\PassOptionsToPackage{usenames,dvipsnames}{xcolor}
\documentclass[twocolumn,twocolappendix]{aastex63}

\pdfoutput=1 
\usepackage{amsmath,amstext}
\usepackage[T1]{fontenc}
\usepackage{apjfonts}
\usepackage{ae,aecompl}
\usepackage[utf8]{inputenc}
\usepackage[flushleft]{threeparttable}

\usepackage{url}
\usepackage{xspace}
\usepackage{lineno}

\usepackage{hyperref}
\usepackage[figure,figure*]{hypcap}
\input{hyperlink-year-only-natbib-patch}
\modulolinenumbers[2]

\newcommand*{\eg}{{\it e.g.}}
\newcommand*{\ie}{{\it i.e.}}
\newcommand*{\etc}{{\it etc.}}
\newcommand*{\etal}{{et~al.}}

\newcommand*{\redmapper}{redMaPPer\xspace}

\def\reply #1{{\color{black}{#1}}}

\definecolor{bleudefrance}{rgb}{0.19, 0.55, 0.91}

\shorttitle{Validating Synthetic Galaxy Catalogs}
\shortauthors{Kovacs  \etal~(LSST~DESC)}

\begin{document}
\title{Validating Synthetic Galaxy Catalogs for Dark Energy Science in the LSST Era}
\author[0000-0002-2545-1989]{Eve~Kovacs}
\affiliation{Argonne National Laboratory, Lemont, IL 60439, USA}

\author[0000-0002-1200-0820]{Yao-Yuan~Mao}
\altaffiliation{NHFP Einstein Fellow}
\affiliation{Department of Physics and Astronomy, Rutgers, The State University of New Jersey, Piscataway, NJ 08854, USA}


\author{Michel Aguena}
\affiliation{Laboratoire d'Annecy de Physique des Particules, Universit\'e Savoie Mont Blanc, CNRS/IN2P3, F-74941 Annecy, France}

\author{Anita~Bahmanyar}
\affiliation{Astronomy and Astrophysics, University of Toronto, Toronto, Ontario M5S 1A1, Canada}

\author{Adam~Broussard}
\affiliation{Department of Physics and Astronomy, Rutgers, The State University of New Jersey, Piscataway, NJ 08854, USA}

\author{James~Butler}
\affiliation{Department of Physics, University of Chicago, Chicago, IL 60637, USA}

\author{Duncan~Campbell}
\affiliation{McWilliams Center for Cosmology, Department of Physics, Carnegie Mellon University, Pittsburgh, PA 15213, USA}

\author{Chihway~Chang}
\affiliation{Department of Astronomy and Astrophysics, University of Chicago, Chicago, IL 60637, USA}
\affiliation{Kavli Institute for Cosmological Physics, University of Chicago, Chicago, IL 60637, USA}

\author[0000-0001-5422-1958]{Shenming~Fu}
\affiliation{Department of Physics, Brown University, 182 Hope Street, Box 1843, Providence, RI 02912, USA}
\affiliation{NSF’s National Optical-Infrared
Astronomy Research Laboratory, 950 North Cherry Avenue, Tucson, AZ
85719, USA}

\author[0000-0003-1468-8232]{Katrin~Heitmann}
\affiliation{Argonne National Laboratory, Lemont, IL 60439, USA}

\author{Danila Korytov}
\affiliation{Argonne National Laboratory, Lemont, IL 60439, USA}

\author[0000-0001-7956-0542]{Fran\c{c}ois Lanusse}
\affiliation{AIM, CEA, CNRS, Universit\'e Paris-Saclay, Universit\'e Paris Diderot, Sorbonne Paris Cit\'e, F-91191 Gif-sur-Yvette, France}

\author[0000-0001-9592-4676]{Patricia~Larsen}
\affiliation{Argonne National Laboratory, Lemont, IL 60439, USA}

\author[0000-0003-2271-1527]{Rachel Mandelbaum}\affiliation{McWilliams Center for Cosmology, Department of Physics, Carnegie Mellon University, Pittsburgh, PA 15213, USA}

\author{Christopher B. Morrison}
\affiliation{Department of Astronomy, University of Washington, Box 351580, Seattle, WA 98195, USA}

\author{Constantin~Payerne}
\affiliation{Laboratoire de Physique Subatomique et de Cosmologie, Universit\'e Grenoble-Alpes, CNRS/IN2P3, 53 avenue des Martyrs, 38026 Grenoble, France}

\author[0000-0002-3645-9652]{Marina~Ricci}
\affiliation{Laboratoire d'Annecy de Physique des Particules, Universit\'e Savoie Mont Blanc, CNRS/IN2P3, F-74941 Annecy, France}

\author{Eli Rykoff}
\affiliation{SLAC National Accelerator Laboratory, Menlo Park, CA 94025, USA}
\affiliation{Kavli Institute for Particle Astrophysics and Cosmology, Stanford University, Stanford, CA  94305, USA}

\author[0000-0003-3136-9532]{F. Javier S\'{a}nchez}
\affiliation{Fermi National Accelerator Laboratory, PO Box 500, Batavia, IL 60510, USA}

\author[0000-0002-1831-1953]{Ignacio Sevilla-Noarbe}
\affiliation{Centro de Investigaciones Energ\'eticas, Medioambientales y Tecnol\'ogicas (CIEMAT), E-28040 Madrid, Spain}

\author{Melanie~Simet}
\affiliation{University of California, Riverside, 900 University Avenue, Riverside, CA 92521, USA}

\author{Chun-Hao~To}
\affiliation{SLAC National Accelerator Laboratory, Menlo Park, CA 94025, USA}
\affiliation{Kavli Institute for Particle Astrophysics and Cosmology, Stanford University, Stanford, CA  94305, USA}

\author{Vinu Vikraman}
\affiliation{Argonne National Laboratory, Lemont, IL 60439, USA}

\author{Rongpu Zhou}
\affiliation{Department of Physics and Astronomy, University of Pittsburgh, Pittsburgh, PA 15260, USA}

\author[0000-0001-8868-0810]{Camille Avestruz}
\affiliation{Department of Physics, University of Michigan, Ann Arbor, MI 48109, USA}
\affiliation{Leinweber Center for Theoretical Physics, University of Michigan, Ann Arbor, MI 48109, USA}

\author{Christophe Benoist}
\affiliation{Université C\^{o}te d'Azur, OCA, CNRS, Lagrange, UMR 7293, CS 34229, 06304, Nice Cedex 4, France}

\author[0000-0001-5501-6008]{Andrew~J.~Benson}
\affiliation{Carnegie Observatories, Pasadena, CA 91101, USA}

\author[0000-0001-7665-5079]{Lindsey Bleem}
\affiliation{Argonne National Laboratory, Lemont, IL 60439, USA}

\author{Aleksandra \'Ciprijanovi\'c}
\affiliation{Fermi National Accelerator Laboratory, PO Box 500, Batavia, IL 60510, USA}

\author[0000-0001-6487-1866]{C\'eline~Combet}
\affiliation{Laboratoire de Physique Subatomique et de Cosmologie, Universit\'e Grenoble-Alpes, CNRS/IN2P3, 53 avenue des Martyrs, 38026 Grenoble, France}

\author[0000-0003-1530-8713]{Eric~Gawiser}
\affiliation{Department of Physics and Astronomy, Rutgers, The State University of New Jersey, Piscataway, NJ 08854, USA}

\author{Shiyuan He}
\affiliation{Institute of Statistics and Big Data, Renmin University of China, Beijing, China}

\author[0000-0002-2704-5028]{Remy Joseph}
\affiliation{Department of Astrophysical Sciences, Princeton University, Princeton, NJ 08544, USA}

\author{Jeffrey A. Newman}
\affiliation{Department of Physics and Astronomy, University of Pittsburgh, Pittsburgh, PA 15260, USA}
\affiliation{Pittsburgh Particle Physics, Astrophysics and Cosmology Center, University of Pittsburgh, Pittsburgh, PA 15260, USA}

\author{Judit Prat}
\affiliation{Department of Physics, University of Chicago, Chicago, IL 60637, USA}

\author[0000-0002-5091-0470]{Samuel Schmidt}
\affiliation{Department of Physics and Astronomy, University of California, Davis, One Shields Ave, Davis, CA, 95616, USA}

\author[0000-0002-8713-3695]{An\v{z}e Slosar}
\affiliation{Physics Department, Brookhaven National Laboratory, Upton, NY 11973, USA }

\author[0000-0001-9789-9646]{Joe Zuntz}
\affiliation{Institute for Astronomy, University of Edinburgh, Edinburgh EH9 3HJ, United Kingdom}

\author{The LSST Dark Energy Science Collaboration}
\noaffiliation

\begin{abstract}
Large simulation efforts are required to provide synthetic galaxy catalogs for ongoing and upcoming cosmology surveys. These extragalactic catalogs are being used for many diverse purposes covering a wide range of scientific topics.
In order to be useful, they must offer realistically complex information about the galaxies they contain. Hence, it is critical to implement a rigorous validation procedure that ensures that the simulated galaxy properties faithfully
capture observations and delivers an assessment of the level of realism attained by the catalog.
We present here a suite of validation tests that have been developed by the
Rubin Observatory Legacy Survey of Space and Time (LSST) Dark Energy Science Collaboration (DESC). We discuss how the inclusion of each test is driven by the scientific targets for static ground-based dark energy science and by the availability of suitable validation data. 
The validation criteria that are used to assess the performance of a catalog are flexible and depend on the science goals. We illustrate the utility of this suite by showing examples for the validation of cosmoDC2, the extragalactic catalog recently released for the LSST~DESC second Data Challenge.
\end{abstract}





\section{Introduction}
\label{sec:intro}

The upcoming Vera C.\@ Rubin Observatory Legacy Survey of Space and Time \citep[LSST;][]{ivezic2019, lsst} is an ambitious ground-based imaging survey covering approximately half the sky and beginning as early as 2023. The advent of this wide and deep data set
will usher in a new era of cosmology characterized by small statistical errors and better control of systematic errors. The Rubin Observatory LSST Dark Energy Science Collaboration (LSST~DESC)\footnote{\url{https://lsstdesc.org}} has been formed to prepare for the wealth of scientific opportunities offered by this large influx of data~\citep{lsst-desc2012}. The collaboration was established with an initial focus on probes of the fundamental properties of dark energy~\citep{weinberg2013}, including measurements of weak lensing, strong lensing, large-scale structure, galaxy clusters and supernovae. Within the collaboration, Working Groups (WGs) have been convened to develop and test a variety of scientific analyses and make forecasts for the precision with which the parameters describing the behavior of dark energy can be constrained~\citep{lsst-srd}. 

These activities would not be possible without an extensive and contemporaneous simulation campaign that is designed to provide synthetic galaxy catalogs with various levels of realism. As with any simulation campaign, it is critical to perform validation to assess the catalog's level of fidelity compared to the real universe and determine the extent of its utility for testing scientific analyses. The validation process relies on a series of tests that compare the catalog data with suitably chosen validation data sets. 
The three considerations that underlie the construction of a test suite are: (1) the scientific goals that will be pursued with the catalog, (2) the availability of suitable validation data and (3) an evaluation of how realistic the catalog must be in order to meet those goals. 

A number of simulated catalogs with various levels of complexity have been produced for ongoing and past surveys~\citep[e.g.,][]{mice2015, alam2017, avila2018, derose2019, shirasaki2019, safonova2020, lin2020, derose2021, wechsler2021}. These catalogs range from large general-purpose catalogs~\citep{mice2015, derose2019, derose2021, wechsler2021} intended to cover a wide range of science targets for surveys such as the Dark Energy Survey (DES), \reply{VISTA and WISE}, to more specialized catalogs intended for specific surveys or analyses. For example, catalogs can be produced to provide synthetic data for  
specific scientific analyses such as the analysis of shear autocorrelations~\citep{shirasaki2019}, or to mimic a specific data set such as those of the Dark Energy Survey BAO sample~\citep{avila2018} or
the Dark Energy Spectroscopic Instrument Bright Galaxy Survey~\citep{safonova2020} \reply {or the emission line galaxy (ELG) sample from the extended Baryon Oscillation Spectroscopic Survey~\citep{lin2020}.} The three considerations listed above enter into the validation efforts that
accompanied the production of these catalogs.

Turning to the test suite that we present here,
the nature of the tests that are of interest to us is driven by our goal of studying dark energy science. The suite must include checks that the statistical distributions of galaxies on the simulated sky are realistic and that the statistical distributions of the properties relevant for dark energy science agree with those that are observed in the real universe. The level of agreement required must be specified for each distribution. 

The range and availability of validation data necessarily has a big impact on the tests that can be performed. Hence
the assembly and curation of an appropriate validation data set is a fundamental step in the construction of any validation test suite. Understanding which data sets can be used to validate which galaxy property and determining what ranges of the property values can be constrained are key factors in constructing the suite.

A specific example that illustrates this point is the validation of synthetic galaxy properties as a function of redshift. Synthetic catalogs are available which provide galaxies out to higher redshifts, $z \gtrsim 1.5$~\citep{derose2019, cosmodc2}, but there are very few \reply{complete, magnitude-limited} observational data samples available  \reply{with sufficient statistics at LSST depths to be useful for validation purposes.
While there are several high-redshift spectroscopic galaxy surveys such as the VIMOS VLT Deep Survey (VVDS)~\citep{vvds2013}, the VIMOS Ultra Deep Survey (VUDS)~\citep{vimos2014} and zCOSMOS~\citep[][]{lilly2007}, these data typically suffer from incompleteness issues and low statistics at  higher redshifts.  Photometric catalogs assembled from COSMOS data~\citep{laigle2016, cosmos2021} suffer from significant outlier rates, particularly at fainter magnitudes, and, due to the template-fitting methods employed to estimate the redshifts, their color and redshift distributions are not unbiased. The ever-present issue of cosmic variance that accompanies small-area survey data may also be a consideration for specific validation tests.
Consequently, the validation of full catalog distributions at high redshift is quite problematic and} we must, in many cases, rely on extrapolations of existing data.

Another aspect of catalog validation is the verification of the catalog contents. Rather than making comparisons with external data, the verification process consists of 
checks to ensure that the simulated catalog delivers the expected results for given quantities. Such checks include tests for the uniqueness of object identifiers, tests of the statistics  of object-property distributions and comparisons of object properties with theoretical predictions based on the input parameters for the underlying cosmological simulations. 

As mentioned above, in addition to developing the validation tests, a set of validation criteria has to be defined. These criteria are informed by the specific science goals that will be pursued using the synthetic catalog. Different goals may provide different requirements for the same validation test and may even evolve with time. In this sense, the criteria can be viewed independently of the test. A test suite could be adapted for other surveys by retaining the tests and validation data sets but altering the
validation criteria to require different levels of fidelity with the observational data.
The entire validation process, which encompasses the tests and the criteria, is in fact the key to understanding how a catalog can be used and what its limitations are.

The validation test suite that we present in this paper provides a set of comparisons of carefully selected measurements from a synthetic galaxy catalog with a curated set of validation data. The tests have been developed in conjunction with LSST~DESC WGs\footnote{\url{https://lsstdesc.org/pages/organization.html}}
and focus on the validation of static galaxy properties. The validation of time-domain properties is discussed elsewhere in~\citet{DC2_survey_paper}. Due to the broad range of dark energy science targets within LSST~DESC, we expect that our validation test suite is representative of the set of tests required to evaluate the realism of synthetic galaxy catalogs targeted for studying dark energy science in optical imaging surveys, both now and in the future. The test suite for spectroscopic surveys would have some overlap with the suite presented here.   

Our suite of tests has been used to validate cosmoDC2 \citep{cosmodc2}\footnote{cosmoDC2 is publicly available from the NERSC website: \url{https://portal.nersc.gov/project/lsst/cosmoDC2/_README.html}}, the extragalactic catalog underlying the LSST~DESC second data challenge, DC2~\citep{DC2_survey_paper} This data challenge is one of two challenges 
that have been instituted by the LSST~DESC in order to assess and quantify the progress towards preparedness for the arrival of LSST data. 
DC2 is an ambitious program to produce a realistic LSST-like data set that can be used for a multitude of science goals such as testing analysis and production pipelines, studying a variety of systematic effects, testing the Rubin LSST Science Pipelines~\citep{rubinpipe2019}, studying time-domain objects and running joint analyses that involve multiple observational probes. Our validation process is an integral part of catalog development and production because (1) the underlying model for the galaxy-halo connection used to populate the synthetic catalog typically needs to be adjusted to meet the validation criteria and (2) the production of the catalog requires significant human effort and computing resources to produce, so a staged and iterative approach is necessary.

In this work, we use DESCQA~\citep{descqa}, which is the automated validation framework developed by LSST~DESC to compare simulated catalogs with a variety of validation data. This framework is quite general in that it ingests sets of catalogs and sets of validation data (which may be observational data or theoretical predictions) and then compares the results, in the form of summary statistics and comparison plots, for a specified set of tests. 
Each test can incorporate validation criteria by producing a numerical score derived from a comparison of selected summary statistics for the catalog and validation data and issuing a pass or fail based on the score.

In summary, we present here a validation test suite designed to evaluate simulated galaxy catalogs that have been targeted for ground-based imaging surveys focused on dark energy science. The suite is based on a curated set of validation data that includes both observational data and theoretical predictions. We demonstrate the utility of the suite with the validation of cosmoDC2 using the DESCQA framework and the DC2 validation criteria supplied by LSST~DESC.

This paper is organized as follows. In \autoref{sec:science}, we discuss the construction of our validation test suite by considering the scientific goals of our planned analyses and the availability of suitable validation data sets. The reader who is primarily interested in the test results for cosmoDC2 can skip to \autoref{sec:tests}.
In \autoref{sec:prelude}, we briefly describe the cosmoDC2 extragalactic catalog and the validation framework DESCQA. Next, in \autoref{sec:tests}, we show examples of the validation tests comprising the suite and discuss the validation criteria used to evaluate the fidelity of cosmoDC2 with the observational data. Finally, we summarize our results and discuss the implications and future work in \autoref{sec:summary}.

\section{Construction of the Validation Test Suite}
\label{sec:science}

Here we discuss the development of the validation test suite that is presented in this paper.  
The scientific goals of the analyses planned by the LSST~DESC  WGs\footnote{\url{https://lsstdesc.org/assets/pdf/docs/DESC_SRM_latest.pdf}} act as drivers for identifying critical properties of the simulated data that must be realistically rendered in the synthetic galaxy catalog. 
Although these scientific goals, and hence the intended uses of the catalog, are varied and complex, a relatively small set of critical properties have emerged from this exercise. We find considerable overlap in the requirements coming from the different analysis WGs.
In the following subsections, for each of these science goals, we  present: (1) the science drivers and the critical properties that must be accurately rendered by the catalog, (2) the data sets that are available to validate these properties and (3) the tests that we construct by matching the available data with the galaxy properties that require validation.

\subsection{Galaxy Number Densities}
\label{sec:science:n}

\subsubsection{Scientific Drivers and Critical Properties}
\label{sec:science:driver:n}
The galaxy number-density distribution is one of the most fundamental properties of a synthetic galaxy catalog. Many targets for dark energy science rely on the assumption that this distribution is realistically rendered. In the real universe, the observed number density of galaxies can be viewed as a multi-dimensional distribution that is dependent on redshift and the spectral energy distribution (SED) of each galaxy, as well as other factors such as Galactic reddening and atmospheric and instrumental effects. In practice, for imaging surveys, the incoming light is integrated over the band passes of the filters in the survey and this high dimensional space is reduced to redshift plus the luminosities in these filters. For LSST, there will be 6 filters $u$, $g$, $r$, $i$, $z$, and $y$. Any validation test suite must incorporate tests that compare the number density as a function of both redshift and observed magnitudes in the available filters.

The primary scientific drivers that require realistic galaxy number densities are large-scale structure (LSS) and weak lensing (WL) analyses. These analyses extract information from measurements of two-point correlation functions such as galaxy clustering, shear auto-correlations and galaxy-density -- shear cross-correlations, all of which are used in the so-called $3\times 2$pt analysis \citep[\eg,][]{des32pt, descollaboration2021dark, kids1000J}. Most importantly, for a single redshift slice, the noise on a correlation function measurement is inversely proportional to number density. If the redshift distribution of galaxies, $N(z)$, is wrong, then the mean redshift of tomographic slices will be wrong, but more subtly, the number of source and lens galaxies will be modulated and the effective redshift, from which the signal in both the auto-correlation and the cross-correlation is derived, will be affected in an unpredictable way.

Another critical ingredient of LSS and WL analyses is an understanding of the accuracy with which photometric redshifts can be determined. This understanding is limited by uncertainties in instrumental calibration and the realism of the distribution of galaxy spectral energy distributions (SEDs) and their evolution with redshift. 
The population of galaxies has evolved as the universe has aged, and the resulting changes in their luminosity functions, combined with observational selection effects, are reflected in the evolution of the observed redshift distribution as a function of magnitude. 
Photometric redshift algorithms estimate galaxy redshifts by comparing observed magnitudes to either model templates \citep[\eg,][]{Benitez:2000} or machine learning algorithms \citep[\eg,][]{Sadeh:16,Izbicki:17} using training sets of observed galaxies with known redshifts \citep{Salvato2019}. These redshift estimates are subject to biases and degeneracies due to the limited information available, \eg~in LSST we will have only six magnitudes from broad-band filters with which to estimate a redshift.  In order to avoid being dominated by systematic uncertainties, we must fully characterize such degeneracies and thus it is vital to have a galaxy population that is reflective of that observed in the real universe: systematic differences between the model galaxy populations and the actual observed populations could introduce biases in the analysis. In addition to one-point distributions, the spatial clustering of galaxies is a critical component in calibrating photo-$z$-derived redshift distributions, so properly modeling the spatial correlations of galaxies is also important (see~\autoref{sec:science:driver:2pt}).


\reply{One of the principal uses of the cosmoDC2 catalog has been to provide the extragalactic catalog inputs for the DC2 image simulation campaign~\citep{DC2_survey_paper}. The simulated images enable a large array of science goals based on the processing and analysis of these images.} One particularly important scientific target that is strongly affected by galaxy number densities is the phenomenon of blending~\citep{Chang2013, Sanchez2021}.
This is the apparent overlap, on the plane of the sky, of the light profiles of different stars and galaxies. Blending is observed, either because objects at different redshifts appear on the same line of sight, or because galaxies are in close physical proximity. Overlapping light profiles in a blended scene therefore affect the measurement of galaxy shapes and magnitudes as well as the number and spatial distribution of galaxies. With $\sim 60\%$ of galaxies found to be affected by blending in the HSC Wide survey~\citep{Bosch2018} and a similar fraction expected in LSST \citep{Newman2019, Sanchez2021}, blending will act as a major source of systematic effects for shear estimations, flux measurements, galaxy number densities and photometric redshifts estimations. In a recent study, \cite{Sanchez2021} showed that $62\%$ of galaxies in LSST will have at least $1\%$ of their flux coming from overlapping, neighbouring sources. This effect, combined with shot noise, decreases the effective number density of galaxies by $18\%$.

Because the occurrence of blending is so high, it is unlikely that WL and LSS analyses can be conducted by removing blended objects. Even with state-of-the-art deblenders \citep[see][for instance]{Melchior2018} to model the light profiles of blended objects, we will have to calibrate the effects of residual blending as well as the effect of the deblender itself on dark energy measurements. The accurate characterization of the performance of these deblending algorithms is a critical component of the systematics analysis and depends on realistic galaxy number densities as a function of luminosity and redshift.

One particularly problematic aspect of deblending is that of unrecognized blends \citep{Dawson2015, Martinet2019}. Most deblending algorithms rely on the detection of each galaxy before their models can be extracted. In cases where galaxies are too close to each other, detection and deblending algorithms cannot determine whether one or more objects are present. Deblending algorithms are likely to erroneously model the scene as one object whose properties (flux, shape, color) will not be representative of the underlying truth. The rate and impact of such unrecognized blends on shear and photometric redshift estimation have to be characterised in the context of LSST, and simulations will play a key role in understanding this effect and mitigating this systematic. Depending on the deblending strategy, it is also possible that modeling errors might introduce biases in the size, shape, flux and colors of deblended galaxies. Simulations will allow us to identify which deblending strategies are expected to be successful and help us understand the impact of these strategies on measurements relevant for higher level science.

\subsubsection{ Galaxy Number-Density Data Sets}
\label{sec:science:data:n}

We now examine the available data sets that can be used to validate galaxy number densities as a function of redshift and magnitude. 
It is critical to have at least one data set that provides a check on the absolute normalization of the galaxy number densities in the catalog. One simple possibility is 
to validate the cumulative number density as a function of magnitude. The construction of such a test requires data from a survey that is both wide and deep. The first requirement suppresses the uncertainty from cosmic variance, while the second
ensures that we can validate the number density of very faint galaxies. 
The data set from the first data release \citep{2018PASJ...70S...8A} of the Deep layer of the Hyper Suprime-Cam (HSC) Subaru Strategic  Program~\citep{2018PASJ...70S...4A} provides the deepest measurements available from an ongoing survey with an area exceeding a few square degrees. 

A convenient way to characterize the behavior of this data set is to obtain power-law fits to the measured cumulative number counts of galaxies in the $g$, $r$, $i$, $z$ and $y$ bands. These fits are valid for CModel~\citep{lupton2001, Bosch2018} magnitudes $\lesssim 25$. The extrapolation of these fits to fainter magnitudes is justified by measurements from deep pencil-beam surveys, which appear to have power-law number counts for CModel magnitudes $\lesssim 28$  \citep[\eg][]{2006AJ....132.1729B}.
We note that although the HSC data have star-galaxy separation selections applied, we expect the systematic errors associated with these selections to grow with fainter magnitudes. Hence, if a stringent validation test for the number density is required, it may be necessary to apply a  more careful extrapolation procedure that takes such systematic effects into account.

Additional verification of our extrapolations can be done by comparing the fitted number counts against the counts obtained from deep pencil beam surveys.  For example, very deep Subaru observations in the Cosmic Evolution Survey (COSMOS) field~\citep{2007ApJS..172...99C} yield a cumulative number density of  150/arcmin$^2$ or $5.4\times 10^5$/deg$^2$ for $I<26.5$, which is quite close to our extrapolated \reply{$i$-band}\footnote{The mean effective wavelengths for COSMOS $I$-band (F814W) and LSST i-band are $\approx 8100$ and $\approx 7500$ \AA, respectively.  The transmission curve for F814W is asymmetrical, so the peak of its transmission occurs at $7500$ \AA} HSC fit of $(5.17 \pm 0.04) \times 10^5$/deg$^2$. \reply{We caution the reader, however, that data from pencil beam surveys are subject to considerable cosmic variance and hence are not as reliable or useful as our extrapolated fits.}

 Observational data for galaxy  redshift  distributions are available from \cite{coil} and \cite{deep2}. These observations are reported as parameterized fits to the shape of $dN/dz$ distributions for a variety of magnitude-limited  samples. The selection cuts for these samples were imposed on the CFHT $R$- and $I$-band magnitudes and range from $18-27$ and $22-24$, respectively, for the works cited above. The parameterized fits have a simple analytic form that can be readily integrated to compute probability distributions over the range of redshifts being validated.
 In order to make meaningful comparisons,
 the cuts used to select the validation data must also be applied to the catalog data. In practice, the filters simulated for a given catalog will be different from those in the validation data, so  depending on the precision required by the validation criteria, it may be necessary to apply correction factors to account for this difference.  
 The fits to the observational data are available out to a redshift of $z \lesssim 1.5$, which as mentioned in~\autoref{sec:intro}, limits the range of the validation that can be performed.
 
 \reply{We reiterate that although there are additional data available from other spectroscopic surveys such as VVDS~\citep{vvds2013}, VUDS~\citep{vimos2014} and zCOSMOS~\citep{lilly2007} and from photometric catalogs that are obtained from COSMOS data~\citep{laigle2016, cosmos2021}, these data are not as suitable for validating the full catalog redshift distribution. For example, VVDS has very few highly-secure redshifts at $z>1$,  VUDS does not provide a complete magnitude-limited sample in redshift ranges relevant for LSST, and, due to the VUDS color selections, the statistics in the redshift range $1 < z < 2$ are low. Similarly, zCOSMOS-deep data is color-selected and does not span the full galaxy population and VANDELS~\citep{vandels2018} targets only red galaxies in the range $1.5 < z < 2.5$. COSMOS data have a photometric redshift outlier rate of $>10$\% for galaxies at high redshift or with $I$-band magnitudes in the range $24 < I < 25$. All of these issues have the potential to confuse any conclusions about the level of agreement between the catalog and validation data.}
 
Two observational data sets  are available to validate galaxy SED distributions. The first is the data set from the Sloan Digital Sky Survey (SDSS) main galaxy sample which provides measurements of the $ugriz$ colors for an apparent magnitude-limited sample from SDSS Data Release (DR) 13 \citep{2017ApJS..233...25A}. We select galaxies from this sample with well defined redshifts\footnote{We used the following SQL query: survey=``SDSS" AND class=``GALAXY" AND z>0 AND zWarning=0} that lie in the range $0.06<z<0.09$,  and use this data set to validate color distributions at low redshifts.

The second data set is a combination of DEEP2 \citep{deep2} and DEEP3 spectroscopic redshifts and Canada-France-Hawaii Telescope Legacy Survey (CFHTLS) \citep{CFHTLS} photometry in $ugriz$ bands; this combined data set is described in  \citet{Zhou2019}. The depth of the CFHTLS-Deep imaging is comparable to that of the LSST 10-year stack and the DEEP2 and DEEP3 spectroscopy is complete to $R_{AB}=24.1$ over the redshift range of $0<z\lesssim 1.4$. The area covered by both CFHTLS-Deep imaging and DEEP2/3 spectroscopic is roughly 0.3 square degrees. We use this data set for validating the colors of galaxies at a redshift of $z \sim 0.9$. Although Subaru $Y$-band data are also included in this data set, they are not used here due to their incomplete and non-uniform spatial coverage. The photometry of both data sets is corrected for (Milky-Way) Galactic extinction before being used for validation.

\reply{Similar to the caveats discussed for the the validation of redshift distributions, the color completeness of data from other spectroscopic and photometric surveys limits their utility for our validation applications.
None of the aforementioned alternative surveys enables an assessment of the full range of galaxy colors at fixed redshift and magnitude.  
In contrast, the DEEP2/DEEP3 data includes objects of all colors at $z>0.7$ by design. It is the single data set, readily available at this time, that is best suited for a validation of the full catalog.  In the future, with additional effort and access to catalogs with well-matched photometry and redshifts, further comparisons based on other data sets could be made in the regimes where those data sets are relatively complete.
}

\subsubsection{Number Density Tests}

 As we have seen in~\autoref{sec:science:data:n}, not all  projections of galaxy number densities in the multi-dimensional space of redshift and available filter magnitudes have corresponding validation data sets available. In order to simplify the test suite and match the available validation data sets, we have chosen to implement tests based on various projections of this multi-dimensional distribution. We include:
\begin{itemize}
    \item the cumulative number density of galaxies as a function of magnitude (\autoref{sec:test:dndmag}),
    \item the probability distribution as a function of redshift for magnitude-limited samples of galaxies (\autoref{sec:test:nz}),
    \item color distributions (\autoref{sec:test:color}):
    \begin{itemize}
    \item the probability distribution as a function of galaxy colors for magnitude-limited samples of galaxies in slices of redshift (\autoref{sec:test:color-1d}),
    \item the 2-dimensional probability distributions of galaxy colors in slices of redshift (\autoref{sec:test:color-color}).
    \end{itemize}
    
\end{itemize}
The first test checks the realism of the absolute number density as a function of magnitude but does not provide a detailed check on the redshift distribution. The cumulative distribution used in this test is sufficient to ensure that the catalog number counts scale approximately as expected with survey depth. A more stringent test would employ differential counts. The second test
checks the realism of the redshift distribution as a function of magnitude but is insensitive to the absolute normalization of the data. The remaining tests check the realism of various projected distributions of galaxy colors and hence provide validation tests for galaxy SEDs. Because galaxy colors and magnitudes are correlated and vary with redshift, a  comprehensive set of tests to validate galaxy SEDs should cover this multi-dimensional color-magnitude space. Many projections of this space are possible. As a first test, we use simple one-point distributions as a basic check of the realism of galaxy colors, but this test alone is not sufficient to determine if a catalog captures the full complexity of the SED distributions in the real universe. Other more complicated color distributions provide further diagnostics for catalog performance. In our suite, we include color-color distribution tests to capture information on the correlations between colors.

\subsection{Galaxy Shapes, Sizes and Morphologies}
\label{sec:science:wl}

\subsubsection{Scientific Drivers and Critical Properties}
\label{sec:science:driver:wl}

The observable quantity for WL is the subtle distortion (or {\it shear}) of observed galaxy shapes due to the deflection of light due to the large-scale matter distribution between those galaxies and the observers \citep{Bartelmann2001}. 
In the two-point correlation function described in \autoref{sec:science:driver:2pt}, as well as other WL statistics, we  estimate shear for an ensemble of galaxies. Operationally, we average galaxy ellipticities (inferred from the light profile) over a sample of galaxies. As the mean ellipticities of most galaxies are more than 10 times larger than the WL-induced shears, one needs to average over a large number of galaxies to suppress this intrinsic {\it shape noise} and reveal the cosmological shear signal. This means that we cannot only use the large bright galaxies for WL measurements. Rather, the small and faint galaxies, due to their large numbers, dominate the galaxy samples used for WL measurements, and present additional challenges for the estimation of shear.   

At the accuracy level required for Stage IV galaxy surveys such as LSST \citep{SRD}, shear estimation is challenging. Multiplicative and additive biases in the shear estimation can arise from each step of the estimation. In particular, the modeling and interpolation of the point spread function (PSF) across the focal plane, the assumed model for the galaxy profiles, and the selection used to arrive at the WL galaxy sample all  have the potential to introduce biases in the shear estimation. A number of approaches have been taken to {\it calibrate} these biases, ranging from image simulations to more empirical approaches using the data themselves~\citep[for a review of systematic effects and  state-of-the-art shear estimation algorithms, see][]{Mandelbaum2018review, sheldon2020}

For WL science, several aspects of the simulations are important. In addition to having realistic number density and magnitude distributions for galaxies, the morphology (size and shape) distributions are also critical. The importance of this can be seen in previous work such as \cite{FenechConti2017}, \cite{Mandelbaum2018} and \cite{Zuntz2018}, where image simulations were specifically developed to calibrate specific shear measurement algorithms. As discussed above, since the universe produces many more small, faint galaxies than large, bright galaxies, any lensing sample will be dominated in number by galaxies close to our minimum cutoffs in size and brightness. As such, an unrealistic model for the size of the galaxies relative to the size of the PSF could result in a mismatch of this cutoff compared to the real world.
Even more recent shear estimation methods that attempt to self-calibrate the shear biases require basic validation against simulations with a reasonable level of complexity in the galaxy population~\citep{maccrann2020}.
On the other hand, the shape of the galaxies is the dominant source of statistical error in shape measurements. The apparent shape of the galaxies is a combination of the intrinsic galaxy shape and measurement errors, both could be redshift and magnitude dependent. Overall, a coherent framework that encompasses all the relevant correlations between the different galaxy properties is essential for testing systematic errors that could couple, \eg, weak lensing and photo-z systematic effects.

Another effect impacting shear measurement is that of objects in close proximity on the plane of the sky, or blending. This contribution has two major effects on shear measurements. Firstly, flux contamination between well-identified close neighbours will lead to biased estimates of galaxy shapes \citep{Dawson2015, Martinet2019, Hoekstra2017}. Secondly, in the regime of unrecognized blends, not only the measured shapes of galaxies will be biased, but the number of galaxies available for weak lensing will be decreased, as a group of galaxies is mistakenly identified as one \citep{Sanchez2021}.

It is expected that blending will not only be a function of density and distance between galaxies, but will also depend on the shapes and alignments of galaxies. For instance, \cite{Dawson2015} showed that unrecognized blending will affect more strongly highly elliptical objects. This is of particular relevance for WL analysis as it probes alignments and ellipticities of galaxies. In order to account for blending properly in WL and other probes, it is paramount that simulations provide a realistic account of the spatial distribution of galaxy shapes.

Since deblending is a promising avenue to mitigate the impact of blending, simulations should serve as a realistic testing ground for algorithms. Deblenders can make various assumptions as to the morphological properties and intrinsic alignments of the objects they intend to model. Providing deblender developers with realistic distributions of shapes and sizes will help refine their algorithms in view of upcoming real images~\citep{sheldon2020}.

\subsubsection{Galaxy Shape, Size and Morphology Data Sets }
\label{sec:science:data:wl}

The available data that we use to validate the sizes of faint galaxies is obtained from parametric, two-dimensional, intensity-profile fits~\citep{2014ApJS..212....5M} to postage stamps of COSMOS galaxies~\citep{cosmos2012} with $i<25.2$.  The fitting procedure used in this work was first presented in \citet{2012MNRAS.421.2277L}. 

For validating galaxy sizes as a function of redshift, we use the measurements of \citet{2014ApJ...788...28V} that combine data from 3D-HST~\citep{3D-HST2012} and CANDELS~\citep{candels2011a, candels2011b} to provide size distributions as a function of rest-frame V-band luminosity in six redshift bins spanning the range $0< z<3$~\citep[see Table 5,][]{2014ApJ...788...28V}. For validating the sizes of the galaxy disk and bulge components, we use the fits provided in \citet{2014ApJS..212....5M}, obtained from COSMOS data, which give estimates of the size distributions for each galaxy component as a function of rest-frame F814W luminosities for four redshift bins spanning the range $0< z< 2$. 

Validation data for the distribution of galaxy ellipticities is available from the COSMOS survey \citep{Joachimi2013}. These data are available for four morphological classes: luminous red galaxies (LRG), disk, early and late type galaxies in the redshift range $0 < z < 2$ and are selected based on $i$-band and absolute $V$-band magnitude cuts of $i < 24$ and $M_V < -19.0$ and $-21.0 < M_V < -17.0$ for the LRG and other morphology classes respectively. The selection cuts to determine the morphology class are quite complex and would be challenging to implement for a synthetic catalog. Instead, as described in \autoref{sec:test:ellipticity}, in order to approximate the morphological selections in the COSMOS sample, we implement selections on $B/T$, the ratio of bulge-to-total fluxes in $i$-band.  

\subsubsection{Size, Shape and Morphology Tests}
\label{sec:science:suite:wl}

The galaxy orientation, size and shape tests that we implement  are comparisons with both theoretical expectations and observational data for the following distributions:
\begin{itemize}
    \item the position angle distribution (\autoref{sec:test:pa}),
    \item the cumulative distribution function (CDF) of galaxy sizes for a faint-galaxy sample (\autoref{sec:test:size}),
    \item the size-luminosity relation as a function of absolute magnitude and redshift (\autoref{sec:test:size-mag})
    \begin{itemize}
        \item for the total light profile (\autoref{sec:test:size-mag-1}),
        \item for the disk- and bulge-component light profiles
        (\autoref{sec:test:size-mag-2}),
    \end{itemize}
    \item the ellipticity distributions for selected morphological classifications and luminosities (\autoref{sec:test:ellipticity}).
\end{itemize}

The first test is designed to check that the position angle of galaxies is randomly distributed with a uniform probability distribution. Although this is an assumption that is normally built into catalogs, it is  sufficiently important to require a test ensuring that the uniformity has propagated to the final catalog.  

The second test ensures that the slope of the galaxy-size distribution near our faint-magnitude resolution cutoff is approximately realistic, so that any selection effects in the catalogs derived from images are similar to what they would be in real data.  

The third and fourth tests check that galaxy sizes scale with luminosity as expected and that the ellipticities of selected sub-samples of galaxies follow the observed distributions, respectively.

\subsection{Galaxy Correlation Functions} 
\label{sec:science:2pt}
\subsubsection{Scientific Drivers and Critical Properties}
\label{sec:science:driver:2pt}

Two-point correlation functions measure the scale- and time-dependent properties of the clustering of galaxies.  In particular, two sets of correlation functions will be particularly well-measured by LSST: the density (or number count) of galaxies, and the correlation between their shapes.  The former, galaxy clustering, uses the galaxy density as a (biased) tracer of the underlying dark matter field.  The latter, cosmic shear, measures the gravitational lensing applied to galaxy populations by intervening matter fields.  In imaging surveys such as LSST, these correlation functions are two-dimensional: they are measured on thick tomographic bins of galaxies determined via photmometric redshifts.

Galaxy clustering measurements provide a very high signal-to-noise but biased probe of the underlying matter field \citep{1978ApJ...221....1D,1980ApJ...236..351D}. The bias is expected to evolve as a function of redshift and scale and this unknown multiplicative uncertainty limits the power of galaxy clustering on its own.

Lensing measurements provide a nearly-unbiased measurement of the underlying matter density, integrated over the line of sight to each galaxy \citep{Mandelbaum2018review}.  Lensing tomography is sensitive to a combination of the growth of structure, since higher mass densities act as stronger lenses, and to the expansion history of the universe, since the expansion modifies the geometry of the observer-lens-source system.

As mentioned in~\autoref{sec:science:driver:n}, the $3\times 2$pt analysis combines two-point clustering measurements with lensing auto- and cross-correlations. The power of this analysis lies in the observation that the unknown galaxy bias and possible systematic photometric redshift errors can be constrained by a combination of measurements~\citep{des32pt}.  Since these systematic effects contribute a large component of the overall error budget in the analysis, we can significantly improve our cosmological constraints using this combination. The three parts of this analysis each place demands on the fidelity of the catalog; errors in any of them will distort the recovered cosmological parameters.

Galaxy clustering correlations, and especially tomographic auto-correlations\footnote{In ``perfect'' surveys, density correlations between different redshift bins would be negligibly small, but in real surveys redshift errors induce measurable correlations.} in simulated catalogs must be as realistic as possible to avoid either spuriously powerful constraints (if the model used is too simple) or incorrect ones. In particular, the galaxy bias prescription used to populate the simulation with galaxies is critical.

Many different prescriptions to describe galaxy bias have been developed \citep{Kaiser:1984sw,1986ApJ...304...15B,Bernardeau:1995ty,2011MNRAS.415..383M, desjacques2018}. The uncertainties in the constraints on this relation strongly increase the errors in the dark energy equation of state or gravitational growth index~\citep{Eriksen:2015hqa}. Thus, having realistic models for galaxy bias is key for studying how to obtain the maximum performance in current and future data analyses. In particular, given the increased statistical power of modern surveys, linear biasing models~\citep{2017arXiv171005045G} are no longer sufficient. Furthermore, since galaxy bias depends on galaxy formation and evolution, it necessarily depends on the galaxy population selected for the analysis. The validation of galaxy bias is therefore a complex and challenging issue. However, a minimal expectation for the catalog data is that the galaxy bias inferred from fitting a linear bias model to the galaxy 2-pt correlation functions shows increasing bias with redshift.  This behavior is expected because the sample of galaxies used to evaluate the 2-pt correlation functions is flux limited and hence includes more faint, low-bias galaxies at low redshift. 

There is less ambiguity about the model that should be used to obtain faithful galaxy lensing from a simulation: it should match the integrated distortion along the line of sight. The practical calculation of this quantity, however, can be computationally demanding.  Errors in going from snapshot information to integrated shear can induce multiplicative biases in overall shear measurement \citep{PhysRevD.95.123503}.  Furthermore, since some cosmic shear statistics probe small physical scales, down to $k \sim 10 h^{-1}$Mpc, the fidelity of the simulation's matter power spectrum on these scales must also be well-understood in order to account for the possible biases that result from uncertainties in the simulations at small scales.

The final component of the  $3\times 2$pt analysis is the cross-correlation of foreground (lens) galaxy positions and background (source) galaxy shear, commonly known as galaxy-galaxy lensing. This measurement probes the average halo profile for a given foreground galaxy sample \citep{Bartelmann2001} and provides a powerful tool for studying the galaxy-halo connection as shown, for example, in~\citet{Mandelbaum2006}. Since this cross-correlation, as the third probe in the $3\times 2$pt scheme, is crucial for breaking degeneracies between cosmological parameters and nuisance parameters (photometric redshift errors, galaxy bias, baryonic effects and intrinsic alignments), it too must be realised to high accuracy in synthetic catalogs.

A final scientific application that is dependent on the fidelity of the large-scale structure of synthetic catalogs is the determination and calibration of redshifts.
Many cosmological analyses rely on detailed knowledge of the redshift distribution for tomographic subsets of the overall sample.  A powerful set of techniques have been developed that use the fact that galaxies are embedded in large scale structure and thus have strong spatial correlations with other objects that are nearby. \citet{Newman:08} presented a method to determine the redshift distribution for a set of objects precisely and accurately by measuring their spatial cross-correlation with a smaller set of objects with known spectroscopic redshifts.  More recent techniques \citep{mcquinn2013, rahman2015, Sanchez:19, rau_et_al_2020} have combined photometric and clustering information into a single redshift estimator.  While powerful, these methods are dependent on complementary estimates of both the weak lensing magnification signal and the evolving galaxy bias of the samples.  In order to model these systematic effects, any simulation must include realistically complex galaxy clustering and magnification signals, as well as a complex galaxy bias evolution model mapping how the galaxies populate dark matter halos.

\subsubsection{Correlation-Function Data Sets}
\label{sec:science:data:2pt}

Several measurements are available for validating galaxy clustering at low redshifts.  From the SDSS DR7 galaxy sample, \cite{tpcf_wang} present results for the angular correlation function $w(\theta)$, the over-abundance of galaxy pairs at an angular separation $\theta$, relative to a random distribution. Table~2 in that work gives $w(\theta)$ for four magnitude limited samples based on SDSS $r$-band magnitudes. The angular separations vary from $\sim 0.005$ to $\sim 10$ degrees. The median redshift for this sample is 0.22. The values of $w(\theta)$ are measured using the Landy-Szalay estimator~\citep{1993ApJ...412...64L} and are in agreement with earlier measurements, but offer much higher precision due to the large size of the galaxy sample and the uniformity of the data.

To check the estimated galaxy bias, $b(z)$, we use the empirical fit obtained by \citet{nicola2020}, which analyzes the clustering of magnitude-limited samples of galaxies in tomographic redshift bins from the first data release of HSC Subaru Strategic Program (HSCSSP). Equation 4.12 from that work provides a simple fitting function for the linear galaxy bias which is given by
\begin{equation}
\label{eq:galaxy_bias}
  b(z) =(b_1*(m_{\rm lim}-24)+b_0)*D(z)^{\alpha},
\end{equation}
where $b_1=-0.0624\pm0.0070$, $b_0=0.8346\pm0.161$, $\alpha=1.30\pm0.19$, $D(z)$ is the cosmology-dependent linear growth factor and $m_{\rm lim}$ is the limiting magnitude of the data sample. \autoref{eq:galaxy_bias} was obtained for an HSCSSP data sample with $m_{\rm lim} = 24.5$.

SDSS data for the projected correlation function, $w_p(r_p)$, is available for low redshifts in the range $0.02 < z < 0.25$ and for various selection cuts on the magnitude and colors of the galaxy sample \citep{zehavi2011}. Tables~C7,  C9, and C10 in \citet{zehavi2011} provide measurements of volume-limited samples of galaxies in absolute $r$-band (Hubble parameter $h=1$) magnitude bins ranging from $-23 < M_r^h < -17$ for the entire sample and $-23 < M_r^h < -19$ for the blue sample and red samples, respectively. Galaxies are separated into blue and red samples by using a magnitude-dependent color cut defined by $(M_g^h- M_r^h)_{\rm{cut}} = 0.21 - 0.03M_r^h$. In order to ensure that the galaxy samples in each magnitude bin are volume limited, bin-specific redshift cuts are applied. All of these cuts must be reproduced for the catalog selections. We reiterate that the absolute magnitude values in \citet{zehavi2011} are quoted for  $h=1$, and need to be corrected using 
\begin{equation}
\label{eq:mag_corr}
M_r = M_r^h + 5\log_{10}h,
\end{equation}
for the cosmology used to simulate the synthetic catalog.

For testing the evolution of galaxy clustering at higher redshifts than those found in SDSS galaxy samples, we use data available from the DEEP2 survey  \citep{mostek13}. The projected auto-correlation functions measured from these data
have been fit to power-laws for stellar-mass limited samples and are valid for redshifts $0.74 < z < 1.05$. \reply{These data are particularly convenient for validating synthetic catalogs because stellar-mass cuts are very straightforward to implement and interpret at the catalog level. Additional validation data for magnitude-limited samples are available from the VIPERS survey~\citep{vipers-wprp2013, marulli2013}. \citet{marulli2013} also present results for stellar-mass limited samples of the data, which have slopes similar to those of the DEEP2 data, but whose normalizations are somewhat higher. }

We now turn to the available data sets and theoretical predictions for validating weak-lensing quantities.
First, we use the predictions for the shear-shear auto-correlations as a function of angular separation to validate the catalog shear-shear signal at large angular scales. These predictions are obtained from linear theory in CAMB~\citep{CAMB} and the Core Cosmology Library \citep[CCL;][]{CCL}.

For validating the galaxy-galaxy lensing signal, there are several reference sources available. A crucial feature in the implementation of any test based on observational data is to ensure that the observed lens galaxy selection can be reproduced in the synthetic catalog thereby checking how the galaxy-halo connection in the synthetic catalog matches with that of data. Our first data set is the galaxy-galaxy lensing signal from the  Baryon Oscillation Spectroscopic Survey (BOSS) LOWZ sample presented in~\citep{Singh2016}. This sample consists of LRGs with $z<0.4$ selected from SDSS DR8. The selection cuts to reproduce this sample are presented in \citet{reid2016} and consist of  color selections based on SDSS $g-r$ and $r-i$ colors, a color-dependent $r$-band magnitude cut and a selection cut on $r$-band magnitudes requiring that $16 < r < 19$. 

A second set of validation data uses the galaxy-galaxy lensing signal from the SDSS Locally Brightest Galaxies (LBG) sample presented in~\citet{Mandelbaum2016}. The LBG sample is defined in~\citet{Anderson2015} and is a subsample of the flux-limited SDSS Main
galaxy sample~\citep{Blanton2005}. This LBG sample has accurate stellar mass and color measurements, which allow us to test synthetic catalogs in these different parameter spaces. The selection cuts that must be reproduced for the catalog lens sample include redshifts in the range $0.03 <z < 0.2$, SDSS $r$-band magnitudes with $r < 17.7$ and stellar masses in the range $\log_{10}(M^*/M_\odot) > 10$. The measurements of CFHTLenS~\citep{Velander2014} also provide further more qualitative checks of the galaxy-galaxy lensing signal at higher redshifts.

\subsubsection{Correlation Function Tests}
\label{sec:science:suite:2pt}

The correlation function tests we implement in our suite are based directly on the available validation data and are as follows:
\begin{itemize}
    \item galaxy-galaxy two-point correlation as a function of angular separation for magnitude limited samples at low redshfit (\autoref{sec:test:angular-cf}),
    \item comparison of galaxy bias with a linear bias model (\autoref{sec:test:galaxy-bias}),
    \item galaxy-galaxy two-point correlation as a function of projected distance (\autoref{sec:test:galaxy-galaxy}):
        \begin{itemize}
            \item for magnitude limited samples at low redshift (\autoref{sec:test:galaxy-galaxy-mag}),
            \item for color-selected samples at low redshift (\autoref{sec:test:galaxy-galaxy-color}),
            \item for samples selected by stellar mass at higher redshift (\autoref{sec:test:galaxy-galaxy-z}),
        \end{itemize}
 \end{itemize}
 
 The first test, which validates the two-point angular correlation function, provides a relatively simple test of galaxy clustering because it does not require the implementation of a complicated set of selection cuts to reproduce the observed data sample. 
 
 The galaxy bias test compares the galaxy angular power spectrum with theoretical predictions to obtain an approximate estimate of the linear bias. 
 
 The third set of tests, which validate the spatial clustering of galaxies as a function of projected distance for different galaxy sub-samples, provides more complicated tests of the two-point clustering because of the selection cuts that must be applied to the catalog in order to mimic those used in the observational data. This means that the test results will depend on the details of these selection cuts and therefore this set of tests constitute more stringent checks of the fidelity of the catalog. 
 
 Turning now to weak-lensing statistics,
 we implement two tests to check the validity of shear correlations:
 \begin{itemize}
 \item shear-shear auto-correlations as a function of redshift (\autoref{sec:test:shear-shear}),
 \item the galaxy-galaxy lensing signal as a function of projected distance (\autoref{sec:test:galaxy-shear}),
    \begin{itemize}
    \item for a mock LOWZ (LRG) galaxy sample (\autoref{sec:test:galaxy-shear-lowz}),
    \item for a mock LBG galaxy sample in bins of stellar mass and color (\autoref{sec:test:galaxy-shear-lbg}).
    \end{itemize}
\end{itemize}
 
The first test is a key validation test for catalogs intended to mimic weak lensing surveys and verifies that the auto-correlations of the cosmological shear values in the synthetic catalog match the theoretical predictions from linear theory for sufficiently large scales.
 
The two galaxy-galaxy lensing tests validate the galaxy-shear correlations for different sub-samples of lower redshift galaxies with $z\lesssim 0.4$. The first test validates the galaxy-halo connection for the arguably important LRG galaxy sample. The second test provides a more stringent validation of the galaxy-halo connection owing to the further subdivisions of the sample by stellar mass and color. 

\subsection{Galaxy Clusters}
\label{sec:science:cl}

\subsubsection{Scientific Drivers and Critical Properties}
\label{sec:science:driver:cl}


Galaxy clusters form in the most massive dark matter halos and are the largest-known gravitationally collapsed objects in the universe. Clusters act as high-density tracers of large scale structure. The number density and spatial distribution of clusters as a function of their masses and redshifts are sensitive probes of cosmology. In particular, the measurement of cluster abundance is sensitive both to the growth of structure and to the geometry of the universe~\citep[and references therein]{allen2011}.

Galaxy clusters are detected from their peculiar observational signature with respect to the background. In the optical and infrared, clusters appear as characteristic spatial concentration of galaxies forming a peak in the redshift distribution, and displaying similar colors and a characteristic luminosity distribution. Their center is taken to be the most likely central galaxy, that is often also the brightest cluster galaxy (BCG). 

The LSST~DESC reference cluster detection algorithm, \redmapper, finds galaxy clusters in optical surveys by identifying over-densities of red-sequence (on a color-magnitude diagram) member galaxies. The \redmapper richness, $\lambda$, is a cluster observable defined as the sum of the membership probabilities of the galaxies within a specified radius and above a luminosity threshold whose scatter in the mass-richness relation is minimized~\citep{rykoff2014}. The membership probability of one galaxy is determined by its position, color and magnitude~\citep {rykoff2012, rykoff2014, rykoff2016}. For each detection, \redmapper outputs a list of potential member galaxies, as well as estimates of the richness, redshift and most likely central galaxy. In order to provide a redshift independent mass estimator, optical cluster finders limit the membership assignment based on  $L_*(z)$, the knee of the luminosity function~\citep{rykoff2016, wazp}.

In surveys, cluster masses are inferred from observables, such as cluster member galaxy counts (\ie,  richness), X-ray luminosities, weak lensing measurements, galaxy velocity dispersions or the flux of the Sunyaev–Zel'dovich (SZ) effect on the Cosmic Microwave Background. Those observables are then linked to the true cluster masses via mass-observable relations (MOR), which need to be calibrated. This is one of the most critical aspects of using clusters as robust cosmological probes.
The MOR calibration often relies on gravitational lensing, since this method does not require assumptions about the cluster dynamical state and is believed to be less biased (at least for high mass systems) than, \eg, hydrostatic mass estimates from X-ray measurements~\citep{vonderlinden2014, mantz2015, becker2011}.
In the optical, the lensing masses are usually obtained by fitting the radial profile of the shear signal of background sources around clusters, by assuming a given form of the underlying radial mass profile. In order to enhance the signal-to-noise, which is low for an individual cluster, the radial shear signal is often stacked, by grouping clusters in bins of richness and redshift.

Another critical element for the use of cluster samples is the determination of their selection function (\ie, how well the detected sample represents the true population). For optical surveys, apart from survey design and detection algorithm specification, the selection function depends not only on the true mass (or observable proxy) and redshift of the systems, but also on many aspects of the physics of clusters (see next paragraph for examples) and the underlying covariance between their measured masses and observables.

The assembly of cluster samples usable for cosmological analyses thus requires knowledge of the detection efficiency, the correct measurement of their observables, an estimation of their masses and redshifts and the evaluation of the sample selection function.
All steps should be extensively tested and validated by using simulated data sets. This is particularly challenging because cluster analyses require many aspects of the synthetic catalog to be as realistic as possible, \eg, member galaxy redshift, luminosity and color distributions, member galaxy spatial distributions and their variation from the field to the inner regions of clusters, the gravitational lensing implementation, the ability to determine high-quality photometric redshifts  and the link between the galaxy distribution and properties and the underlying dark matter distributions.

In this analysis we focus on validating the most important cluster properties responsible for their detection and mass estimation in the future LSST data. In particular, we base many of our requirements on the use of the red-sequence Matched-filter Probabilistic Percolation cluster-finder (\redmapper) detection algorithm \citep{rykoff2014, rykoff2016}.  
However, as we test for generic cluster properties, our validation scheme should also enable tests of other detection methods.

\subsubsection{Galaxy Cluster Validation Data }
\label{sec:science:data:cl}

For the validation of synthetic clusters, we use both observational data samples and parametrizations measured from observed data or other simulations. We include measurements of observed clusters, simulated halos and cluster-member galaxies.

Polynomial fits to measurements of the mean $g-r$, $r-i$ and $i-z$ colors, as a function of redshift, for red-sequence galaxies observed by DES~\citep{rykoff2016} provide a convenient summary of the expected dependence of the color of cluster-member galaxies with redshift.\footnote{These fits were also used in the cosmoDC2 production pipeline as described in~\citet{cosmodc2}.} We also assemble a validation data sample of 2661 \redmapper clusters by selecting clusters from the publicly available \redmapper catalog (version 6.4) for DES Y1 data~\footnote{ \url{https://des.ncsa.illinois.edu/releases/y1a1/key-catalogs/key-redmapper}}. For each cluster, this catalog provides a redshift and color information and membership probabilites for the cluster members.  In selecting this sample, we impose a richness cut of $\lambda > 30$ and we assign a median color to the cluster by computing the median color for all cluster member galaxies whose membership probability is greater than 70\%.

The  evolution of the luminosity of member galaxies as a function of redshift, which impacts the detection efficiency for cluster members, requires an estimation of the evolution of $L_*(z)$ (the knee of the luminosity function).  We derive the corresponding apparent magnitude $m_*(z)$ (assuming that LSST $i$-band is used for the detection) from the passive evolution of a burst galaxy with a formation redshift $z=3$, taken from the PEGASE2 library~\citep[\texttt{burst\_sc86\_zo.sed},][]{Fio97}. The relationship is calibrated using the value of ${\rm K}^{\star}$($z=0.25$) (the value of $m_*(z)$ in K band) derived by \citet{2006ApJ...650L..99L} from an observed cluster sample. We use this estimate as a prediction for $m_*(z)$.

In order to estimate the luminosity distribution of member galaxies within galaxy clusters, we begin with the \redmapper v5.10 cluster catalog derived from the SDSS DR8 data set, which covers approximately  $10405$  $\rm{deg}^2$ with $i$-band depth of $\sim 20.9$ \citep{Rozo15}. We select clusters with redshift ($z_\lambda$) less than $0.3$ to avoid the selection effects from the SDSS survey depth and richness $\lambda$ greater than $5$ to maximize the signal to noise. This results in a catalog consisting of $80507$ clusters. We then compute the conditional luminosity function following the method described in \citet{to2020}.


For the validation of cluster galaxy-density profiles, we use the parametrization of galaxy density-profiles from \cite{2017MNRAS.467.4015H}, measured on a set of SZ selected galaxy clusters. High mass SZ selected cluster samples are known to present high purity and to be close to pure mass selection~\citep{sptpol2020a, sptpol2020b, atacama2021}. The mean galaxy-density profiles in these clusters can thus be compared to that of massive halos in simulations. The clusters used in \cite{2017MNRAS.467.4015H} are detected from the South Pole Telescope survey~\citep{bleem2015} and their galaxy density profiles are measured from optical data taken during the science verification phase of the Dark Energy Survey~\citep{jarvis2015}. The evolution of the profiles with mass and redshift are studied by using NFW parametrizations~\citep{1997ApJ...490..493N} of the profiles.

The data sets available for validating the mass-richness relation of galaxy clusters are obtained from the DES (Year 1 and Science Verification) and the SDSS surveys. 
\cite{mcclintock2019} use the DES Y1 data set ($\sim$1500 deg$^2$), with \redmapper depth $0.2L_*$ at $z=0.7$, to obtain $> 6500$ clusters with richness $\lambda>20$.
They divide these \redmapper clusters 
into $4\times3$ richness ($\lambda\geq20$) and redshift ($0.2\leq z \leq0.65$) bins and measure the mean masses in these bins via the stacked weak lensing signal.  The modeled mass-richness relation (\ie, the expectation value of the halo mass at a given richness and redshift) is determined to be:
\begin{equation}
    \langle M_{\rm 200m}|\lambda, z\rangle = M_0 \left(\frac{\lambda}{\lambda_0}\right)^F \left(\frac{1+z}{1+z_0}\right)^G,
    \label{eq:mass-richness_relation}
\end{equation}
where $M_0 = [3.081 \pm 0.075(\rm stat) \pm 0.133(\rm sys)] \cdot10^{14} M_\odot$, 
$F = 1.356 \pm 0.051 (\rm stat) \pm 0.008 (\rm sys)$, 
$G = - 0.30 \pm 0.30 (\rm stat) \pm 0.06 (\rm sys)$. \autoref{eq:mass-richness_relation} uses the halo mass definition $M_{\rm 200m}$, which denotes the spherical overdensity (SO) mass within a sphere whose mean internal density is 200 times the mean matter density of the universe at redshift $z$.
The pivot values $\lambda_0 = 40$ and $z_0 = 0.35$ are close to the median values for the \citet{mcclintock2019} cluster sample. For the DES SV and SDSS surveys, in order to facilitate easy comparisons of the data,  \citet{mcclintock2019} provide corrections for the impact of analysis details that could cause the definitions of richness to differ among the surveys. We use these corrected results as additional validation data.


A comparison of the halo masses recovered from measured weak-lensing quantities with those obtained directly from a synthetic catalog provides a consistency test that, ideally, would result in an exact match between the two quantities for any given halo. In reality, there will be scatter in this relationship and the amount of scatter will depend on the realism of the catalog and the underlying systematic effects that may or may not have been included. This type of test does not require any external data, but does require a number of theoretical inputs to obtain best-fit mass estimates from, \eg, excess surface-density profiles. For this case, theoretical predictions for the excess surface-density are obtained using the DESC  CLMM\footnote{\url{https://github.com/LSSTDESC/CLMM} [github.com]} library assuming an NFW dark-matter profile \citep{1997ApJ...490..493N} and a concentration-mass relation \citep{Duffy2008}.  We note that such an analysis delivers estimates of $M_{\rm 200m}$, whereas the synthetic catalog may use an alternate mass definition such as a Friends-Of-Friends (FoF) halo mass. In order to obtain the most meaningful consistency check, it is necessary either to estimate the effect of using different mass definitions or, if possible, to convert the masses from one definition to another.

Another proxy for the mass of a galaxy cluster is the velocity dispersion of member galaxies. \cite{evrard2008} present an estimate of the bulk virial scaling relation for dark matter halos that is determined from an ensemble of simulations. They determine that the dark matter velocity dispersion, $\sigma_{DM}(M,z)$,  in halos with masses larger than $10^{14} h^{-1}\text{M}_\odot$ scales with the halo mass as 

\begin{equation}
\sigma_{DM} = \sigma_{DM,15}\left(\frac{h_{100}(z)M_{\rm 200c}}{10^{15}\textup{M}_\odot}\right)^\alpha,
\label{eq:vel_disp}
\end{equation}
where $\sigma_{DM,15}$ = 1082.9 $\pm$ 4.0 km/sec, $\alpha$ = 0.3361 $\pm$ .0026, $h_{100}(z)$ denotes the dimensionless Hubble parameter normalized by 100 ${\rm km}~{\rm sec}^{-1} {\rm Mpc}^{-1}$ and $M_{\rm 200c}$ denotes the SO mass defined as the enclosed mass within a sphere whose mean internal density is 200 times the critical density of the universe at redshift $z$.
The dark matter velocity dispersion is related to the galaxy velocity dispersion via a dimensionless factor, the velocity bias, which is defined as $\sigma_{gal}/\sigma_{DM}$. The values of velocity bias can be determined from simulations and are scattered in the range 0.9 to 1.1, with averages that vary from 0.96 at high redshift to 1.04 at low redshift~\citep{evrard2008}. A more recent study~\citep{gifford2013} has determined the velocity bias for a number of semi-analytic galaxy models and for a variety of galaxy samples with varying fractions of red, blue, bright and dim galaxies. The values of the bias in most of these cases vary between 0.95 and 1.05 except for samples with low fractions of red galaxies which have biases as high as 1.1.  
For simplicity, we choose the velocity bias to be equal to 1 for validation test in this paper.

\subsubsection{Galaxy Cluster Tests}
\label{sec:science:suite:cl}

Based on the requirements presented in \autoref{sec:science:driver:cl} and the available validation data sets, we develop 7 validation tests for clusters. These are based either on the properties of their galaxies: 
\begin{itemize}
    \item the color-redshift relation of central galaxies (\autoref{sec:test:red-sequence}),
    \item the redshift evolution of member-galaxy magnitudes in comparison to $m_*$, (\ie, $L_*$ converted to the corresponding $i$-band apparent magnitude\footnote{See \autoref{sec:science:data:cl}}) (\autoref{sec:test:cl_mag_star}),
    \item the shape of the cluster galaxy conditional luminosity function (\autoref{sec:test:clf}),
    \item the shape of the number-density profile of member galaxies (\autoref{sec:test:cl_density_profile}));
\end{itemize}
or on the connection between the true halo mass and different observables:
\begin{itemize}
    \item the mass-richness relation for \redmapper selected clusters (\autoref{sec:test:mass_richness}),
    \item the weak lensing cluster profiles and the associated measured mass (\autoref{sec:test:cl_shear}),
    \item the velocity-dispersion--halo-mass relation (\autoref{sec:test:velocity_disp}).
\end{itemize}

The ability to run \redmapper to identify clusters in a synthetic catalog is a non-trivial requirement that demands the existence of a clearly identifiable red sequence with tightly constrained scatter on a color-magnitude diagram for galaxies in halos with masses $M_{halo} \gtrsim  10^{13} M_\odot/h$. The first test checks for the presence of the red sequence and is a minimal requirement for a synthetic cluster catalog intended for use by optical surveys.

As described in \autoref{sec:science:data:cl}, the evolution of the luminosity of member galaxies with redshift  impacts the performance of optical cluster finders. Therefore, to test for the consistency of cluster detection as a function of redshift, the second test checks how well the redshift evolution of halo-member magnitudes is traced by $m_*$.  
This test ensures that a richness estimation based on $m_*$ is redshift independent and is also an important requirement for alternative optical cluster finders such as WaZP \citep{wazp}.

The performance of \redmapper also depends on the relative luminosity distribution of cluster central and satellite galaxies. Thus, in order to ensure that the performance of \redmapper on the synthetic catalog is sufficiently realistic, it is important to check the fidelity of the luminosity functions against the observed data. Ideally, we want to compare the luminosity function in bins of halo masses; however, halo masses are not an observable in the real data. Instead, in the third test, we measure the luminosity function of centrals and satellites in bins of richness ($\lambda$), which serves as a low-scatter halo mass proxy for \redmapper clusters.

The  fourth test checks the number-density profile of member galaxies, which is one of the drivers of cluster detection~\citep[see, \eg,][]{2019A&A...627A..23E}. The distribution of member galaxies not only impacts the detection but also the richness estimation~\citep{rykoff2014, rykoff2016}. Moreover, it can be used as a proxy of the underlying dark matter distribution~\citep{guo2012, Augustsson2010} and  
at larger radii, it is also important for testing weak-lensing mass measurements and possible splash-back effects.

Apart from the cluster member properties, it is also crucial to validate the link between cluster observables and mass.
The fifth test provides a way to investigate this link by determining the cluster mass-richness relation in the synthetic catalog  and comparing it to that measured with \redmapper\ catalogs from, \eg, DES (see \autoref{sec:science:data:cl})
Gravitational lensing will be the primary probe of mass for LSST clusters, it is therefore mandatory to test the validity of the shear profiles around massive halos. Moreover, the sixth test then naturally arises from  examining the relation between the true halo mass and the estimate of the mass recovered from weak lensing measurements   

Finally, as mentioned in \autoref{sec:science:data:cl} the velocity dispersion of a dark matter halo acts as another proxy for its mass and therefore the seventh test which examines the validity of the velocity dispersion-mass relation thus provides another significant validity test for cluster cosmological analyses.

\subsection{Emission Line Galaxies}
\label{sec:science:elg}

\subsubsection{Scientific Drivers and Critical Properties}
\label{sec:science:driver:emf}

Nebular emission lines in galaxies are a critical tracer of many elements of galaxy physics, including gas phase metallicity, star formation rate, and dust content \citep{Runco2021}.  Additionally, they have been found to have a significant effect on galaxy colors used for photometric redshift fitting \citep{Csornyei2021}, indicating that improper emission line modeling could lead to a reduction on photometric redshift precision and accuracy.  Measurements of large-scale structure (LSS) with LSST will rely heavily on the ability to recover galaxy redshifts accurately from photometry using photometric redshift fitting, as galaxy clustering analyses require splitting the sample into narrow redshift bins to perform angular clustering analysis~\citep{nicola2020}.

Typically, so-called red sequence galaxies are used for such fits because the Balmer break -- the strong decrease in flux density at wavelengths bluer than rest-frame $4000$~{\AA} -- greatly increases the precision of photometric redshift fits \citep{gladders2000, Rozo2016, Stoughton2002}.  However, an alternative analysis that is being actively pursued within LSST~DESC focuses on another sub-population of galaxies that could yield high-quality photometric redshifts by utilizing strong emission line features, rather than continuum features.  We refer to such objects as extreme emission line (EEL) galaxies.  Because emission lines are isolated to very specific wavelengths, it is possible that EEL galaxies could yield comparable or higher redshift precision than red sequence galaxies, allowing for improved measurements of galaxy clustering by reducing the redshift width of redshift bins and reducing contamination between adjacent bins.   Indeed, \cite{Csornyei2021} find that galaxy emission lines often have a more pronounced effect on galaxy colors than photometric uncertainties. As a result, galaxy catalogs with realistic emission-line properties are crucial to the development of this analysis.  



\subsubsection{Emission Line Data Sets}
Data for emission line ratio validation are sourced from SDSS DR12 \citep{Alam2015}.  SDSS DR12 provides data from the culmination of SDSS-III observations, which includes data from BOSS~\citep{Dawson2013}.  BOSS provides optical spectra for over 2 million galaxies across 10,000 deg$^2$ at $z<0.7$.  We select galaxies with non-zero measured continuum and emission line fluxes in H$\alpha$, H$\beta$, [OIII]$_{4959, 5007}$, and [OII]$_{3727, 3729}$ to generate a catalog of 87,949 objects at $z\lesssim 0.4$.  While this SDSS catalog covers a much smaller range in redshift than that of the LSST, we can use it broadly as a test of the shapes of emission line ratio distributions.

\subsubsection{Emission-Line Tests}

We consider a combination of hydrogen (H$\beta$) and oxygen ([OII]$_{3727, 3729}$, [OIII]$_{4959, 5007}$) emission lines for our emission line ratio test.  These emission lines are considered because (1) they are relatively bright in $ugrizy$ bands at $z \lesssim 1.6$ and (2) ratios of different combinations of emission line fluxes can function as proxies for physical properties of interest. The ratio between $F_\mathrm{H\alpha}$ and $F_\mathrm{H\beta}$, for example, probes dust attenuation of nebular light.  Meanwhile, $F_\mathrm{[OIII]}/F_\mathrm{[OII]}$ traces the ionization parameter and is commonly known as $O_{23}$ \citep{Nakajima2013}.  Oxygen and hydrogen lines can be used together to calculate $R_{23}$, or $(F_\mathrm{[OIII]} + F_\mathrm{[OII]}/F_\mathrm{[H\beta]})$, which is a tracer of metallicity \citep{Lilly2003}.

\subsection{Stellar Mass Function }
\label{sec:science:smf}

\subsubsection{Scientific Drivers and Critical Properties}
\label{sec:science:driver:smf}

The stellar mass function (SMF), which is defined as the number density of galaxies as a function of their stellar mass, is one of the most fundamental products of a synthetic galaxy catalog and acts as an important probe of the galaxy-halo connection. Although the SMF was not identified by the LSST~DESC WGs as a critical static property requiring a specific validation criterion, it is strongly correlated with the behavior of other observable properties across cosmic time and therefore should be validated. 

The stellar mass of a galaxy is assembled over time, with each burst of star formation contributing an evolving component to its SED, as described by stellar population synthesis models \citep{bruzual2003}. In fact, these models form the basis for the inference of galaxy stellar masses from observations of their SEDs and are a critical component in determining the SMF from observed data. Recently, \citet{chaves-montero} have demonstrated in model universes that galaxy colors are largely regulated by their recent star-formation history and that variations in these histories are the dominant influence on the distributions of galaxy colors. Thus there is a direct, but complex, connection between the SMF and the SEDs, which are the fundamental galaxy observables that are sampled by optical surveys. The SMF at a given redshift provides a summary statistic for the galaxy content of the universe at a given time, and the evolution of the SMF as a function of redshift traces the evolution of the galaxy population and its properties over the history of the universe \citep{behroozi_etal18}. If the SMF as a function of redshift does not agree with the observational data then the luminosity and color distributions of synthetic galaxies can, at most, be right only at one redshift.

The distributions of stellar masses and star-formation rates directly impact the simulation of time-domain objects such as supernovae and strong lenses. 
Volumetric supernova (SN) rates depend on the star-formation rates of their host galaxies. The properties of Type Ia SN, which are of cosmological interest, are known to depend on their host environments and are strongly correlated with the stellar mass of their host galaxies \citep{sullivan2010, kelly2010, lampeitl2010}. Furthermore, Type Ia SN occur more frequently in galaxies with high stellar masses so the properties of these galaxies are of particular relevance for SN science. 
Although we do not discuss the validation of time-domain objects in this paper, from the above discussion it is clear that an accurate SMF over the range of redshifts of interest is a crucial driver for enabling realistic simulations of the host assignments of SN. 

\subsubsection{Stellar Mass Function Data Sets}
\label{sec:science:data:smf}

The data available for the validation of the SMF comes from two sources. \cite{moustakas2013} use data from the PRIMUS and SDSS surveys to measure the SMF in redshift bins with widths varying from $0.1$ to $0.2$ and spanning the redshift range $0 < z < 1$. Tables 3 and 4 in \citet{moustakas2013} present SMF measurements over a wide range of stellar-mass bins spanning the range $8.8 < \log(M/M_\odot h_{70}^{-2}) < 12.0$. These data are reasonably consistent with earlier measurements, given the differences in how the stellar masses were defined in some of the earlier studies. \reply{More recent high-redshift ($0.5 < z < 1$) measurements of the SMF from the VIPERS survey~\citep{vipers-smf2013} are almost a factor two lower than the PRIMUS results and lie on the lower boundary of earlier measurements for all redshift bins. We do not include these data in our validation test.  }

Data from the BOSS DR12 CMASS galaxy sample are available to validate the SMF for galaxies in the redshift range $0.4< z <0.7$. This data sample includes specific selection cuts on their SDSS $g$, $r$, $i$ band-magnitudes and colors as described in \citet{reid2016}. These cuts, given by:  
\begin{enumerate}
\item $d_\perp > 0.55$ where $d_\perp = (r-i) - (g-r)/8$,
\item $i < 19.86 + 1.6(d_\perp - 0.8)$,
\item $17.5 < i < 19.9$,
\item $(r-i) < 2$,
\end{enumerate}
select a sample of galaxies with high stellar masses and can be used to validate a CMASS-like sample in the synthetic catalog.

\subsubsection{Stellar Mass Function Tests}
\label{sec:science:suite:smf}

Our test suite includes two tests of the SMF that are based on the available validation data described in \autoref{sec:science:data:smf}: 
\begin{itemize}
    \item the SMF in redshift bins matching those of the PRIMUS data (\autoref{sec:test:smf-primus}), and
    \item the stellar mass distribution for a mock CMASS galaxy sample (\autoref{sec:test:smf-cmass}). 
\end{itemize}
The first test checks the redshift dependence of the SMF for redshifts in the range $0 < z <1$. The second test narrows the selection to a bright galaxy sample that provides a specific validation test for high stellar-mass galaxies.

\subsection{Catalog Verification}
\label{sec:science:suite:readiness}

As mentioned in \autoref{sec:intro}, verification tests ensure that the {\it contents} of the catalog have been correctly produced.
Synthetic catalogs provide data for a variety of analysis needs,  
one of the most important of which is the extragalactic input for image simulations. This is a particularly costly use of the catalog both in terms of computer time and human effort. 
It is therefore critical to ensure that values of the galaxy properties supplied by the catalog are finite, have reasonable distributions that lie within acceptable physical ranges, that object identifiers are unique and that quantities that should be related to each other in specific ways are related as expected. Our validation test suite therefore includes a set of tests that we dub {\it readiness} tests (\autoref{sec:test:readiness}), to verify that the catalog contents are as expected. 

\subsection{Summary}

In~\autoref{tab:tests}, we present a summary of the tests comprising the validation test suite that we have discussed in this section.  The tests are grouped by the type of test (one-point distributions, two-point correlations, relations). The science drivers relevant for each test and the section showing results for the cosmoDC2 catalog are also listed, along with the sources of validation data for each test. The tests listed in italics have quantitative validation criteria. 

\reply{We provide a complementary summary of the tests in the validation suite in~\autoref{tab:testsummary}, where we have listed the selection cuts that have been applied to the data for each test, as well as the validation criteria that have been implemented. We also summarize the results obtained for the application of these tests to the cosmoDC2 catalog, which will be discussed in~\autoref{sec:tests}. The selections of redshift, magnitudes, colors and other quantities listed in this table define the ranges of these quantities that are possible to validate with the currently available observational data sets. Although there are significant amounts of observational data in regions outside the ranges listed in \autoref{tab:testsummary}, they are problematic to use for validation purposes due to the difficulty of understanding the selection effects and hence the completeness of the data. As previously mentioned, the validation criteria listed in \autoref{tab:testsummary} have been developed by the LSST~DESC WGs, and are specific to the validation of the cosmoDC2 catalog for LSST science. In most cases, we have required qualitative agreement between the catalog and validation data over all regimes where suitable validation data exists. A few tests have more quantitative validation criteria and the motivations for these are discussed in \autoref{sec:tests}.}

\begin{table*}[t]
\begin{center}
\begin{tabular}{llllll}
Test                  & Section                      & Science Drivers   & Validation Data Sources\\ 
\hline \hline
\textbf{verification tests} & &  & \\
readiness  & \ref{sec:test:readiness} &  & \\
\hline
\textbf{one-point distributions} & & & \\
\it{cumulative number density}    & \ref{sec:test:dndmag}        & WL, BL, LSS  & \citet{2018PASJ...70S...8A, 2006AJ....132.1729B}; \\
 & & & \citet{2007ApJS..172...99C}  \\
redshift              & \ref{sec:test:nz}            & PZ, LSS, BL   &  \citet{coil, deep2}\\
\it{color (SDSS data)}                 & \ref{sec:test:color-1d}         & PZ, CL, BL   & \citet{2017ApJS..233...25A}  \\
 \it{color (DEEP2 data)} & & & \citet{deep2, CFHTLS}; \\
  & & & \citet{Zhou2019} \\
\it{position angle}       & \ref{sec:test:pa}            & WL   &   \citet{scipy} (uniform distributions) \\
\it{size}                  & \ref{sec:test:size}          & WL, BL  &   \citet{2014ApJS..212....5M, 2012MNRAS.421.2277L} \\
ellipticity           & \ref{sec:test:ellipticity}            & WL, BL   &  \citet{Joachimi2013} \\
stellar mass (redshift dependence)         & \ref{sec:test:smf-primus}         & SN, SL, CL & \citet{moustakas2013}\\
stellar mass (mock CMASS sample)       & \ref{sec:test:smf-cmass}         & SN, SL, CL &  \citet{reid2016}\\
\hline
\textbf{two-point correlations} & &  &\\
galaxy--galaxy angular  & \ref{sec:test:angular-cf} & LSS, PZ, BL  & \citet{tpcf_wang}\\
\it{galaxy--density (bias)}     & \ref{sec:test:galaxy-bias}   & LSS, BL &   \citet{nicola2020, CCL} \\

galaxy--galaxy projected distance       & \ref{sec:test:galaxy-galaxy-mag} & WL, LSS, PZ,  & \citet{zehavi2011}\\
(luminosity dependence) & & BL & \\
galaxy--galaxy projected distance      & \ref{sec:test:galaxy-galaxy-color} & WL, LSS, PZ,  & \citet{zehavi2011}\\
(color dependence) & & BL & \\
\it{galaxy--galaxy projected distance}    & \ref{sec:test:galaxy-galaxy-z} & WL, LSS, PZ,  & \citet{mostek13}\\
\it{(redshift dependence)} & & BL & \\
shear--shear          & \ref{sec:test:shear-shear}   & WL    &  \citet{CAMB, CCL} \\

galaxy--shear (mock LOWZ sample)      & \ref{sec:test:galaxy-shear-lowz}  & WL   &  \citet{Singh2015} \\
galaxy--shear  (mock LBG sample)      & \ref{sec:test:galaxy-shear-lbg}  & WL   &  \citet{Mandelbaum2016}\\
\hline
\textbf{relations} (binned statistics) & & & \\
\it{color--color}          & \ref{sec:test:color-color}   & PZ, CL  & \citet{deep2, 2017ApJS..233...25A}; \\
& & & \citet{CFHTLS, Zhou2019} \\
size--luminosity (total light profile)      & \ref{sec:test:size-mag-1}      & WL, BL   &   \citet{2014ApJ...788...28V} \\
size--luminosity (disk and bulge)      & \ref{sec:test:size-mag-2}      & WL, BL   &  \citet{2014ApJS..212....5M} \\
color--redshift (red sequence) & \ref{sec:test:red-sequence} & PZ, CL  & \citet{rykoff2016} \\
cluster-member magnitude--redshift & \ref{sec:test:cl_mag_star} & CL  & \citet{Fio97, 2006ApJ...650L..99L}\\ 
magnitude--halo mass (CLF) & \ref{sec:test:clf}      & CL   &  \citet{Rozo15, to2020} \\
galaxy-density profile--halo mass & 
\ref{sec:test:cl_density_profile} & CL  & \citet{2017MNRAS.467.4015H} \\
halo mass--richness & \ref{sec:test:mass_richness} & CL & \citet{mcclintock2019}\\
cluster weak-lensing masses & \ref{sec:test:cl_shear} & CL & \citet{1997ApJ...490..493N, Duffy2008} \\
velocity dispersion--halo mass & \ref{sec:test:velocity_disp}      & CL    & \citet{evrard2008} \\
\it{emission-line ratios} & \ref{sec:test:em-line}  & PZ, LSS  &  \citet{Alam2015} \\
\hline
\end{tabular}
\end{center}
\caption{Summary table of tests, science drivers and validation data sources for the tests in our test suite. The acronyms  used for the science drivers are: PZ: photometric redshift determination; WL: weak lensing; BL: blending; SL: strong lensing; CL: cluster cosmology; LSS: large-scale structure clustering; SN: supernovae. The sections in the paper in which each test is described in detail are also given. The tests listed in italics are those with specific quantitative validation criteria supplied by the LSST~DESC WGs. See  \autoref{tab:testsummary} for a complementary summary of the data selections, more detailed validation criteria and cosmoDC2 results for each of the validation tests.}
\label{tab:tests}
\end{table*}

\begin{table*}[t]
\movetableright=-1in
\begin{threeparttable}
\begin{tabular}{l|l|l|l|l}
\hline
Test              & Redshift  & Other Selections   & Validation Criteria & CosmoDC2 \\ 
 & Selection & & & Status \\
\hline \hline
\textbf{verification tests}  & & & & \\
readiness  & none &  none & visual inspection &  \\
\hline
\textbf{one-point distributions} & & & & \\
cumulative number density   & none &  $\rm{mag} < 27.5$  & within $\pm 40$\% of HSC\tnote{a} & $g$\tnote{b}, $r$, $i$, $z$ pass \\
redshift          & $0.0 < z < 1.5$ &  $18 < r < 27 $   & qualitative agreement & fail for $r<= 21$\\
color (SDSS data)   & $0.05 <z<0.1$  & $r < 17.7$ & Cram\'{e}r-von Mises  & $r-i$ passes\\
color (DEEP2 data)   & $0.85 <z<0.95$ & $r< 24$ & statistic: $w < .05$ & $g-r$ passes\\
position angle    & none & none & uniform  & pass \\
size           & none & $i < 25.2 $\tnote{c} & $1 < dN/dx < 4$  & pass \\
ellipticity          & $0.0 < z < 2.0$ & $i < 24$, $-21.0 < V < -17.0$\tnote{d} & qualitative agreement & pass \\
stellar mass (redshift dependence) & $0.0 < z < 1.0$ & none & qualitative agreement & pass\\
stellar mass (mock CMASS sample) & $0.4 < z < 0.7$ & $17.5 < i < 19.9$, $r-i <2$\tnote{e} &qualitative agreement &  pass\\
\hline
\textbf{two-point correlations} & &  & & \\
galaxy--galaxy angular  & none & $17 < r < 21$  & qualitative agreement & pass \\
galaxy--density (bias)  &$0.2 < z< 1.2$ & $i< 25.3$  & qualitative agreement & pass\\
galaxy--galaxy projected distance        & none & $-23 < M_r^h < -19$ &  qualitative agreement & pass \\
(luminosity dependence) & &  &  & \\
galaxy--galaxy projected distance      & none & $-23 < M_r^h < -19$  &  qualitative agreement & pass (red galaxies) \\
(color dependence) &  & $(M_g^h - M_r^h)_{\rm{cut}} = 0.21 - 0.03M_r^h$  & & fail (blue galaxies)\\
galaxy--galaxy projected distance   & $0.74 <z< 1.05$ & $\log_{10}(M^*/M_\odot) > 10.5, 10.8$ & $\chi^2 < 2$ for & pass \\
(redshift dependence)  & & &  $1-10$ $h^{-1}$Mpc & \\
shear--shear          & $ 0.5 < z < 3.0$ & $M_r < -19$  &  qualitative agreement  & pass\\
galaxy--shear (mock LOWZ sample)    & $0.4 < z < 1.0$ &  $16<r< 19$\tnote{f} &   qualitative agreement & pass for $r_p > 0.4 \rm{Mpc}/h$ \\
galaxy--shear  (mock LBG sample)  &  $0.03 < z <0.2$ & $r<17.7$, $\log_{10}(M^*/M_\odot) > 10$  &  qualitative agreement & pass (red galaxies); \\
 & & & & fail (blue galaxies)\\
\hline
\textbf{relations} (binned statistics)  & & & & \\
color--color          & $0.1 < z <0.3$ & $r < 24$  & Wasserstein distance & fail \\
size--luminosity (total light profile)  & $0.0 < z< 2.5$  & none  & qualitative agreement  & pass\\
size--luminosity (disk and bulge)   & $0.0 < z < 2.0$  &  none & qualitative agreement & pass \\
color--redshift (red sequence) & $0.0 < z < 1.5$  &  $M_{\rm halo} > 10^{13.5} M_\odot$\tnote{g} &  qualitative agreement & pass\\
cluster-member magnitude--redshift & $0.0 < z < 1.5$ & $M_{\rm halo} > 10^{13.2} M_\odot$ &  qualitative agreement & pass\\ 
magnitude--halo mass (CLF)  & $0.1 < z < 0.2$ & $5 < \lambda <100$\tnote{h} &  qualitative agreement & pass\\
galaxy-density profile--halo mass  & $\langle z\rangle$=0.46 & 
$\langle M\rangle$=$6\times 10^{14} M_{\odot}$, $i<22$ & qualitative agreement  &  pass\\
halo mass--richness  & $0.3 < z < 0.4$ & $M_{FoF} > 5.10^{13}$, $\lambda > 20$\tnote{i} &  qualitative agreement & pass \\
cluster weak-lensing masses  & $0.1 < z < 0.7$ & $M_{FoF} > 10^{14} M_\odot$, $i<25$\tnote{j} &  qualitative agreement & pass\\
velocity dispersion--halo mass  & none & $M_{FoF} > 10^{14} M_\odot$,  & qualitative agreement & pass \\
 & &  $M^* > 10^9 M_\odot$, $N_g >=5$ & & \\
emission-line ratios  & $0.0 < z < 0.7$ & $r<19.5$ $(z<0.4)$ & 2d K-S test & fail \\
              &       &  $i<19.9$  $(z>0.4)$ & & \\
\hline
\end{tabular}
\begin{tablenotes}
\footnotesize
\item[a] Extrapolated fit to HSC data in magnitude range $24.0 < \rm{mag} < 27.5$
\item[b] $g$-band passes almost all of the required range. See \autoref{sec:test:dndmag} for details.
\item[c] Galaxies are further selected such that their size is less than the median size of the sample. See \autoref{sec:test:size}.
\item[d] Additional selections on $B/T$ are described in \autoref{sec:test:ellipticity}.
\item[e] Additional color and magnitude selections are described in \autoref{sec:science:data:smf}.
\item[f] Additional color selections are described in~\citet{reid2016}.
\item[g] Galaxies are further selected to be on the red sequence. See \autoref{sec:test:red-sequence}.
\item[h] Test is run on \redmapper-selected clusters from cosmoDC2
\item[i] Additional selections are described in \autoref{sec:test:mass_richness}
\item[j] Additional selections are described in \autoref{sec:test:cl_shear}
\end{tablenotes}
\caption{Summary table of tests showing the redshift and other selection cuts applied to the catalog data, the validation criteria and the test status for the cosmoDC2 catalog. The selections listed in this table span the ranges of availability of suitable validation data.  See \autoref{tab:tests} for a complementary summary of the science drivers, the validation data sources and the paper section describing each test.}
\label{tab:testsummary}
\end{threeparttable}

\end{table*}

\section{Simulated Catalogs and Validation Test Framework}
\label{sec:prelude}

In this section, we briefly describe the catalog and the underlying simulation we employ to demonstrate our validation procedure, as well as the DESCQA validation framework which we use to develop and run all of the validation tests shown in \autoref{sec:tests}.


\subsection{Simulations}
\label{subsec:sims}
The cosmoDC2 catalog~\citep{cosmodc2, hearin2020}, which was produced for the LSST~DESC DC2, is based on a trillion-particle, ($4.225$~Gpc)$^3$ cosmological, gravity-only N-body simulation, the `Outer Rim'~\citep{outerrim2019}. \reply{The mass resolution of this simulation is $2.6\cdot 10^9 $M$_\odot$.}
The underlying cosmological model parameters are similar to those of WMAP7 and the simulation identifies halos using a parallel FoF-based halo finder with a linking length of b= 0.168 and the requirement of a minimum of 20 particles per halo. \reply{Using a recent determination of the stellar-to-halo-mass relation~\citep{girelli2020}, we expect that the stellar mass resolution of a catalog based on this simulation to be of the order of a few $\times 10^8$M$_\odot$.}
The cosmoDC2 catalog covers $440$~deg$^2$ of sky area to a redshift of $z=3$ and is complete to a magnitude depth of 28 in the LSST $r$-band. Each galaxy is characterized by a multitude of galaxy properties including stellar mass, morphology, SEDs, broadband filters magnitudes, host halo information and weak lensing shear. The catalog has been constructed using a new hybrid technique that combines data-based empirical approaches with semi-analytic galaxy modeling. The empirical model establishes a restricted set of fundamental galaxy properties, which are then completed by adding the properties of matching galaxies obtained from the semi-analytic model (SAM). In this approach, the empirical model is tuned to match the observational data and the SAM galaxies function as a library from which to draw the large and complex set of properties required by DC2. The method captures the highly nonlinear correlations between galaxy properties that are built into the SAM but preserves agreement with observational data by selective sampling of the SAM library. The underlying assumption in the approach is that the quantities tuned in the empirical model are sufficiently correlated with the quantities available from the SAM so that the statistical distributions of the latter will be realistic.  Our validation suite tests the validity of this assumption. 

The empirical model used for cosmoDC2 is based on UniverseMachine~\citep{behroozi_etal18} and is augmented
with additional rest-frame magnitude and color modeling.
\reply{Since the parameters of UniverseMachine are optimized using two-point correlation function data, the model is able to preserve stellar-mass and star-formation-rate correlations as a function of environment.}
This model is applied to Outer Rim halos to populate halo lightcones with galaxies.
At this stage, the galaxies have redshifts, positions, stellar masses, star-formation rates, LSST $r$-band rest-frame magnitudes and $g-r$ and $r-i$ colors. Then, these galaxies are matched to those that have been obtained by running the Galacticus SAM~\citep{benson_2010b} on a small companion simulation to Outer Rim. \reply{Owing to the computing time required to run the SAM, this companion simulation is about 1600 times smaller than Outer Rim ($360$~Mpc)$^3$.} The matching is done on the rest-frame magnitudes and colors. The properties of the matched SAM galaxies supply the full complement of required properties including LSST and SDSS rest-frame and observer-frame magnitudes, as well as coarse-grained SEDs obtained from a set of top-hat filters spanning a wavelength range from 100 nm to 2 $\mu$m. All of the magnitudes are available separately for disk and bulge components, with and without host-galaxy extinction corrections and with and without emission-line corrections. After the match, we use additional empirical models, based on properties obtained from the SAM galaxies (\eg, the bulge-to-total ratio), to provide values for galaxy shapes, orientations and sizes. 

In general, the observational data sets that are used to tune the galaxy properties in the catalog are different from those used in the validation tests. The exceptions are the PRIMUS data for the stellar mass function~\citep{moustakas2013} and the SDSS data for the two point correlation functions~\citep{zehavi2011}, which were used to tune the UniverseMachine empirical model.

The weak-lensing quantities in the catalog are constructed from the particle-lightcone data from the Outer Rim simulation by projecting particles onto a series of mass sheets (lens planes) and performing a full ray-tracing calculation to produce maps of the weak-lensing distortions. The surface densities on the mass sheets are computed on a HEALPix\footnote{\url{https://sourceforge.net/projects/healpix/}}~\citep{healpix} grid with Nside=4096. The mass sheets have a median width of approximately $114~\rm{Mpc}$.
Ray-tracing follows photon paths backwards in time from an observer grid to a source plane and adding deflections that depend on the surface density of particles at each lens plane between the observer and the source. The shears and convergence for each galaxy are obtained by first shifting the galaxy to its observed position and then interpolating the source map to that position.

\subsection{The Validation Framework: DESCQA}
\label{subsec:descqa}

The DESCQA validation framework~\citep{descqa} provides an efficient, flexible and convenient infrastructure that allows developers to produce and run a wide range of validation tests.\footnote{\url{https://github.com/LSSTDESC/descqa} [github.com]} The streamlined workflow enables non-expert users to run the available tests very simply. The framework is composed of three principal parts: (1) a catalog reader interface, (2) a validation test interface and (3) a web interface. A main script\footnote{This script was originally called the master script in \citet{descqa}} runs the selected catalogs and tests and outputs the test results.  These results are viewed on a dedicated web page that is generated by the web-interface portion of the framework. Each test typically produces comparison plots of the catalog and validation data and delivers a status and an optional score that is the result of applying a test-specific  metric comparing the catalog and validation data. 

The catalog reader interface employs the Python package \texttt{GCRCatalogs}\footnote{\url{https://github.com/LSSTDESC/gcr-catalogs} [github.com]} to supply a unified interface for ingesting a diverse range of catalogs. \texttt{GCRCatalogs} in turn uses the Generic Catalog Reader\footnote{\url{https://github.com/yymao/generic-catalog-reader} [github.com]} (GCR)
base class to provide many convenient features for filtering the data, which renders it especially suited for use in the validation test suite.

 The validation test interface contains a class (the \texttt{BaseValidationClass}) which provides a unified infrastructure for constructing a test.  Test developers need only to write a module that runs the desired test on an arbitrary catalog and provide an input configuration file that contains any required auxiliary information. The latter can include paths to the validation data, selection cuts required to match the catalog to the validation data and plotting and processing options that must be passed to the test.

\section{Validation Test Results for cosmoDC2}
\label{sec:tests}

We now provide a concrete use case of applying our validation test suite to assess a synthetic catalog. In this section, we present the results of running the tests summarised in \autoref{tab:tests} on the  cosmoDC2 extragalactic catalog. For each test, we describe the implementation details and discuss the 
validation criteria set by the relevant LSST~DESC WGs.

\subsection{Readiness Tests }
\label{sec:test:readiness}

The readiness tests provide a range of general checks that enable a very simple first quality check of the catalog data. These quality checks can be easily summarized and include various statistics evaluated on the distributions that are included in the test. Statistics include minimum and maximum values, means and standard deviations, fractions of zero values, fractions of infinite values \etc, as well as warnings for outlying values that exceed the expected ranges. Other checks are implemented to check if data columns are repeated, that all galaxy\_ids are unique, that the halo\_ids of central galaxies are unique and that the expected relationships between galaxy properties are found. For example, the relationship between the galaxy ellipticities and the lengths of the major and minor axes and between the shears, magnification and convergence are tested. The configuration file for the tests allows the user to choose which quantities are checked and to set values for the expected ranges in which the statistics for each distribution should lie. In order to illustrate the capabilities of these tests, in \autoref{app:ready} we show selected screen shots from the DESCQA web interface for the test outputs produced for the cosmoDC2 catalog. \reply{Problems with the catalog data are readily identified by visual inspection of the test outputs.}

\subsection{Cumulative Galaxy Number Density}
\label{sec:test:dndmag}

\begin{figure}[!ht]
  \centering \includegraphics[width=8.5cm]{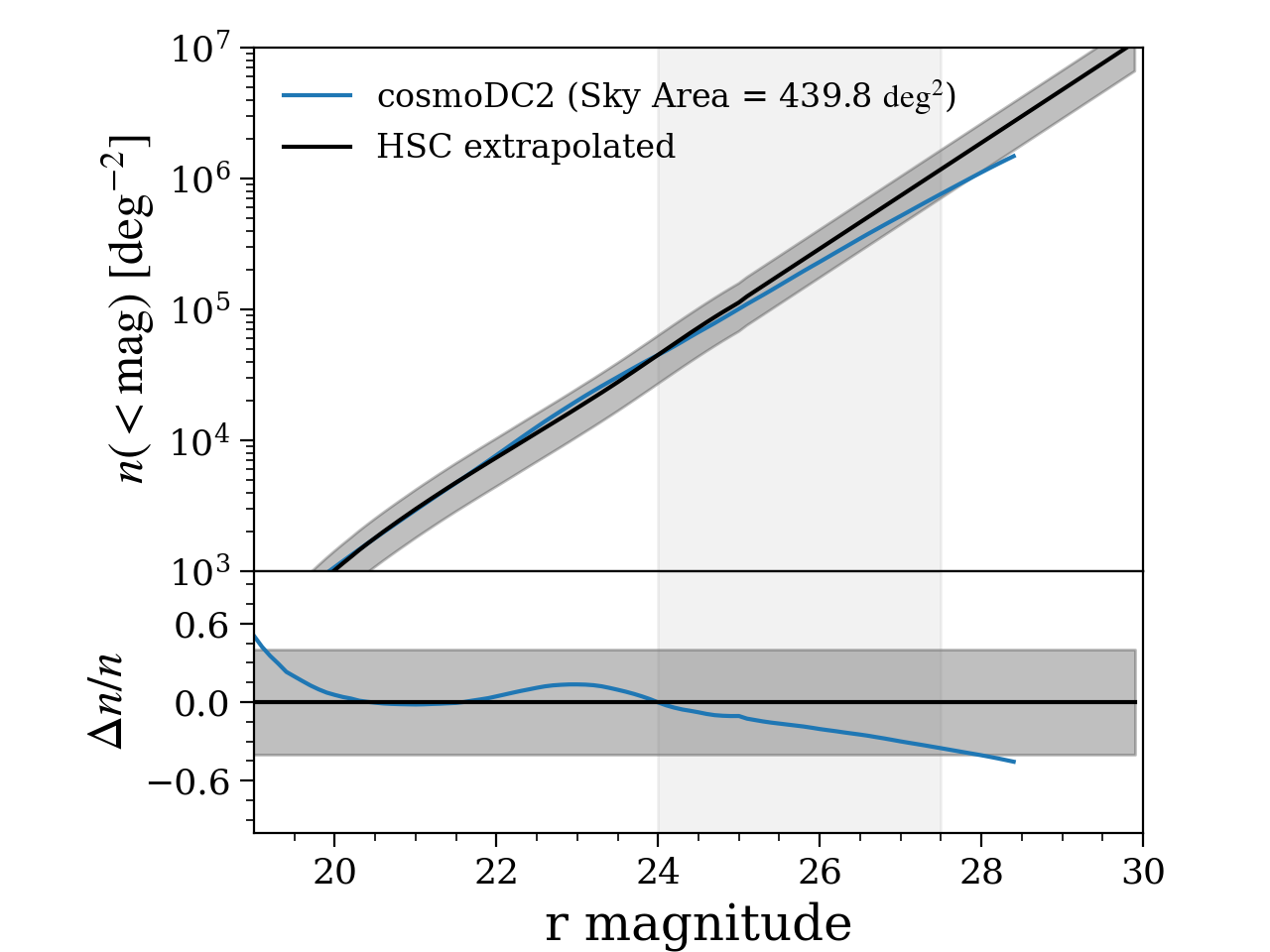}
  \caption{Top panel: Observed cumulative $r$-band number counts per square degree as a function of magnitude from cosmoDC2 (blue) and extrapolated from
  the HSC survey (black) (see text for more details). The grey shaded band shows a $\pm 40\%$ uncertainty around the HSC extrapolation. The vertical shaded region shows the magnitude range  within which the two curves are compared. Bottom panel: Relative difference between the two curves in the top panel. \label{fig:dndmag}}
\end{figure}

In~\autoref{fig:dndmag}, we compare the cosmoDC2 galaxy number counts with the HSC extrapolations described in~\autoref{sec:science:data:n}. The \reply{top panel} shows the cumulative number counts per square degree for galaxies with $r$-band magnitude less than the value of the abscissa, as a function of that value. The validation suite contains similar tests for $g$-, $i$-, $z$- and $y$-band magnitudes and the results for all of these tests are comparable to the one shown here. 

The grey-shaded band around the HSC extrapolated curve shows the region corresponding to a $\pm 40\%$ fractional difference in the number counts. The validation criteria set by the LSST~DESC WL and LSS groups required that the number counts for cosmoDC2 lie within this grey band in the magnitude range $24 < r <27.5$, which is indicated by the vertical shade band in the figure. This loose validation criterion is designed to check that the catalog has reasonably realistic cumulative number counts at the expected depths of the LSST WL and LSS data samples.
The number counts of galaxies fainter than $r=27.5$ are not critical for the scientific analyses planned for cosmoDC2, since such galaxies are generally below the detection limits of the LSST survey. We note, however, that there is a limitation in the comparison between the HSC extrapolation and the cosmoDC2 data. The HSC galaxy detections come from measurements of images for which blending can impact the number of detections and their fluxes \reply{(because deblended objects can potentially add to or fall out of the sample)}, so both the shape and magnitude of the extrapolation could be affected by undetected blends. The loose validation criteria for this test partially compensate for this additional uncertainty.

In the validation region, the maximum fractional difference between the catalog data and HSC extrapolation is determined and the catalog passes the test if this difference is below the specified tolerance (0.4 in our case). The maximum fractional differences for cosmoDC2 for $g$-, $r$- $i$-, $z$- and $y$-bands are 0.49, 0.34, 0.17, 0.10, and 0.27, respectively.
We note that cosmoDC2 passes this important validation test in every band except $g$-band, where the passing criterion is only marginally exceeded due to a small deficit in the number density above $g=27$. We do not expect that this minor deficit will significantly affect any of our verification tests. 

In \autoref{fig:dndmag}, the decrease in the cosmoDC2 number counts at fainter  magnitudes is an effect of the finite mass resolution of the underlying N-body simulation. Very faint galaxies reside in halos whose mass is too small to be resolved in the simulation. In order for cosmoDC2 to pass this test, it was necessary to add synthetic ultra-faint galaxies to the catalog to boost the number counts for magnitudes $\gtrsim 27$ as described in \citet{cosmodc2}. The required number of such galaxies was estimated by using a low-mass power-law extrapolation of the halo mass function to determine the number of missing halos and then populating each of these with a single galaxy. Without these additional galaxies, cosmoDC2 would have fallen well below the validation criteria shown in \autoref{fig:dndmag}. This test provides an example of why it is critical that validation testing is an integral part of the catalog production process. In this case, it was necessary to alter the production methodology so that the science goals for the catalog could be realized. 

\subsection{Redshift Distributions}
\label{sec:test:nz}

\begin{figure}
  \centering \includegraphics[width=8.5cm]{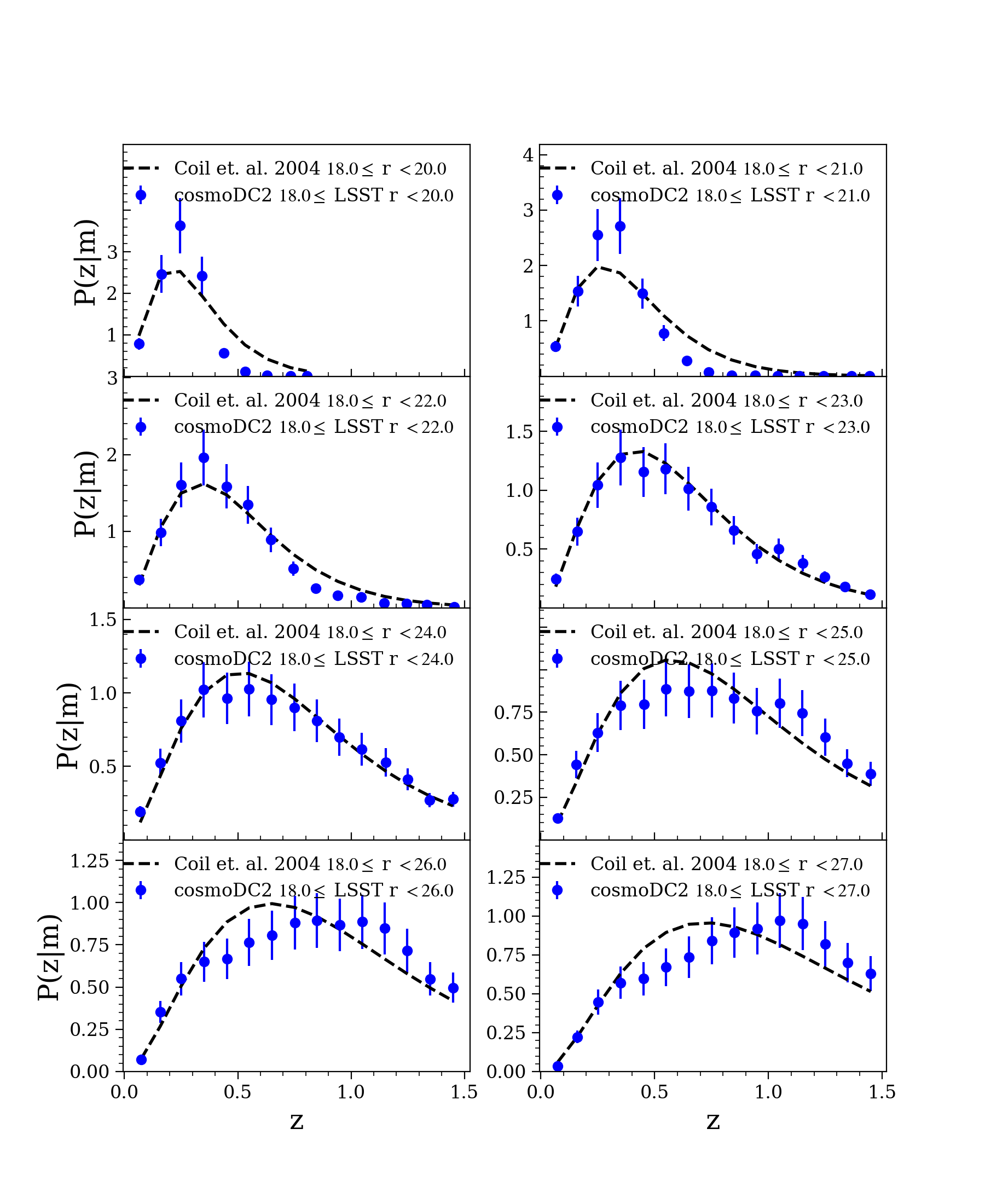}
  \caption{Redshift distribution of cosmoDC2 galaxies compared with fits to DEEP2 data for  a redshift range of $0 < z <1.5$ for eight magnitude bins. The selection cuts are on LSST $r$-band magnitudes with values ranging from 18 to 27 as indicated in the legends. The fits to the DEEP2 data are valid for $z\lesssim 1.5$. }
  \label{fig:nz}
\end{figure}

 In~\autoref{fig:nz}, we show the comparison between a $\sim$60 deg$^2$ patch of cosmoDC2 and the DEEP2 data fits from \citet{coil} as described in~\autoref{sec:science:data:n} for eight magnitude bins in the range $18 < r <27$. These fits to the DEEP2 data are valid for $z\lesssim 1.5$. Our test suite includes several similar tests for different combinations of magnitude cuts on both $r$- and $i$-band magnitudes. (A similar test for magnitude-limited galaxy samples with $r < 22 - 24$ is shown in Figure~13 of \citealt{cosmodc2}). We note that we have imposed the magnitude selection cuts on the LSST filter magnitudes in the catalog data, with no additional corrections to account for the differences between LSST and CFHT filters. 

 A $\chi^2$ test between the catalog and validation data is a convenient way to quantify the level of agreement between the two data sets. 
 A summary score for the test is computed by calculating the average of the reduced $\chi^2$ between the catalog data and observed fits for each of the magnitude-limited samples.
 Some care is required in implementing this test due to the smallness of the statistical errors in the binned catalog data. These statistical errors do not properly account for the errors due to cosmic variance, particularly for catalogs covering only small sky areas. Instead, we implement a jackknife procedure where we estimate the error, $\sigma_{i}$, on the $i$-th redshift bin by excluding regions of RA and Dec from the catalog data using a k-means\footnote{https://scikit-learn.org/stable/} algorithm. The elements of the covariance matrix are then given by
\begin{equation}
    \sigma^2_{ij} = \frac{N_{\rm{jack}}-1}{N_{\rm{jack}}} \sum_k (\bar{N}_i - N^k_i)\cdot(\bar{N}_j - N^k_j),
\end{equation}
where $N_{\rm{jack}}$ denotes the number of jackknife regions and $\bar{N}_i$ and $N^k_i$ denote the number of galaxies in the $i$-th redshift bin for the full sample and for the sample with the $k$-th region excluded, respectively.  The computation of covariances is lengthy so this test is faster if run over smaller sky areas (\eg, $\sim$100 deg$^2$). For these smaller areas, the above covariance matrix is often not invertible due to instabilities in the off-diagonal matrix elements, so we use only the diagonal elements in the $\chi^2$ computation. Consequently, the $\chi^2$ may be underestimated.

Except for the brightest galaxy samples ($r < 21$) at high redshift, the cosmoDC2 redshift distributions are in reasonable agreement with the DEEP2 fits, and we obtain $\chi^2/\rm{d.o.f}$ values of 79210, 8085, 13.3, .53, .62, .93, .96 and 2.2 for $r$-band magnitude ranges between 18 and 20 - 27, respectively. The large $\chi^2$ values for the   brightest galaxies shown in the upper two panels of \autoref{fig:nz} are due to a deficit of these galaxies at redshifts $z \gtrsim 0.6$. The validation criteria for cosmoDC2 did not specify a quantitative limit for  $\chi^2/\rm{d.o.f}$, so redshift distributions that are broadly in agreement with the observational data are considered acceptable. For cosmoDC2, all but the brighter galaxies with $r < 21$ and redshifts in the range $0.6 \lesssim z \lesssim 1$ approximately follow the expected distributions. 

\subsection{Color Distributions}
\label{sec:test:color}

\subsubsection{1-d Color Distributions}
\label{sec:test:color-1d}

\begin{figure}[htp]
  \centering \includegraphics[width=8.5cm]{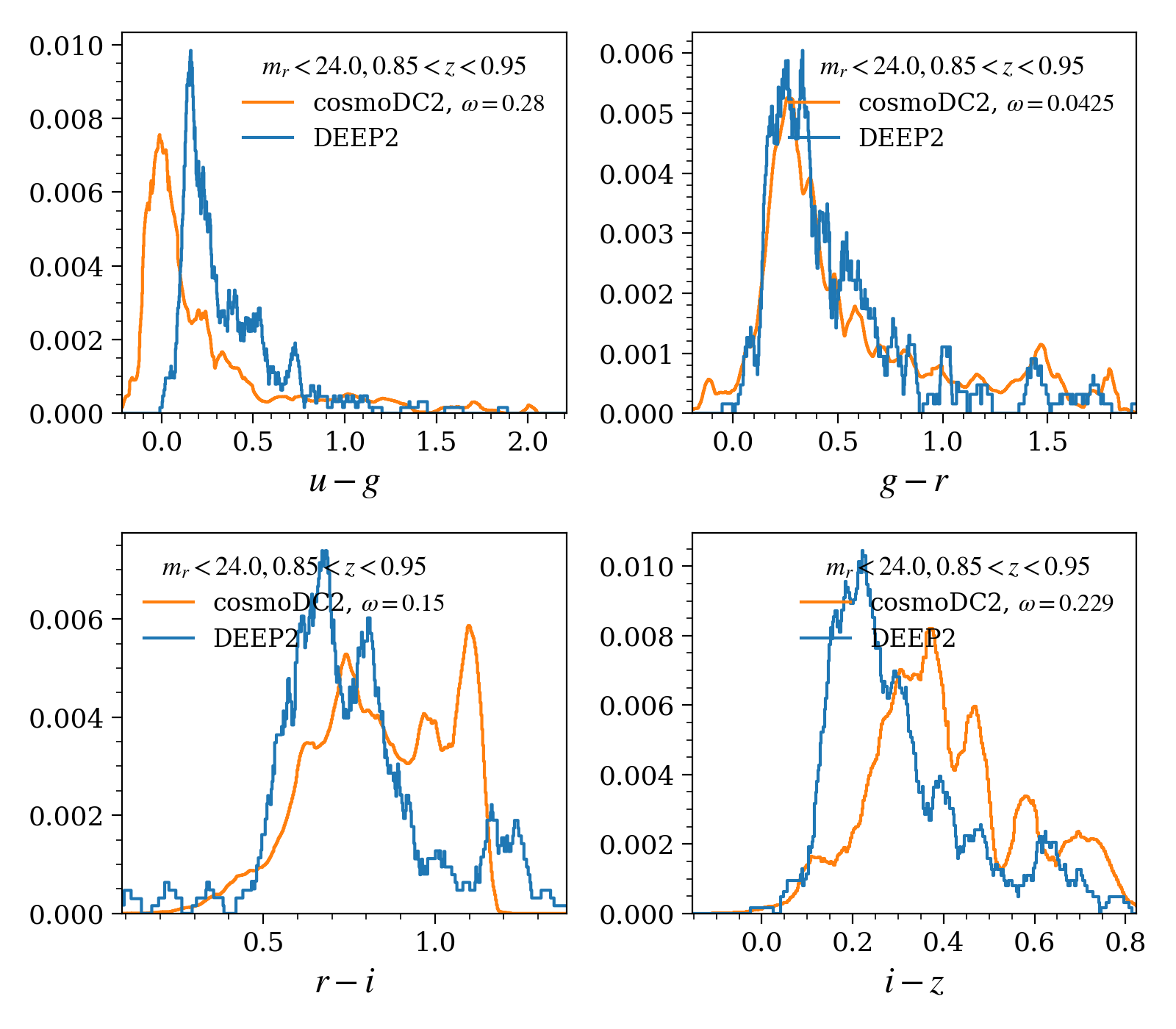}
  \caption{Color distributions of cosmoDC2 galaxies compared with those of DEEP2 data for $u-g$, $g-r$, $r-i$ and $i-z$ colors. The cosmoDC2 galaxies comprise a magnitude- and redshift-limited sample with $r$-band magnitude $<24$ and a redshift range of $0.85 < z < 0.95$.
  The SDSS filter magnitudes in cosmoDC2 have been corrected to compensate for the differences with the DEEP2 CFHT filters.}
  \label{fig:color}
\end{figure}

The 1-d color-distribution test in our suite is similar to the color test described in \cite{descqa}, but has been adapted for use on a lightcone catalog. For cosmoDC2, we select galaxies within a specified redshift range and having apparent magnitudes brighter than a specified limit, and compare the resulting color distributions to those of observed galaxies, selected using the same cuts. It should be noted that although we only compare the distributions of colors, the magnitude limits on the catalog samples implies that it is only possible to pass the test if the magnitude distributions are also sufficiently realistic. In~\autoref{fig:color}, we show an example of the color test for high-redshift galaxies. For this example, cosmoDC2 galaxies have $r$-band magnitudes $<24$ and  redshifts in the range $0.85 < z < 0.95$ and are compared with DEEP2 data. The suite contains a similar test comparing synthetic galaxies having $r$-band magnitudes $<17.7$ and redshifts in the range $0.05 < z < 0.10$ with SDSS data. The results for this test for the cosmoDC2 catalog are shown in Figure 14 in \citet{cosmodc2}.

As described in \citet{descqa}, we obtain a quantitative estimate of the level of difference between the validation data and the synthetic color distributions, by calculating the Cram\'{e}r-von Mises (CvM) statistic, $\omega$~\citep{anderson62}. This statistic provides a non-parametric method to test if two data samples are drawn from the same underlying distribution by measuring the average $L^2$ distance between the two cumulative distribution functions. The values of $\omega$ in~\autoref{fig:color} are 0.28, 0.0425, 0.15 and 0.229 for $u-g$, $g-r$, $r-i$ and $i-z$ color distributions, respectively. Lower values for this statistic correspond to better agreement between the distributions and values of $\omega \lesssim 0.05$ indicate that the two distributions are in very good agreement. Only the $g-r$ distribution satisfies this stringent criterion.

For DC2, no quantitative requirement has been given for $\omega$. As can be seen from~\autoref{fig:color}, the cosmoDC2 color distributions, except for $g-r$, do not agree well with the validation data. There are notable differences, particularly in the numbers of redder galaxies. As discussed in \citet{cosmodc2}, in order to facilitate the study of cluster systematic effects, the empirical color model for cosmoDC2 has been tuned to enhance the efficiency with which the \redmapper algorithm is able to identify synthetic clusters. This tuning results in an enhancement, particularly evident in the $r-i$ color distribution, of the red fraction in cosmoDC2.

\subsubsection{Color--Color Distributions }
\label{sec:test:color-color}

We have expanded the test suite to include 2-dimensional distributions as well. It is well known that, unlike the CvM statistic, there is not a well-defined metric to compare multivariate datasets in a unique way. In this case, we have instead used the so-called Wasserstein metric~\citep[in Russian]{dobrushin1970definition} measured across several random directions in color-color space, where the distributions are projected. These different measurements are averaged and provided as additional outputs for our color-color plot comparisons, which include a heatmap versus contour plot visualization as well. We  implemented a kernel two sample test from \citet{gretton2012kernel} as an additional cross-check.

\begin{figure}
  \centering \includegraphics[width=8.5cm]{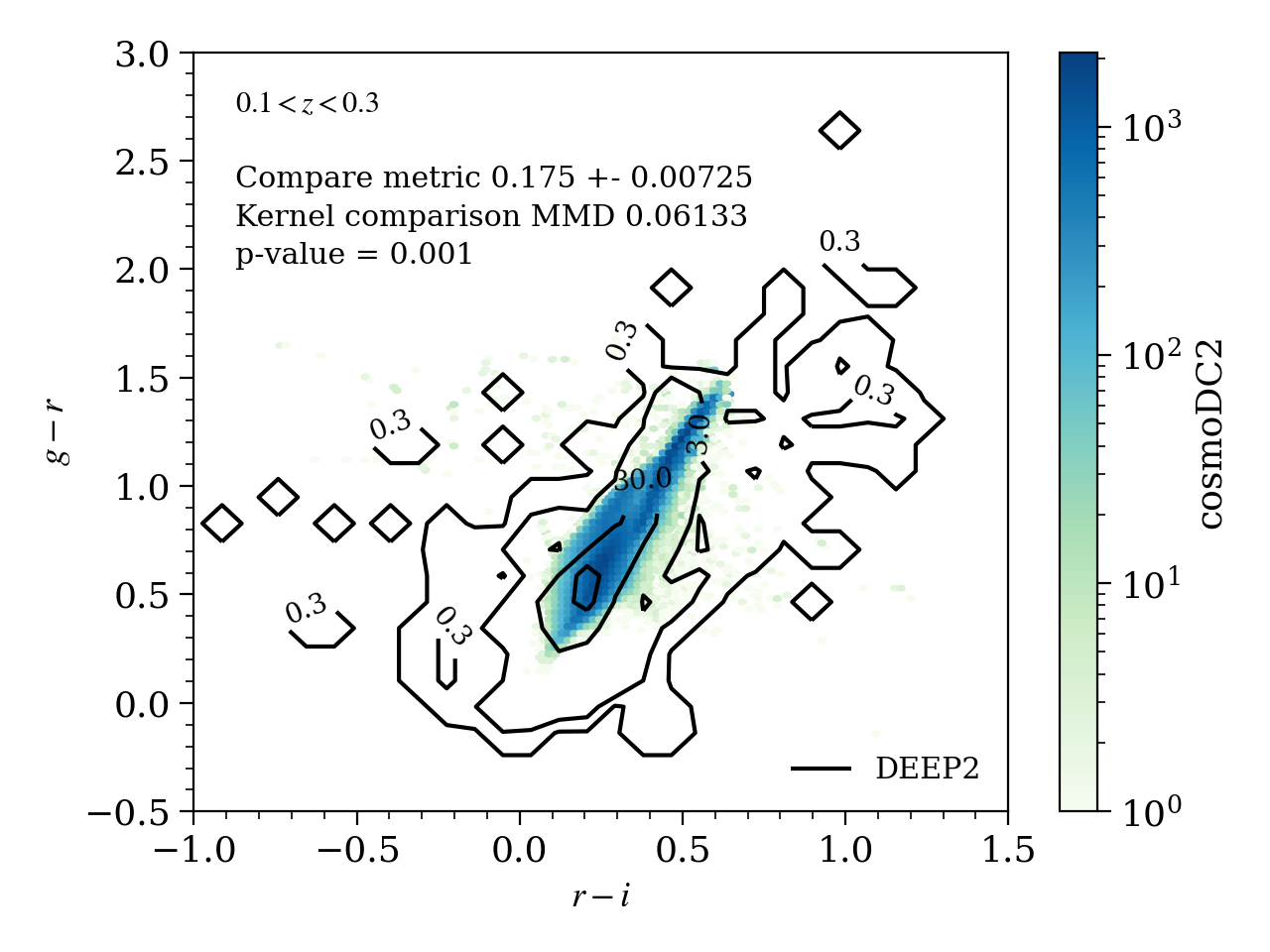}
  \caption{The 2 dimensional distribution of $g-r$ (y-axis) and $r-i$ (x-axis) cosmoDC2 colors compared with those of the DEEP2 data for the redshift range $0.1 < z < 0.3$. The color bar and the black contours shows the scales for the densities of cosmoDC2 and DEEP2 galaxies, respectively. The black contours are shown at density levels of 0.3, 3.0, 30.0 and 100.0. The results for the comparison metrics discussed in the text are given in the legend. }
  \label{fig:color-color}
\end{figure}

In \autoref{fig:color-color}, we show the two dimensional distribution of $g-r$ and $r-i$  cosmoDC2 colors compared with that of the DEEP2 data for the redshift range $0.1 < z < 0.3$. The``Compare" metric refers to the averaged Wasserstein distances, which in this case indicate a non-zero bias is present in this color space. The kernel comparison MMD (Maximum Mean Discrepancy) test also shows that the difference between the two distributions is significant (p value is 0.001 for both coming from the same underlying true distribution).

\subsection{Position Angle Distribution }
\label{sec:test:pa}
 
\begin{figure}
  \centering \includegraphics[width=8.5cm]{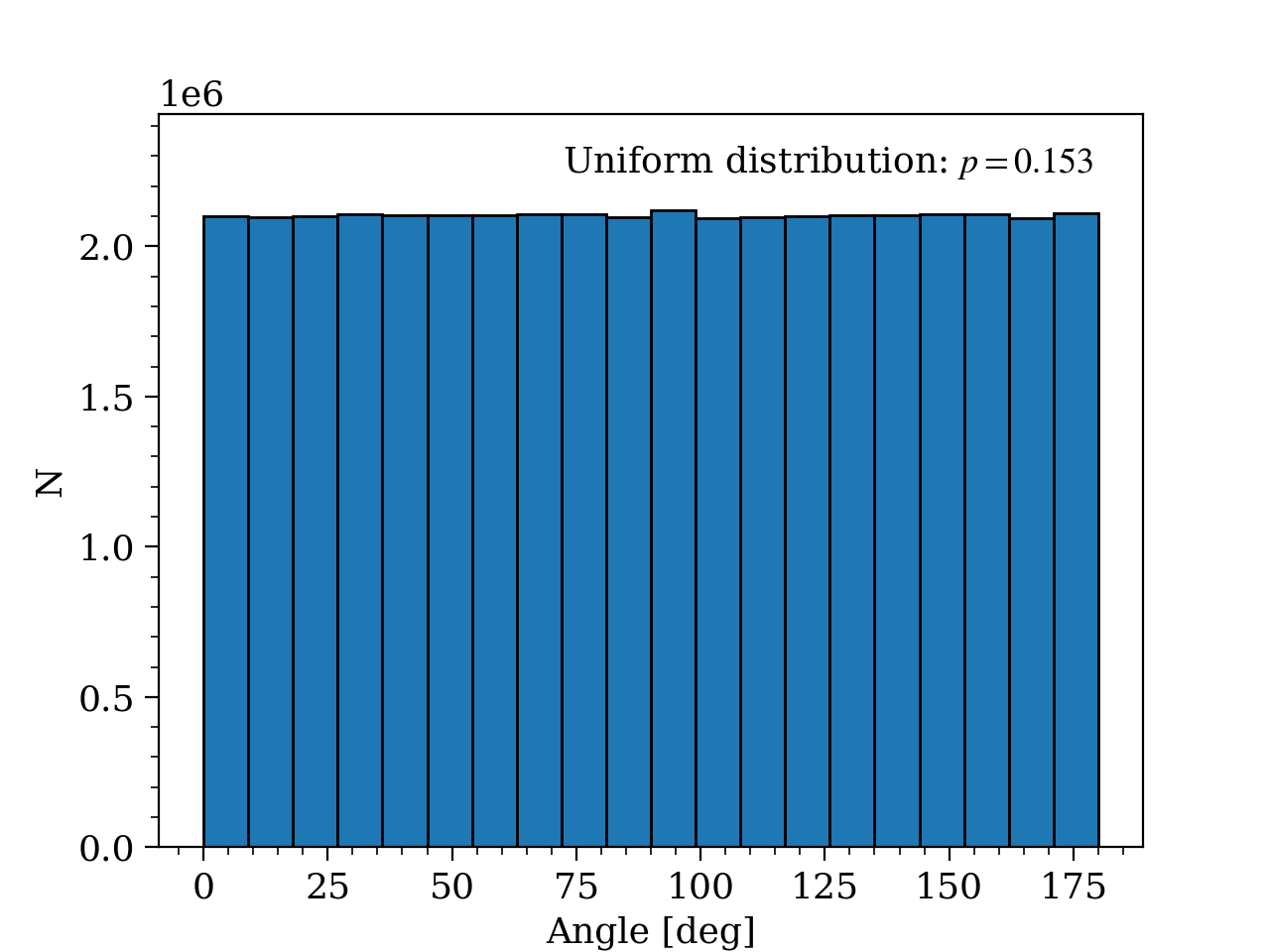}
  \caption{The distribution of position angles for cosmoDC2. The $p$-value statistic is obtained by using the two-sample K--S test to compare a uniformly down-sampled distribution of catalog position angles to a uniform distribution. The validation criterion requires that $p>0.05$.} 
  \label{fig:pos-angle}
\end{figure}

In \autoref{fig:pos-angle}, we show the galaxy position-angle distribution for 60~deg$^2$ of cosmoDC2.   The uniformity of the distribution can be verified by visual inspection. This test also provides a more quantitative measure of the uniformity by
using a statistic based on the two-sample Kolmogorov--Smirnov (K--S) test, as implemented in NumPy~\citep{numpy}, to compare the galaxy position-angle distribution to a uniform distribution.  The default requirement on the $p$-value of the test statistic was chosen to be $p\ge 0.05$. However, as we discuss below, the $p$-value is not a particularly reliable or robust statistic, so we prefer to use visual inspection of the distribution to verify its uniformity.

Historically, this test was developed for a small catalog for which the $p$-value statistic provided a reasonably robust measure of uniformity. However, for the large sample sizes that are obtained with catalogs such as cosmoDC2, the $p$-value statistic is sensitive to the sample size and we find that its value decreases with increasing sample size. When running this test on larger catalogs it is necessary to down-sample the distribution uniformly to remain in a sample-size regime where the $p$-values are reasonably consistent. We tested a variety of catalog sizes and down-sampling factors to determine that the size $N_\theta$ of the position-angle array used in the K--S test should satisfy $N_\theta \lesssim 5\times10^6$.

\subsection{Size Distribution}
\label{sec:test:size}

The modeling of galaxy sizes in cosmoDC2 is described at length in \citet{cosmodc2}. Here, we summarize the main ingredients. The size of the disk and bulge component of each galaxy is obtained by using the fits to the mean relation for size versus SDSS absolute $r$-band magnitude presented in \citet{zhang2019} for $B/T < 0.5$ and  $B/T > 0.5$, respectively. A scatter about the mean relation of 0.2 dex and 0.15 dex~\citep{somerville2018} for the disk and bulge components, respectively is introduced in the size assignments. Redshift dependence is included by imposing a smooth decrease in size with redshift such that mean sizes of both disks and bulges are reduced by a factor of two at $z= 1$. In cosmoDC2, an approximate prescription given by the luminosity-weighted sum of the disk- and bulge-component sizes is used to estimate the total size of each galaxy and serves as a proxy for other methods of determining size.
Since the galaxy sizes in cosmoDC2 are derived from empirical data, 
the tests described in this section and \autoref{sec:test:size-mag} therefore constitute a check on the cosmoDC2 modeling procedure and verify that the mean relationships used are consistent with those derived from other data sets and that the redshift dependence and the scatter are reasonable.

For the size-distribution test presented here, a comparison with the available validation data requires an estimate of the half-light radius, $R_{\rm eff}$, which is the size parameter characterizing the S\'{e}rsic (single-component) profile of the galaxy. Since this parameter is not available in cosmoDC2, we use instead the approximation for $R_{\rm eff}$ described above, but note that it does not give the same result as fitting a single Sersic profile to the sum of the disk and bulge profiles.

\begin{figure}
  \centering \includegraphics[width=8.5cm]{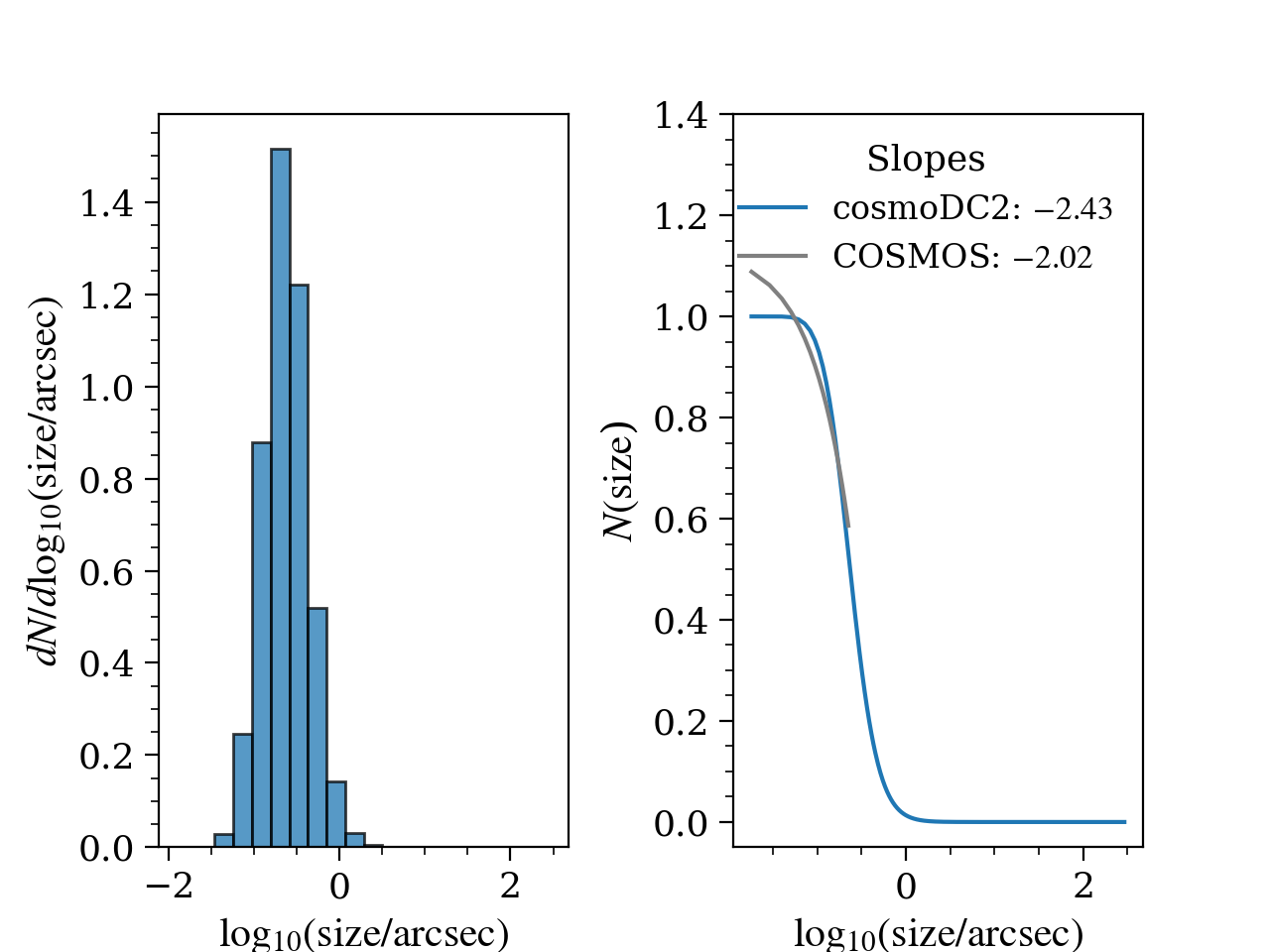}
  \caption{True (unlensed) size distribution of small cosmoDC2  galaxies (left panel)  and the comparison of cosmoDC2 galaxy sizes (blue) with a similar sample of COSMOS galaxy sizes (grey) (right panel). 
  The sizes of cosmoDC2 galaxies are estimated by computing the luminosity-weighted sums of the disk and bulge component sizes. Sizes are given in arcseconds.
  CosmoDC2 galaxies are selected such that their $i$-band magnitude $<25.2$ and small galaxies are defined such that their size is smaller than the median size of the sample. 
  In the right panel, the values of the slopes for linear fits to $dN/dx$ for cosmoDC2 and COSMOS data are given in the legend. The validation criterion for this test is that the slopes agree to within a factor of 2.}
  \label{fig:size}
\end{figure}

In~\autoref{fig:size}, we show the size distribution for cosmoDC2 galaxies (left panel) and a comparison of the sizes for cosmoDC2 and COSMOS galaxies (right panel). Sizes are measured in units of arcseconds. To mimic the COSMOS data selection, 
we apply an $i<25.2$ magnitude cut to the cosmoDC2 galaxies. We measure the cumulative distribution function (CDF) of these galaxy sizes in thin bins of $R_{\rm eff}$, with a variable cutoff to match the validation data set, and use the slope of the CDF as our comparison statistic.

In order to ensure we are probing the the small-size slope only, we discard all galaxies with a size greater than the median size. We then bin both data sets in optimally spaced size bins and measure the number of cosmoDC2 and COSMOS galaxies in each bin. We fit a line to both $dN/dx$ measurements and compare the slopes of the best-fit line.  This linear description is not very accurate, so we require simply that the two slopes agree within a factor of 2. This criterion is limited by the quality of the metric more than the quality of the data or the simulations. Since our sensitivity to this slope is low, and the fidelity of galaxy sizes is also checked by the size-luminosity relation (see \autoref{sec:test:size-mag}), this loose criterion is sufficient for DC2 analyses.

\subsection{Size--Luminosity Relations }
\label{sec:test:size-mag}

For these tests we compare the observed size-luminosity relations as described in \autoref{sec:science:data:wl}, which have been derived from the total light profiles (single components) for 3D-HST and CANDELS data and the disk and bulge light profiles (two components) for COSMOS data.
For the simulated catalog samples, we mimic the redshift selections in the observed data. We also convert the absolute magnitudes of galaxies to luminosities as described in detail below. As discussed in \autoref{sec:test:size}, the total size of a cosmoDC2 galaxy is obtained by computing the luminosity-weighted sum of the disk and bulge components. We remind the reader that this is an approximate prescription for estimating galaxy size and does not give the same result as fitting a single Sersic profile to the sum of the disk and bulge profiles.

\subsubsection{Total Light Profile}
\label{sec:test:size-mag-1}

\begin{figure}
  \centering \includegraphics[width=8.5cm]{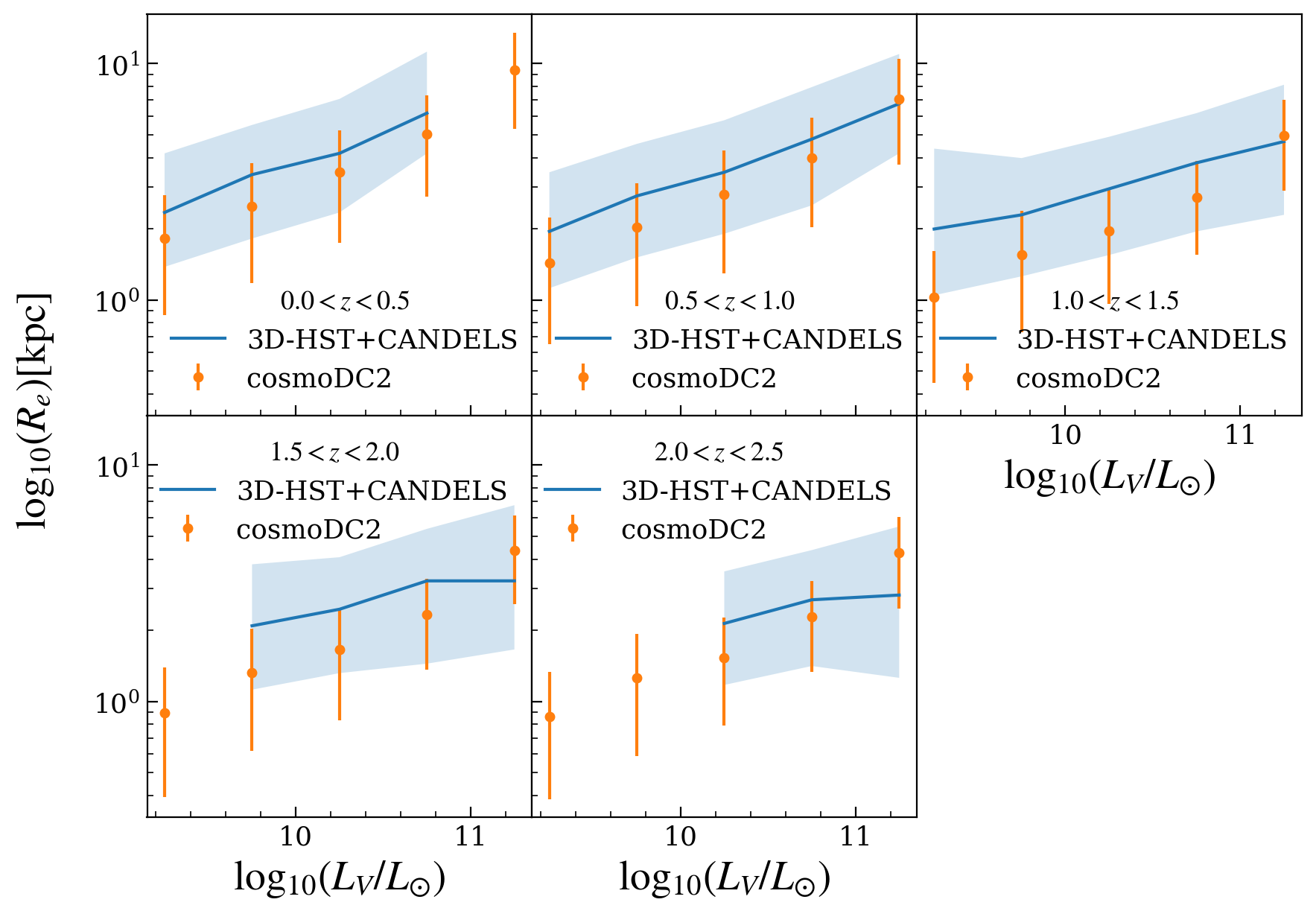}
  \caption{The size-luminosity relation for cosmoDC2 galaxies (orange points) compared to 3D-HST + CANDELS data (blue solid lines) for five redshift ranges as indicated in each panel. The shaded bands show the $1-\sigma$ population uncertainties in the data.
  }
  \label{fig:size_lum1}
\end{figure}

For the comparison of sizes based on the total light profile, we use $M_V$, the cosmoDC2 absolute $V$-band magnitudes to obtain luminosities using the relation $\log_{10}(L/L_\odot) = (4.80 - M_V)/2.5$~\citep[Table 3] {willmer2018}\footnote{4.80 is the absolute AB magnitude of the sun measured in $V$-band}.  In~\autoref{fig:size_lum1}, we show the size-luminosity relation for cosmoDC2 (orange points) and the mean relation for the 3D-HST + CANDELS data (blue lines), with the $1-\sigma$ population uncertainties in the validation data depicted by blue shaded bands. The catalog data agree with the mean validation data within the uncertainties.

\subsubsection{Disk and Bulge Light Profiles}
\label{sec:test:size-mag-2}

\begin{figure}
  \centering \includegraphics[width=8.5cm]{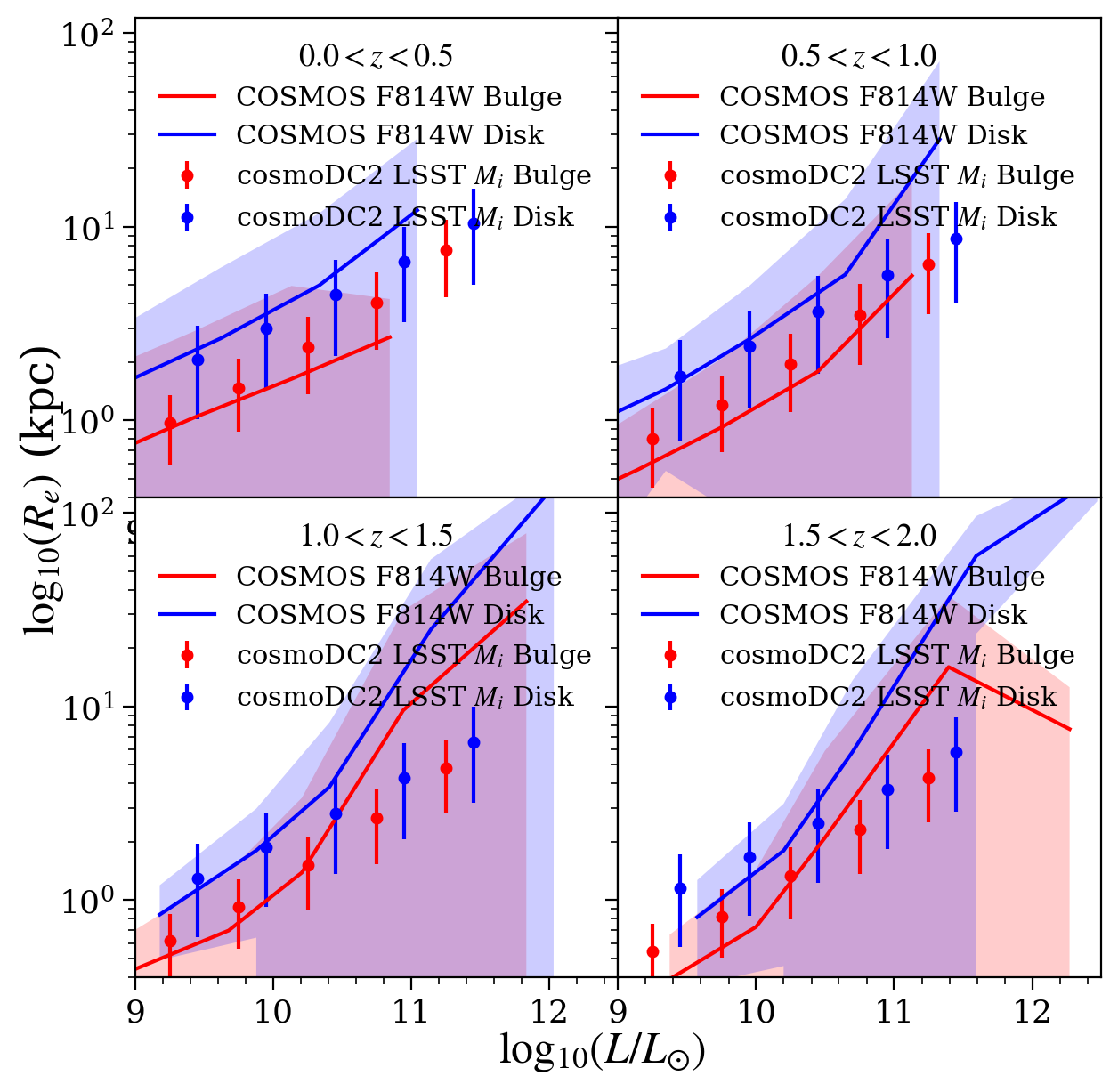}
  \caption{The size-luminosity relation for the bulge (blue) and disk (red) components of cosmoDC2 galaxies (points) compared to COSMOS data (solid lines) for four redshift bins as indicated in each panel. The shaded bands show the $1-\sigma$ population scatter in the COSMOS data. We convert the absolute  $i$-band magnitudes for the catalog data to units of solar luminosities as described in the text.
  }
  \label{fig:size_lum2}
\end{figure}

For size estimates based on the disk and bulge components of the light profile, we use
the cosmoDC2 LSST  absolute $i$-band magnitudes to obtain luminosities using the relation $\log_{10}(L/L_\odot) = (4.52 - M_i)/2.5$~\citep[Table 3] {willmer2018}. Here, we have assumed that LSST $i$-band and COSMOS F814W are approximately equivalent. In~\autoref{fig:size_lum2} we show the disk and bulge data in blue and red, respectively.  The cosmoDC2 and the COSMOS data are shown as points and lines, respectively. The $1-\sigma$ population scatter in the validation data is depicted by shaded bands. Overall, the catalog data are in good agreement with the mean relationship for the validation data, with some deviations for bright galaxies at redshifts $z > 1$. The deviations are within the measured uncertainites, which unfortunately, are quite large.

\subsection{Ellipticity Distribution }
\label{sec:test:ellipticity}

\begin{figure}
  \centering \includegraphics[width=8.5cm]{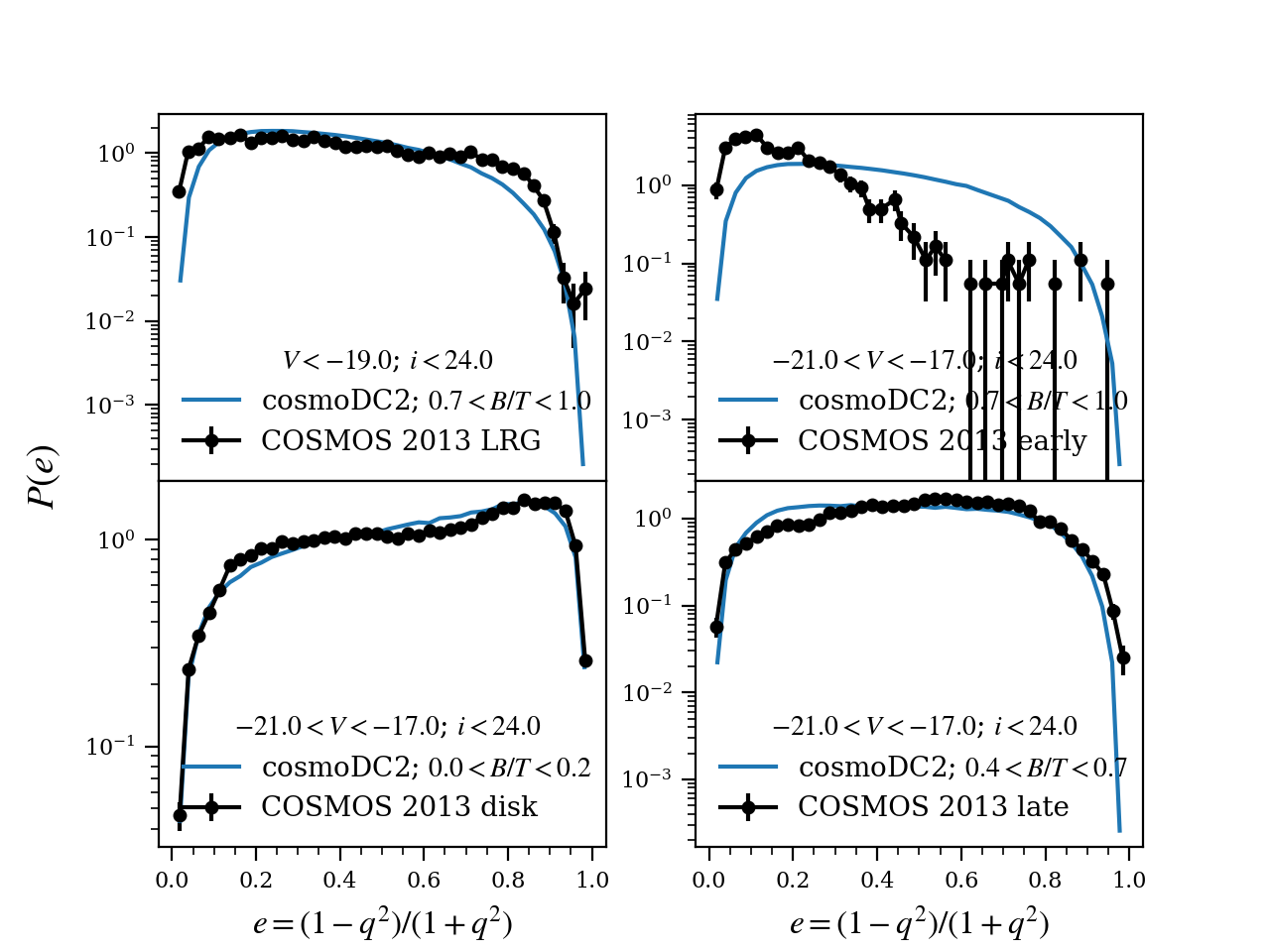}
  \caption{A comparison of the ellipticity distributions for magnitude-limited samples of cosmoDC2 galaxies and COSMOS galaxies. Since the COSMOS data are used to assign ellipticities to cosmoDC2 galaxies, this test is a consistency check that the assigned values match the intended distributions. 
  }
  \label{fig:ellipticity}
\end{figure}

We compare the ellipticity distributions of cosmoDC2 galaxies with data from the COSMOS survey~\citep{Joachimi2013}. In~\autoref{fig:ellipticity} we show distributions for the four morphology selections and $V$-band magnitude ranges given in \cite{Joachimi2013}.  The morphologies of the cosmoDC2 galaxies are matched to those of the COSMOS data by selecting subsamples with $i$-band bulge-to-total ratios ($B/T$) given by $0.7 < B/T < 1.0$,  $0.0 < B/T < 0.2$ and $0.4 < B/T < 0.7$ for LRG, early, disk and late types, respectively. The magnitudes of the ellipticities for cosmoDC2 galaxies are defined as the $r$-band luminosity-weighted average of the disk and bulge ellipticities.  This prescription for estimating the ellipticities provides an approximation for measuring the ellipticity from the shape of the total light profile of the galaxy. The disk and bulge ellipticities in cosmoDC2 are drawn from distributions that have been tuned to match the COSMOS data as described in \cite{cosmodc2}. Hence this test amounts to a consistency check that the assignment is consistent with the expected distributions. From \autoref{fig:ellipticity}, we see that all the distributions agree well except for early type galaxies. This is probably due to the fact that our selection on $B/T$ is too crude to match the more complicated morphological selections for early type galaxies in the COSMOS data.

\subsection{Galaxy-Galaxy Angular Correlation Function}
\label{sec:test:angular-cf}

In this test we use the \texttt{TreeCorr}\footnote{\url{ https://rmjarvis.github.io/TreeCorr/_build/html/index.html}} package~\citep{treecorr} to compute the two-point angular correlation function, $w(\theta)$, the over-abundance of galaxy pairs at some angular separation $\theta$ relative to a random distribution. We apply selection cuts to the catalog sample to mimic measurements from \cite{tpcf_wang} of $w(\theta)$ in bins of SDSS $r$-band magnitudes in the range $17 < r < 21$. The computation of $w(\theta)$ uses the estimator of \citet{1993ApJ...412...64L}. This test has been described previously in \citet{cosmodc2}, so we do not reproduce it here. The results for the test showed that the amplitudes and slopes of the clustering signals for the catalog data were in reasonable agreement with the validation data,  except at very small angular scales ($\theta \lesssim 0.4^\circ$), where the catalog clustering signal was slightly steeper than the SDSS data.

\subsection{Galaxy Bias }
\label{sec:test:galaxy-bias}

For this test, we divide the catalog galaxy sample into tomographic redshift bins in the range $0.2 < z < 1.2$ and select galaxies whose observed $i$-band magnitudes satisfy $i < 25.3$, which is  the expected $i$-band limiting magnitude for the 10-year LSST gold sample~\citep{SRD}. For each tomographic bin, the redshift distribution is determined and the angular correlation function is computed using \texttt{TreeCorr} as in \autoref{sec:test:angular-cf}. Next, the values of the cosmological parameters used in the simulated catalog are input to \texttt{CCL}~\citep{CCL} to compute the theoretical prediction for the angular correlation function. A linear bias model is independently fit in each tomographic bin 
by minimizing $\chi^{2}_{b}$, which is defined as:
\begin{equation}
\label{eq:bias_fit}
\chi^{2}_{b}=\sum_{i=1,N_\text{bins}}\frac{\left(w_\text{data}(\theta_{i})-b \,  w_\text{th}(\theta_{i})\right)^2}{\left(\Delta w(\theta_{i})\right)^2}.
\end{equation}
Here $\Delta w(\theta)$ are the error bars reported by \texttt{TreeCorr} (using 80 jackknife patches) and $N_\text{bins}$ is the number of angular bins considered in each tomographic bin. We note that in \autoref{eq:bias_fit}, we are neglecting significant correlations between the $\theta$ bins.
The range of angles included in the fits are $0.01^\circ  < \theta < 0.3^\circ-0.6^\circ$, depending on the tomographic bin (with the smallest fit range corresponding to the highest redshift range).

\begin{figure}
  \centering \includegraphics[width=8.5cm]{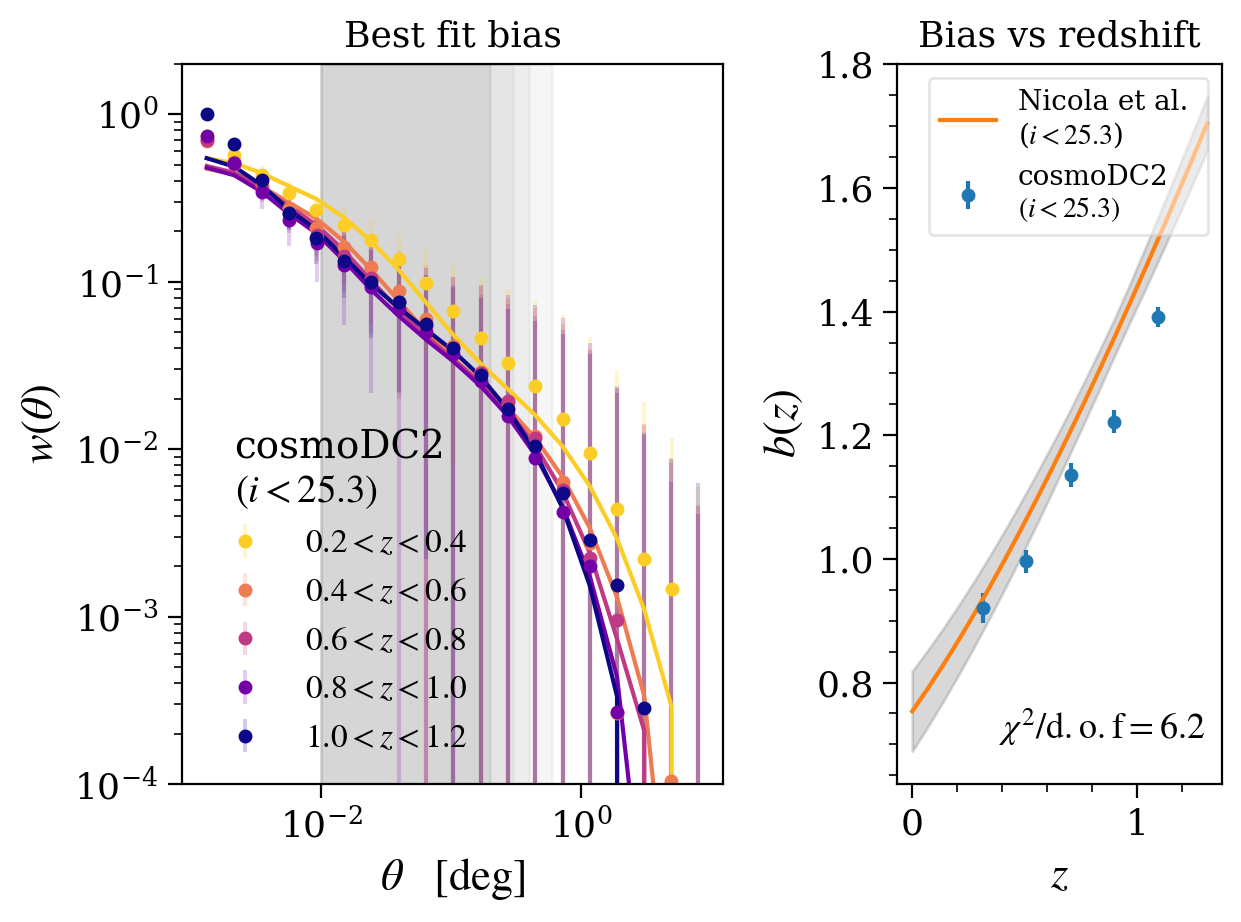}
  \caption{The left panel shows the two-point correlation function evaluated for the series of redshift bins indicated in the legend. The points show the cosmoDC2 data and the solid lines show the theoretical predictions, as described in the text. Error bars are computed with \texttt{TreeCorr} using 80 jackknife patches. We note that, due to the large correlations between $\theta$ bins that are caused by
  cosmic variance, the error bars are far larger than the point-to-point scatter.
  The grey shaded bands show the range of $\theta$ values over which the bias is fit. These ranges decrease with increasing redshift. The right panel shows the estimated linear bias as a function of the mean redshift of each tomographic bin compared to the empirical fit in \autoref{eq:galaxy_bias}~\citep{nicola2020} with $m_{\rm lim}=25.3$, which is the expected limiting magnitude of the LSST 10-year gold sample~\citep{SRD}.  The $\chi^2$ agreement between the estimated bias and the empirical fit is $\chi^2/d.o.f = 6.2$. Although the estimated bias is lower in normalization and slope than the empirical fit, the two data sets are in qualitative agreement, which is sufficient to satisfy the loose validation criterion for this test.
  }
  \label{fig:galaxy_bias}
\end{figure}

We show the results for the correlation function and the fitted bias parameters in the left and right panels of \autoref{fig:galaxy_bias}, respectively. In the right panel, we compare our bias parameters to the empirical fit given in \autoref{eq:galaxy_bias}~\citep{nicola2020} with $m_{\rm lim}=25.3$.  The shaded grey band shows the uncertainty in the fit.  We see that both the normalization and the slope of our estimated linear bias are low compared to the empirical fit for redshifts in the range $z \gtrsim 0.5$.  We evaluate the $\chi^2$ agreement between the estimated bias and the empirical fit and find that $\chi^2/d.o.f=6.2$. 
We also tried changing the limiting magnitude of our galaxy sample to $m_{\rm lim}=24.5$ and obtained qualitatively similar results (lower slope and normalization for the estimated bias) with $\chi^2/d.o.f=12$.

We emphasize that the estimates of the bias shown here have some caveats. We have assumed a linear galaxy bias for all tomographic bins, which may not be a correct assumption for this catalog over all scales used. Furthermore, although we attempted to reproduce the data selection in 
 \citet{nicola2020}, there may be some residual selection effects at these faint magnitudes which have not been taken into account. Therefore, we impose a loose validation criterion that the estimated catalog bias be in qualitative agreement with the empirical fit of \citet{nicola2020}, which we consider to be met in this case. In the future, with a more careful analysis of the uncertainties in our procedure and in the validation data, we expect to impose a more stringent validation criterion for this test.

\subsection{Galaxy--Galaxy Projected Correlation Function}
\label{sec:test:galaxy-galaxy}

\subsubsection{Luminosity Dependence}
\label{sec:test:galaxy-galaxy-mag}

\begin{figure}
  \centering \includegraphics[width=8cm]{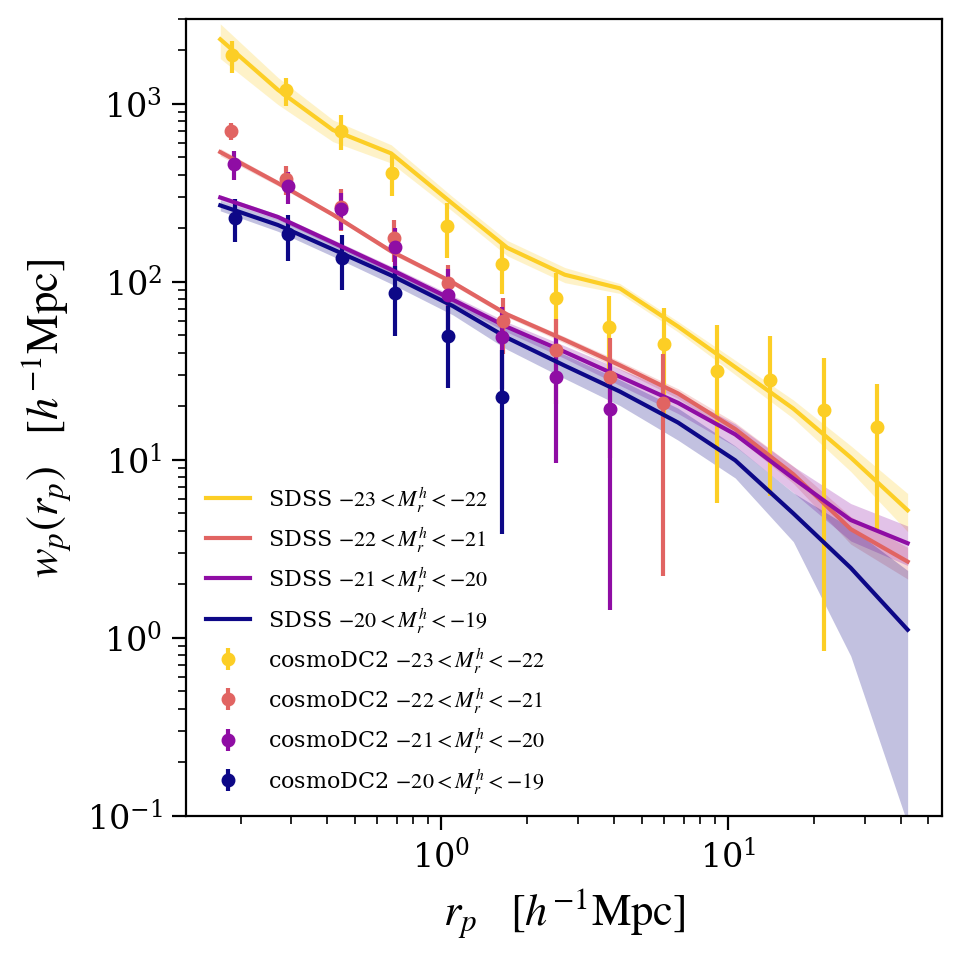}
  \caption{Two-point auto-correlation as a function of projected distance between galaxy pairs for cosmoDC2 galaxies (points) compared to SDSS data (lines) for bins of SDSS $r$-band absolute ($h=1$) magnitudes as indicated in the legend.
  The magnitude bin limits are converted to the cosmoDC2 cosmology using \autoref{eq:mag_corr}. Additional redshift cuts are applied to the catalog data to mimic the redshift selections used in the validation data. The uncertainties for the catalog data are computed using a jackknife estimation procedure (see text). Catalog data with very large uncertainties are not shown in the figure.
  Uncertainties in the observational data for each magnitude bin are indicated by the shaded bands.}
  \label{fig:test:wprp_sdss_mag}
\end{figure}

In \autoref{fig:test:wprp_sdss_mag} we show the two-point  auto-correlation function, computed using \texttt{TreeCorr}, as a function of the projected distance $r_p$ between galaxy pairs for cosmoDC2 (points) and SDSS data (lines) for bins of absolute SDSS $r$-band magnitudes in the range $-23 < M_r^h < -18$. For the cosmoDC2 data, we apply the redshift cuts listed in Table 1 of \citet{zehavi2011} to each magnitude bin to mimic the redshift selections that have been applied to the validation data. The errors shown for the catalog data are computed using $20$ patches for the jackknife covariance-estimation option supplied by \texttt{TreeCorr}. Due to the limited statistics available for the cosmoDC2 data, we do not show results for the magnitude bin $-18 < M_r^{h} < -17$, or for some of the larger $r_p$ values. We also note that it is critical to convert the magnitude values to the cosmoDC2 cosmology using \autoref{eq:mag_corr} (results in a shift of -0.74 in the bin limits). The uncertainties in the SDSS data are indicated by the shaded bands.

For this test, the agreement between the catalog and validation data is reasonably good for the brightest galaxies and gets slightly worse as the galaxy sample becomes fainter. The sample sizes for the cosmoDC2 galaxies passing the requirements for this test are quite small, numbering only 3000 to 6000 galaxies in each magnitude bin, so the statistics are limited. Indeed, for the magnitude bins in the range $-17 < M-r < -19$, the statistical uncertainties in the catalog data are too large to provide a meaningful comparison with the validation data and are not shown. We note that, owing to the tight selections on both magnitudes and redshifts for each bin, this test is much more stringent than the angular correlation function test discussed in \autoref{sec:test:angular-cf}. Nevertheless, the clustering signals from cosmoDC2 have qualitatively the right amplitudes and slopes and scale as expected with galaxy brightness. 

We remark that it is very difficult to obtain good agreement for clustering amplitudes between any mock catalog and the observational data without extensive and computationally expensive tuning of the catalog. For example, because of these modeling and computational challenges, SAMs have only recently 
begun to use correlation functions on an equal footing with SMFs~\citep{vanDaalen2016}. Bearing in mind that the clustering amplitudes in cosmoDC2 were not tuned beyond the level of agreement inherited from the underlying empirical model~\citep{cosmodc2, behroozi_etal18}, we should not expect better than qualitative agreement with the validation data. In fact, the agreement is quite good, given the available statistics.

\subsubsection{Color dependence }
\label{sec:test:galaxy-galaxy-color}

\begin{figure*}
  \centering \includegraphics[width=7.0in]{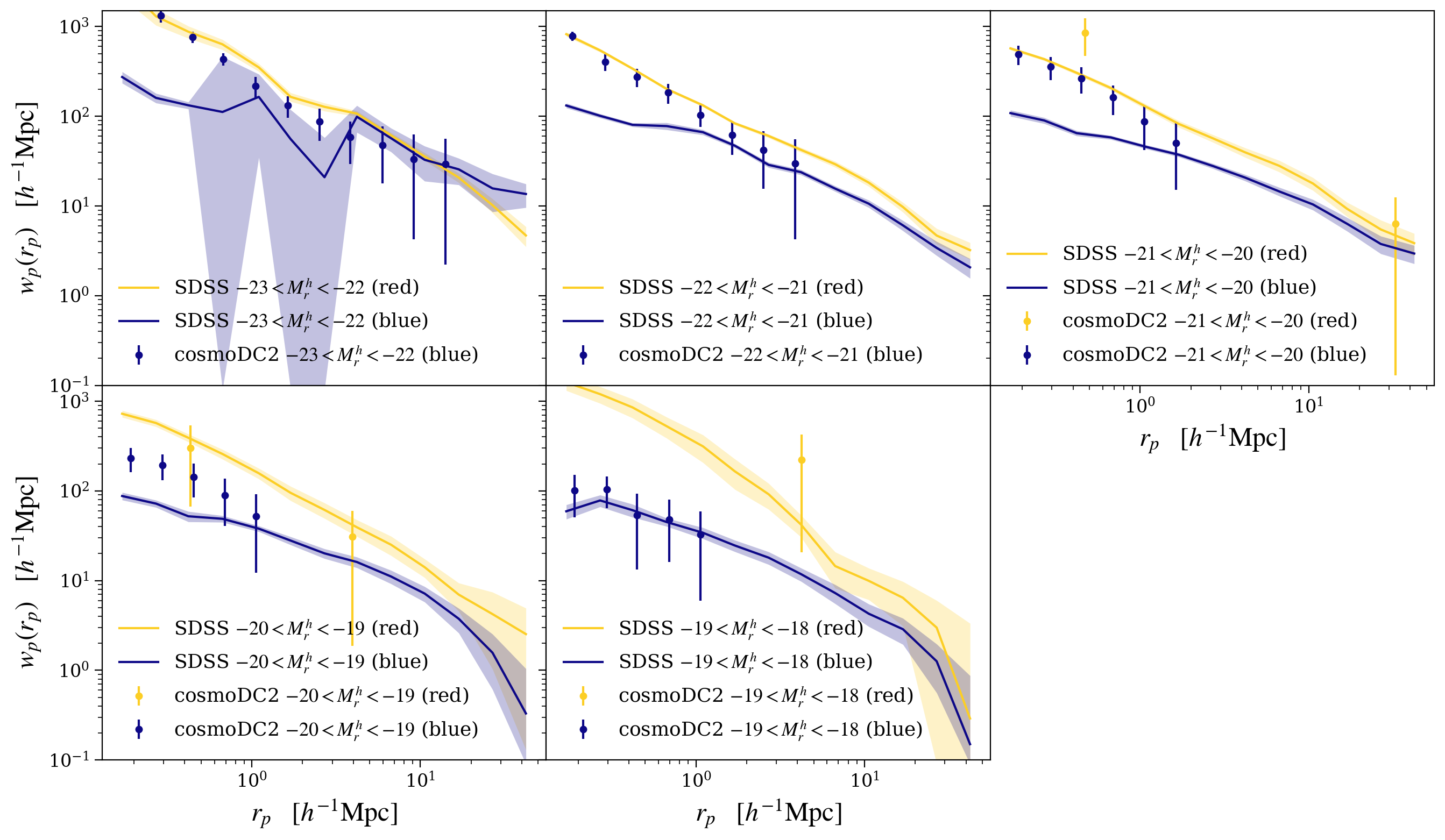}
  \caption{As for \autoref{fig:test:wprp_sdss_mag}, but with an additional color selection to divide the sample into red and blue galaxy sub-samples. For each bin, the magnitude-dependent color cut is given by $(g-r)_{\rm{cut}}^h = 0.21 - 0.03M_r^h$, with blue (red) galaxies falling below (above) the specified limit. We note that there are no cosmoDC2 galaxies that are classified as red for $M_r^h < -22$. The uncertainties in the catalog data are computed using 10 jackknife resamplings (see text). Catalog data points with very large uncertainties are not shown.
  }
  \label{fig:test:wprp_sdss_color}
\end{figure*}

In \autoref{fig:test:wprp_sdss_color} we show the two-point  auto-correlation function, computed as described in \autoref{sec:test:galaxy-galaxy-mag}, but with an additional color selection applied to separate the sample into red and blue galaxy subsamples. For each bin, as described in \autoref{sec:science:data:2pt} and \citet{zehavi2011}, the magnitude-dependent color cut is given by $(M_g^h - M_r^h)_{\rm{cut}} = 0.21 - 0.03M_r^h$, with blue (red) galaxies falling below (above) the specified limit. The errors on the catalog data are estimated with \texttt{TreeCorr} using 10 jackknife patches. Catalog data points with very large uncertainties are omitted from the plot.

Owing to the additional color cut, this validation test is more demanding than the luminosity-dependent test described in the previous section. Overall the agreement between the catalog and the validation data is poor.
From the figure, we see that cosmoDC2 has very few galaxies that pass the red-galaxy color selection, suggesting that the red fraction of bright galaxies at low redshift is underestimated. There are no cosmoDC2 galaxies that are classified as red for $M_r^h < -22$. Furthermore, the clustering signal for the brighter blue catalog samples follows more closely the validation data for the corresponding red samples. Only for the faintest magnitude bin shown, $-19 < M_r^{h} < -18$, do we find better agreement between the catalog and validation data for the blue sample. The available statistics for the red samples are extremely low but are in qualitative agreement with the observational data.  As for the test discussed in the previous section, it is very challenging to produce a catalog that agrees well with the validation data for this test. 

\subsubsection{Clustering at Higher Redshift} 
\label{sec:test:galaxy-galaxy-z}
\begin{figure}
  \centering \includegraphics[width=8.5cm]{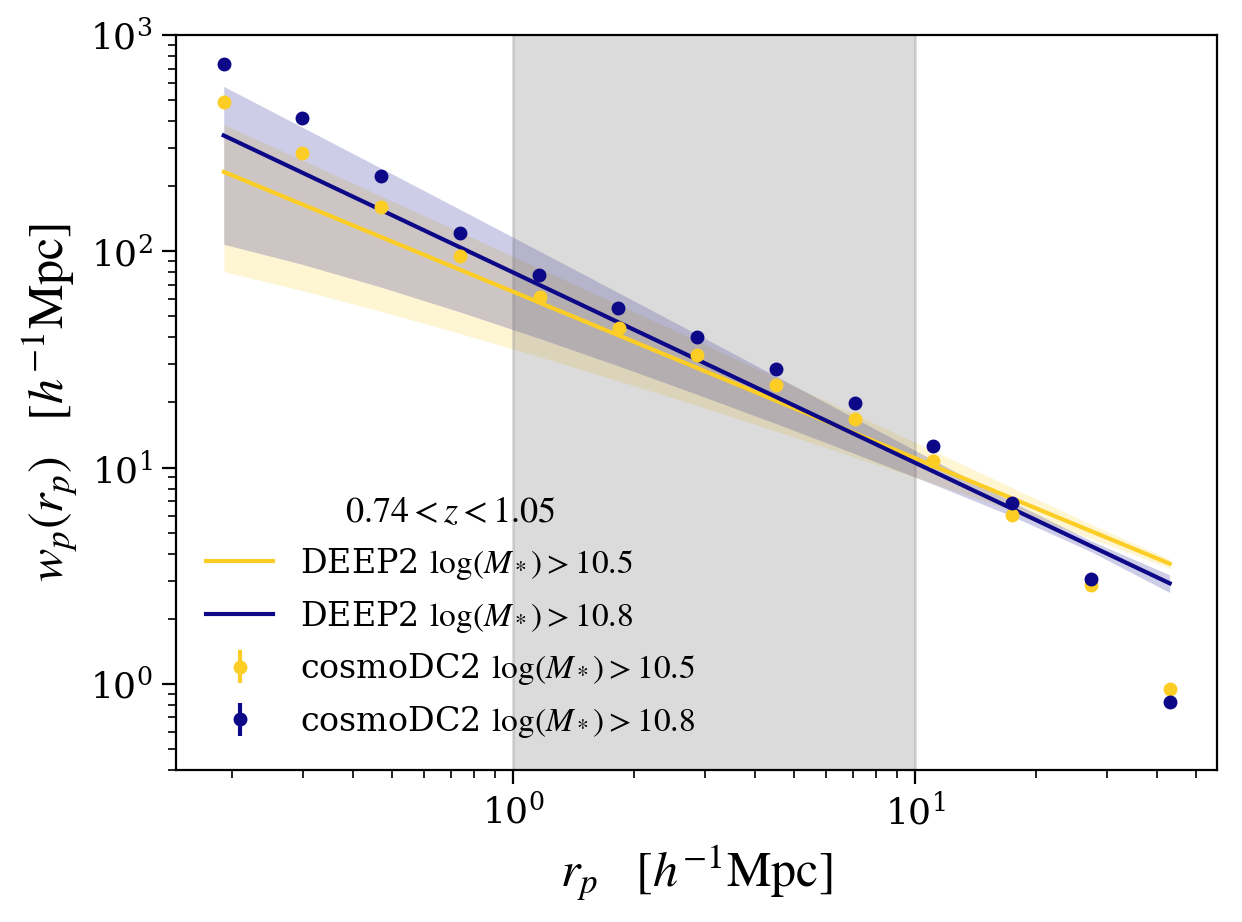}
  \caption{Projected auto-correlation functions for cosmoDC2 galaxies compared with power-law fits to DEEP2 data for two stellar-mass limited samples with thresholds $M^*>10.5$ (yellow) and $M^*>10.8$ (blue). Errors in the fits are shown by the colored shaded bands. The redshifts of the selected cosmoDC2 and DEEP2 galaxies lie in the range $0.74 < z < 1.05$. The vertical grey shaded band shows the validation region.}
  \label{fig:tpcf_high-z}
\end{figure}

For this test, we wish to validate the two-point clustering signal at higher redshifts.
In~\autoref{fig:tpcf_high-z}, we compare the projected auto-correlations for two stellar-mass limited samples of cosmoDC2 galaxies with the best fit power law values obtained from the measurements of the projected correlation functions for DEEP2 data from Table 1 in \citet{mostek13}.
The stellar-mass thresholds are $\log_{10}(M^*/M_\odot)>10.5$ and $\log_{10}(M^*/M_\odot)>10.8$ and the redshifts of the selected galaxies lie in the range $0.74 < z < 1.05$. The colored shaded bands show the errors on the fits for each mass-limited sample. The vertical grey shaded band shows the validation region, $1-10$ $h^{-1}$Mpc. The metric computed is the average $\chi^2$ per data point measured from the simulations using the errors from the DEEP2 power-law fits over the validation region. The value for this test is 0.72. The test passes if the average $\chi^2$ is $<2$. 

\subsection{Shear--Shear Correlation }
\label{sec:test:shear-shear}

\begin{figure*}
  \centering \includegraphics[width=7.0in]{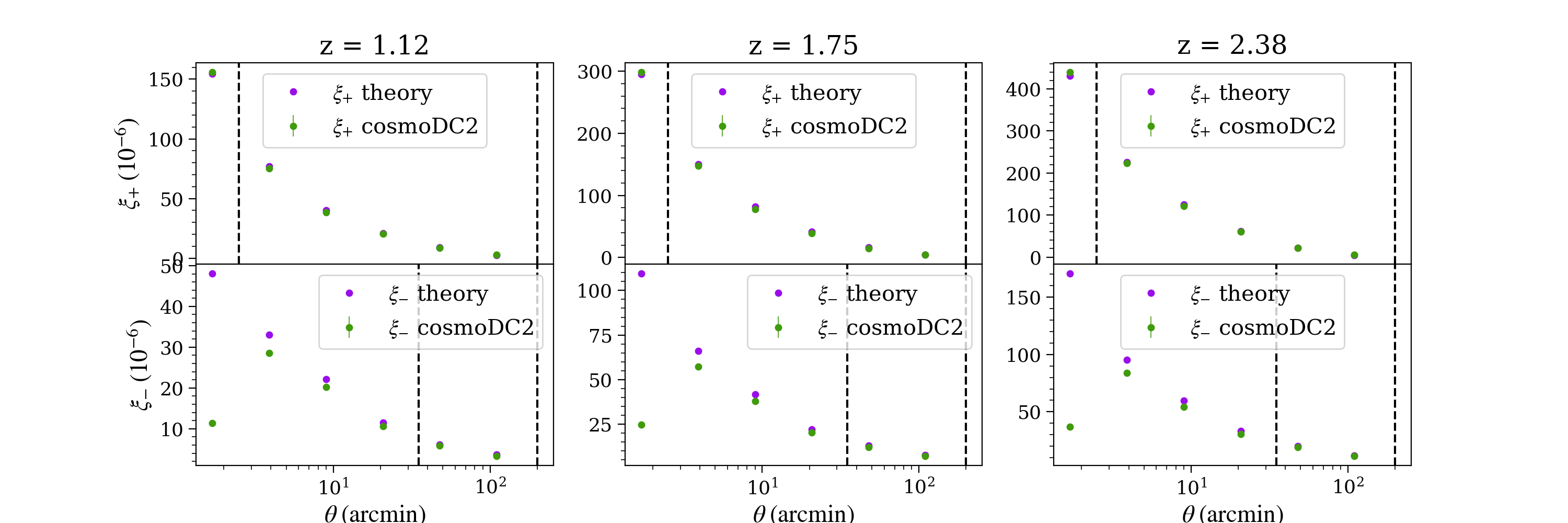}
  \caption{A comparison of the $\xi_+$ (upper panels) and $\xi_-$ (lower panels) statistics for the catalog data (green) with the theoretical predictions (purple) as a function of angular separation on the sky for 3 redshift bins (increasing in redshift from left to right) as given in title of each panel. The vertical dashed lines indicate the expected range of angular separations over which the comparison is valid~\citep{DES_Y1_WL}. The fall-off evident in the values of $\xi_-$ at small scales is due to the effects of pixelization used in the computation of the shears.}
  \label{fig:shear-shear}
\end{figure*}

The shear-shear test in our suite compares the $\xi_+$ and $\xi_-$ statistics using \texttt{TreeCorr} with the theoretical predictions based on the cosmology of the input simulation. An example of this test is shown in \autoref{fig:shear-shear} for 3 tomographic redshift bins.

We extract catalog values for a number of evenly-spaced redshift bins that are obtained from user-defined inputs. In order to mimic observed galaxy samples (by removing very faint galaxies), we apply an absolute magnitude cut requiring that the SDSS $r$-band magnitudes satisfy $M_r^{\rm SDSS} < -19$. Once the galaxy samples are selected, \texttt{TreeCorr} is run to compute their shear auto-correlations. The theoretical predictions for $\xi_\pm$, given the computed redshift distribution of the galaxy samples in each tomographic bin, are generated using \texttt{CAMB}~\citep{CAMB} to compute the lensing power power spectra and \texttt{CCL}~\citep{CCL} to perform the real space projection.

For any simulated catalog there are anticipated limitations in this comparison depending on the input catalog. These can include a smoothing at small scales due to the pixelization of the lensing map, a drop off at large scales due to limited sky area, and cosmic variance differences on small patches and/or low redshifts. In \autoref{fig:shear-shear} one can see the shear auto-correlations and assess the level of agreement with the theory at a range of redshifts and  angular scales. The vertical dashed lines in the figure indicate the regions of angular separation where we expect good agreement with the theoretical predictions~\citep{DES_Y1_WL}. We note that a sign convention mismatch will nullify $\xi_-$
and can be caught on visual inspection. 

\subsection{Galaxy--Shear Correlation  }
\label{sec:test:galaxy-shear}

Our test suite includes galaxy-shear correlation tests for the galaxy-galaxy lensing signal. These tests compare the catalog and observational data for the excess surface mass density, $\Delta\Sigma_+(r_p)$, measured at a given projected distance $r_p$ from a sample of foreground lens galaxies.  $\Delta\Sigma_+(r_p)$ is related to $\langle \gamma_{+}^{l,s}(r_p)\rangle$, the average tangential shear of foreground lens and background source galaxy pairs ($l,s$), separated by $r_p$, as follows: 
\begin{equation}
\Delta\Sigma_+(r_p) = \Sigma_{\rm crit}^{l,s}
\langle\gamma_{+}^{l,s}(r_p)\rangle, 
\label{eq:delta_sigma}
\end{equation}
where
\begin{equation}
 \gamma_{+}^{l,s} = -[\gamma_{1}\cos(2 \phi_{s})+\gamma_{2}\sin(2 \phi_{s})].
\label{eq:gammat}
\end{equation}
Here, $\phi_s$ is the angle of the vector connecting the projected lens and source galaxy positions, $(\gamma_1, \gamma_2)$ are the measured shear components for the source galaxy\footnote{Note that for this test and the cluster weak-lensing mass test in \autoref{sec:test:cl_shear} we are not using reduced shear, which would be the measured quantity.} and
$\Sigma_{\rm crit}^{l,s}$ is the critical surface density of the lens-source plane defined as
\begin{equation}
    \Sigma_{{\rm crit}}^{l,s} = \frac{c^2}{4 \pi G} \frac{D_{s}}{D_{l} D_{ls}},
    \label{eq:critical_surface_dens}
\end{equation}
where $D_l$, $D_s$ and $D_{ls}$ are the angular diameter distance to the lens, to the source, and between the lens and the source, respectively. 

The galaxy-galaxy lensing tests described in this section are designed to be qualitative checks rather than quantitative pass-fail tests. That is, we use the tests to determine for which regimes of the available parameter space the DC2 simulations can be trusted. We describe the tests in detail in the following two subsections, but summarize the main conclusions for both tests here. For cosmoDC2, we conclude that the weak-lensing quantities are not reliable below scales of $r_p \lesssim 0.3 h^{-1}{\rm Mpc}$, but otherwise, the galaxy-halo connection is consistent with observations over the range of magnitude, stellar-mass and color selections that have been probed by the tests. 
This result is a highly non-trivial finding since the catalog production procedure did not tune to achieve this result.

\subsubsection{The SDSS LOWZ Sample}
\label{sec:test:galaxy-shear-lowz}

 In~\citet{cosmodc2}, we presented results for the comparison between cosmoDC2 data and the SDSS LOWZ measurements~\citep{Singh2015} of $\Delta\Sigma_+(r_p)$. The lens galaxy sample was selected using the data cuts described in \autoref{sec:science:data:2pt} and the source galaxies were selected by requiring their redshifts to lie in the ranges $0.4 < z < 1.0$.  We found good agreement between the catalog and the validation data for $r_p \gtrsim 0.4~h^{-1}{\rm Mpc}$ and refer the reader to \citet{cosmodc2} for further discussion of this test.
 
\subsubsection{The SDSS Locally Bright Galaxy Sample}
\label{sec:test:galaxy-shear-lbg}

\begin{figure}
  \centering \includegraphics[width=0.99\columnwidth]{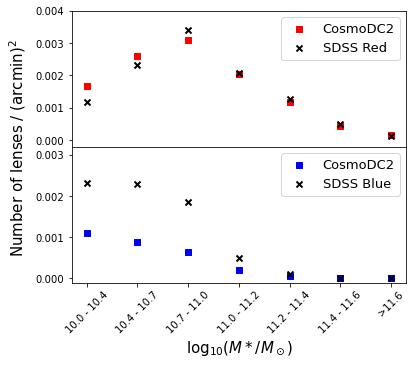}
  \caption{Comparison of the number density of lenses for cosmoDC2 (squares) with the SDSS data (crosses) from \citet{Mandelbaum2016}. Number densities are compared for bins of $\log_{10}(M^*/M_\odot)$ as indicated on the x-axis. Galaxies are split into red (upper panel) and blue (lower panel) samples according to whether their $M_g-M_r$ color is above or below $0.7$}
  \label{fig:gg_number_density}
\end{figure}

Here, we present comparisons to the measurements in \citet{Mandelbaum2016} for the SDSS Locally Brightest Galaxy sample.
We first define the lens selection to follow as closely as possible the selections used for the observed data samples. In \citet{Mandelbaum2016}, the lenses are binned into a number of stellar mass, luminosity and color bins. The selections include galaxies with redshifts in the range
$0.03 <z < 0.2$, SDSS $r$-band magnitudes with $r < 17.7$ and stellar masses  $ \log_{10}(M^*/M_\odot) > 10$. Red and blue lenses are selected by requiring that their $M_g-M_r$ color is above or below 0.7, respectively. In \autoref{fig:gg_number_density}, we compare the number density of the selected lens galaxies from cosmoDC2 with the observed data for bins of $\log_{10}(M^*/M_\odot)$ in the ranges indicated in the figure. The trend in the number counts of red galaxies as a function of stellar mass traces the observed data quite well, especially for the higher stellar-mass bins. The blue galaxy samples are less well matched, and for the lower stellar-mass bins, there are about a factor of 2 fewer lens galaxies in the simulations compared to data.

\begin{figure}
  \centering \includegraphics[width=0.99\columnwidth]{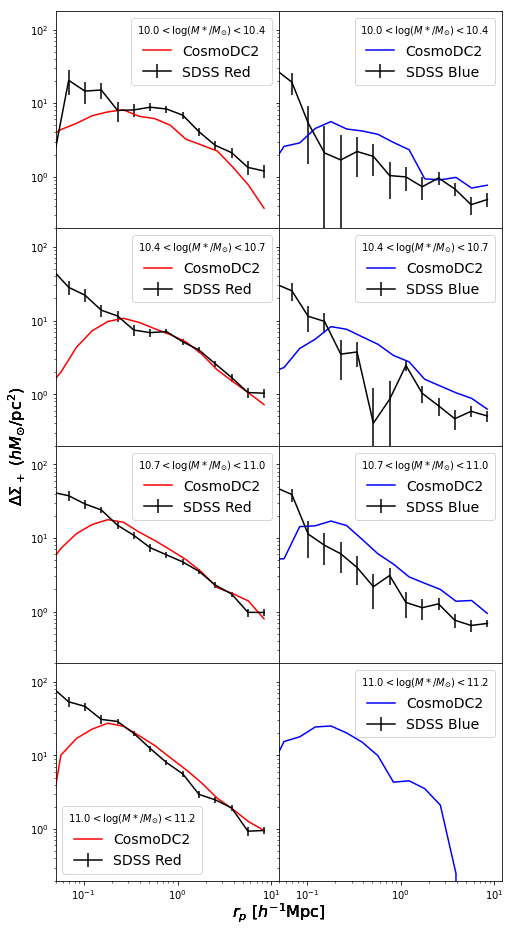}
  \caption{Comparison of the galaxy-galaxy lensing signal $\Delta \Sigma_+$ (see text) for cosmoDC2 (colored lines) with
  the SDSS data (black points) from \citet{Mandelbaum2016} for selected bins of stellar mass. Red and blue galaxy samples are shown in the left and right panels, respectively. Observational data are not available for the blue sample in the stellar-mass bins with  $\log_{10}(M^*/M_\odot) > 11$. The fall-off evident in the cosmoDC2 values of $\Delta\Sigma_+$ at small scales is due to the effects of pixelization used in the computation of the shears.}
  \label{fig:delta_sigma_red_blue}
\end{figure}

 We calculate the average tangential shear signal using the true shear from source galaxies which are selected such that they have $r < 22$ and lie in a redshift range behind the lens sample given by $0.2 < z < 1.0$. We down-sample from these selections as needed to speed up the calculation.
 In \autoref{fig:delta_sigma_red_blue} we show the comparison of cosmoDC2 with the observed data for the four lowest stellar mass bins in \autoref{fig:gg_number_density}. For the three highest stellar mass bins, there is no observational data available for the blue sample. Meanwhile, the comparisons for the red sample for these higher mass bins are very similar to those shown in \autoref{fig:delta_sigma_red_blue}. The agreement between the catalog signal and the observed data for the red sample is quite good for scales $r_p \gtrsim .2-.3 h^{-1}{\rm Mpc}$. The suppression of the catalog signal below these scales is due to the pixelization used in the ray-tracing calculation of the shear maps~\citep{cosmodc2}. The catalog signal overshoots the observed data by a factor of a few in amplitude for the blue galaxy sample, but has approximately the correct slope with $r_p$.

\subsection{Red-Sequence Color }
\label{sec:test:red-sequence}

\begin{figure}
  \centering \includegraphics[width=8.5cm]{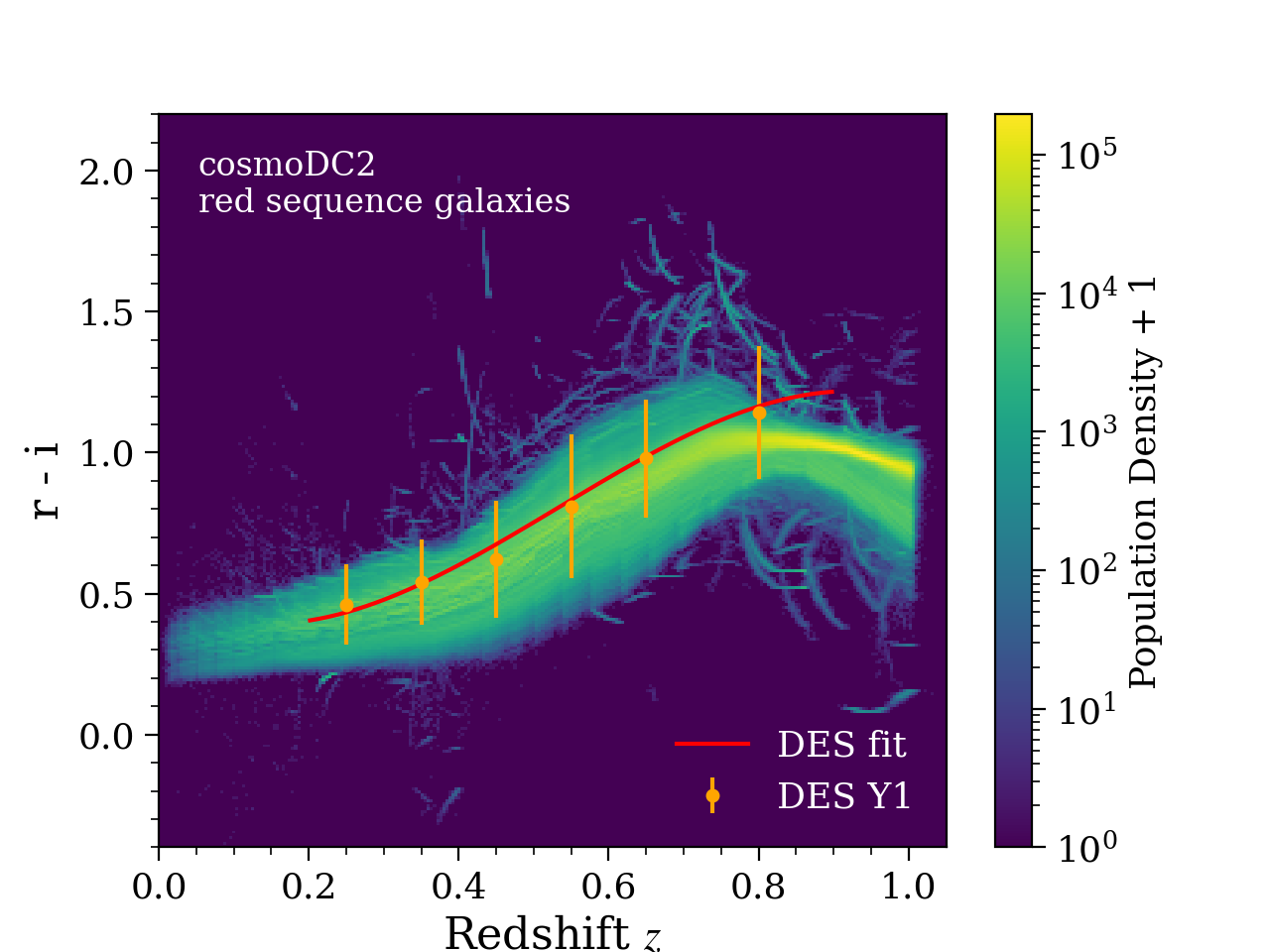}
  \centering \includegraphics[width=8.5cm]{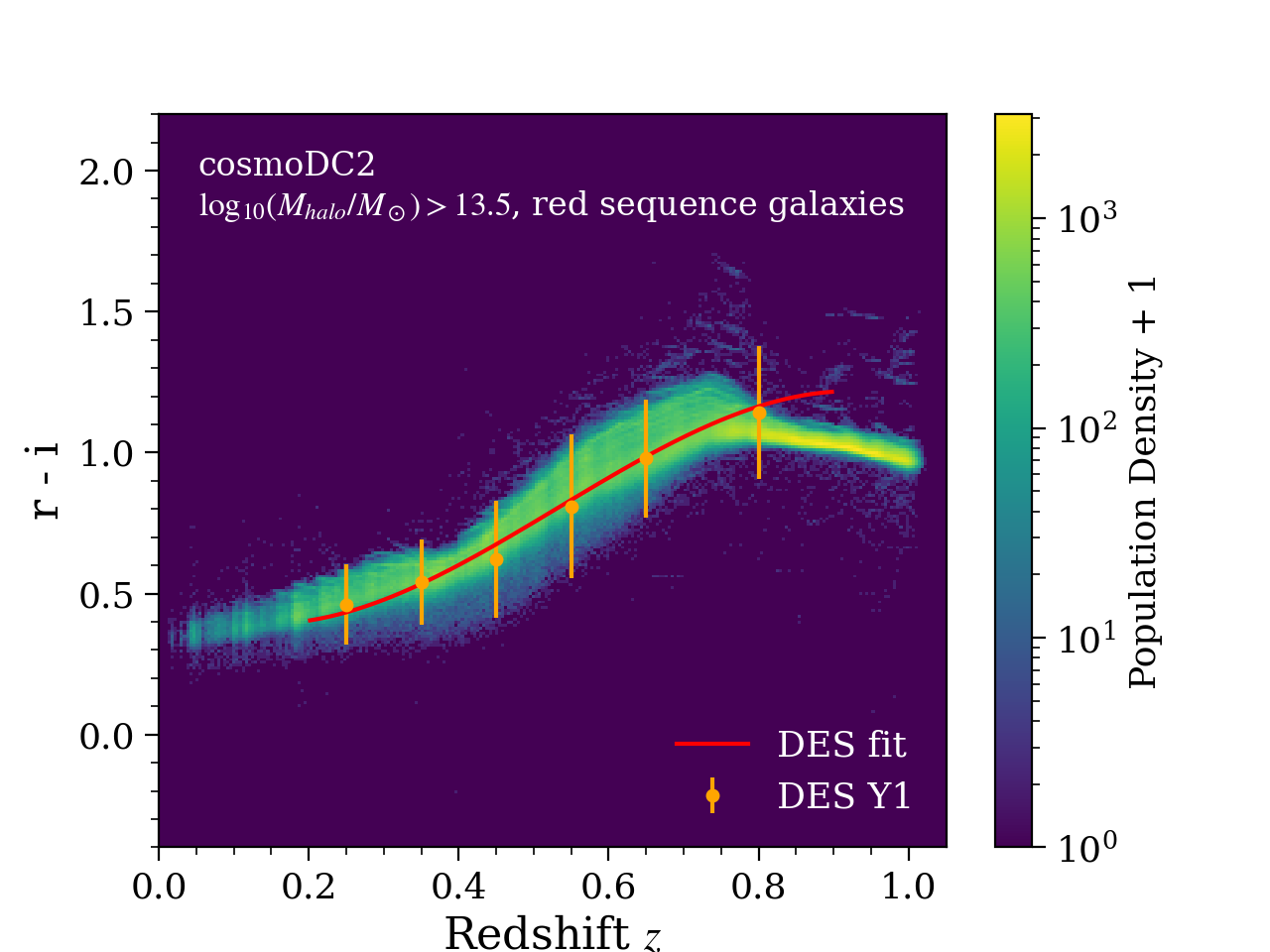}
  \caption{The LSST $r-i$ color as a function of redshift for cosmoDC2 red-sequence field and cluster galaxies (top) and red-sequence cluster galaxies (bottom). Galaxies are flagged as being on the red-sequence if they are selected from the red peaks in the empirical model for rest-frame galaxy colors~\citep{cosmodc2}. Cluster galaxies are defined as those residing in halos with mass $M_{\rm{halo}} > 10^{13.5} M_{\odot}$, whereas field galaxies reside in lower mass halos. The scatter in the color-redshift relationship for cluster galaxies is reduced compared to that for field galaxies due to an additional constraint applied to the observer-frame colors in the cosmoDC2 galaxy model (see text). The red solid line shows a polynomial fit to the mean observed red sequence in DES data and the orange points show the average median color for clusters selected from the DES Y1 \redmapper catalog as described in \autoref{sec:science:data:cl}.
  }
  \label{fig:rs}
\end{figure}

In~\autoref{fig:rs} we show the color-redshift relation for cosmoDC2 red-sequence field and cluster galaxies (top panel) and red-sequence cluster galaxies (bottom panel). Galaxies are flagged as being on the red-sequence if they are selected from the red distribution in the model for rest-frame galaxy colors~\citep{cosmodc2}. Cluster galaxies are defined as those residing in halos with mass $M_{\rm halo} > 10^{13.5} M_\odot$. For these cluster galaxies, the cosmoDC2 galaxy model imposes additional constraints on the observer-frame colors such that they match the mean and scatter for the red-sequence color distributions observed by DES~\citep{rykoff2016}. A polynomial fit to this mean relationship is shown by the red solid line in \autoref{fig:rs}. The orange points show the mean of the median color of cluster members for clusters selected from the DES Y1 \redmapper catalog as described in \autoref{sec:science:data:cl}. Hence this test acts as a verification test for that matching algorithm and ensures that the red 
sequence in the catalog is clearly visible and sufficiently narrow to enable optical red-sequence cluster finders to be run successfully.

\subsection{Cluster Member-Galaxy Magnitude Evolution}
\label{sec:test:cl_mag_star}

\begin{figure}
  \centering \includegraphics{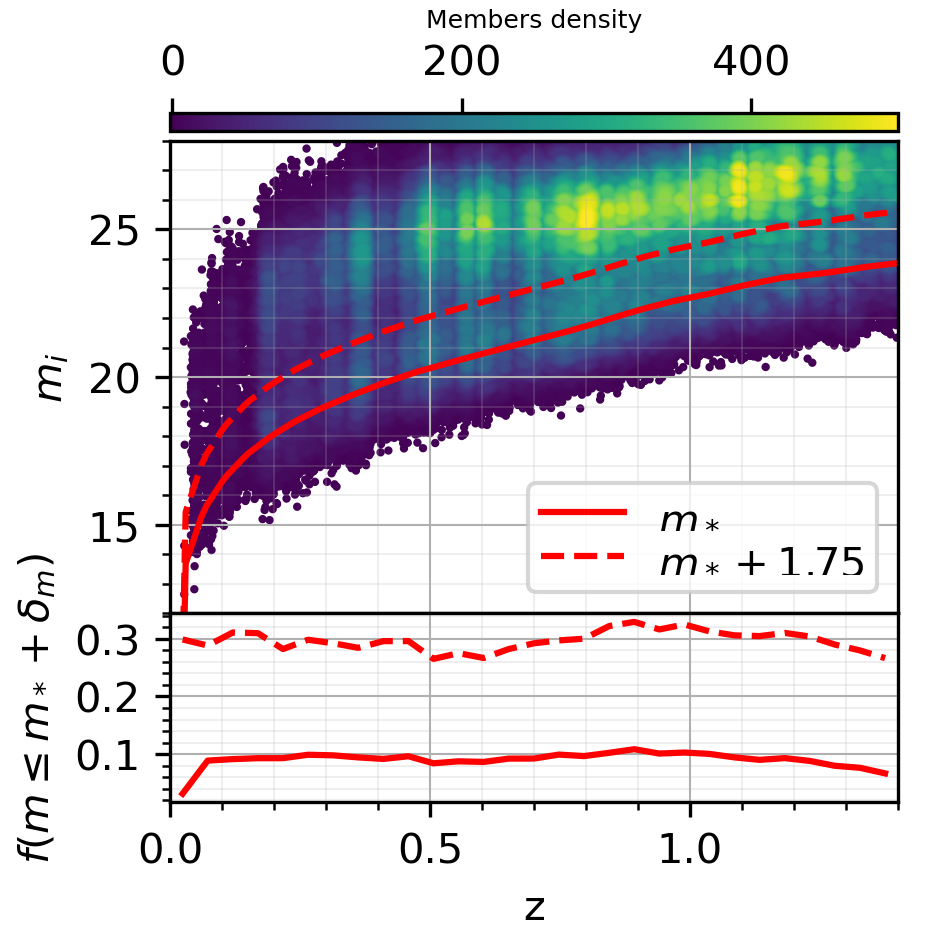}
  \caption{Distribution of halo member magnitudes as a function of redshift. The red lines are defined by the characteristic magnitude marking the knee of the luminosity function $m_*$ (see \autoref{sec:science:data:cl}).
  The bottom panel shows the fraction of cluster members brighter than $m_*$ and $m_*+1.75$, which is equivalent to a cut in luminosity of $0.2L_*$.
  }
  \label{fig:cl_mstar}
\end{figure}

In the top panel of~\autoref{fig:cl_mstar} we show the distribution of the $i$-band magnitudes of cosmoDC2 galaxies residing in halos with mass $M_{\rm{halo}} > 10^{13.2} M_{\odot}$ as a function of their redshift.
We see that the evolution of magnitude is being traced by $m_*$.
This is quantified in the bottom panel, where we compute the fraction of member galaxies brighter than $m_*+\delta_m$, where we have chosen $\delta_m=1.75$. The plot illustrates that
richness estimated using $m_*+\delta_m$ cuts will be redshift independent.
The same result is obtained by restricting the sample to contain only central galaxies
and all centrals are found to be brighter than than $m_*+1.75$, which is equivalent to a cut in luminosity of $0.2L_*$.
We also verified that over 90\% of central galaxies are brighter than $m_*$.

\subsection{Conditional Luminosity Function }
\label{sec:test:clf}

\begin{figure*}
  \centering \includegraphics[width=6.5in]{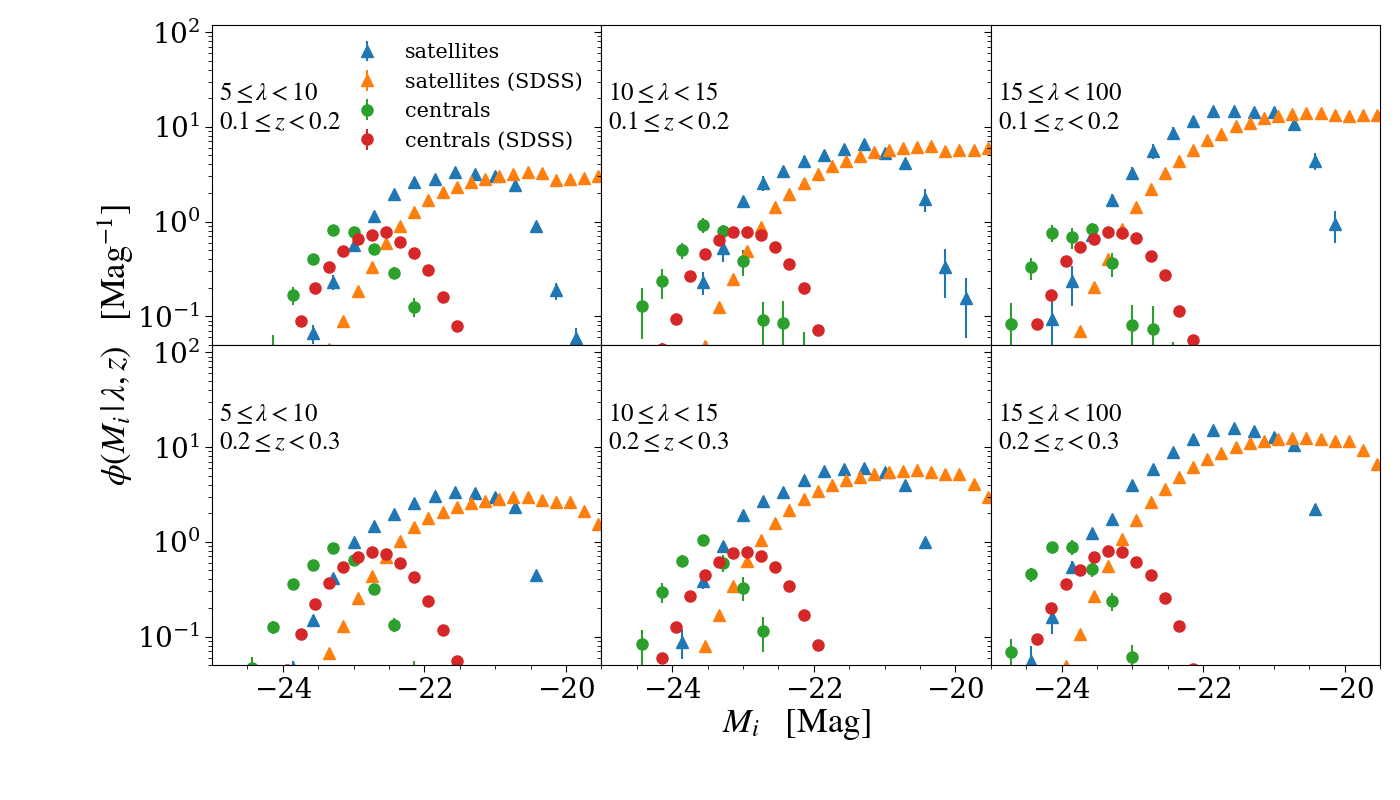}
  \caption{The conditional luminosity function for central galaxies (circles) and satellite galaxies (triangles) for \redmapper selected clusters in cosmoDC2 (green and blue points, respectively) and the SDSS data (red and orange points, respectively). Different panels show different richness and redshift bins as indicated in the legend in each panel. Richness increases from top to bottom and redshift increases from left to right. Errorbars show the $1\sigma$ errors estimated from 50 jackknife resamplings.}
  \label{fig:cluster_clf}
\end{figure*}

For this test, we first run the \redmapper algorithm on dark matter halo centers in the simulation to avoid the confusion of associating clusters to dark matter halos. We then measure the luminosity function of centrals and satellites and compare the measurement to the conditional luminosity function (CLF) measured in SDSS DR8 \redmapper cluster sample~\citep{to2020}. In bins of richness and redshift, we check whether central galaxies in the synthetic catalog are much brighter or fainter relative to satellites. The relative brightness of centrals and satellites is closely related to the performance of \redmapper cluster finder. 
The comparison of the catalog and observational data for both centrals and satellites is shown in \autoref{fig:cluster_clf} for various bins of richness and redshift in the ranges $5 \leq \lambda <100$ and $0.1 \leq z < 0.3$, respectively.  Assuming the clusters found in the synthetic catalog and the real data have a similar mass-richness relation, we find that the relative brightness of centrals and satellites in the synthetic catalog is similar to that in the real data except for the faintest satellite galaxies.

\subsection{Galaxy Density Profiles around Clusters}
\label{sec:test:cl_density_profile}

In order to test the spatial distribution of cluster satellite galaxies in cosmoDC2, we measured the member-galaxy density profiles around massive halos and compare it to that of observed clusters from \cite{2017MNRAS.467.4015H}. In this work, as discussed in \autoref{sec:science:data:cl}, the authors investigate the galaxy density profiles of individual SZ galaxy clusters by using NFW parametrizations to characterize their evolution with mass and redshift.

For this test, we selected a sample of cosmoDC2 halos such that the mean mass and redshift of the sample is equal to the pivot mass and redshift in \cite{2017MNRAS.467.4015H} ($\langle z\rangle$=0.46 and $
\langle M\rangle$=$6\times 10^{14} M_{\odot}$), assuming $M_{\rm FoF} \approx M_{\rm 200c}$. (In  \autoref{fig:M200c_Mfof_OR} in \autoref{app:mass_conversion}, we show that for cosmoDC2 halos with mass $>10^{13} h^{-1}{\rm M_\odot}$ and $z < 1$, the average value of the FoF to SO mass ratio, $\langle M_{\rm 200c}/M_{\rm FoF}\rangle = 0.92$  which is sufficiently close to 1 for the purposes of the test). This selection yields 39 halos.

We then selected galaxies whose $i$-band magnitudes are brighter than $i=22$, to approximately reproduce the selection in \cite{2017MNRAS.467.4015H}. We measured the profiles using two methods: (1) including only galaxies belonging to the clusters and (2) including all galaxies along the line of sight, but performing a background subtraction. We estimate  the background by averaging the density of galaxies in annuli of 6 to 7 Mpc around each cluster.

To construct the stacked profile, we measure the projected distance of included galaxies from the halo center (defined as the position of the central galaxy) and stack their number in linear radial bins. The projected density is then simply obtained by dividing by the surface area corresponding to each radial bin and subtracting the background density, if we are including all galaxies in the measurement.
The resulting profiles are then compared to that obtained in \cite{2017MNRAS.467.4015H} by using their measured NFW parametrizations  and setting the mass and redshift to that of the mean values for clusters in the sample. We also estimate the scaled radius $R_{\rm 200c}$, at the same mass and redshift values, assuming $M_{\rm FoF} \approx M_{\rm 200c}$.

\begin{figure}
  \centering 
  \includegraphics[width=0.99\columnwidth]{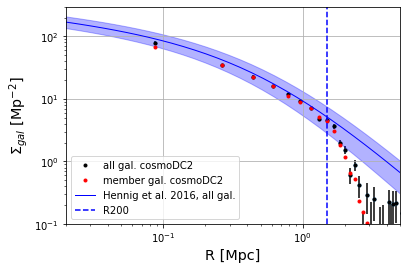}
  \caption{Projected density profile of galaxies around clusters as a function of radial separation from cluster centre. The points show the resulting stacked profiles for cosmoDC2 clusters with $0.4<z<0.5$ and $M_{\rm FoF}>4\times10^{14} M_{\odot}$. Galaxies have been selected to have $i$-band magnitudes brighter than $i = 22$. The red points show the profiles for member galaxies associated with the clusters and the black points show the background subtracted profiles including all galaxies along the line of sight.
  The blue line and shaded region show the NFW parametrization and associated uncertainty  expected for a cluster with a mass and redshift equal to the mean mass and redshift of the cosmoDC2 sample, as measured by \cite{2017MNRAS.467.4015H}. The dashed blue line indicates an estimate of the scaled radius $R_{\rm 200c}$, measured at the mean redshift and true halo mass (FoF masses). The error bars are computed by assuming a Poisson distribution for the galaxy counts.  }
\label{fig:cl_profile}
\end{figure}

\autoref{fig:cl_profile} shows the results of the comparison. We can see that the density profiles in cosmoDC2 agree well with the observed one for a radius less than $R_{\rm 200c}$. At higher radii, the signal drops considerably with respect to the validation data. While this can be expected for member galaxies in the simulation due to the method used to associate galaxies within halos, the fact that the background-subtracted profiles follow the same trend is yet to be understood. More work is thus needed to validate the distribution of galaxies at the outskirts of clusters. We also note that, in cosmoDC2, the galaxy distributions do not follow the exact dark matter distribution based on the halo's shape, but rather a truncated NFW distribution, with random (spherical) symmetry. 
This feature is expected to have minor impact on the current detection and richness measurements because for \redmapper the maximum radius for assigning membership is less than $R_{\rm 200c}$~\citep[see, \eg,][]{rykoff2016}

We comment, however, that for testing weak-lensing mass measurements, it is important to have the correct number density of cluster members at larger radii because such interloper foreground galaxies can significantly dilute the shear signal if they are mistaken for background galaxies \cite[\eg,][]{medezinski18,varga19}. 
Moreover, one major source of bias for optical cluster cosmological analyses is due to projection effects~\citep[see, \eg,][]{mcclintock2019} from uncorrelated structures and from the effects of halo tri-axiality. Therefore, it is crucial to measure the correlation between richness and WL observables up to large radii in simulations and to understand any biases  that result from using the galaxy distribution as a tracer for the dark matter distribution.

\subsection{The Mass-Richness Relation} 
\label{sec:test:mass_richness}

\begin{figure}
  \centering 
  \includegraphics[width=9.0cm]{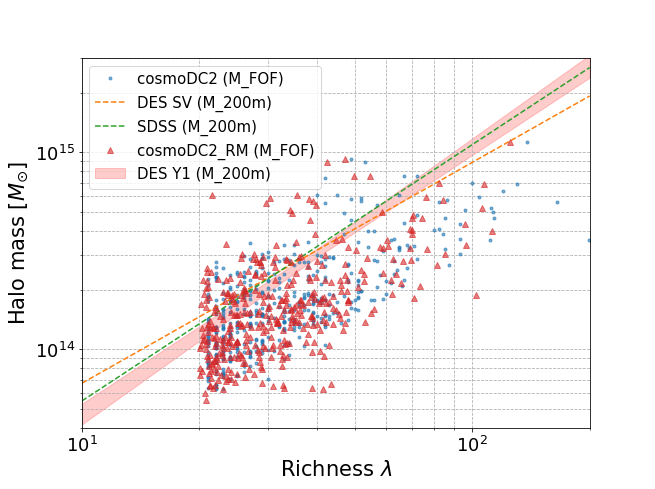}
 
  \caption{Mass-richness relation of cosmoDC2 galaxy clusters compared with the measured relation from DES Y1 (red shaded band), DES SV (red dashed line), and SDSS (green dashed line). The blue points indicate halos with radial distance, richness, color, and magnitude cuts that approximate the thresholds in DES Y1, and the corresponding galaxy count is used as richness. The red triangles indicate halos that are bijectively matched between cosmoDC2 \redmapper and extragalactic catalogs, and the richness comes from \redmapper. The halo-mass definition used for the observed data is $M_{\rm 200m}$, which is approximately equivalent to $M_{\rm FoF}$, the halo-mass definition used in the cosmoDC2 simulation (see text and \autoref{app:mass_conversion}). }
\label{fig:mass-rich}
\end{figure}

In \autoref{fig:mass-rich} we show the mass-richness relation in cosmoDC2 versus observations using two methods to identify clusters and determine their richness. In both cases, we select clusters from cosmoDC2 with halo masses $M_{\rm FoF}>5\cdot10^{13}M_{\odot}$, (which corresponds to $\lambda\sim10$ in \autoref{eq:mass-richness_relation}) and redshift $0.3<z<0.4$ ($\sim z_0=0.35$). Here 
$M_{\rm FoF}$ denotes the mass determined by the FoF algorithm that was used to find halos in the underlying simulation. We now summarize the details of the two methods:

First, we try to reproduce the results of running the \redmapper algorithm by imposing some simple selection cuts on the cosmoDC2 catalog. 
We select clusters with $\lambda > 20$, counting only galaxies within a radial distance of $(\lambda/100)^{0.2}h^{-1} {\rm Mpc}$ from the cluster center. We require that galaxies be on the red sequence (by requiring their red-sequence flag for $r-i$ and $g-r$ be true) and that their magnitudes are brighter than $0.2L*$ in the $z$-band. These selections approximate the thresholds in \citet{mcclintock2019}.  We define ``richness'' as the galaxy count in each halo after these cuts to select the member galaxies have been applied.  The results are shown by the blue points in \autoref{fig:mass-rich}.

For the second method, we match geometrically the positions of objects in the cosmoDC2 extragalactic catalog with those in the catalog obtained by running \redmapper on cosmoDC2. We use a bijective nearest match within a cylinder that has a radius $R_{FoF}=\{3M_{\rm FoF}/[800\pi\rho_c(z)]\}^{1/3}$ and a depth $2\delta_z (1+z)$, where $\delta_z=0.05$. This enables us to assign a halo mass to the objects identified by \redmapper. The richness is obtained from the  cosmoDC2 \redmapper catalog and is required to have a lower limit of 20. The results are shown by the red triangles in \autoref{fig:mass-rich}.

We see that the first method, in general, gives similar richness to the second method and that the catalog data are scattered about the observed relation and have similar slopes.
We note that the mass definitions for the halo masses, $M_{\rm FoF}$ and $M_{\rm 200m}$, in cosmoDC2 and the observational data respectively, are not identical. In \autoref{app:mass_conversion} we show that for the cosmoDC2 catalog, the expected ratio between these two definitions is $M_{\rm 200m}/M_{\rm FoF}\approx 1.08$ (see \autoref{fig:M200m_Mfof_OR}). Hence we expect the red triangles and blue points from cosmoDC2 to be scattered about the observed mass-richness relationship but biased slightly low. We conclude that mass-richness relationship for \redmapper selected clusters in cosmoDC2 are in qualitative agreement with the expectations from observations.



\subsection{Cluster Weak-lensing Masses }
\label{sec:test:cl_shear}

We investigate the weak lensing mass determination of galaxy clusters using the cosmoDC2 shear signal. First, we extract background source galaxy catalogs from 3300 cosmoDC2 clusters for $M_{\rm FoF}$ ranging from $10^{14}$ to $10^{15}\ \rm{M}_\odot$, with redshift $0.1 < z_\mathrm{cl} < 0.7$. To build the background galaxy catalogs, we select galaxies with $i-$band magnitude $<\ 25$, redshifts $>z_\mathrm{cl} + 0.1$ and  within a constant angular window of 0.36 deg$^2$ around the position of the central galaxy of the cluster. The average number density of background galaxies is $\bar{n} = 23\  \mathrm{arcmin}^{-2}$. We measure average cluster masses by stacking selected clusters in true halo mass bins and cluster redshift bins. Following \autoref{eq:delta_sigma} and using a weighted average of the tangential shear, we estimate the stacked excess surface density by summing over all lens-source ($l,s$) pairs as:
\begin{equation}
     \widehat{\Delta\Sigma}_+(r_p) = \frac{1}{\sum\limits_{l,s = 1} w_{l,s}}
     \sum\limits_{l,s= 1}w_{l,s}\Sigma_{{\rm crit}}^{l,s}\gamma_+^{l,s}, 
     \label{eq:deltasigma_stack}
\end{equation}
 where $\gamma_+^{l,s}$ and $\Sigma_{{\rm crit}}^{l,s}$ are defined in \autoref{eq:gammat} and \autoref{eq:critical_surface_dens} respectively and
  $(\gamma_1, \gamma_2)$ are the measured shear components provided in cosmoDC2 for each source galaxy. The lens-source pairs in \autoref{eq:deltasigma_stack} are identified in bins of $r_p$ defined using $r_p = D_l  \theta_{ls} $, where $\theta_{ls}$ is the angular separation between the lens and source on the sky.
The weights $w_{l,s} = \left(\Sigma_{\rm crit}^{l,s}\right)^{-2}$ are the optimal weights that formally reduce the signal-to-noise ratio of the estimator $\widehat{\Delta\Sigma}_+$ in the shot-noise limited regime~\citep{sheldon2004, 2018MNRAS.478.4277S}. These weights have the effect of down-weighting lens-source pairs that are close in redshift, for which the shear induced deformation of background galaxy shape is weak. For consistency, we also estimate the cross lensing signal $\widehat{\Delta\Sigma}_\times$ where $\gamma_+^{l,s}$ is replaced in \autoref{eq:deltasigma_stack}  by $\gamma_\times^{l,s} = \gamma_1\sin(2\phi_\mathrm{s}) - \gamma_2\cos(2\phi_\mathrm{s})$.  Due to symmetry, this quantity should be consistent with zero for the ensemble. We estimate the bootstrap covariance matrix for $\Delta \Sigma_{+}$ and $\Delta \Sigma_{\times}$  by using 100 random samplings with replacement for each $M_{\rm FoF}$ and $z$ bin. The lensing signal is estimated in 20 logarithmic radial bins from 0.2 to 5~Mpc from the projected cluster center.
\begin{figure}
  \centering \includegraphics[width=8cm]{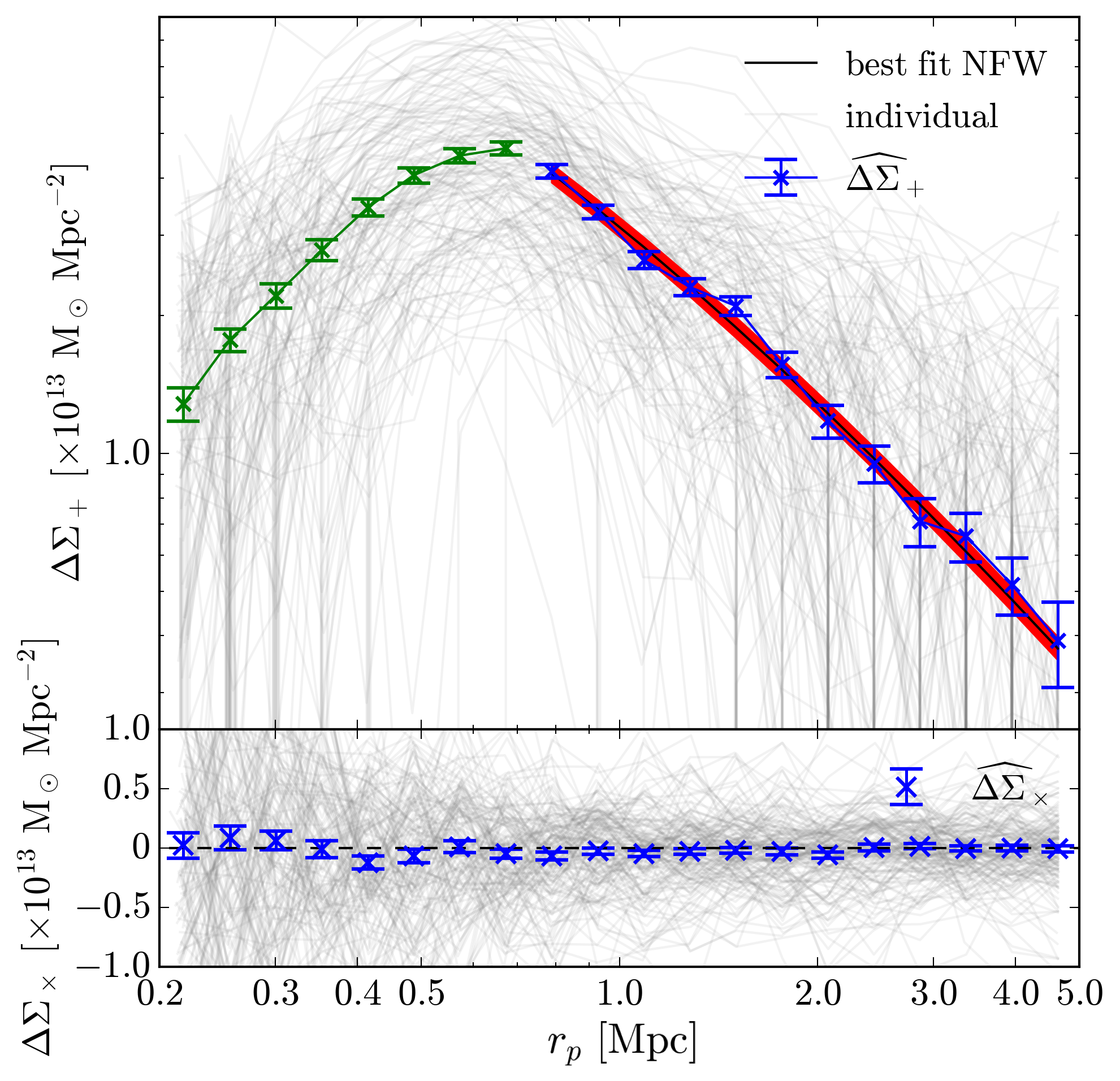}
  \caption{Estimation of $\widehat{\Delta\Sigma}_+$ and $\widehat{\Delta\Sigma}_\times$ (blue crosses) for 151 selected clusters in the $M_{\rm FoF}$ bin $[10^{14}, 1.2\times 10^{14}]\ \rm{M}_\odot$ and redshift bin $[0.3, 0.4]$. The signal attenuation due to the limited resolution of the shear maps is denoted in green crosses for $\widehat{\Delta\Sigma}_+$. The estimates of the shear signal for each cluster of the stack are represented in grey solid lines, where the lens index l is fixed in \autoref{eq:deltasigma_stack}. The best fit reconstructed mass is $M_{\rm 200m} = 1.31^{+0.03}_{-0.03} \times 10^{14}\ \rm{M}_\odot$ ($\chi^2$/dof = 0.93). The NFW excess surface density predicted for the best fit mass is indicated by the black solid line, with the $3\sigma$ interval shown by the red shaded band. Error bars are given by the square roots of diagonal components of the bootstrap covariance matrix.}
  \label{fig:DeltaSigma_Stacking}
\end{figure}
 
 \autoref{fig:DeltaSigma_Stacking} shows the individual (light grey) and stacked (symbols) $\Delta\Sigma_+$ profiles in a given $M_{\rm FoF}$ bin $[10^{14}, 1.2\times 10^{14}]\ \rm{M}_\odot$ and redshift bin $[0.3, 0.4]$. The cross component (lower panel), is consistent with zero, as expected. The tangential component (upper panel) increases towards the central region of the cluster, before dropping off in the innermost regions. The region of attenuated signal was similarly noted in \citet{cosmodc2}, when looking at galaxy-galaxy lensing, and is explained by the resolution to which the shear is estimated in cosmoDC2. We therefore exclude this inner region (green symbols) and proceed with fitting the cluster mass on the remaining radial range (blue symbols). The shear signal for low mass clusters and for $z_{\rm cl} > 1$ is significantly degraded, due to limited $i < 25$ galaxy number density at these redshifts.
 
\begin{figure}
  \centering \includegraphics[width=8.0cm]{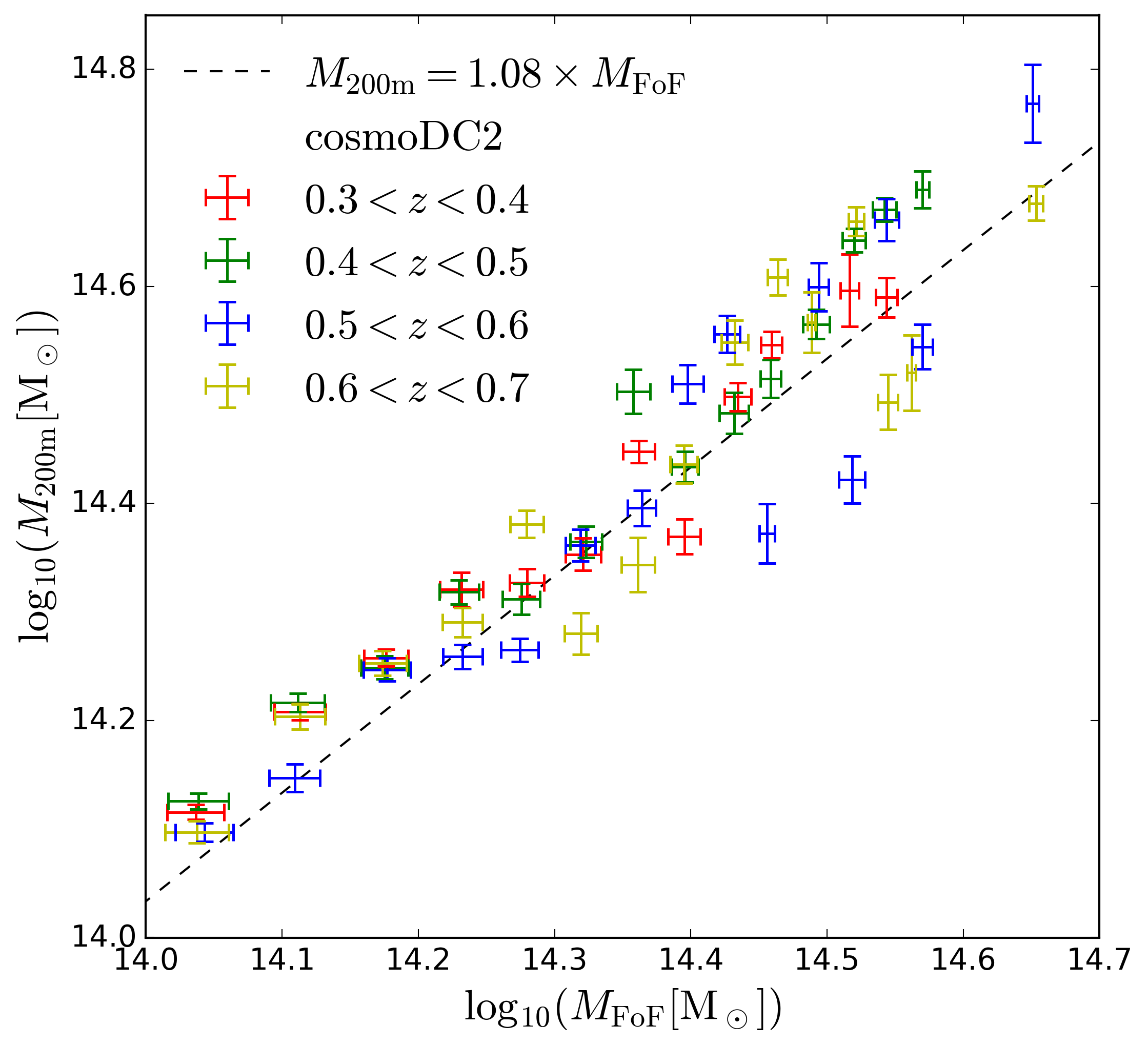}
  \caption{Validation test for DC2 mass determination using shear observable; The markers represent the reconstructed $M_{\rm 200m}$ mass for several true $M_{\rm FoF}$ mass bins and redshift bins. The different colors corresponds to the four redshifts bins. The dashed line corresponds to $M_{\rm 200m}/M_{\rm FoF} = 1.08$, which is the expected mean relationship between $M_{\rm 200m}$ and $M_{\rm FoF}$ that is derived in \autoref{app:mass_conversion}. The errors on the $x$ and $y$-axes are the r.m.s. of the halo masses in each mass bin and the weak-lensing mass reconstruction errors, respectively.}
  \label{fig:M200m_Mfof_cosmoDC2}
\end{figure}

 We assume the halo dark matter density is modeled by a NFW profile~\citep{1997ApJ...490..493N} that depends on the cluster mass $M_{\rm 200m}$ and its concentration $c_{\rm 200m}$. We predict the $\Delta\Sigma_+$ profile using the CLMM library~\citep{aguena2021clmm}, and assuming the mass-concentration relation of \citet{Duffy2008}. The corresponding $M_{\rm 200m}$ is fitted by likelihood maximization using the diagonal terms of the bootstrap covariance matrix. The model using the best fit cluster mass is shown in solid red in \autoref{fig:DeltaSigma_Stacking} for this particular bin of true halo mass and redshift. We repeat the same analysis in a set of equal-size true mass bins $\Delta M_{\rm FoF} = 0.2\times10^{14} \rm{M}_\odot$ from $10^{14}$ to $10^{15} \rm{M}_\odot$ and for four different redshift bins $\Delta z = 0.1$ for $0.3 < z_{\rm cl} < 0.7$. The reconstructed masses $M_{\rm 200m}$ thus obtained are shown in  \autoref{fig:M200m_Mfof_cosmoDC2}. 
 The dashed line in the figure shows the average relationship, derived from simulations and discussed in \autoref{app:mass_conversion}, between $M_{\rm 200m}$ and $M_{\rm FoF}$. The vertical and horizontal bars show the errors on the weak lensing mass reconstruction and the r.m.s. of the halos in each mass bin, respectively.
 This test provides a verification that the values of the weak-lensing masses $M_{\rm 200m}$ and the true halo masses $M_{\rm FoF}$ in cosmoDC2 are qualitatively well correlated and demonstrates that the shear signal in cosmoDC2 behaves as expected. The derived values of $M_{\rm 200m}$ follow approximately the expected relationship with $M_{\rm FoF}$. The same analysis was repeated without assuming a mass-concentration relation and gives comparable results with more scatter in the mass determination. 

\subsection{Velocity Dispersion in Clusters}
\label{sec:test:velocity_disp}

\begin{figure}
  \centering \includegraphics[width=8.5cm]{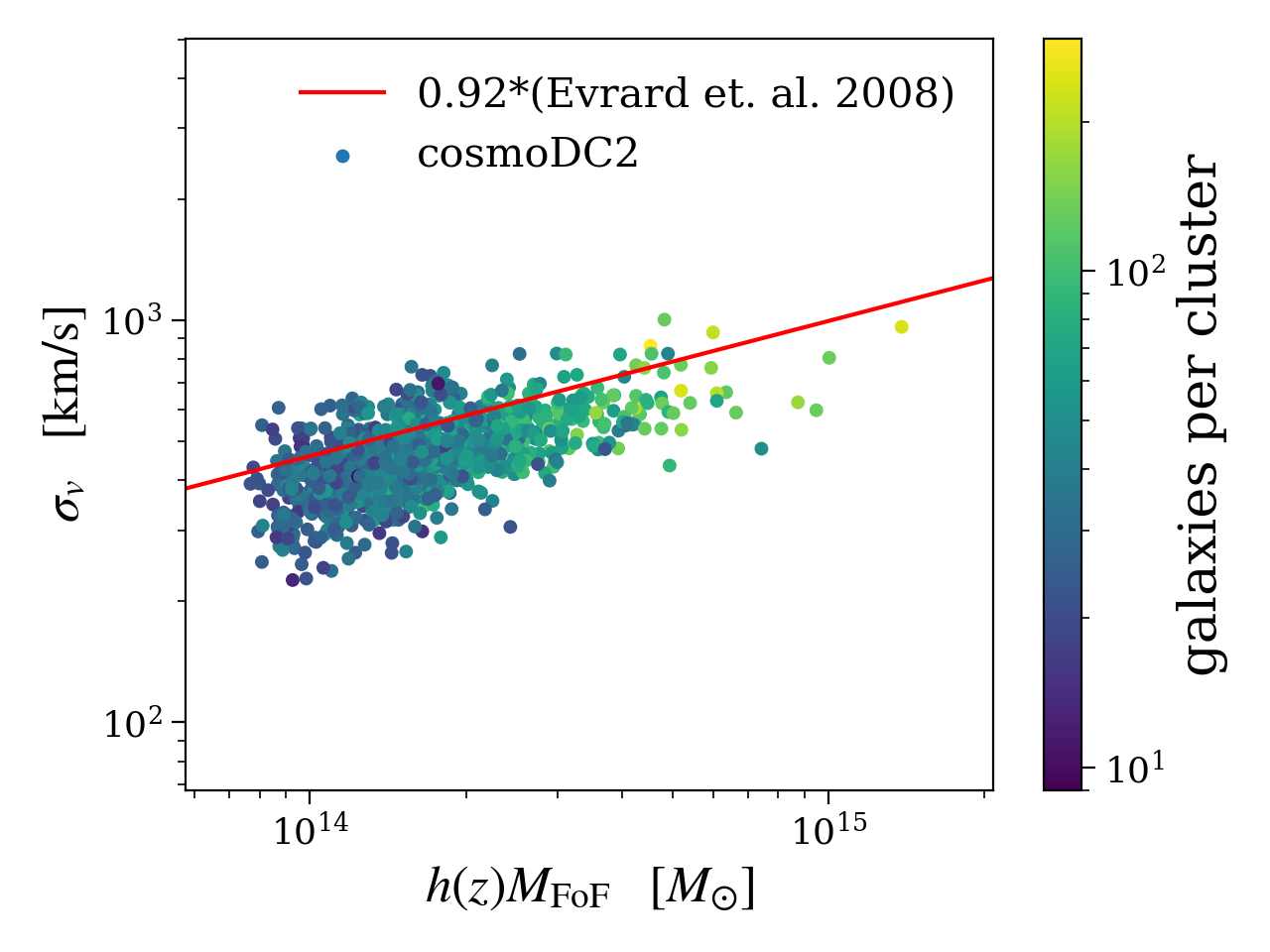}
  \caption{Velocity dispersion of member galaxies within a given cosmoDC2 cluster (points) as a function of the cluster's parent halo mass, which has been scaled by the dimensionless Hubble parameter $h(z)$ evaluated at the median of the redshifts for the member galaxies in the cluster. The points are colored according to the number of member galaxies in each cluster. The red line shows the observed scaling relation from \citet{evrard2008}  (\autoref{eq:vel_disp}), which has been scaled by a factor of 0.92 to account for the difference in the halo mass definition between $M_{\rm 200c}$ and $M_{\rm FoF}$ (see text and \autoref{app:mass_conversion}).}
  \label{fig:virial}
\end{figure}

In \autoref{fig:virial}, we present the velocity dispersions of member galaxies of selected clusters as a function of their parent halo mass and compare them to the mean relationship measured in \citet{evrard2008} and given in \autoref{eq:vel_disp}. We select galaxy clusters with parent halos masses $>10^{14}M_\odot$, member galaxies with stellar masses  $> 10^{9}M_\odot$ and we include only clusters whose resulting member-galaxy counts are $> 5$. The velocity dispersion is computed as the biweight scale (square root of the biweight midvariance\footnote{\url{https://docs.astropy.org/en/stable/api/astropy.stats.biweight.biweight_scale.html}}) of the velocity magnitudes of the selected member galaxies. We select this statistic rather than the conventional standard deviation because it is a more robust statistic for determining the standard deviation of a distribution with outliers. In order to match the cosmoDC2 FoF masses with the mass in relationship from \citet{evrard2008} which is derived for $M_{\rm 200c}$ we scale \autoref{eq:vel_disp} by 0.92, which is the mean of the ratios of $M_{\rm 200c}/M_{\rm FoF}$ measured in the Outer Rim simulation and discussed in \autoref{app:mass_conversion} (see \autoref{fig:M200c_Mfof_OR}). To account for the factor of $h(z)$ in \autoref{eq:vel_disp}, for each cluster in \autoref{fig:virial}, we scale the cosmoDC2 FoF halo mass by the dimensionless Hubble parameter $h(z)$ evaluated at the median of the redshifts for the member galaxies in the cluster.
We see that the velocity dispersions of the selected cosmoDC2 clusters are in qualitative agreement with the expected scaling relation from the observational data. In the future, a more quantitative comparison  would require a measurement of the SO masses for the synthetic catalog or a conversion from FoF masses to SO masses using either measured SO concentrations or an approximate concentration-mass relation.

\subsection{Emission-Line Ratios}
\label{sec:test:em-line}

\begin{figure*}
  \centering \includegraphics[width=6.5in]{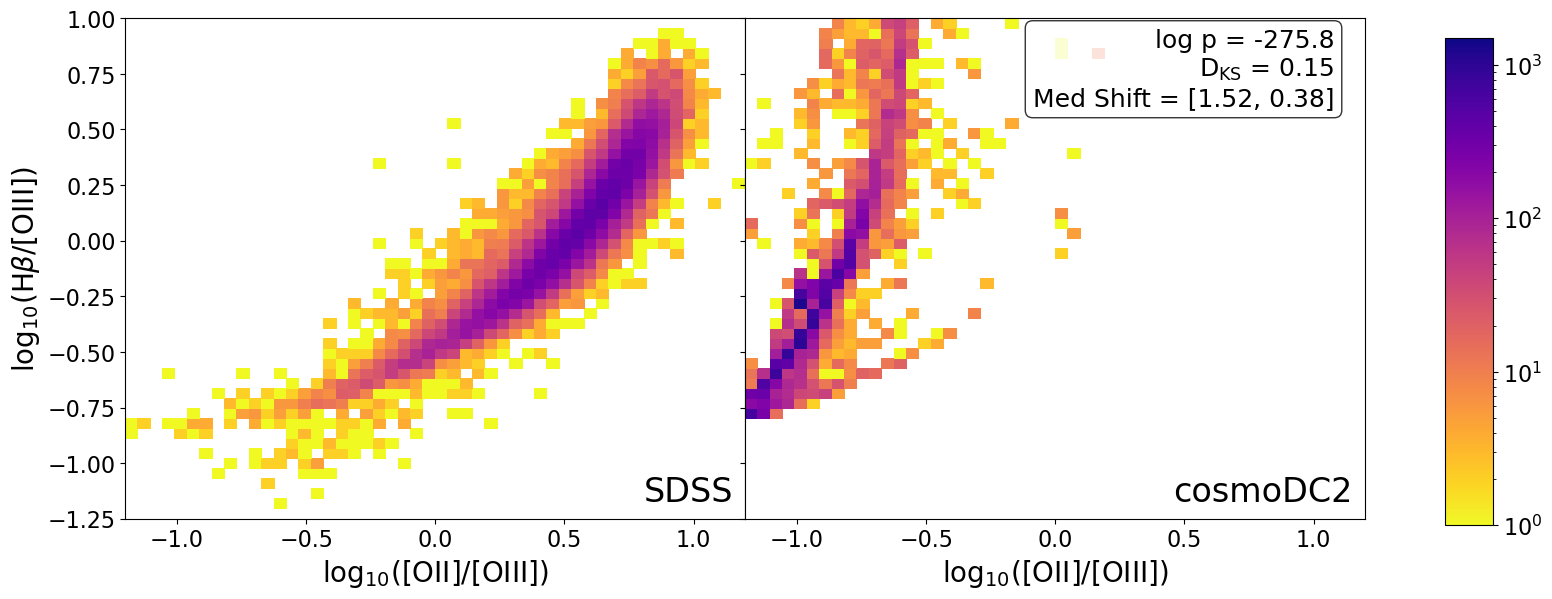}
  \caption{Emission-line ratio distributions for $\log(\rm H\beta/OIII)$ versus $\log(\rm OII/OIII)$ for SDSS galaxies (left panel) compared with cosmoDC2 (right panel).
  }
  \label{fig:em}
\end{figure*}

\autoref{fig:em} shows the distribution of log emission line ratios for $\mathrm H\beta/[\mathrm{OIII}]$ and $\mathrm{[OII]/[OIII]}$.  We plot these results for the SDSS comparison set in the left panel and cosmoDC2 in the right panel. \reply{CosmoDC2 galaxies are selected to lie in the redshift range $0.0 < z < 0.7$. Galaxies in the range $0.0 < z < 0.4$ are further required to have $r$-band magnitudes $<19.5$, while those in the range $0.4 < z < 0.7$ are required to have $i$-band magnitudes $<19.9$. These magnitude cuts correspond to the published SDSS spectroscopic detection limits~\citep{bolton2012}. The distributions are compared using a two-dimensional KS test. We find that we need to limit the sample sizes for both the catalog and the validation data to random subsamples of 30000 in order to prevent the KS test from becoming too discriminatory.} We report the log $p$-value for the KS test, the corresponding $D$ value, and the shift in distribution medians in the upper left corner of the right panel.  While the SDSS data set does not span the same range of redshifts that the LSST will probe, we can already see many artifacts in the cosmoDC2 emission-line ratio distribution.  Additionally, it is apparent that the mode of the  two-dimensional emission-line ratio distribution for cosmoDC2 in the right panel is shifted to lower values of both line ratios in comparison to those of the SDSS sample on the left. This shift could  be indicating an over-representation of $\mathrm{[OIII]}$ luminosity in cosmoDC2.  In the future, this emission-line ratio test will require an observed comparison sample that more accurately represents the redshift range probed by LSST. 

\subsection{Stellar Mass Function }
\label{sec:test:mstar}

\subsubsection{Redshift Dependence}
\label{sec:test:smf-primus}

\begin{figure}
  \centering \includegraphics[width=8.5cm]{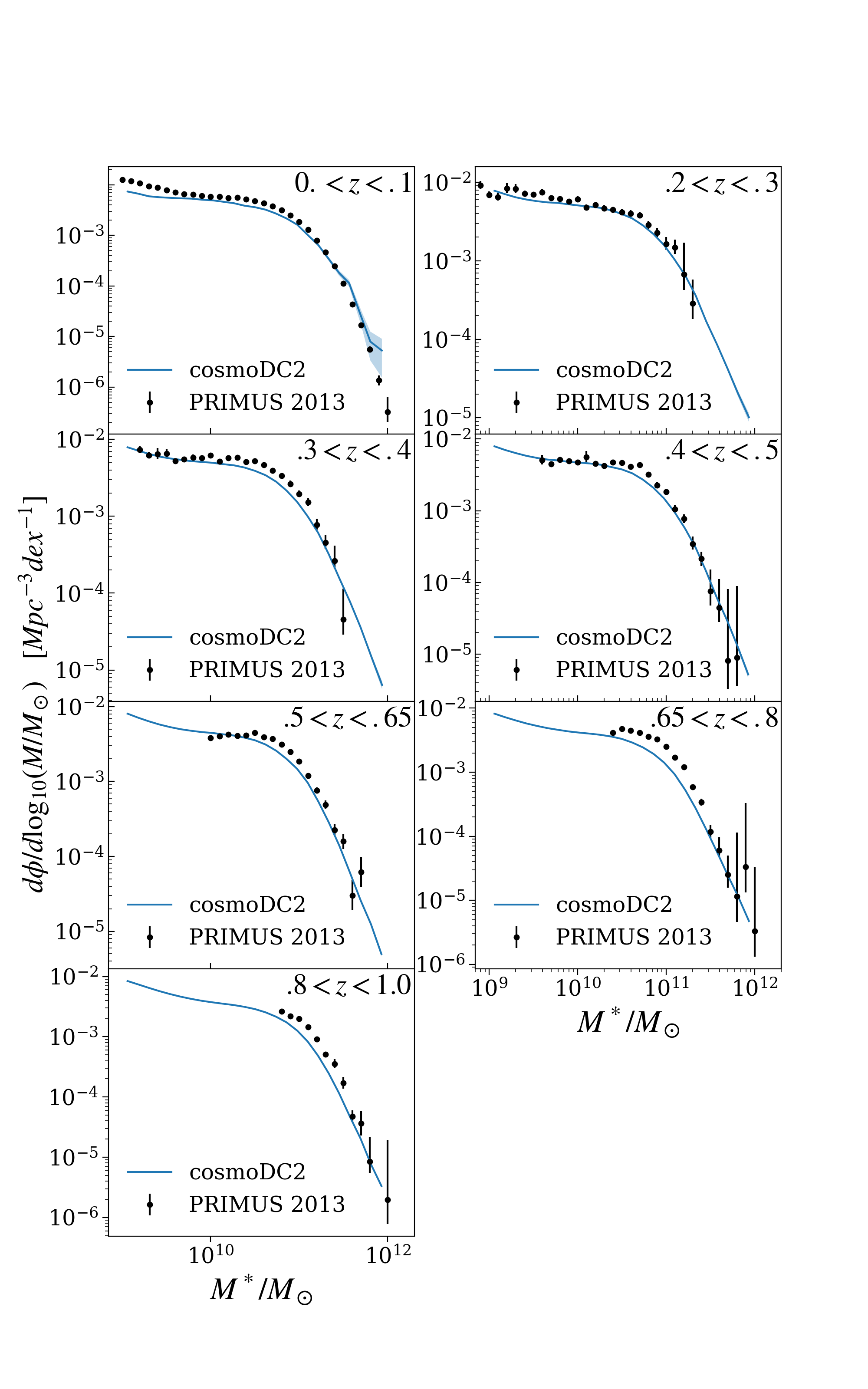}
  \caption{A comparison of the stellar mass function for cosmoDC2 (blue solid lines) with PRIMUS data~\citep[black points]{moustakas2013} for five redshift ranges as indicated in each panel.
  }
  \label{fig:smf}
\end{figure}

In~\autoref{fig:smf} we show a comparison of the stellar mass function for cosmoDC2 galaxies with PRIMUS data~\citep{moustakas2013} for selected redshift slices in the range $0 < z < 1$. Overall, the agreement between the catalog and observational data is good, but not perfect, over a wide range in redshift. The most discrepant regions in stellar mass and redshift, where the catalog data falls below the PRIMUS data, are $M^*/M_\odot < 10^9$ and $0 < z < 0.1$ and $M^*/M_\odot \gtrsim 10^{11}$ and $z > 0.5$. The level of agreement shown here is obtained without explicit tuning of the catalog data to the SMF. Instead, the underlying empirical model used for cosmoDC2 employed resampling of galaxies and their stellar masses from a simulation with somewhat different cosmology, resolution and size~\citep{cosmodc2, behroozi_etal18}, which has been tuned to match the observed SMF. The resampling is performed in narrow bins of halo mass. Hence, provided that the resampling procedure is unbiased, we expect that the distributions of stellar masses in cosmoDC2 will follow those of the observational data. This test thus validates that no significant distortions of the stellar-mass function have been introduced by the hybrid catalog-production algorithm.

\subsubsection{SMF for a CMASS Galaxy Sample}
\label{sec:test:smf-cmass}

\begin{figure}
  \centering \includegraphics[width=8.5cm]{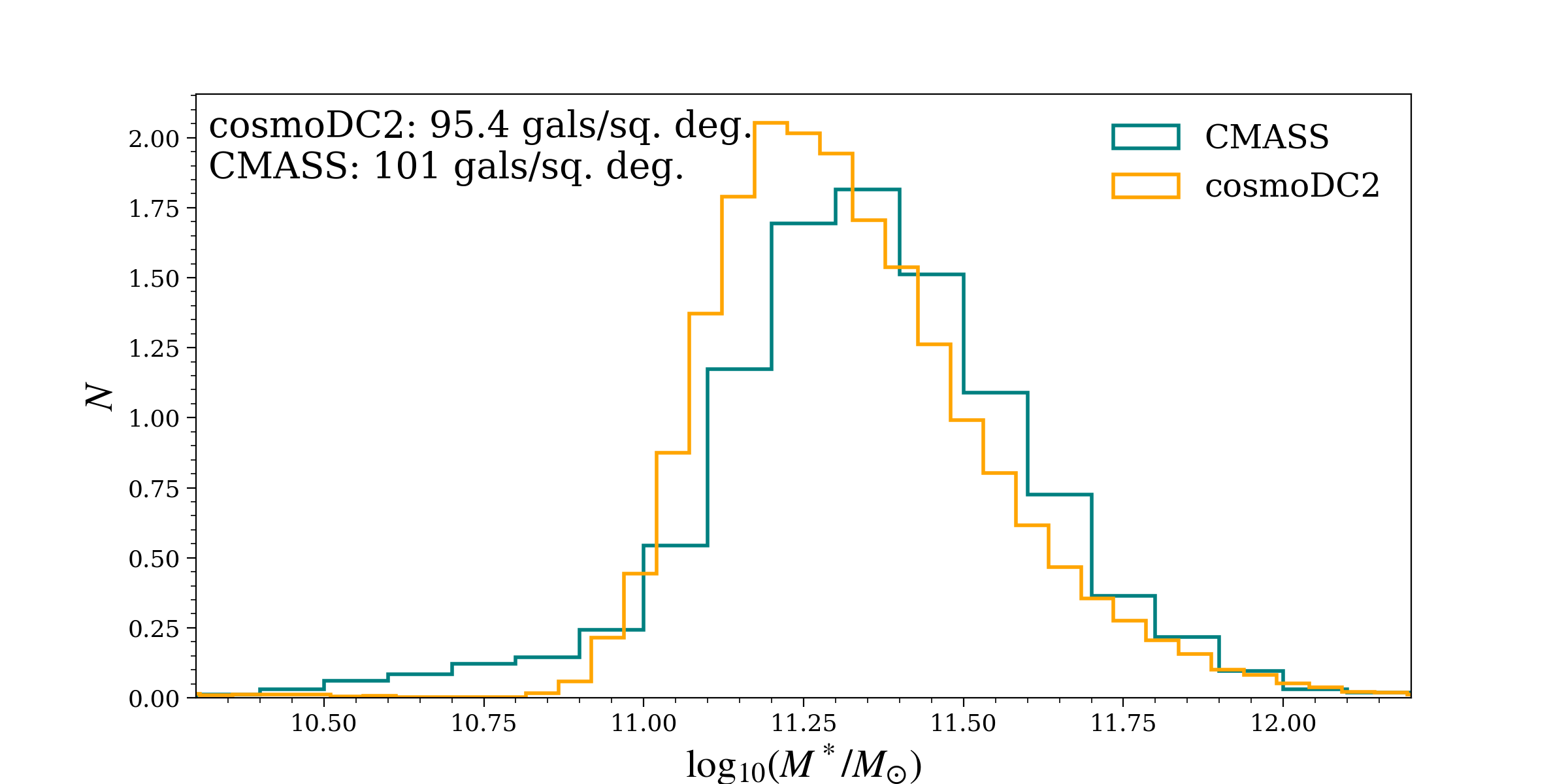}
  \caption{A comparison of the stellar mass distributions for a sample of cosmoDC2 (orange) and CMASS (blue) galaxies. The cosmoDC2 galaxies have been selected to mimic the CMASS data sample using the cuts listed in \autoref{sec:science:data:smf}. The galaxy number densities for the cosmoDC2 and CMASS samples are indicated in the figure and are in good agreement, thus providing a non-trivial test for galaxy colors in addition to the test of the stellar mass distribution. 
  }
  \label{fig:cmass}
\end{figure}

In \autoref{fig:cmass}, we show the stellar mass distribution of a sample of cosmoDC2 galaxies obtained by selecting galaxies in the redshift range $0.4 < z < 0.7$ and applying the color and magnitude cuts listed in \autoref{sec:science:data:smf}. These selection cuts reproduce the CMASS galaxy sample~\citep{reid2016} and are designed to select a sample of galaxies with high stellar masses.
The peak of the resulting stellar-mass distribution for cosmoDC2 galaxies lies in the range $10^{11} M_{\odot}$ - $10^{11.5} M_{\odot}$, which is in qualitative agreement with the location of the peak for CMASS galaxies.  The galaxy number density for the cosmoDC2 sample is 95.4 galaxies per square degree which agrees well with the number density of 101 galaxies per square degree for the CMASS sample~\citep[see Table 2]{reid2016}. This test therefore also provides a non-trivial test for galaxy colors, since color cuts are imposed to select the CMASS galaxy sample.

\section{Summary and Discussion}
\label{sec:summary}

In this paper we have presented a test suite for validating synthetic catalogs that have been developed primarily for evaluating dark energy science targets based on data from optical surveys. Generating synthetic catalogs for these surveys requires a large amount of human effort and consumes significant computer resources. Crucially, these catalogs must be sufficiently realistic to ensure that they are useful for testing survey and data analysis pipelines. Therefore it is imperative to validate them against observational data and theoretical expectations. It is only by carrying out these validation steps that catalog users can in fact assess how useful the catalogs will be for studying their science case of interest. This in turn implies that prospective users have specified their science goals and have developed some validation criteria that the catalogs must meet in order to fulfill those goals. Different surveys will require different levels of fidelity to the real universe. Our validation framework can accommodate a flexible set of validation tests and validation criteria. 

We discussed a broad range of scientific goals for the cosmological analyses of data from optical surveys and identified the key galaxy properties that are required to be rendered accurately by the synthetic catalogs. These properties include the number density of galaxies in a multi-dimensional space of redshift, survey filter magnitudes and colors, galaxy size and morphology distributions and galaxy clustering, as well as the properties of more specialized galaxy samples such as galaxy clusters. They form the building blocks of the validation test suite. Key constraints for further development of the tests are the availability of suitable validation data. We showed that due to the limited availability of observational data, only certain ranges of the synthetic galaxy properties can actually be compared with observations and that only specific projections of properties that have a complex structure in a multi-dimensional feature space are generally available. In the future, more observational data, particularly at high redshifts, will be needed to extend the range over which synthetic catalogs can be validated.

As part of this paper, we have assembled a curated set of validation data that can be used by optical surveys to perform catalog validation in the future. We then constructed a test suite based on the science requirements and the available data, which we summarized in \autoref{tab:tests}. This summary table should serve as a useful guideline for other optical surveys doing cosmological analyses. 

We illustrated the utility of our test suite by showing examples of more than 25 tests run on the LSST~DESC cosmoDC2 synthetic catalog. The wide range of test results that are described in \autoref{sec:tests} paint a picture of the overall fidelity of the catalog to the real universe. The tests also point out the regions of phase space where the catalog does not perform so well and gives users an idea of the limitations of the catalog. Only five of the tests had specific validation critera (cumulative number density, position angle, size and galaxy-galaxy projected correlation function at high redshift). The catalog passed all of these tests. The remaining tests showed varying levels of agreement with the validation data.

In general, we find that the catalog fidelity is reduced for tests that require sub-selections of galaxy samples, such as restrictions in both the magnitude range and color of the sample being tested. This is because more tightly selected galaxy sub-samples place increasing demands on the fidelity of the galaxy-halo connection model. Several tests showed discrepancies for faint galaxies.
Curiously, bright galaxy samples perform quite well in some tests (stellar mass distributions, galaxy-shear correlations) and less well in others (higher $z$ distributions, galaxy-galaxy projected correlations for color selected sub-samples). Only the $g-r$ one-point color distribution is in good qualitative agreement with the data but none of the two-point color distributions agree well. The tests on cluster member galaxies show good qualitative agreement with the observed data but more quantitative comparisons require more precise conversions between the various halo mass definitions. The differences between the simulated and observed galaxy profiles in the outskirts of clusters need further investigation. 

We emphasized the importance of running the catalog-validation program concurrently with the catalog production. Some of the intended uses of the synthetic catalog, such as image simulations, are extremely costly and cannot be easily rerun if problems with the catalog surface at a later date. Ideally, the catalog production process should be flexible enough to
incorporate changes when problems are discovered. Similarly, the test suite and the framework in which it is run should also easily accommodate changes or additions to the tests. Both of these features of the catalog-production and DESCQA validation pipelines were heavily used during the production and validation of cosmoDC2. 

One example of this interplay is the validation test for the presence of a prominent red-sequence of galaxies, which in early iterations of the cosmoDC2 catalog, could not be identified by the \redmapper algorithm. Since cosmoDC2 is built partly on an empirically-tuned model, it was possible to change the model parameters to enhance the red sequence to solve this problem. Another example is the development of the readiness tests during catalog production when it was discovered that outlying values in some galaxy properties were causing the image simulation code to crash. Rapid development and deployment of the readiness tests were made possible by the flexibility of the DESCQA framework that we use to run our test suite.

We anticipate that our validation test suite will be expanded in the future as users pursue more detailed analyses of the catalog data and begin prototyping new scientific analyses. Indeed, many of the tests of cluster member galaxies shown in this paper have been developed only recently as the specialized \redmapper-selected galaxy sample became available.  By the same token, as new observational data becomes available, the range of properties that can be validated by existing tests can also be expanded. Both of these additions to the test suite extend the utility of the catalog. Another area of development concerns the improvement of validation criteria. For many of the tests in our suite, we do not have quantitative validation criteria which the catalog is required to pass. With increasing analysis of the catalog data, users will be able to provide additional and more stringent validation criteria to ensure that future catalogs will better meet their needs.

 Our test suite highlights the areas in which the catalog needs improvement and provides valuable feedback for the catalog producers. It is clear that better methods for generating galaxy SEDs are required. Such methods should include techniques to match higher-point color distributions to incorporate correlations between colors and to obtain more realistic distributions of emission-lines. The results of our tests provide guidance on how to change the galaxy-halo connection model to improve the rendering of galaxy properties in the synthetic catalog. Ideally, the validation procedure is an integral part of the catalog production pipeline, so that the model parameters of the galaxy-halo connection model can be optimized against the validation data before a final version of the catalog is produced. Our test suite provides the data and a guide to the kinds of tests that need to be included in an optimized production pipeline in the future.


\section*{Acknowledgements}
\label{sec:acknowledgements}


This paper has undergone internal review in the LSST Dark Energy Science Collaboration. We thank the internal reviewers Ian Dell'Antonio, Alex Drlica-Wagner and  Andr\'es Plazas Malag\'on. We also thank Peter Brown for reading the manuscript and providing comments and Andrew Hearin for many useful discussions on the results of the validation tests. In addition, we thank Minghao Yue for providing feedback that helped us to identify and correct a bug in the ellipticity distributions in cosmoDC2. We also wish to thank the paper referee for a thorough and thoughtful review and for suggesting a number of changes to improve the clarity and readability of the paper.

The contributions from the authors are listed below. \\
Eve~Kovacs: Co-developed the validation framework, developed and worked on many validation tests, contributed text to the paper and led the writing of the paper.\\
Yao-Yuan Mao:  Co-developed the validation framework, worked on several validation tests to integrate them into the framework, and contributed text to the paper.\\
Michel~Aguena: Developed the magnitude evolution test for consistency of galaxy cluster detection.\\
Anita Bahmanyar: Developed the validation test for the stellar-mass distribution and number density of the CMASS galaxy sample.\\
Adam Broussard: Developed emission line luminosity validation tests and contributed corresponding text to the paper.\\
James Butler: Developed the cluster velocity dispersion test.\\
Duncan Campbell: Developed the cumulative number-density test.\\
Chihway~Chang: Contributed to validation of galaxy-galaxy lensing and edited text related to lensing validation tests.\\
Shenming Fu: Contributed to the validation of galaxy cluster mass-richness relation and edited part of the text related to this test.\\
Katrin Heitmann: Contributed to the text of the paper, helped to develop the overall structure of the paper, provided comments and edits to several subsections of the paper.\\
Danila Korytov: Developed the color-redshift validation tests.\\
Francois Lanusse: Contributed to validation of galaxy-galaxy lensing.\\
Patricia Larsen:  Developed the shear-shear correlation test and contributed to the associated text.\\
Rachel Mandelbaum: Contributed to development of several tests related to lensing or use of HSC data; provided feedback on paper text.\\
Christopher Morrison: Developed the high-redshift clustering test and contributed to the associated text.\\
Constantin Payerne: Contributed to the  galaxy cluster mass reconstruction test using weak gravitational lensing.\\
Marina Ricci: Developed and contributed to the development of several tests related to clusters, and contributed to write sections pertaining to clusters.\\
Eli Rykoff: Provided the redMaPPer catalog that is used in the galaxy cluster tests.\\
F.~Javier Sanchez: Developed galaxy bias test and edited part of the relevant text.\\
Ignacio~Sevilla-Noarbe: Developed and analyzed the color-color test, and edited the associated text.\\
Melanie Simet: Developed the galaxy size test and contributed to the text describing that test.\\
Chun-Hao To: Developed the conditional luminosity function test for redMaPPer clusters and contributed texts describing the test.\\
Vinu Vikraman: Co-developed the projected correlation function and luminosity dependent galaxy size tests and contributed to the text describing these tests.\\
Rongpu Zhou: Developed the color distribution test and supplied the color validation data.



The DESC acknowledges ongoing support from the Institut National de Physique Nucl\'eaire et de Physique des Particules in France; the Science \& Technology Facilities Council in the United Kingdom; and the Department of Energy, the National Science Foundation, and the LSST Corporation in the United States.  DESC uses resources of the IN2P3 Computing Center (CC-IN2P3--Lyon/Villeurbanne - France) funded by the Centre National de la Recherche Scientifique; the National Energy Research Scientific Computing Center, a DOE Office of Science User Facility supported by the Office of Science of the U.S.\ Department of Energy under Contract No.\ DE-AC02-05CH11231; STFC DiRAC HPC Facilities, funded by UK BIS National E-infrastructure capital grants; and the UK particle physics grid, supported by the GridPP Collaboration.  This work was performed in part under DOE Contract DE-AC02-76SF00515.



We thank the developers and maintainers of the following software used at various stages of our catalog-testing suite: 
IPython \citep{ipython},
Jupyter (\href{https://jupyter.org}),
NumPy \citep{numpy},
SciPy \citep{scipy},
h5py (\href{http://www.h5py.org}{h5py.org}),
Matplotlib \citep{matplotlib},
Astropy \citep{astropy},
Healpix \citep{healpix},
CAMB \citep{CAMB},
Colossus \citep{colossus_soft},
DESCQA \citep{descqa-v2.0.0-0.4.5},
Treecorr \citep{treecorr}

\bibliographystyle{yahapj}
\bibliography{references,software}

\begin{thebibliography}{}
\providecommand\natexlab[1]{#1}
\providecommand\JournalTitle[1]{#1}

\bibitem[{{Aguena} {et~al.}(2021{\natexlab{a}}){Aguena}, {Avestruz}, {Combet},
  {Fu}, {Herbonnet}, {Malz}, {Penna-Lima}, {Ricci}, {Vitenti}, {Baumont},
  {Fan}, {Fong}, {Ho}, {Kirby}, {Payerne}, {Boutigny}, {Lee}, {Liu},
  {McClintock}, {Miyatake}, {Sif{\'o}n}, {von der Linden}, {Wu}, {Yoon}, \&
  {The LSST Dark Energy Science Collaboration}}]{aguena2021clmm}
{Aguena}, M., {Avestruz}, C., {Combet}, C., {et~al.} 2021{\natexlab{a}},
  \JournalTitle{arXiv e-prints}, arXiv:2107.10857

\bibitem[{{Aguena} {et~al.}(2021{\natexlab{b}}){Aguena}, {Benoist}, {da Costa},
  {Ogando}, {Gschwend}, {Sampaio-Santos}, {Lima}, {Maia}, {Allam}, {Avila},
  {Bacon}, {Bertin}, {Bhargava}, {Brooks}, {Carnero Rosell}, {Carrasco Kind},
  {Carretero}, {Costanzi}, {De Vicente}, {Desai}, {Diehl}, {Doel}, {Everett},
  {Evrard}, {Ferrero}, {Fert{\'e}}, {Flaugher}, {Fosalba}, {Frieman},
  {Garc{\'\i}a-Bellido}, {Giles}, {Gruendl}, {Gutierrez}, {Hinton},
  {Hollowood}, {Honscheid}, {James}, {Jeltema}, {Kuehn}, {Kuropatkin}, {Lahav},
  {Melchior}, {Miquel}, {Morgan}, {Palmese}, {Paz-Chinch{\'o}n}, {Plazas},
  {Romer}, {Sanchez}, {Santiago}, {Schubnell}, {Serrano}, {Sevilla-Noarbe},
  {Smith}, {Soares-Santos}, {Suchyta}, {Tarle}, {To}, {Tucker}, \&
  {Wilkinson}}]{wazp}
{Aguena}, M., {Benoist}, C., {da Costa}, L.~N., {et~al.} 2021{\natexlab{b}},
  \href{http://dx.doi.org/10.1093/mnras/stab264}{\JournalTitle{\mnras}, 502,
  4435}

\bibitem[{{Agustsson}(2010)}]{Augustsson2010}
{Agustsson}, I. 2010, in American Astronomical Society Meeting Abstracts, Vol.
  215, American Astronomical Society Meeting Abstracts \#215, 375.04

\bibitem[{{Aihara} {et~al.}(2018{\natexlab{a}}){Aihara}, {Armstrong},
  {Bickerton}, {Bosch}, {Coupon}, {Furusawa}, {Hayashi}, {Ikeda}, {Kamata},
  {Karoji}, {Kawanomoto}, {Koike}, {Komiyama}, {Lang}, {Lupton}, {Mineo},
  {Miyatake}, {Miyazaki}, {Morokuma}, {Obuchi}, {Oishi}, {Okura}, {Price},
  {Takata}, {Tanaka}, {Tanaka}, {Tanaka}, {Uchida}, {Uraguchi}, {Utsumi},
  {Wang}, {Yamada}, {Yamanoi}, {Yasuda}, {Arimoto}, {Chiba}, {Finet},
  {Fujimori}, {Fujimoto}, {Furusawa}, {Goto}, {Goulding}, {Gunn}, {Harikane},
  {Hattori}, {Hayashi}, {He{\l}miniak}, {Higuchi}, {Hikage}, {Ho}, {Hsieh},
  {Huang}, {Huang}, {Imanishi}, {Iwata}, {Jaelani}, {Jian}, {Kashikawa},
  {Katayama}, {Kojima}, {Konno}, {Koshida}, {Kusakabe}, {Leauthaud}, {Lee},
  {Lin}, {Lin}, {Mandelbaum}, {Matsuoka}, {Medezinski}, {Miyama}, {Momose},
  {More}, {More}, {Mukae}, {Murata}, {Murayama}, {Nagao}, {Nakata}, {Niida},
  {Niikura}, {Nishizawa}, {Oguri}, {Okabe}, {Ono}, {Onodera}, {Onoue}, {Ouchi},
  {Pyo}, {Shibuya}, {Shimasaku}, {Simet}, {Speagle}, {Spergel}, {Strauss},
  {Sugahara}, {Sugiyama}, {Suto}, {Suzuki}, {Tait}, {Takada}, {Terai}, {Toba},
  {Turner}, {Uchiyama}, {Umetsu}, {Urata}, {Usuda}, {Yeh}, \&
  {Yuma}}]{2018PASJ...70S...8A}
{Aihara}, H., {Armstrong}, R., {Bickerton}, S., {et~al.} 2018{\natexlab{a}},
  \href{http://dx.doi.org/10.1093/pasj/psx081}{\JournalTitle{\pasj}, 70, S8}

\bibitem[{{Aihara} {et~al.}(2018{\natexlab{b}}){Aihara}, {Arimoto},
  {Armstrong}, {Arnouts}, {Bahcall}, {Bickerton}, {Bosch}, {Bundy}, {Capak},
  {Chan}, {Chiba}, {Coupon}, {Egami}, {Enoki}, {Finet}, {Fujimori}, {Fujimoto},
  {Furusawa}, {Furusawa}, {Goto}, {Goulding}, {Greco}, {Greene}, {Gunn},
  {Hamana}, {Harikane}, {Hashimoto}, {Hattori}, {Hayashi}, {Hayashi},
  {He{\l}miniak}, {Higuchi}, {Hikage}, {Ho}, {Hsieh}, {Huang}, {Huang},
  {Ikeda}, {Imanishi}, {Inoue}, {Iwasawa}, {Iwata}, {Jaelani}, {Jian},
  {Kamata}, {Karoji}, {Kashikawa}, {Katayama}, {Kawanomoto}, {Kayo}, {Koda},
  {Koike}, {Kojima}, {Komiyama}, {Konno}, {Koshida}, {Koyama}, {Kusakabe},
  {Leauthaud}, {Lee}, {Lin}, {Lin}, {Lupton}, {Mandelbaum}, {Matsuoka},
  {Medezinski}, {Mineo}, {Miyama}, {Miyatake}, {Miyazaki}, {Momose}, {More},
  {More}, {Moritani}, {Moriya}, {Morokuma}, {Mukae}, {Murata}, {Murayama},
  {Nagao}, {Nakata}, {Niida}, {Niikura}, {Nishizawa}, {Obuchi}, {Oguri},
  {Oishi}, {Okabe}, {Okamoto}, {Okura}, {Ono}, {Onodera}, {Onoue}, {Osato},
  {Ouchi}, {Price}, {Pyo}, {Sako}, {Sawicki}, {Shibuya}, {Shimasaku},
  {Shimono}, {Shirasaki}, {Silverman}, {Simet}, {Speagle}, {Spergel},
  {Strauss}, {Sugahara}, {Sugiyama}, {Suto}, {Suyu}, {Suzuki}, {Tait},
  {Takada}, {Takata}, {Tamura}, {Tanaka}, {Tanaka}, {Tanaka}, {Tanaka},
  {Terai}, {Terashima}, {Toba}, {Tominaga}, {Toshikawa}, {Turner}, {Uchida},
  {Uchiyama}, {Umetsu}, {Uraguchi}, {Urata}, {Usuda}, {Utsumi}, {Wang}, {Wang},
  {Wong}, {Yabe}, {Yamada}, {Yamanoi}, {Yasuda}, {Yeh}, {Yonehara}, \&
  {Yuma}}]{2018PASJ...70S...4A}
{Aihara}, H., {Arimoto}, N., {Armstrong}, R., {et~al.} 2018{\natexlab{b}},
  \href{http://dx.doi.org/10.1093/pasj/psx066}{\JournalTitle{\pasj}, 70, S4}

\bibitem[{{Alam} {et~al.}(2015){Alam}, {Albareti}, {Allende Prieto}, {Anders},
  {Anderson}, {Anderton}, {Andrews}, {Armengaud}, {Aubourg}, {Bailey}, {Basu},
  {Bautista}, {Beaton}, {Beers}, {Bender}, {Berlind}, {Beutler}, {Bhardwaj},
  {Bird}, {Bizyaev}, {Blake}, {Blanton}, {Blomqvist}, {Bochanski}, {Bolton},
  {Bovy}, {Shelden Bradley}, {Brandt}, {Brauer}, {Brinkmann}, {Brown},
  {Brownstein}, {Burden}, {Burtin}, {Busca}, {Cai}, {Capozzi}, {Carnero
  Rosell}, {Carr}, {Carrera}, {Chambers}, {Chaplin}, {Chen}, {Chiappini},
  {Chojnowski}, {Chuang}, {Clerc}, {Comparat}, {Covey}, {Croft}, {Cuesta},
  {Cunha}, {da Costa}, {Da Rio}, {Davenport}, {Dawson}, {De Lee}, {Delubac},
  {Deshpande}, {Dhital}, {Dutra-Ferreira}, {Dwelly}, {Ealet}, {Ebelke},
  {Edmondson}, {Eisenstein}, {Ellsworth}, {Elsworth}, {Epstein}, {Eracleous},
  {Escoffier}, {Esposito}, {Evans}, {Fan}, {Fern{\'a}ndez-Alvar}, {Feuillet},
  {Filiz Ak}, {Finley}, {Finoguenov}, {Flaherty}, {Fleming}, {Font-Ribera},
  {Foster}, {Frinchaboy}, {Galbraith-Frew}, {Garc{\'\i}a},
  {Garc{\'\i}a-Hern{\'a}ndez}, {Garc{\'\i}a P{\'e}rez}, {Gaulme}, {Ge},
  {G{\'e}nova-Santos}, {Georgakakis}, {Ghezzi}, {Gillespie}, {Girardi},
  {Goddard}, {Gontcho}, {Gonz{\'a}lez Hern{\'a}ndez}, {Grebel}, {Green},
  {Grieb}, {Grieves}, {Gunn}, {Guo}, {Harding}, {Hasselquist}, {Hawley},
  {Hayden}, {Hearty}, {Hekker}, {Ho}, {Hogg}, {Holley-Bockelmann}, {Holtzman},
  {Honscheid}, {Huber}, {Huehnerhoff}, {Ivans}, {Jiang}, {Johnson},
  {Kinemuchi}, {Kirkby}, {Kitaura}, {Klaene}, {Knapp}, {Kneib}, {Koenig},
  {Lam}, {Lan}, {Lang}, {Laurent}, {Le Goff}, {Leauthaud}, {Lee}, {Lee},
  {Licquia}, {Liu}, {Long}, {L{\'o}pez-Corredoira}, {Lorenzo-Oliveira},
  {Lucatello}, {Lundgren}, {Lupton}, {Mack}, {Mahadevan}, {Maia}, {Majewski},
  {Malanushenko}, {Malanushenko}, {Manchado}, {Manera}, {Mao}, {Maraston},
  {Marchwinski}, {Margala}, {Martell}, {Martig}, {Masters}, {Mathur},
  {McBride}, {McGehee}, {McGreer}, {McMahon}, {M{\'e}nard}, {Menzel},
  {Merloni}, {M{\'e}sz{\'a}ros}, {Miller}, {Miralda-Escud{\'e}}, {Miyatake},
  {Montero-Dorta}, {More}, {Morganson}, {Morice-Atkinson}, {Morrison},
  {Mosser}, {Muna}, {Myers}, {Nandra}, {Newman}, {Neyrinck}, {Nguyen},
  {Nichol}, {Nidever}, {Noterdaeme}, {Nuza}, {O'Connell}, {O'Connell},
  {O'Connell}, {Ogando}, {Olmstead}, {Oravetz}, {Oravetz}, {Osumi}, {Owen},
  {Padgett}, {Padmanabhan}, {Paegert}, {Palanque-Delabrouille}, {Pan},
  {Parejko}, {P{\^a}ris}, {Park}, {Pattarakijwanich}, {Pellejero-Ibanez},
  {Pepper}, {Percival}, {P{\'e}rez-Fournon}, {P{\textasciiacute}rez-Ra`fols},
  {Petitjean}, {Pieri}, {Pinsonneault}, {Porto de Mello}, {Prada}, {Prakash},
  {Price-Whelan}, {Protopapas}, {Raddick}, {Rahman}, {Reid}, {Rich}, {Rix},
  {Robin}, {Rockosi}, {Rodrigues}, {Rodr{\'\i}guez-Torres}, {Roe}, {Ross},
  {Ross}, {Rossi}, {Ruan}, {Rubi{\~n}o-Mart{\'\i}n}, {Rykoff},
  {Salazar-Albornoz}, {Salvato}, {Samushia}, {S{\'a}nchez}, {Santiago},
  {Sayres}, {Schiavon}, {Schlegel}, {Schmidt}, {Schneider}, {Schultheis},
  {Schwope}, {Sc{\'o}ccola}, {Scott}, {Sellgren}, {Seo}, {Serenelli}, {Shane},
  {Shen}, {Shetrone}, {Shu}, {Silva Aguirre}, {Sivarani}, {Skrutskie},
  {Slosar}, {Smith}, {Sobreira}, {Souto}, {Stassun}, {Steinmetz}, {Stello},
  {Strauss}, {Streblyanska}, {Suzuki}, {Swanson}, {Tan}, {Tayar}, {Terrien},
  {Thakar}, {Thomas}, {Thomas}, {Thompson}, {Tinker}, {Tojeiro}, {Troup},
  {Vargas-Maga{\~n}a}, {Vazquez}, {Verde}, {Viel}, {Vogt}, {Wake}, {Wang},
  {Weaver}, {Weinberg}, {Weiner}, {White}, {Wilson}, {Wisniewski},
  {Wood-Vasey}, {Ye`che}, {York}, {Zakamska}, {Zamora}, {Zasowski}, {Zehavi},
  {Zhao}, {Zheng}, {Zhou}, {Zhou}, {Zou}, \& {Zhu}}]{Alam2015}
{Alam}, S., {Albareti}, F.~D., {Allende Prieto}, C., {et~al.} 2015,
  \href{http://dx.doi.org/10.1088/0067-0049/219/1/12}{\JournalTitle{\apjs},
  219, 12}

\bibitem[{{Alam} {et~al.}(2017){Alam}, {Ata}, {Bailey}, {Beutler}, {Bizyaev},
  {Blazek}, {Bolton}, {Brownstein}, {Burden}, {Chuang}, {Comparat}, {Cuesta},
  {Dawson}, {Eisenstein}, {Escoffier}, {Gil-Mar{\'\i}n}, {Grieb}, {Hand}, {Ho},
  {Kinemuchi}, {Kirkby}, {Kitaura}, {Malanushenko}, {Malanushenko}, {Maraston},
  {McBride}, {Nichol}, {Olmstead}, {Oravetz}, {Padmanabhan},
  {Palanque-Delabrouille}, {Pan}, {Pellejero-Ibanez}, {Percival}, {Petitjean},
  {Prada}, {Price-Whelan}, {Reid}, {Rodr{\'\i}guez-Torres}, {Roe}, {Ross},
  {Ross}, {Rossi}, {Rubi{\~n}o-Mart{\'\i}n}, {Saito}, {Salazar-Albornoz},
  {Samushia}, {S{\'a}nchez}, {Satpathy}, {Schlegel}, {Schneider},
  {Sc{\'o}ccola}, {Seo}, {Sheldon}, {Simmons}, {Slosar}, {Strauss}, {Swanson},
  {Thomas}, {Tinker}, {Tojeiro}, {Maga{\~n}a}, {Vazquez}, {Verde}, {Wake},
  {Wang}, {Weinberg}, {White}, {Wood-Vasey}, {Y{\`e}che}, {Zehavi}, {Zhai}, \&
  {Zhao}}]{alam2017}
{Alam}, S., {Ata}, M., {Bailey}, S., {et~al.} 2017,
  \href{http://dx.doi.org/10.1093/mnras/stx721}{\JournalTitle{\mnras}, 470,
  2617}

\bibitem[{{Albareti} {et~al.}(2017){Albareti}, {Allende Prieto}, {Almeida},
  {Anders}, {Anderson}, {Andrews}, {Arag{\'o}n-Salamanca},
  {Argudo-Fern{\'a}ndez}, {Armengaud}, {Aubourg}, \&
  et~al.}]{2017ApJS..233...25A}
{Albareti}, F.~D., {Allende Prieto}, C., {Almeida}, A., {et~al.} 2017,
  \href{http://dx.doi.org/10.3847/1538-4365/aa8992}{\JournalTitle{\apjs}, 233,
  25}

\bibitem[{{Allen} {et~al.}(2011){Allen}, {Evrard}, \& {Mantz}}]{allen2011}
{Allen}, S.~W., {Evrard}, A.~E., \& {Mantz}, A.~B. 2011,
  \href{http://dx.doi.org/10.1146/annurev-astro-081710-102514}{\JournalTitle{\araa},
  49, 409}

\bibitem[{{Anderson} {et~al.}(2015){Anderson}, {Gaspari}, {White}, {Wang}, \&
  {Dai}}]{Anderson2015}
{Anderson}, M.~E., {Gaspari}, M., {White}, S. D.~M., {Wang}, W., \& {Dai}, X.
  2015, \href{http://dx.doi.org/10.1093/mnras/stv437}{\JournalTitle{\mnras},
  449, 3806}

\bibitem[{Anderson(1962)}]{anderson62}
Anderson, T.~W. 1962,
  \href{http://dx.doi.org/10.1214/aoms/1177704477}{\JournalTitle{Ann. Math.
  Statist.}, 33, 1148}

\bibitem[{{Astropy Collaboration} {et~al.}(2013){Astropy Collaboration},
  {Robitaille}, {Tollerud}, {Greenfield}, {Droettboom}, {Bray}, {Aldcroft},
  {Davis}, {Ginsburg}, {Price-Whelan}, {Kerzendorf}, {Conley}, {Crighton},
  {Barbary}, {Muna}, {Ferguson}, {Grollier}, {Parikh}, {Nair}, {Unther},
  {Deil}, {Woillez}, {Conseil}, {Kramer}, {Turner}, {Singer}, {Fox}, {Weaver},
  {Zabalza}, {Edwards}, {Azalee Bostroem}, {Burke}, {Casey}, {Crawford},
  {Dencheva}, {Ely}, {Jenness}, {Labrie}, {Lim}, {Pierfederici}, {Pontzen},
  {Ptak}, {Refsdal}, {Servillat}, \& {Streicher}}]{astropy}
{Astropy Collaboration}, {Robitaille}, T.~P., {Tollerud}, E.~J., {et~al.} 2013,
  \href{http://dx.doi.org/10.1051/0004-6361/201322068}{\JournalTitle{\aap},
  558, A33}

\bibitem[{{Avila} {et~al.}(2018){Avila}, {Crocce}, {Ross},
  {Garc{\'\i}a-Bellido}, {Percival}, {Banik}, {Camacho}, {Kokron}, {Chan},
  {Andrade-Oliveira}, {Gomes}, {Gomes}, {Lima}, {Rosenfeld}, {Salvador},
  {Friedrich}, {Abdalla}, {Annis}, {Benoit-L{\'e}vy}, {Bertin}, {Brooks},
  {Carrasco Kind}, {Carretero}, {Castander}, {Cunha}, {da Costa}, {Davis}, {De
  Vicente}, {Doel}, {Fosalba}, {Frieman}, {Gerdes}, {Gruen}, {Gruendl},
  {Gutierrez}, {Hartley}, {Hollowood}, {Honscheid}, {James}, {Kuehn},
  {Kuropatkin}, {Miquel}, {Plazas}, {Sanchez}, {Scarpine}, {Schindler},
  {Schubnell}, {Sevilla-Noarbe}, {Smith}, {Sobreira}, {Suchyta}, {Swanson},
  {Tarle}, {Thomas}, {Walker}, \& {Dark Energy Survey
  Collaboration}}]{avila2018}
{Avila}, S., {Crocce}, M., {Ross}, A.~J., {et~al.} 2018,
  \href{http://dx.doi.org/10.1093/mnras/sty1389}{\JournalTitle{\mnras}, 479,
  94}

\bibitem[{{Bardeen} {et~al.}(1986){Bardeen}, {Bond}, {Kaiser}, \&
  {Szalay}}]{1986ApJ...304...15B}
{Bardeen}, J.~M., {Bond}, J.~R., {Kaiser}, N., \& {Szalay}, A.~S. 1986,
  \href{http://dx.doi.org/10.1086/164143}{\JournalTitle{\apj}, 304, 15}

\bibitem[{{Bartelmann} \& {Schneider}(2001)}]{Bartelmann2001}
{Bartelmann}, M., \& {Schneider}, P. 2001,
  \href{http://dx.doi.org/10.1016/S0370-1573(00)00082-X}{\JournalTitle{\physrep},
  340, 291}

\bibitem[{{Becker} \& {Kravtsov}(2011)}]{becker2011}
{Becker}, M.~R., \& {Kravtsov}, A.~V. 2011,
  \href{http://dx.doi.org/10.1088/0004-637X/740/1/25}{\JournalTitle{\apj}, 740,
  25}

\bibitem[{{Beckwith} {et~al.}(2006){Beckwith}, {Stiavelli}, {Koekemoer},
  {Caldwell}, {Ferguson}, {Hook}, {Lucas}, {Bergeron}, {Corbin}, {Jogee},
  {Panagia}, {Robberto}, {Royle}, {Somerville}, \&
  {Sosey}}]{2006AJ....132.1729B}
{Beckwith}, S.~V.~W., {Stiavelli}, M., {Koekemoer}, A.~M., {et~al.} 2006,
  \href{http://dx.doi.org/10.1086/507302}{\JournalTitle{\aj}, 132, 1729}

\bibitem[{{Behroozi} {et~al.}(2019){Behroozi}, {Wechsler}, {Hearin}, \&
  {Conroy}}]{behroozi_etal18}
{Behroozi}, P., {Wechsler}, R.~H., {Hearin}, A.~P., \& {Conroy}, C. 2019,
  \href{http://dx.doi.org/10.1093/mnras/stz1182}{\JournalTitle{\mnras}, 488,
  3143}

\bibitem[{{Ben{\'\i}tez}(2000)}]{Benitez:2000}
{Ben{\'\i}tez}, N. 2000,
  \href{http://dx.doi.org/10.1086/308947}{\JournalTitle{\apj}, 536, 571}

\bibitem[{{Benson}(2012)}]{benson_2010b}
{Benson}, A.~J. 2012,
  \href{http://dx.doi.org/10.1016/j.newast.2011.07.004}{\JournalTitle{\na}, 17,
  175}

\bibitem[{{Bernardeau}(1996)}]{Bernardeau:1995ty}
{Bernardeau}, F. 1996, \JournalTitle{\aap}, 312, 11

\bibitem[{{Blanton} {et~al.}(2005){Blanton}, {Schlegel}, {Strauss},
  {Brinkmann}, {Finkbeiner}, {Fukugita}, {Gunn}, {Hogg}, {Ivezi{\'c}}, {Knapp},
  {Lupton}, {Munn}, {Schneider}, {Tegmark}, \& {Zehavi}}]{Blanton2005}
{Blanton}, M.~R., {Schlegel}, D.~J., {Strauss}, M.~A., {et~al.} 2005,
  \href{http://dx.doi.org/10.1086/429803}{\JournalTitle{\aj}, 129, 2562}

\bibitem[{{Bleem} {et~al.}(2015){Bleem}, {Stalder}, {de Haan}, {Aird}, {Allen},
  {Applegate}, {Ashby}, {Bautz}, {Bayliss}, {Benson}, {Bocquet}, {Brodwin},
  {Carlstrom}, {Chang}, {Chiu}, {Cho}, {Clocchiatti}, {Crawford}, {Crites},
  {Desai}, {Dietrich}, {Dobbs}, {Foley}, {Forman}, {George}, {Gladders},
  {Gonzalez}, {Halverson}, {Hennig}, {Hoekstra}, {Holder}, {Holzapfel},
  {Hrubes}, {Jones}, {Keisler}, {Knox}, {Lee}, {Leitch}, {Liu}, {Lueker},
  {Luong-Van}, {Mantz}, {Marrone}, {McDonald}, {McMahon}, {Meyer}, {Mocanu},
  {Mohr}, {Murray}, {Padin}, {Pryke}, {Reichardt}, {Rest}, {Ruel}, {Ruhl},
  {Saliwanchik}, {Saro}, {Sayre}, {Schaffer}, {Schrabback}, {Shirokoff},
  {Song}, {Spieler}, {Stanford}, {Staniszewski}, {Stark}, {Story}, {Stubbs},
  {Vanderlinde}, {Vieira}, {Vikhlinin}, {Williamson}, {Zahn}, \&
  {Zenteno}}]{bleem2015}
{Bleem}, L.~E., {Stalder}, B., {de Haan}, T., {et~al.} 2015,
  \href{http://dx.doi.org/10.1088/0067-0049/216/2/27}{\JournalTitle{\apjs},
  216, 27}

\bibitem[{{Bleem} {et~al.}(2020){Bleem}, {Bocquet}, {Stalder}, {Gladders},
  {Ade}, {Allen}, {Anderson}, {Annis}, {Ashby}, {Austermann}, {Avila}, {Avva},
  {Bayliss}, {Beall}, {Bechtol}, {Bender}, {Benson}, {Bertin}, {Bianchini},
  {Blake}, {Brodwin}, {Brooks}, {Buckley-Geer}, {Burke}, {Carlstrom}, {Rosell},
  {Carrasco Kind}, {Carretero}, {Chang}, {Chiang}, {Citron}, {Moran},
  {Costanzi}, {Crawford}, {Crites}, {da Costa}, {de Haan}, {De Vicente},
  {Desai}, {Diehl}, {Dietrich}, {Dobbs}, {Eifler}, {Everett}, {Flaugher},
  {Floyd}, {Frieman}, {Gallicchio}, {Garc{\'\i}a-Bellido}, {George}, {Gerdes},
  {Gilbert}, {Gruen}, {Gruendl}, {Gschwend}, {Gupta}, {Gutierrez}, {Halverson},
  {Harrington}, {Henning}, {Heymans}, {Holder}, {Hollowood}, {Holzapfel},
  {Honscheid}, {Hrubes}, {Huang}, {Hubmayr}, {Irwin}, {James}, {Jeltema},
  {Joudaki}, {Khullar}, {Klein}, {Knox}, {Kuropatkin}, {Lee}, {Li}, {Lidman},
  {Lowitz}, {MacCrann}, {Mahler}, {Maia}, {Marshall}, {McDonald}, {McMahon},
  {Melchior}, {Menanteau}, {Meyer}, {Miquel}, {Mocanu}, {Mohr}, {Montgomery},
  {Nadolski}, {Natoli}, {Nibarger}, {Noble}, {Novosad}, {Padin}, {Palmese},
  {Parkinson}, {Patil}, {Paz-Chinch{\'o}n}, {Plazas}, {Pryke}, {Ramachandra},
  {Reichardt}, {Remolina Gonz{\'a}lez}, {Romer}, {Roodman}, {Ruhl}, {Rykoff},
  {Saliwanchik}, {Sanchez}, {Saro}, {Sayre}, {Schaffer}, {Schrabback},
  {Serrano}, {Sharon}, {Sievers}, {Smecher}, {Smith}, {Soares-Santos}, {Stark},
  {Story}, {Suchyta}, {Tarle}, {Tucker}, {Vanderlinde}, {Veach}, {Vieira},
  {Wang}, {Weller}, {Whitehorn}, {Wu}, {Yefremenko}, \& {Zhang}}]{sptpol2020a}
{Bleem}, L.~E., {Bocquet}, S., {Stalder}, B., {et~al.} 2020,
  \href{http://dx.doi.org/10.3847/1538-4365/ab6993}{\JournalTitle{\apjs}, 247,
  25}

\bibitem[{{Bolton} {et~al.}(2012){Bolton}, {Schlegel}, {Aubourg}, {Bailey},
  {Bhardwaj}, {Brownstein}, {Burles}, {Chen}, {Dawson}, {Eisenstein}, {Gunn},
  {Knapp}, {Loomis}, {Lupton}, {Maraston}, {Muna}, {Myers}, {Olmstead},
  {Padmanabhan}, {P{\^a}ris}, {Percival}, {Petitjean}, {Rockosi}, {Ross},
  {Schneider}, {Shu}, {Strauss}, {Thomas}, {Tremonti}, {Wake}, {Weaver}, \&
  {Wood-Vasey}}]{bolton2012}
{Bolton}, A.~S., {Schlegel}, D.~J., {Aubourg}, {\'E}., {et~al.} 2012,
  \href{http://dx.doi.org/10.1088/0004-6256/144/5/144}{\JournalTitle{\aj}, 144,
  144}

\bibitem[{{Bosch} {et~al.}(2018){Bosch}, {Armstrong}, {Bickerton}, {Furusawa},
  {Ikeda}, {Koike}, {Lupton}, {Mineo}, {Price}, {Takata}, {Tanaka}, {Yasuda},
  {AlSayyad}, {Becker}, {Coulton}, {Coupon}, {Garmilla}, {Huang}, {Krughoff},
  {Lang}, {Leauthaud}, {Lim}, {Lust}, {MacArthur}, {Mand elbaum}, {Miyatake},
  {Miyazaki}, {Murata}, {More}, {Okura}, {Owen}, {Swinbank}, {Strauss},
  {Yamada}, \& {Yamanoi}}]{Bosch2018}
{Bosch}, J., {Armstrong}, R., {Bickerton}, S., {et~al.} 2018,
  \href{http://dx.doi.org/10.1093/pasj/psx080}{\JournalTitle{\pasj}, 70, S5}

\bibitem[{{Bosch} {et~al.}(2019){Bosch}, {AlSayyad}, {Armstrong}, {Bellm},
  {Chiang}, {Eggl}, {Findeisen}, {Fisher-Levine}, {Guy}, {Guyonnet},
  {Ivezi{\'c}}, {Jenness}, {Kov{\'a}cs}, {Krughoff}, {Lupton}, {Lust},
  {MacArthur}, {Meyers}, {Moolekamp}, {Morrison}, {Morton}, {O'Mullane},
  {Parejko}, {Plazas}, {Price}, {Rawls}, {Reed}, {Schellart}, {Slater},
  {Sullivan}, {Swinbank}, {Taranu}, {Waters}, \& {Wood-Vasey}}]{rubinpipe2019}
{Bosch}, J., {AlSayyad}, Y., {Armstrong}, R., {et~al.} 2019, in Astronomical
  Society of the Pacific Conference Series, Vol. 523, Astronomical Data
  Analysis Software and Systems XXVII, ed. P.~J. {Teuben}, M.~W. {Pound}, B.~A.
  {Thomas}, \& E.~M. {Warner}, 521

\bibitem[{{Brammer} {et~al.}(2012){Brammer}, {van Dokkum}, {Franx},
  {Fumagalli}, {Patel}, {Rix}, {Skelton}, {Kriek}, {Nelson}, {Schmidt},
  {Bezanson}, {da Cunha}, {Erb}, {Fan}, {F{\"o}rster Schreiber}, {Illingworth},
  {Labb{\'e}}, {Leja}, {Lundgren}, {Magee}, {Marchesini}, {McCarthy},
  {Momcheva}, {Muzzin}, {Quadri}, {Steidel}, {Tal}, {Wake}, {Whitaker}, \&
  {Williams}}]{3D-HST2012}
{Brammer}, G.~B., {van Dokkum}, P.~G., {Franx}, M., {et~al.} 2012,
  \href{http://dx.doi.org/10.1088/0067-0049/200/2/13}{\JournalTitle{\apjs},
  200, 13}

\bibitem[{{Bruzual} \& {Charlot}(2003)}]{bruzual2003}
{Bruzual}, G., \& {Charlot}, S. 2003,
  \href{http://dx.doi.org/10.1046/j.1365-8711.2003.06897.x}{\JournalTitle{\mnras},
  344, 1000}

\bibitem[{{Capak} {et~al.}(2007){Capak}, {Aussel}, {Ajiki}, {McCracken},
  {Mobasher}, {Scoville}, {Shopbell}, {Taniguchi}, {Thompson}, {Tribiano},
  {Sasaki}, {Blain}, {Brusa}, {Carilli}, {Comastri}, {Carollo}, {Cassata},
  {Colbert}, {Ellis}, {Elvis}, {Giavalisco}, {Green}, {Guzzo}, {Hasinger},
  {Ilbert}, {Impey}, {Jahnke}, {Kartaltepe}, {Kneib}, {Koda}, {Koekemoer},
  {Komiyama}, {Leauthaud}, {Le Fevre}, {Lilly}, {Liu}, {Massey}, {Miyazaki},
  {Murayama}, {Nagao}, {Peacock}, {Pickles}, {Porciani}, {Renzini}, {Rhodes},
  {Rich}, {Salvato}, {Sanders}, {Scarlata}, {Schiminovich}, {Schinnerer},
  {Scodeggio}, {Sheth}, {Shioya}, {Tasca}, {Taylor}, {Yan}, \&
  {Zamorani}}]{2007ApJS..172...99C}
{Capak}, P., {Aussel}, H., {Ajiki}, M., {et~al.} 2007,
  \href{http://dx.doi.org/10.1086/519081}{\JournalTitle{\apjs}, 172, 99}

\bibitem[{{Chang} {et~al.}(2013){Chang}, {Jarvis}, {Jain}, {Kahn}, {Kirkby},
  {Connolly}, {Krughoff}, {Peng}, \& {Peterson}}]{Chang2013}
{Chang}, C., {Jarvis}, M., {Jain}, B., {et~al.} 2013,
  \href{http://dx.doi.org/10.1093/mnras/stt1156}{\JournalTitle{\mnras}, 434,
  2121}

\bibitem[{{Chaves-Montero} \& {Hearin}(2020)}]{chaves-montero}
{Chaves-Montero}, J., \& {Hearin}, A. 2020,
  \href{http://dx.doi.org/10.1093/mnras/staa1230}{\JournalTitle{\mnras}, 495,
  2088}

\bibitem[{{Child} {et~al.}(2018){Child}, {Habib}, {Heitmann}, {Frontiere},
  {Finkel}, {Pope}, \& {Morozov}}]{child2018}
{Child}, H.~L., {Habib}, S., {Heitmann}, K., {et~al.} 2018,
  \href{http://dx.doi.org/10.3847/1538-4357/aabf95}{\JournalTitle{\apj}, 859,
  55}

\bibitem[{{Chisari} {et~al.}(2019){Chisari}, {Alonso}, {Krause}, {Leonard},
  {Bull}, {Neveu}, {Villarreal}, {Singh}, {McClintock}, {Ellison}, {Du},
  {Zuntz}, {Mead}, {Joudaki}, {Lorenz}, {Tr{\"o}ster}, {Sanchez}, {Lanusse},
  {Ishak}, {Hlozek}, {Blazek}, {Campagne}, {Almoubayyed}, {Eifler}, {Kirby},
  {Kirkby}, {Plaszczynski}, {Slosar}, {Vrastil}, {Wagoner}, \& {LSST Dark
  Energy Science Collaboration}}]{CCL}
{Chisari}, N.~E., {Alonso}, D., {Krause}, E., {et~al.} 2019,
  \href{http://dx.doi.org/10.3847/1538-4365/ab1658}{\JournalTitle{\apjs}, 242,
  2}

\bibitem[{{Coil} {et~al.}(2004){Coil}, {Newman}, {Kaiser}, {Davis}, {Ma},
  {Kocevski}, \& {Koo}}]{coil}
{Coil}, A.~L., {Newman}, J.~A., {Kaiser}, N., {et~al.} 2004,
  \href{http://dx.doi.org/10.1086/425676}{\JournalTitle{\apj}, 617, 765}

\bibitem[{{Crocce} {et~al.}(2015){Crocce}, {Castander}, {Gazta{\~n}aga},
  {Fosalba}, \& {Carretero}}]{mice2015}
{Crocce}, M., {Castander}, F.~J., {Gazta{\~n}aga}, E., {Fosalba}, P., \&
  {Carretero}, J. 2015,
  \href{http://dx.doi.org/10.1093/mnras/stv1708}{\JournalTitle{\mnras}, 453,
  1513}

\bibitem[{{Cs{\"o}rnyei} {et~al.}(2021){Cs{\"o}rnyei}, {Dobos}, \&
  {Csabai}}]{Csornyei2021}
{Cs{\"o}rnyei}, G., {Dobos}, L., \& {Csabai}, I. 2021,
  \href{http://dx.doi.org/10.1093/mnras/stab261}{\JournalTitle{\mnras}},
  \href{http://arxiv.org/abs/2101.11368}{{\sffamily arXiv:2101.11368
  [astro-ph.GA]}}

\bibitem[{{Davidzon} {et~al.}(2013){Davidzon}, {Bolzonella}, {Coupon},
  {Ilbert}, {Arnouts}, {de la Torre}, {Fritz}, {De Lucia}, {Iovino}, {Granett},
  {Zamorani}, {Guzzo}, {Abbas}, {Adami}, {Bel}, {Bottini}, {Branchini},
  {Cappi}, {Cucciati}, {Franzetti}, {Fumana}, {Garilli}, {Krywult}, {Le Brun},
  {Le F{\`e}vre}, {Maccagni}, {Ma{\l}ek}, {Marulli}, {McCracken}, {Paioro},
  {Peacock}, {Polletta}, {Pollo}, {Schlagenhaufer}, {Scodeggio}, {Tasca},
  {Tojeiro}, {Vergani}, {Zanichelli}, {Burden}, {Di Porto}, {Marchetti},
  {Marinoni}, {Mellier}, {Moscardini}, {Moutard}, {Nichol}, {Percival},
  {Phleps}, \& {Wolk}}]{vipers-smf2013}
{Davidzon}, I., {Bolzonella}, M., {Coupon}, J., {et~al.} 2013,
  \href{http://dx.doi.org/10.1051/0004-6361/201321511}{\JournalTitle{\aap},
  558, A23}

\bibitem[{{Davis} {et~al.}(1978){Davis}, {Geller}, \&
  {Huchra}}]{1978ApJ...221....1D}
{Davis}, M., {Geller}, M.~J., \& {Huchra}, J. 1978,
  \href{http://dx.doi.org/10.1086/156000}{\JournalTitle{\apj}, 221, 1}

\bibitem[{{Dawson} {et~al.}(2013){Dawson}, {Schlegel}, {Ahn}, {Anderson},
  {Aubourg}, {Bailey}, {Barkhouser}, {Bautista}, {Beifiori}, {Berlind},
  {Bhardwaj}, {Bizyaev}, {Blake}, {Blanton}, {Blomqvist}, {Bolton}, {Borde},
  {Bovy}, {Brandt}, {Brewington}, {Brinkmann}, {Brown}, {Brownstein}, {Bundy},
  {Busca}, {Carithers}, {Carnero}, {Carr}, {Chen}, {Comparat}, {Connolly},
  {Cope}, {Croft}, {Cuesta}, {da Costa}, {Davenport}, {Delubac}, {de Putter},
  {Dhital}, {Ealet}, {Ebelke}, {Eisenstein}, {Escoffier}, {Fan}, {Filiz Ak},
  {Finley}, {Font-Ribera}, {G{\'e}nova-Santos}, {Gunn}, {Guo}, {Haggard},
  {Hall}, {Hamilton}, {Harris}, {Harris}, {Ho}, {Hogg}, {Holder}, {Honscheid},
  {Huehnerhoff}, {Jordan}, {Jordan}, {Kauffmann}, {Kazin}, {Kirkby}, {Klaene},
  {Kneib}, {Le Goff}, {Lee}, {Long}, {Loomis}, {Lundgren}, {Lupton}, {Maia},
  {Makler}, {Malanushenko}, {Malanushenko}, {Mandelbaum}, {Manera}, {Maraston},
  {Margala}, {Masters}, {McBride}, {McDonald}, {McGreer}, {McMahon}, {Mena},
  {Miralda-Escud{\'e}}, {Montero-Dorta}, {Montesano}, {Muna}, {Myers},
  {Naugle}, {Nichol}, {Noterdaeme}, {Nuza}, {Olmstead}, {Oravetz}, {Oravetz},
  {Owen}, {Padmanabhan}, {Palanque-Delabrouille}, {Pan}, {Parejko},
  {P{\^a}ris}, {Percival}, {P{\'e}rez-Fournon}, {P{\'e}rez-R{\`a}fols},
  {Petitjean}, {Pfaffenberger}, {Pforr}, {Pieri}, {Prada}, {Price-Whelan},
  {Raddick}, {Rebolo}, {Rich}, {Richards}, {Rockosi}, {Roe}, {Ross}, {Ross},
  {Rossi}, {Rubi{\~n}o-Martin}, {Samushia}, {S{\'a}nchez}, {Sayres}, {Schmidt},
  {Schneider}, {Sc{\'o}ccola}, {Seo}, {Shelden}, {Sheldon}, {Shen}, {Shu},
  {Slosar}, {Smee}, {Snedden}, {Stauffer}, {Steele}, {Strauss}, {Streblyanska},
  {Suzuki}, {Swanson}, {Tal}, {Tanaka}, {Thomas}, {Tinker}, {Tojeiro},
  {Tremonti}, {Vargas Maga{\~n}a}, {Verde}, {Viel}, {Wake}, {Watson}, {Weaver},
  {Weinberg}, {Weiner}, {West}, {White}, {Wood-Vasey}, {Yeche}, {Zehavi},
  {Zhao}, \& {Zheng}}]{Dawson2013}
{Dawson}, K.~S., {Schlegel}, D.~J., {Ahn}, C.~P., {et~al.} 2013,
  \href{http://dx.doi.org/10.1088/0004-6256/145/1/10}{\JournalTitle{\aj}, 145,
  10}

\bibitem[{Dawson {et~al.}(2015)Dawson, Schneider, Tyson, \& Jee}]{Dawson2015}
Dawson, W.~A., Schneider, M.~D., Tyson, J.~A., \& Jee, M.~J. 2015,
  \href{http://dx.doi.org/10.3847/0004-637x/816/1/11}{\JournalTitle{The
  Astrophysical Journal}, 816, 11}

\bibitem[{{de la Torre} {et~al.}(2013){de la Torre}, {Guzzo}, {Peacock},
  {Branchini}, {Iovino}, {Granett}, {Abbas}, {Adami}, {Arnouts}, {Bel},
  {Bolzonella}, {Bottini}, {Cappi}, {Coupon}, {Cucciati}, {Davidzon}, {De
  Lucia}, {Fritz}, {Franzetti}, {Fumana}, {Garilli}, {Ilbert}, {Krywult}, {Le
  Brun}, {Le F{\`e}vre}, {Maccagni}, {Ma{\l}ek}, {Marulli}, {McCracken},
  {Moscardini}, {Paioro}, {Percival}, {Polletta}, {Pollo}, {Schlagenhaufer},
  {Scodeggio}, {Tasca}, {Tojeiro}, {Vergani}, {Zanichelli}, {Burden}, {Di
  Porto}, {Marchetti}, {Marinoni}, {Mellier}, {Monaco}, {Nichol}, {Phleps},
  {Wolk}, \& {Zamorani}}]{vipers-wprp2013}
{de la Torre}, S., {Guzzo}, L., {Peacock}, J.~A., {et~al.} 2013,
  \href{http://dx.doi.org/10.1051/0004-6361/201321463}{\JournalTitle{\aap},
  557, A54}

\bibitem[{{DeRose} {et~al.}(2019){DeRose}, {Wechsler}, {Becker}, {Busha},
  {Rykoff}, {MacCrann}, {Erickson}, {Evrard}, {Kravtsov}, {Gruen}, {Allam},
  {Avila}, {Bridle}, {Brooks}, {Buckley-Geer}, {Carnero Rosell}, {Carrasco
  Kind}, {Carretero}, {Castander}, {Cawthon}, {Crocce}, {da Costa}, {Davis},
  {De Vicente}, {Dietrich}, {Doel}, {Drlica-Wagner}, {Fosalba}, {Frieman},
  {Garcia-Bellido}, {Gutierrez}, {Hartley}, {Hollowood}, {Hoyle}, {James},
  {Krause}, {Kuehn}, {Kuropatkin}, {Lima}, {Maia}, {Menanteau}, {Miller},
  {Miquel}, {Ogando}, {Plazas Malag{\'o}n}, {Romer}, {Sanchez}, {Schindler},
  {Serrano}, {Sevilla-Noarbe}, {Smith}, {Suchyta}, {Swanson}, {Tarle}, \&
  {Vikram}}]{derose2019}
{DeRose}, J., {Wechsler}, R.~H., {Becker}, M.~R., {et~al.} 2019,
  \JournalTitle{arXiv e-prints}, arXiv:1901.02401

\bibitem[{{DeRose} {et~al.}(2021){DeRose}, {Wechsler}, {Becker}, {Rykoff},
  {Pandey}, {MacCrann}, {Amon}, {Myles}, {Krause}, {Gruen}, {Jain}, {Troxel},
  {Prat}, {Alarcon}, {S{\'a}nchez}, {Blazek}, {Crocce}, {Giannini}, {Gatti},
  {Bernstein}, {Zuntz}, {Dodelson}, {Fang}, {Friedrich}, {Secco},
  {Elvin-Poole}, {Everett}, {Choi}, {Harrison}, {Cordero}, {Rodriguez-Monroy},
  {McCullough}, {Cawthon}, {Chen}, {Alves}, {Camacho}, {Campos}, {Diehl},
  {Drlica-Wagner}, {Eifler}, {Fosalba}, {Huang}, {Porredon}, {Raveri},
  {Rosenfeld}, {Ross}, {Sanchez}, {Sheldon}, {Yanny}, {Yin}, {Aguena}, {Allam},
  {Andrade-Oliveira}, {Annis}, {Avila}, {Bacon}, {Bechtol}, {Bhargava},
  {Brooks}, {Buckley-Geer}, {Burke}, {Carnero Rosell}, {Carrasco Kind},
  {Chang}, {Costanzi}, {da Costa}, {Pereira}, {De Vicente}, {Desai},
  {Dietrich}, {Doel}, {Eckert}, {Evrard}, {Ferrero}, {Fert{\'e}}, {Flaugher},
  {Frieman}, {Garc{\'\i}a-Bellido}, {Gaztanaga}, {Giannantonio}, {Gruendl},
  {Gschwend}, {Gutierrez}, {Hartley}, {Hinton}, {Hollowood}, {Honscheid},
  {Huff}, {Huterer}, {James}, {Kuehn}, {Kuropatkin}, {Lahav}, {Lima}, {Maia},
  {Marshall}, {Melchior}, {Menanteau}, {Miquel}, {Mohr}, {Morgan}, {Palmese},
  {Paz-Chinch{\'o}n}, {Pieres}, {Plazas Malag{\'o}n}, {Sanchez}, {Scarpine},
  {Serrano}, {Sevilla-Noarbe}, {Smith}, {Soares-Santos}, {Suchyta}, {Tarle},
  {Thomas}, {To}, {Varga}, \& {Zhang}}]{derose2021}
---. 2021, \JournalTitle{arXiv e-prints}, arXiv:2105.13547

\bibitem[{{DES Collaboration}(2018)}]{des32pt}
{DES Collaboration}. 2018,
  \href{http://dx.doi.org/10.1103/PhysRevD.98.043526}{\JournalTitle{\prd}, 98,
  043526}

\bibitem[{{DES Collaboration}(2021)}]{descollaboration2021dark}
---. 2021, Dark Energy Survey Year 3 Results: Cosmological Constraints from
  Galaxy Clustering and Weak Lensing,
  \href{http://arxiv.org/abs/2105.13549}{{\sffamily arXiv:2105.13549
  [astro-ph.CO]}}

\bibitem[{{Desjacques} {et~al.}(2018){Desjacques}, {Jeong}, \&
  {Schmidt}}]{desjacques2018}
{Desjacques}, V., {Jeong}, D., \& {Schmidt}, F. 2018,
  \href{http://dx.doi.org/10.1016/j.physrep.2017.12.002}{\JournalTitle{\physrep},
  733, 1}

\bibitem[{{Diemer}(2015)}]{colossus_soft}
{Diemer}, B. 2015, {Colossus: COsmology, haLO, and large-Scale StrUcture toolS}

\bibitem[{{Diemer}(2018)}]{colossus}
---. 2018,
  \href{http://dx.doi.org/10.3847/1538-4365/aaee8c}{\JournalTitle{\apjs}, 239,
  35}

\bibitem[{Dobrushin(1970)}]{dobrushin1970definition}
Dobrushin, R.~L. 1970, \JournalTitle{Teoriya Veroyatnostei i ee Primeneniya},
  15, 469

\bibitem[{{Dressler}(1980)}]{1980ApJ...236..351D}
{Dressler}, A. 1980,
  \href{http://dx.doi.org/10.1086/157753}{\JournalTitle{\apj}, 236, 351}

\bibitem[{Duffy {et~al.}(2008)Duffy, Schaye, Kay, \& Dalla~Vecchia}]{Duffy2008}
Duffy, A.~R., Schaye, J., Kay, S.~T., \& Dalla~Vecchia, C. 2008,
  \JournalTitle{Monthly Notices of the Royal Astronomical Society: Letters}

\bibitem[{{Eriksen} \& {Gaztanaga}(2015)}]{Eriksen:2015hqa}
{Eriksen}, M., \& {Gaztanaga}, E. 2015, \JournalTitle{ArXiv e-prints},
  arXiv:1508.00035

\bibitem[{{Euclid Collaboration}(2019)}]{2019A&A...627A..23E}
{Euclid Collaboration}. 2019,
  \href{http://dx.doi.org/10.1051/0004-6361/201935088}{\JournalTitle{\aap},
  627, A23}

\bibitem[{{Evrard} {et~al.}(2008){Evrard}, {Bialek}, {Busha}, {White}, {Habib},
  {Heitmann}, {Warren}, {Rasia}, {Tormen}, {Moscardini}, {Power}, {Jenkins},
  {Gao}, {Frenk}, {Springel}, {White}, \& {Diemand}}]{evrard2008}
{Evrard}, A.~E., {Bialek}, J., {Busha}, M., {et~al.} 2008,
  \href{http://dx.doi.org/10.1086/521616}{\JournalTitle{\apj}, 672, 122}

\bibitem[{{Fenech Conti} {et~al.}(2017){Fenech Conti}, {Herbonnet}, {Hoekstra},
  {Merten}, {Miller}, \& {Viola}}]{FenechConti2017}
{Fenech Conti}, I., {Herbonnet}, R., {Hoekstra}, H., {et~al.} 2017,
  \href{http://dx.doi.org/10.1093/mnras/stx200}{\JournalTitle{\mnras}, 467,
  1627}

\bibitem[{{Fioc} \& {Rocca-Volmerange}(1997)}]{Fio97}
{Fioc}, M., \& {Rocca-Volmerange}, B. 1997, \JournalTitle{\aap}, 326, 950

\bibitem[{{Gifford} {et~al.}(2013){Gifford}, {Miller}, \& {Kern}}]{gifford2013}
{Gifford}, D., {Miller}, C., \& {Kern}, N. 2013,
  \href{http://dx.doi.org/10.1088/0004-637X/773/2/116}{\JournalTitle{\apj},
  773, 116}

\bibitem[{{Girelli} {et~al.}(2020){Girelli}, {Pozzetti}, {Bolzonella},
  {Giocoli}, {Marulli}, \& {Baldi}}]{girelli2020}
{Girelli}, G., {Pozzetti}, L., {Bolzonella}, M., {et~al.} 2020,
  \href{http://dx.doi.org/10.1051/0004-6361/201936329}{\JournalTitle{\aap},
  634, A135}

\bibitem[{{Gladders} \& {Yee}(2000)}]{gladders2000}
{Gladders}, M.~D., \& {Yee}, H.~K.~C. 2000,
  \href{http://dx.doi.org/10.1086/301557}{\JournalTitle{\aj}, 120, 2148}

\bibitem[{{G{\'o}rski} {et~al.}(2005){G{\'o}rski}, {Hivon}, {Banday},
  {Wandelt}, {Hansen}, {Reinecke}, \& {Bartelmann}}]{healpix}
{G{\'o}rski}, K.~M., {Hivon}, E., {Banday}, A.~J., {et~al.} 2005,
  \href{http://dx.doi.org/10.1086/427976}{\JournalTitle{\apj}, 622, 759}

\bibitem[{Gretton {et~al.}(2012)Gretton, Borgwardt, Rasch, Sch{\"o}lkopf, \&
  Smola}]{gretton2012kernel}
Gretton, A., Borgwardt, K.~M., Rasch, M.~J., Sch{\"o}lkopf, B., \& Smola, A.
  2012, \JournalTitle{The Journal of Machine Learning Research}, 13, 723

\bibitem[{{Grogin} {et~al.}(2011){Grogin}, {Kocevski}, {Faber}, {Ferguson},
  {Koekemoer}, {Riess}, {Acquaviva}, {Alexander}, {Almaini}, {Ashby}, {Barden},
  {Bell}, {Bournaud}, {Brown}, {Caputi}, {Casertano}, {Cassata}, {Castellano},
  {Challis}, {Chary}, {Cheung}, {Cirasuolo}, {Conselice}, {Roshan Cooray},
  {Croton}, {Daddi}, {Dahlen}, {Dav{\'e}}, {de Mello}, {Dekel}, {Dickinson},
  {Dolch}, {Donley}, {Dunlop}, {Dutton}, {Elbaz}, {Fazio}, {Filippenko},
  {Finkelstein}, {Fontana}, {Gardner}, {Garnavich}, {Gawiser}, {Giavalisco},
  {Grazian}, {Guo}, {Hathi}, {H{\"a}ussler}, {Hopkins}, {Huang}, {Huang},
  {Jha}, {Kartaltepe}, {Kirshner}, {Koo}, {Lai}, {Lee}, {Li}, {Lotz}, {Lucas},
  {Madau}, {McCarthy}, {McGrath}, {McIntosh}, {McLure}, {Mobasher},
  {Moustakas}, {Mozena}, {Nandra}, {Newman}, {Niemi}, {Noeske}, {Papovich},
  {Pentericci}, {Pope}, {Primack}, {Rajan}, {Ravindranath}, {Reddy}, {Renzini},
  {Rix}, {Robaina}, {Rodney}, {Rosario}, {Rosati}, {Salimbeni}, {Scarlata},
  {Siana}, {Simard}, {Smidt}, {Somerville}, {Spinrad}, {Straughn}, {Strolger},
  {Telford}, {Teplitz}, {Trump}, {van der Wel}, {Villforth}, {Wechsler},
  {Weiner}, {Wiklind}, {Wild}, {Wilson}, {Wuyts}, {Yan}, \&
  {Yun}}]{candels2011b}
{Grogin}, N.~A., {Kocevski}, D.~D., {Faber}, S.~M., {et~al.} 2011,
  \href{http://dx.doi.org/10.1088/0067-0049/197/2/35}{\JournalTitle{\apjs},
  197, 35}

\bibitem[{{Gruen} {et~al.}(2018){Gruen}, {Friedrich}, {Krause}, {DeRose},
  {Cawthon}, {Davis}, {Elvin-Poole}, {Rykoff}, {Wechsler}, {Alarcon},
  {Bernstein}, {Blazek}, {Chang}, {Clampitt}, {Crocce}, {De Vicente}, {Gatti},
  {Gill}, {Hartley}, {Hilbert}, {Hoyle}, {Jain}, {Jarvis}, {Lahav}, {MacCrann},
  {McClintock}, {Prat}, {Rollins}, {Ross}, {Rozo}, {Samuroff}, {S{\'a}nchez},
  {Sheldon}, {Troxel}, {Zuntz}, {Abbott}, {Abdalla}, {Allam}, {Annis},
  {Bechtol}, {Benoit-L{\'e}vy}, {Bertin}, {Bridle}, {Brooks}, {Buckley-Geer},
  {Carnero Rosell}, {Carrasco Kind}, {Carretero}, {Cunha}, {D'Andrea}, {da
  Costa}, {Desai}, {Diehl}, {Dietrich}, {Doel}, {Drlica-Wagner}, {Fernandez},
  {Flaugher}, {Fosalba}, {Frieman}, {Garc{\'\i}a-Bellido}, {Gaztanaga},
  {Giannantonio}, {Gruendl}, {Gschwend}, {Gutierrez}, {Honscheid}, {James},
  {Jeltema}, {Kuehn}, {Kuropatkin}, {Lima}, {March}, {Marshall}, {Martini},
  {Melchior}, {Menanteau}, {Miquel}, {Mohr}, {Plazas}, {Roodman}, {Sanchez},
  {Scarpine}, {Schubnell}, {Sevilla-Noarbe}, {Smith}, {Smith}, {Soares-Santos},
  {Sobreira}, {Swanson}, {Tarle}, {Thomas}, {Vikram}, {Walker}, {Weller},
  {Zhang}, \& {DES Collaboration}}]{2017arXiv171005045G}
{Gruen}, D., {Friedrich}, O., {Krause}, E., {et~al.} 2018,
  \href{http://dx.doi.org/10.1103/PhysRevD.98.023507}{\JournalTitle{\prd}, 98,
  023507}

\bibitem[{{Guo} {et~al.}(2012){Guo}, {Cole}, {Eke}, \& {Frenk}}]{guo2012}
{Guo}, Q., {Cole}, S., {Eke}, V., \& {Frenk}, C. 2012,
  \href{http://dx.doi.org/10.1111/j.1365-2966.2012.21882.x}{\JournalTitle{\mnras},
  427, 428}

\bibitem[{{Hearin} {et~al.}(2020){Hearin}, {Korytov}, {Kovacs}, {Benson},
  {Aung}, {Bradshaw}, {Campbell}, \& {LSST Dark Energy Science
  Collaboration}}]{hearin2020}
{Hearin}, A., {Korytov}, D., {Kovacs}, E., {et~al.} 2020,
  \href{http://dx.doi.org/10.1093/mnras/staa1495}{\JournalTitle{\mnras}, 495,
  5040}

\bibitem[{{Heitmann} {et~al.}(2019){Heitmann}, {Finkel}, {Pope}, {Morozov},
  {Frontiere}, {Habib}, {Rangel}, {Uram}, {Korytov}, {Child}, {Flender},
  {Insley}, \& {Rizzi}}]{outerrim2019}
{Heitmann}, K., {Finkel}, H., {Pope}, A., {et~al.} 2019,
  \href{http://dx.doi.org/10.3847/1538-4365/ab4da1}{\JournalTitle{\apjs}, 245,
  16}

\bibitem[{{Hennig} {et~al.}(2017){Hennig}, {Mohr}, {Zenteno}, {Desai},
  {Dietrich}, {Bocquet}, {Strazzullo}, {Saro}, {Abbott}, {Abdalla}, {Bayliss},
  {Benoit-L{\'e}vy}, {Bernstein}, {Bertin}, {Brooks}, {Capasso}, {Capozzi},
  {Carnero}, {Carrasco Kind}, {Carretero}, {Chiu}, {D'Andrea}, {daCosta},
  {Diehl}, {Doel}, {Eifler}, {Evrard}, {Fausti-Neto}, {Fosalba}, {Frieman},
  {Gangkofner}, {Gonzalez}, {Gruen}, {Gruendl}, {Gupta}, {Gutierrez},
  {Honscheid}, {Hlavacek-Larrondo}, {James}, {Kuehn}, {Kuropatkin}, {Lahav},
  {March}, {Marshall}, {Martini}, {McDonald}, {Melchior}, {Miller}, {Miquel},
  {Neilsen}, {Nord}, {Ogando}, {Plazas}, {Reichardt}, {Romer}, {Rozo},
  {Rykoff}, {Sanchez}, {Santiago}, {Schubnell}, {Sevilla-Noarbe}, {Smith},
  {Soares-Santos}, {Sobreira}, {Stalder}, {Stanford}, {Suchyta}, {Swanson},
  {Tarle}, {Thomas}, {Vikram}, {Walker}, \& {Zhang}}]{2017MNRAS.467.4015H}
{Hennig}, C., {Mohr}, J.~J., {Zenteno}, A., {et~al.} 2017,
  \href{http://dx.doi.org/10.1093/mnras/stx175}{\JournalTitle{\mnras}, 467,
  4015}

\bibitem[{{Hilton} {et~al.}(2021){Hilton}, {Sif{\'o}n}, {Naess},
  {Madhavacheril}, {Oguri}, {Rozo}, {Rykoff}, {Abbott}, {Adhikari}, {Aguena},
  {Aiola}, {Allam}, {Amodeo}, {Amon}, {Annis}, {Ansarinejad}, {Aros-Bunster},
  {Austermann}, {Avila}, {Bacon}, {Battaglia}, {Beall}, {Becker}, {Bernstein},
  {Bertin}, {Bhandarkar}, {Bhargava}, {Bond}, {Brooks}, {Burke}, {Calabrese},
  {Carrasco Kind}, {Carretero}, {Choi}, {Choi}, {Conselice}, {da Costa},
  {Costanzi}, {Crichton}, {Crowley}, {D{\"u}nner}, {Denison}, {Devlin},
  {Dicker}, {Diehl}, {Dietrich}, {Doel}, {Duff}, {Duivenvoorden}, {Dunkley},
  {Everett}, {Ferraro}, {Ferrero}, {Fert{\'e}}, {Flaugher}, {Frieman},
  {Gallardo}, {Garc{\'\i}a-Bellido}, {Gaztanaga}, {Gerdes}, {Giles}, {Golec},
  {Gralla}, {Grandis}, {Gruen}, {Gruendl}, {Gschwend}, {Gutierrez}, {Han},
  {Hartley}, {Hasselfield}, {Hill}, {Hilton}, {Hincks}, {Hinton}, {Ho},
  {Honscheid}, {Hoyle}, {Hubmayr}, {Huffenberger}, {Hughes}, {Jaelani}, {Jain},
  {James}, {Jeltema}, {Kent}, {Knowles}, {Koopman}, {Kuehn}, {Lahav}, {Lima},
  {Lin}, {Lokken}, {Loubser}, {MacCrann}, {Maia}, {Marriage}, {Martin},
  {McMahon}, {Melchior}, {Menanteau}, {Miquel}, {Miyatake}, {Moodley},
  {Morgan}, {Mroczkowski}, {Nati}, {Newburgh}, {Niemack}, {Nishizawa},
  {Ogando}, {Orlowski-Scherer}, {Page}, {Palmese}, {Partridge},
  {Paz-Chinch{\'o}n}, {Phakathi}, {Plazas}, {Robertson}, {Romer}, {Carnero
  Rosell}, {Salatino}, {Sanchez}, {Schaan}, {Schillaci}, {Sehgal}, {Serrano},
  {Shin}, {Simon}, {Smith}, {Soares-Santos}, {Spergel}, {Staggs}, {Storer},
  {Suchyta}, {Swanson}, {Tarle}, {Thomas}, {To}, {Trac}, {Ullom}, {Vale}, {Van
  Lanen}, {Vavagiakis}, {De Vicente}, {Wilkinson}, {Wollack}, {Xu}, \&
  {Zhang}}]{atacama2021}
{Hilton}, M., {Sif{\'o}n}, C., {Naess}, S., {et~al.} 2021,
  \href{http://dx.doi.org/10.3847/1538-4365/abd023}{\JournalTitle{\apjs}, 253,
  3}

\bibitem[{Hoekstra {et~al.}(2017)Hoekstra, Viola, \& Herbonnet}]{Hoekstra2017}
Hoekstra, H., Viola, M., \& Herbonnet, R. 2017,
  \href{http://dx.doi.org/10.1093/mnras/stx724}{\JournalTitle{Monthly Notices
  of the Royal Astronomical Society}, 468, 3295–3311}

\bibitem[{{Huang} {et~al.}(2020){Huang}, {Bleem}, {Stalder}, {Ade}, {Allen},
  {Anderson}, {Austermann}, {Avva}, {Beall}, {Bender}, {Benson}, {Bianchini},
  {Bocquet}, {Brodwin}, {Carlstrom}, {Chang}, {Chiang}, {Citron}, {Moran},
  {Crawford}, {Crites}, {Haan}, {Dobbs}, {Everett}, {Floyd}, {Gallicchio},
  {George}, {Gilbert}, {Gladders}, {Guns}, {Gupta}, {Halverson}, {Harrington},
  {Henning}, {Hilton}, {Holder}, {Holzapfel}, {Hrubes}, {Hubmayr}, {Irwin},
  {Khullar}, {Knox}, {Lee}, {Li}, {Lowitz}, {McDonald}, {McMahon}, {Meyer},
  {Mocanu}, {Montgomery}, {Nadolski}, {Natoli}, {Nibarger}, {Noble}, {Novosad},
  {Padin}, {Patil}, {Pryke}, {Reichardt}, {Ruhl}, {Saliwanchik}, {Saro},
  {Sayre}, {Schaffer}, {Sharon}, {Sievers}, {Smecher}, {Stark}, {Story},
  {Tucker}, {Vanderlinde}, {Veach}, {Vieira}, {Wang}, {Whitehorn}, {Wu}, \&
  {Yefremenko}}]{sptpol2020b}
{Huang}, N., {Bleem}, L.~E., {Stalder}, B., {et~al.} 2020,
  \href{http://dx.doi.org/10.3847/1538-3881/ab6a96}{\JournalTitle{\aj}, 159,
  110}

\bibitem[{{Hudelot} {et~al.}(2012){Hudelot}, {Cuillandre}, {Withington},
  {Goranova}, {McCracken}, {Magnard}, {Mellier}, {Regnault}, {Betoule},
  {Aussel}, {Kavelaars}, {Fernique}, {Bonnarel}, {Ochsenbein}, \&
  {Ilbert}}]{CFHTLS}
{Hudelot}, P., {Cuillandre}, J.-C., {Withington}, K., {et~al.} 2012,
  \JournalTitle{VizieR Online Data Catalog}, 2317,
  \url{http://irfu.cea.fr/Images/astImg/3226/T0007-ExecSummary.pdf}

\bibitem[{Hunter(2007)}]{matplotlib}
Hunter, J.~D. 2007,
  \href{http://dx.doi.org/10.1109/MCSE.2007.55}{\JournalTitle{Computing in
  Science Engineering}, 9, 90}

\bibitem[{{Ivezi{\'c}} {et~al.}(2019){Ivezi{\'c}}, {Kahn}, {Tyson}, {Abel},
  {Acosta}, {Allsman}, {Alonso}, {AlSayyad}, {Anderson}, {Andrew}, {Angel},
  {Angeli}, {Ansari}, {Antilogus}, {Araujo}, {Armstrong}, {Arndt}, {Astier},
  {Aubourg}, {Auza}, {Axelrod}, {Bard}, {Barr}, {Barrau}, {Bartlett}, {Bauer},
  {Bauman}, {Baumont}, {Bechtol}, {Bechtol}, {Becker}, {Becla}, {Beldica},
  {Bellavia}, {Bianco}, {Biswas}, {Blanc}, {Blazek}, {Blandford}, {Bloom},
  {Bogart}, {Bond}, {Booth}, {Borgland}, {Borne}, {Bosch}, {Boutigny},
  {Brackett}, {Bradshaw}, {Brandt}, {Brown}, {Bullock}, {Burchat}, {Burke},
  {Cagnoli}, {Calabrese}, {Callahan}, {Callen}, {Carlin}, {Carlson},
  {Chandrasekharan}, {Charles-Emerson}, {Chesley}, {Cheu}, {Chiang}, {Chiang},
  {Chirino}, {Chow}, {Ciardi}, {Claver}, {Cohen-Tanugi}, {Cockrum}, {Coles},
  {Connolly}, {Cook}, {Cooray}, {Covey}, {Cribbs}, {Cui}, {Cutri}, {Daly},
  {Daniel}, {Daruich}, {Daubard}, {Daues}, {Dawson}, {Delgado}, {Dellapenna},
  {de Peyster}, {de Val-Borro}, {Digel}, {Doherty}, {Dubois},
  {Dubois-Felsmann}, {Durech}, {Economou}, {Eifler}, {Eracleous}, {Emmons},
  {Fausti Neto}, {Ferguson}, {Figueroa}, {Fisher-Levine}, {Focke}, {Foss},
  {Frank}, {Freemon}, {Gangler}, {Gawiser}, {Geary}, {Gee}, {Geha}, {Gessner},
  {Gibson}, {Gilmore}, {Glanzman}, {Glick}, {Goldina}, {Goldstein}, {Goodenow},
  {Graham}, {Gressler}, {Gris}, {Guy}, {Guyonnet}, {Haller}, {Harris},
  {Hascall}, {Haupt}, {Hernandez}, {Herrmann}, {Hileman}, {Hoblitt}, {Hodgson},
  {Hogan}, {Howard}, {Huang}, {Huffer}, {Ingraham}, {Innes}, {Jacoby}, {Jain},
  {Jammes}, {Jee}, {Jenness}, {Jernigan}, {Jevremovi{\'c}}, {Johns}, {Johnson},
  {Johnson}, {Jones}, {Juramy-Gilles}, {Juri{\'c}}, {Kalirai}, {Kallivayalil},
  {Kalmbach}, {Kantor}, {Karst}, {Kasliwal}, {Kelly}, {Kessler}, {Kinnison},
  {Kirkby}, {Knox}, {Kotov}, {Krabbendam}, {Krughoff}, {Kub{\'a}nek},
  {Kuczewski}, {Kulkarni}, {Ku}, {Kurita}, {Lage}, {Lambert}, {Lange},
  {Langton}, {Le Guillou}, {Levine}, {Liang}, {Lim}, {Lintott}, {Long},
  {Lopez}, {Lotz}, {Lupton}, {Lust}, {MacArthur}, {Mahabal}, {Mandelbaum},
  {Markiewicz}, {Marsh}, {Marshall}, {Marshall}, {May}, {McKercher}, {McQueen},
  {Meyers}, {Migliore}, {Miller}, {Mills}, {Miraval}, {Moeyens}, {Moolekamp},
  {Monet}, {Moniez}, {Monkewitz}, {Montgomery}, {Morrison}, {Mueller},
  {Muller}, {Mu{\~n}oz Arancibia}, {Neill}, {Newbry}, {Nief}, {Nomerotski},
  {Nordby}, {O'Connor}, {Oliver}, {Olivier}, {Olsen}, {O'Mullane}, {Ortiz},
  {Osier}, {Owen}, {Pain}, {Palecek}, {Parejko}, {Parsons}, {Pease},
  {Peterson}, {Peterson}, {Petravick}, {Libby Petrick}, {Petry},
  {Pierfederici}, {Pietrowicz}, {Pike}, {Pinto}, {Plante}, {Plate}, {Plutchak},
  {Price}, {Prouza}, {Radeka}, {Rajagopal}, {Rasmussen}, {Regnault}, {Reil},
  {Reiss}, {Reuter}, {Ridgway}, {Riot}, {Ritz}, {Robinson}, {Roby}, {Roodman},
  {Rosing}, {Roucelle}, {Rumore}, {Russo}, {Saha}, {Sassolas}, {Schalk},
  {Schellart}, {Schindler}, {Schmidt}, {Schneider}, {Schneider}, {Schoening},
  {Schumacher}, {Schwamb}, {Sebag}, {Selvy}, {Sembroski}, {Seppala}, {Serio},
  {Serrano}, {Shaw}, {Shipsey}, {Sick}, {Silvestri}, {Slater}, {Smith},
  {Smith}, {Sobhani}, {Soldahl}, {Storrie-Lombardi}, {Stover}, {Strauss},
  {Street}, {Stubbs}, {Sullivan}, {Sweeney}, {Swinbank}, {Szalay}, {Takacs},
  {Tether}, {Thaler}, {Thayer}, {Thomas}, {Thornton}, {Thukral}, {Tice},
  {Trilling}, {Turri}, {Van Berg}, {Vanden Berk}, {Vetter}, {Virieux},
  {Vucina}, {Wahl}, {Walkowicz}, {Walsh}, {Walter}, {Wang}, {Wang}, {Warner},
  {Wiecha}, {Willman}, {Winters}, {Wittman}, {Wolff}, {Wood-Vasey}, {Wu},
  {Xin}, {Yoachim}, \& {Zhan}}]{ivezic2019}
{Ivezi{\'c}}, {\v{Z}}., {Kahn}, S.~M., {Tyson}, J.~A., {et~al.} 2019,
  \href{http://dx.doi.org/10.3847/1538-4357/ab042c}{\JournalTitle{\apj}, 873,
  111}

\bibitem[{Izbicki \& Lee(2017)}]{Izbicki:17}
Izbicki, R., \& Lee, A.~B. 2017,
  \href{http://dx.doi.org/10.1214/17-EJS1302}{\JournalTitle{Electron. J.
  Statist.}, 11, 2800}

\bibitem[{{Jarvis}(2015)}]{treecorr}
{Jarvis}, M. 2015, {TreeCorr: Two-point correlation functions}, Astrophysics
  Source Code Library

\bibitem[{{Jarvis} {et~al.}(2015){Jarvis}, {Sheldon}, {Zuntz}, {Bridle},
  {Kacprzak}, \& {Dark Energy Survey Collaboration}}]{jarvis2015}
{Jarvis}, M., {Sheldon}, E., {Zuntz}, J., {et~al.} 2015, in APS Meeting
  Abstracts, Vol. 2015, APS April Meeting Abstracts, Y2.005

\bibitem[{{Joachimi} {et~al.}(2013){Joachimi}, {Semboloni}, {Bett}, {Hartlap},
  {Hilbert}, {Hoekstra}, {Schneider}, \& {Schrabback}}]{Joachimi2013}
{Joachimi}, B., {Semboloni}, E., {Bett}, P.~E., {et~al.} 2013,
  \href{http://dx.doi.org/10.1093/mnras/stt172}{\JournalTitle{\mnras}, 431,
  477}

\bibitem[{{Joachimi} {et~al.}(2021){Joachimi}, {Lin}, {Asgari}, {Tr{\"o}ster},
  {Heymans}, {Hildebrandt}, {K{\"o}hlinger}, {S{\'a}nchez}, {Wright},
  {Bilicki}, {Blake}, {van den Busch}, {Crocce}, {Dvornik}, {Erben}, {Getman},
  {Giblin}, {Hoekstra}, {Kannawadi}, {Kuijken}, {Napolitano}, {Schneider},
  {Scoccimarro}, {Sellentin}, {Shan}, {von Wietersheim-Kramsta}, \&
  {Zuntz}}]{kids1000J}
{Joachimi}, B., {Lin}, C.~A., {Asgari}, M., {et~al.} 2021,
  \href{http://dx.doi.org/10.1051/0004-6361/202038831}{\JournalTitle{\aap},
  646, A129}

\bibitem[{Jones {et~al.}(2001--)Jones, Oliphant, Peterson, {et~al.}}]{scipy}
Jones, E., Oliphant, T., Peterson, P., {et~al.} 2001--, {SciPy}: Open source
  scientific tools for {Python}, [Online;
  \href{http://www.scipy.org/}{scipy.org}]

\bibitem[{{Kaiser}(1984)}]{Kaiser:1984sw}
{Kaiser}, N. 1984, \href{http://dx.doi.org/10.1086/184341}{\JournalTitle{\apj},
  284, L9}

\bibitem[{{Kelly} {et~al.}(2010){Kelly}, {Hicken}, {Burke}, {Mandel}, \&
  {Kirshner}}]{kelly2010}
{Kelly}, P.~L., {Hicken}, M., {Burke}, D.~L., {Mandel}, K.~S., \& {Kirshner},
  R.~P. 2010,
  \href{http://dx.doi.org/10.1088/0004-637X/715/2/743}{\JournalTitle{\apj},
  715, 743}

\bibitem[{{Koekemoer} {et~al.}(2011){Koekemoer}, {Faber}, {Ferguson}, {Grogin},
  {Kocevski}, {Koo}, {Lai}, {Lotz}, {Lucas}, {McGrath}, {Ogaz}, {Rajan},
  {Riess}, {Rodney}, {Strolger}, {Casertano}, {Castellano}, {Dahlen},
  {Dickinson}, {Dolch}, {Fontana}, {Giavalisco}, {Grazian}, {Guo}, {Hathi},
  {Huang}, {van der Wel}, {Yan}, {Acquaviva}, {Alexander}, {Almaini}, {Ashby},
  {Barden}, {Bell}, {Bournaud}, {Brown}, {Caputi}, {Cassata}, {Challis},
  {Chary}, {Cheung}, {Cirasuolo}, {Conselice}, {Roshan Cooray}, {Croton},
  {Daddi}, {Dav{\'e}}, {de Mello}, {de Ravel}, {Dekel}, {Donley}, {Dunlop},
  {Dutton}, {Elbaz}, {Fazio}, {Filippenko}, {Finkelstein}, {Frazer}, {Gardner},
  {Garnavich}, {Gawiser}, {Gruetzbauch}, {Hartley}, {H{\"a}ussler},
  {Herrington}, {Hopkins}, {Huang}, {Jha}, {Johnson}, {Kartaltepe},
  {Khostovan}, {Kirshner}, {Lani}, {Lee}, {Li}, {Madau}, {McCarthy},
  {McIntosh}, {McLure}, {McPartland}, {Mobasher}, {Moreira}, {Mortlock},
  {Moustakas}, {Mozena}, {Nandra}, {Newman}, {Nielsen}, {Niemi}, {Noeske},
  {Papovich}, {Pentericci}, {Pope}, {Primack}, {Ravindranath}, {Reddy},
  {Renzini}, {Rix}, {Robaina}, {Rosario}, {Rosati}, {Salimbeni}, {Scarlata},
  {Siana}, {Simard}, {Smidt}, {Snyder}, {Somerville}, {Spinrad}, {Straughn},
  {Telford}, {Teplitz}, {Trump}, {Vargas}, {Villforth}, {Wagner}, {Wandro},
  {Wechsler}, {Weiner}, {Wiklind}, {Wild}, {Wilson}, {Wuyts}, \&
  {Yun}}]{candels2011a}
{Koekemoer}, A.~M., {Faber}, S.~M., {Ferguson}, H.~C., {et~al.} 2011,
  \href{http://dx.doi.org/10.1088/0067-0049/197/2/36}{\JournalTitle{\apjs},
  197, 36}

\bibitem[{{Korytov} {et~al.}(2019){Korytov}, {Hearin}, {Kovacs}, {Larsen},
  {Rangel}, {Hollowed}, {Benson}, {Heitmann}, {Mao}, {Bahmanyar}, {Chang},
  {Campbell}, {DeRose}, {Finkel}, {Frontiere}, {Gawiser}, {Habib}, {Joachimi},
  {Lanusse}, {Li}, {Mandelbaum}, {Morrison}, {Newman}, {Pope}, {Rykoff},
  {Simet}, {To}, {Vikraman}, {Wechsler}, {White}, \& {(The LSST Dark Energy
  Science Collaboration}}]{cosmodc2}
{Korytov}, D., {Hearin}, A., {Kovacs}, E., {et~al.} 2019,
  \href{http://dx.doi.org/10.3847/1538-4365/ab510c}{\JournalTitle{\apjs}, 245,
  26}

\bibitem[{{Lackner} \& {Gunn}(2012)}]{2012MNRAS.421.2277L}
{Lackner}, C.~N., \& {Gunn}, J.~E. 2012,
  \href{http://dx.doi.org/10.1111/j.1365-2966.2012.20450.x}{\JournalTitle{\mnras},
  421, 2277}

\bibitem[{{Laigle} {et~al.}(2016){Laigle}, {McCracken}, {Ilbert}, {Hsieh},
  {Davidzon}, {Capak}, {Hasinger}, {Silverman}, {Pichon}, {Coupon}, {Aussel},
  {Le Borgne}, {Caputi}, {Cassata}, {Chang}, {Civano}, {Dunlop}, {Fynbo},
  {Kartaltepe}, {Koekemoer}, {Le F{\`e}vre}, {Le Floc'h}, {Leauthaud}, {Lilly},
  {Lin}, {Marchesi}, {Milvang-Jensen}, {Salvato}, {Sanders}, {Scoville},
  {Smolcic}, {Stockmann}, {Taniguchi}, {Tasca}, {Toft}, {Vaccari}, \&
  {Zabl}}]{laigle2016}
{Laigle}, C., {McCracken}, H.~J., {Ilbert}, O., {et~al.} 2016,
  \href{http://dx.doi.org/10.3847/0067-0049/224/2/24}{\JournalTitle{\apjs},
  224, 24}

\bibitem[{{Lampeitl} {et~al.}(2010){Lampeitl}, {Smith}, {Nichol}, {Bassett},
  {Cinabro}, {Dilday}, {Foley}, {Frieman}, {Garnavich}, {Goobar}, {Im}, {Jha},
  {Marriner}, {Miquel}, {Nordin}, {{\"O}stman}, {Riess}, {Sako}, {Schneider},
  {Sollerman}, \& {Stritzinger}}]{lampeitl2010}
{Lampeitl}, H., {Smith}, M., {Nichol}, R.~C., {et~al.} 2010,
  \href{http://dx.doi.org/10.1088/0004-637X/722/1/566}{\JournalTitle{\apj},
  722, 566}

\bibitem[{{Landy} \& {Szalay}(1993)}]{1993ApJ...412...64L}
{Landy}, S.~D., \& {Szalay}, A.~S. 1993,
  \href{http://dx.doi.org/10.1086/172900}{\JournalTitle{\apj}, 412, 64}

\bibitem[{{Le F{\`e}vre} {et~al.}(2013){Le F{\`e}vre}, {Cassata}, {Cucciati},
  {Garilli}, {Ilbert}, {Le Brun}, {Maccagni}, {Moreau}, {Scodeggio}, {Tresse},
  {Zamorani}, {Adami}, {Arnouts}, {Bardelli}, {Bolzonella}, {Bondi},
  {Bongiorno}, {Bottini}, {Cappi}, {Charlot}, {Ciliegi}, {Contini}, {de la
  Torre}, {Foucaud}, {Franzetti}, {Gavignaud}, {Guzzo}, {Iovino}, {Lemaux},
  {L{\'o}pez-Sanjuan}, {McCracken}, {Marano}, {Marinoni}, {Mazure}, {Mellier},
  {Merighi}, {Merluzzi}, {Paltani}, {Pell{\`o}}, {Pollo}, {Pozzetti},
  {Scaramella}, {Tasca}, {Vergani}, {Vettolani}, {Zanichelli}, \&
  {Zucca}}]{vvds2013}
{Le F{\`e}vre}, O., {Cassata}, P., {Cucciati}, O., {et~al.} 2013,
  \href{http://dx.doi.org/10.1051/0004-6361/201322179}{\JournalTitle{\aap},
  559, A14}

\bibitem[{{Le F{\`e}vre} {et~al.}(2014){Le F{\`e}vre}, {Adami}, {Arnouts},
  {Bardelli}, {Bolzonella}, {Bondi}, {Bongiorno}, {Bottini}, {Cappi},
  {Cassata}, {Charlot}, {Ciliegi}, {Contini}, {Cucciati}, {de la Torre},
  {Foucaud}, {Franzetti}, {Garilli}, {Gavignaud}, {Guzzo}, {Ilbert}, {Iovino},
  {Le Brun}, {Lemaux}, {L{\'o}pez-Sanjuan}, {Maccagni}, {McCracken}, {Marano},
  {Marinoni}, {Mazure}, {Mellier}, {Merighi}, {Merluzzi}, {Moreau}, {Paltani},
  {Pell{\`o}}, {Pollo}, {Pozzetti}, {Scaramella}, {Scodeggio}, {Tasca},
  {Tresse}, {Vergani}, {Vettolani}, {Zamorani}, {Zanichelli}, \&
  {Zucca}}]{vimos2014}
{Le F{\`e}vre}, O., {Adami}, C., {Arnouts}, S., {et~al.} 2014,
  \JournalTitle{The Messenger}, 155, 33

\bibitem[{Lewis {et~al.}(2000)Lewis, Challinor, \& Lasenby}]{CAMB}
Lewis, A., Challinor, A., \& Lasenby, A. 2000,
  \href{http://dx.doi.org/10.1086/309179}{\JournalTitle{\apj}, 538, 473}

\bibitem[{{Lilly} {et~al.}(2003){Lilly}, {Carollo}, \& {Stockton}}]{Lilly2003}
{Lilly}, S.~J., {Carollo}, C.~M., \& {Stockton}, A.~N. 2003,
  \href{http://dx.doi.org/10.1086/378389}{\JournalTitle{\apj}, 597, 730}

\bibitem[{{Lilly} {et~al.}(2007){Lilly}, {Le F{\`e}vre}, {Renzini}, {Zamorani},
  {Scodeggio}, {Contini}, {Carollo}, {Hasinger}, {Kneib}, {Iovino}, {Le Brun},
  {Maier}, {Mainieri}, {Mignoli}, {Silverman}, {Tasca}, {Bolzonella},
  {Bongiorno}, {Bottini}, {Capak}, {Caputi}, {Cimatti}, {Cucciati}, {Daddi},
  {Feldmann}, {Franzetti}, {Garilli}, {Guzzo}, {Ilbert}, {Kampczyk}, {Kovac},
  {Lamareille}, {Leauthaud}, {Le Borgne}, {McCracken}, {Marinoni}, {Pello},
  {Ricciardelli}, {Scarlata}, {Vergani}, {Sanders}, {Schinnerer}, {Scoville},
  {Taniguchi}, {Arnouts}, {Aussel}, {Bardelli}, {Brusa}, {Cappi}, {Ciliegi},
  {Finoguenov}, {Foucaud}, {Franceschini}, {Halliday}, {Impey}, {Knobel},
  {Koekemoer}, {Kurk}, {Maccagni}, {Maddox}, {Marano}, {Marconi}, {Meneux},
  {Mobasher}, {Moreau}, {Peacock}, {Porciani}, {Pozzetti}, {Scaramella},
  {Schiminovich}, {Shopbell}, {Smail}, {Thompson}, {Tresse}, {Vettolani},
  {Zanichelli}, \& {Zucca}}]{lilly2007}
{Lilly}, S.~J., {Le F{\`e}vre}, O., {Renzini}, A., {et~al.} 2007,
  \href{http://dx.doi.org/10.1086/516589}{\JournalTitle{\apjs}, 172, 70}

\bibitem[{{Lin} {et~al.}(2020){Lin}, {Tinker}, {Klypin}, {Prada}, {Blanton},
  {Comparat}, {Dawson}, {de Mattia}, {du Mas des Bourboux}, {Percival},
  {Raichoor}, {Rossi}, {Smith}, \& {Zhao}}]{lin2020}
{Lin}, S., {Tinker}, J.~L., {Klypin}, A., {et~al.} 2020,
  \href{http://dx.doi.org/10.1093/mnras/staa2571}{\JournalTitle{\mnras}, 498,
  5251}

\bibitem[{{Lin} {et~al.}(2006){Lin}, {Mohr}, {Gonzalez}, \&
  {Stanford}}]{2006ApJ...650L..99L}
{Lin}, Y.-T., {Mohr}, J.~J., {Gonzalez}, A.~H., \& {Stanford}, S.~A. 2006,
  \href{http://dx.doi.org/10.1086/508940}{\JournalTitle{\apjl}, 650, L99}

\bibitem[{{LSST Dark Energy Science Collaboration}(2012)}]{lsst-desc2012}
{LSST Dark Energy Science Collaboration}. 2012, \JournalTitle{arXiv e-prints},
  arXiv:1211.0310

\bibitem[{{LSST Dark Energy Science Collaboration}(2017)}]{descqa-v2.0.0-0.4.5}
---. 2017, {DESCQA}, v2.0.0-0.4.5.
  \href{https://doi.org/10.5281/zenodo.1120294}{doi:10.5281/zenodo.1120294}

\bibitem[{{LSST Dark Energy Science Collaboration}(2020)}]{DC2_survey_paper}
---. 2020, \JournalTitle{arXiv e-prints}, arXiv:2010.05926

\bibitem[{{LSST Science Collaboration} {et~al.}(2009){LSST Science
  Collaboration}, {Abell}, {Allison}, {Anderson}, {Andrew}, {Angel}, {Armus},
  {Arnett}, {Asztalos}, {Axelrod}, \& et~al.}]{lsst}
{LSST Science Collaboration}, {Abell}, P.~A., {Allison}, J., {et~al.} 2009,
  \JournalTitle{arXiv e-prints},
  \href{http://arxiv.org/abs/0912.0201}{{\sffamily arXiv:0912.0201
  [astro-ph.IM]}}

\bibitem[{{Luki{\'c}} {et~al.}(2009){Luki{\'c}}, {Reed}, {Habib}, \&
  {Heitmann}}]{lukic2009}
{Luki{\'c}}, Z., {Reed}, D., {Habib}, S., \& {Heitmann}, K. 2009,
  \href{http://dx.doi.org/10.1088/0004-637X/692/1/217}{\JournalTitle{\apj},
  692, 217}

\bibitem[{{Lupton} {et~al.}(2001){Lupton}, {Gunn}, {Ivezi{\'c}}, {Knapp}, \&
  {Kent}}]{lupton2001}
{Lupton}, R., {Gunn}, J.~E., {Ivezi{\'c}}, Z., {Knapp}, G.~R., \& {Kent}, S.
  2001, in Astronomical Society of the Pacific Conference Series, Vol. 238,
  Astronomical Data Analysis Software and Systems X, ed. J.~{Harnden}, F.~R.,
  F.~A. {Primini}, \& H.~E. {Payne}, 269

\bibitem[{{MacCrann} {et~al.}(2020){MacCrann}, {Becker}, {McCullough}, {Amon},
  {Gruen}, {Jarvis}, {Choi}, {Troxel}, {Sheldon}, {Yanny}, {Herner},
  {Dodelson}, {Zuntz}, {Eckert}, {Rollins}, {Varga}, {Bernstein}, {Gruendl},
  {Harrison}, {Hartley}, {Sevilla-Noarbe}, {Pieres}, {Bridle}, {Myles},
  {Alarcon}, {Everett}, {S{\'a}nchez}, {Huff}, {Tarsitano}, {Gatti}, {Secco},
  {Abbott}, {Aguena}, {Allam}, {Annis}, {Bacon}, {Bertin}, {Brooks}, {Burke},
  {Carnero Rosell}, {Carrasco Kind}, {Carretero}, {Costanzi}, {Crocce},
  {Pereira}, {De Vicente}, {Desai}, {Diehl}, {Dietrich}, {Doel}, {Eifler},
  {Ferrero}, {Fert{\'e}}, {Flaugher}, {Fosalba}, {Frieman},
  {Garc{\'\i}a-Bellido}, {Gaztanaga}, {Gerdes}, {Giannantonio}, {Gschwend},
  {Gutierrez}, {Hinton}, {Hollowood}, {Honscheid}, {James}, {Lahav}, {Lima},
  {Maia}, {March}, {Marshall}, {Martini}, {Melchior}, {Menanteau}, {Miquel},
  {Mohr}, {Morgan}, {Muir}, {Ogando}, {Palmese}, {Paz-Chinch{\'o}n}, {Plazas},
  {Rodriguez-Monroy}, {Roodman}, {Samuroff}, {Sanchez}, {Scarpine}, {Serrano},
  {Smith}, {Soares-Santos}, {Suchyta}, {Swanson}, {Tarle}, {Thomas}, {To}, \&
  {Wilkinson}}]{maccrann2020}
{MacCrann}, N., {Becker}, M.~R., {McCullough}, J., {et~al.} 2020,
  \JournalTitle{arXiv e-prints}, arXiv:2012.08567

\bibitem[{{Mandelbaum}(2018)}]{Mandelbaum2018review}
{Mandelbaum}, R. 2018,
  \href{http://dx.doi.org/10.1146/annurev-astro-081817-051928}{\JournalTitle{\araa},
  56, 393}

\bibitem[{{Mandelbaum} {et~al.}(2012){Mandelbaum}, {Hirata}, {Leauthaud},
  {Massey}, \& {Rhodes}}]{cosmos2012}
{Mandelbaum}, R., {Hirata}, C.~M., {Leauthaud}, A., {Massey}, R.~J., \&
  {Rhodes}, J. 2012,
  \href{http://dx.doi.org/10.1111/j.1365-2966.2011.20138.x}{\JournalTitle{\mnras},
  420, 1518}

\bibitem[{{Mandelbaum} {et~al.}(2006){Mandelbaum}, {Seljak}, {Kauffmann},
  {Hirata}, \& {Brinkmann}}]{Mandelbaum2006}
{Mandelbaum}, R., {Seljak}, U., {Kauffmann}, G., {Hirata}, C.~M., \&
  {Brinkmann}, J. 2006,
  \href{http://dx.doi.org/10.1111/j.1365-2966.2006.10156.x}{\JournalTitle{\mnras},
  368, 715}

\bibitem[{{Mandelbaum} {et~al.}(2016){Mandelbaum}, {Wang}, {Zu}, {White},
  {Henriques}, \& {More}}]{Mandelbaum2016}
{Mandelbaum}, R., {Wang}, W., {Zu}, Y., {et~al.} 2016,
  \href{http://dx.doi.org/10.1093/mnras/stw188}{\JournalTitle{\mnras}, 457,
  3200}

\bibitem[{{Mandelbaum} {et~al.}(2014){Mandelbaum}, {Rowe}, {Bosch}, {Chang},
  {Courbin}, {Gill}, {Jarvis}, {Kannawadi}, {Kacprzak}, \&
  {Lackner}}]{2014ApJS..212....5M}
{Mandelbaum}, R., {Rowe}, B., {Bosch}, J., {et~al.} 2014,
  \href{http://dx.doi.org/10.1088/0067-0049/212/1/5}{\JournalTitle{\apjs}, 212,
  5}

\bibitem[{{Mandelbaum} {et~al.}(2018){Mandelbaum}, {Lanusse}, {Leauthaud},
  {Armstrong}, {Simet}, {Miyatake}, {Meyers}, {Bosch}, {Murata}, {Miyazaki}, \&
  {Tanaka}}]{Mandelbaum2018}
{Mandelbaum}, R., {Lanusse}, F., {Leauthaud}, A., {et~al.} 2018,
  \href{http://dx.doi.org/10.1093/mnras/sty2420}{\JournalTitle{\mnras}, 481,
  3170}

\bibitem[{{Manera} \& {Gazta{\~n}aga}(2011)}]{2011MNRAS.415..383M}
{Manera}, M., \& {Gazta{\~n}aga}, E. 2011,
  \href{http://dx.doi.org/10.1111/j.1365-2966.2011.18705.x}{\JournalTitle{\mnras},
  415, 383}

\bibitem[{{Mantz} {et~al.}(2015){Mantz}, {von der Linden}, {Allen},
  {Applegate}, {Kelly}, {Morris}, {Rapetti}, {Schmidt}, {Adhikari}, {Allen},
  {Burchat}, {Burke}, {Cataneo}, {Donovan}, {Ebeling}, {Shandera}, \&
  {Wright}}]{mantz2015}
{Mantz}, A.~B., {von der Linden}, A., {Allen}, S.~W., {et~al.} 2015,
  \href{http://dx.doi.org/10.1093/mnras/stu2096}{\JournalTitle{\mnras}, 446,
  2205}

\bibitem[{{Mao} {et~al.}(2018){Mao}, {Kovacs}, {Heitmann}, {Uram}, {Benson},
  {Campbell}, {Cora}, {DeRose}, {Di Matteo}, {Habib}, {Hearin}, {Bryce
  Kalmbach}, {Krughoff}, {Lanusse}, {Luki{\'c}}, {Mandelbaum}, {Newman},
  {Padilla}, {Paillas}, {Pope}, {Ricker}, {Ruiz}, {Tenneti},
  {Vega-Mart{\'\i}nez}, {Wechsler}, {Zhou}, {Zu}, \& {LSST Dark Energy Science
  Collaboration}}]{descqa}
{Mao}, Y.-Y., {Kovacs}, E., {Heitmann}, K., {et~al.} 2018,
  \href{http://dx.doi.org/10.3847/1538-4365/aaa6c3}{\JournalTitle{\apjs}, 234,
  36}

\bibitem[{Martinet {et~al.}(2019)Martinet, Schrabback, Hoekstra, Tewes,
  Herbonnet, Schneider, Hernandez-Martin, Taylor, Brinchmann, \&
  et~al.}]{Martinet2019}
Martinet, N., Schrabback, T., Hoekstra, H., {et~al.} 2019,
  \href{http://dx.doi.org/10.1051/0004-6361/201935187}{\JournalTitle{Astronomy
  \& Astrophysics}, 627, A59}

\bibitem[{{Marulli} {et~al.}(2013){Marulli}, {Bolzonella}, {Branchini},
  {Davidzon}, {de la Torre}, {Granett}, {Guzzo}, {Iovino}, {Moscardini},
  {Pollo}, {Abbas}, {Adami}, {Arnouts}, {Bel}, {Bottini}, {Cappi}, {Coupon},
  {Cucciati}, {De Lucia}, {Fritz}, {Franzetti}, {Fumana}, {Garilli}, {Ilbert},
  {Krywult}, {Le Brun}, {Le F{\`e}vre}, {Maccagni}, {Ma{\l}ek}, {McCracken},
  {Paioro}, {Polletta}, {Schlagenhaufer}, {Scodeggio}, {Tasca}, {Tojeiro},
  {Vergani}, {Zanichelli}, {Burden}, {Di Porto}, {Marchetti}, {Marinoni},
  {Mellier}, {Nichol}, {Peacock}, {Percival}, {Phleps}, {Wolk}, \&
  {Zamorani}}]{marulli2013}
{Marulli}, F., {Bolzonella}, M., {Branchini}, E., {et~al.} 2013,
  \href{http://dx.doi.org/10.1051/0004-6361/201321476}{\JournalTitle{\aap},
  557, A17}

\bibitem[{{McClintock} {et~al.}(2019){McClintock}, {Varga}, {Gruen}, {Rozo},
  {Rykoff}, {Shin}, {Melchior}, {DeRose}, {Seitz}, {Dietrich}, {Sheldon},
  {Zhang}, {von der Linden}, {Jeltema}, {Mantz}, {Romer}, {Allen}, {Becker},
  {Bermeo}, {Bhargava}, {Costanzi}, {Everett}, {Farahi}, {Hamaus}, {Hartley},
  {Hollowood}, {Hoyle}, {Israel}, {Li}, {MacCrann}, {Morris}, {Palmese},
  {Plazas}, {Pollina}, {Rau}, {Simet}, {Soares-Santos}, {Troxel}, {Vergara
  Cervantes}, {Wechsler}, {Zuntz}, {Abbott}, {Abdalla}, {Allam}, {Annis},
  {Avila}, {Bridle}, {Brooks}, {Burke}, {Carnero Rosell}, {Carrasco Kind},
  {Carretero}, {Castander}, {Crocce}, {Cunha}, {D'Andrea}, {da Costa}, {Davis},
  {De Vicente}, {Diehl}, {Doel}, {Drlica-Wagner}, {Evrard}, {Flaugher},
  {Fosalba}, {Frieman}, {Garc{\'\i}a-Bellido}, {Gaztanaga}, {Gerdes},
  {Giannantonio}, {Gruendl}, {Gutierrez}, {Honscheid}, {James}, {Kirk},
  {Krause}, {Kuehn}, {Lahav}, {Li}, {Lima}, {March}, {Marshall}, {Menanteau},
  {Miquel}, {Mohr}, {Nord}, {Ogando}, {Roodman}, {Sanchez}, {Scarpine},
  {Schindler}, {Sevilla-Noarbe}, {Smith}, {Smith}, {Sobreira}, {Suchyta},
  {Swanson}, {Tarle}, {Tucker}, {Vikram}, {Walker}, {Weller}, \& {DES
  Collaboration}}]{mcclintock2019}
{McClintock}, T., {Varga}, T.~N., {Gruen}, D., {et~al.} 2019,
  \href{http://dx.doi.org/10.1093/mnras/sty2711}{\JournalTitle{\mnras}, 482,
  1352}

\bibitem[{{McQuinn} \& {White}(2013)}]{mcquinn2013}
{McQuinn}, M., \& {White}, M. 2013,
  \href{http://dx.doi.org/10.1093/mnras/stt914}{\JournalTitle{\mnras}, 433,
  2857}

\bibitem[{{Medezinski} {et~al.}(2018){Medezinski}, {Oguri}, {Nishizawa},
  {Speagle}, {Miyatake}, {Umetsu}, {Leauthaud}, {Murata}, {Mandelbaum},
  {Sif{\'o}n}, {Strauss}, {Huang}, {Simet}, {Okabe}, {Tanaka}, \&
  {Komiyama}}]{medezinski18}
{Medezinski}, E., {Oguri}, M., {Nishizawa}, A.~J., {et~al.} 2018,
  \href{http://dx.doi.org/10.1093/pasj/psy009}{\JournalTitle{\pasj}, 70, 30}

\bibitem[{{Melchior} {et~al.}(2018){Melchior}, {Moolekamp}, {Jerdee},
  {Armstrong}, {Sun}, {Bosch}, \& {Lupton}}]{Melchior2018}
{Melchior}, P., {Moolekamp}, F., {Jerdee}, M., {et~al.} 2018,
  \href{http://dx.doi.org/10.1016/j.ascom.2018.07.001}{\JournalTitle{Astronomy
  and Computing}, 24, 129}

\bibitem[{{Mostek} {et~al.}(2013){Mostek}, {Coil}, {Cooper}, {Davis}, {Newman},
  \& {Weiner}}]{mostek13}
{Mostek}, N., {Coil}, A.~L., {Cooper}, M., {et~al.} 2013,
  \href{http://dx.doi.org/10.1088/0004-637X/767/1/89}{\JournalTitle{\apj}, 767,
  89}

\bibitem[{{Moustakas} {et~al.}(2013){Moustakas}, {Coil}, {Aird}, {Blanton},
  {Cool}, {Eisenstein}, {Mendez}, {Wong}, {Zhu}, \& {Arnouts}}]{moustakas2013}
{Moustakas}, J., {Coil}, A.~L., {Aird}, J., {et~al.} 2013,
  \href{http://dx.doi.org/10.1088/0004-637X/767/1/50}{\JournalTitle{\apj}, 767,
  50}

\bibitem[{{Nakajima} {et~al.}(2013){Nakajima}, {Ouchi}, {Shimasaku},
  {Hashimoto}, {Ono}, \& {Lee}}]{Nakajima2013}
{Nakajima}, K., {Ouchi}, M., {Shimasaku}, K., {et~al.} 2013,
  \href{http://dx.doi.org/10.1088/0004-637X/769/1/3}{\JournalTitle{\apj}, 769,
  3}

\bibitem[{{Navarro} {et~al.}(1997){Navarro}, {Frenk}, \&
  {White}}]{1997ApJ...490..493N}
{Navarro}, J.~F., {Frenk}, C.~S., \& {White}, S. D.~M. 1997,
  \href{http://dx.doi.org/10.1086/304888}{\JournalTitle{\apj}, 490, 493}

\bibitem[{{Newman} {et~al.}(2019){Newman}, {Blazek}, {Chisari}, {Clowe},
  {Dell'Antonio}, {Gawiser}, {Hlo{\v{z}}ek}, {Kim}, {von der Linden},
  {Lochner}, {Mand elbaum}, {Medezinski}, {Melchior}, {Newman}, {S{\'a}nchez},
  {Schmidt}, {Singh}, \& {LSST Dark Energy Science Collaboration}}]{Newman2019}
{Newman}, J., {Blazek}, J., {Chisari}, N.~E., {et~al.} 2019,
  \JournalTitle{\baas}, 51, 358

\bibitem[{{Newman}(2008)}]{Newman:08}
{Newman}, J.~A. 2008,
  \href{http://dx.doi.org/10.1086/589982}{\JournalTitle{\apj}, 684, 88}

\bibitem[{{Newman} {et~al.}(2013){Newman}, {Cooper}, {Davis}, {Faber}, {Coil},
  {Guhathakurta}, {Koo}, {Phillips}, {Conroy}, {Dutton}, {Finkbeiner}, {Gerke},
  {Rosario}, {Weiner}, {Willmer}, {Yan}, {Harker}, {Kassin}, {Konidaris},
  {Lai}, {Madgwick}, {Noeske}, {Wirth}, {Connolly}, {Kaiser}, {Kirby},
  {Lemaux}, {Lin}, {Lotz}, {Luppino}, {Marinoni}, {Matthews}, {Metevier}, \&
  {Schiavon}}]{deep2}
{Newman}, J.~A., {Cooper}, M.~C., {Davis}, M., {et~al.} 2013,
  \href{http://dx.doi.org/10.1088/0067-0049/208/1/5}{\JournalTitle{\apjs}, 208,
  5}

\bibitem[{{Nicola} {et~al.}(2020){Nicola}, {Alonso}, {S{\'a}nchez}, {Slosar},
  {Awan}, {Broussard}, {Dunkley}, {Gawiser}, {Gomes}, {Mandelbaum}, {Miyatake},
  {Newman}, {Sevilla-Noarbe}, {Skinner}, \& {Wagoner}}]{nicola2020}
{Nicola}, A., {Alonso}, D., {S{\'a}nchez}, J., {et~al.} 2020,
  \href{http://dx.doi.org/10.1088/1475-7516/2020/03/044}{\JournalTitle{\jcap},
  2020, 044}

\bibitem[{{Pentericci} {et~al.}(2018){Pentericci}, {McLure}, {Garilli},
  {Cucciati}, {Franzetti}, {Iovino}, {Amorin}, {Bolzonella}, {Bongiorno},
  {Carnall}, {Castellano}, {Cimatti}, {Cirasuolo}, {Cullen}, {De Barros},
  {Dunlop}, {Elbaz}, {Finkelstein}, {Fontana}, {Fontanot}, {Fumana},
  {Gargiulo}, {Guaita}, {Hartley}, {Jarvis}, {Juneau}, {Karman}, {Maccagni},
  {Marchi}, {Marmol-Queralto}, {Nandra}, {Pompei}, {Pozzetti}, {Scodeggio},
  {Sommariva}, {Talia}, {Almaini}, {Balestra}, {Bardelli}, {Bell}, {Bourne},
  {Bowler}, {Brusa}, {Buitrago}, {Caputi}, {Cassata}, {Charlot}, {Citro},
  {Cresci}, {Cristiani}, {Curtis-Lake}, {Dickinson}, {Fazio}, {Ferguson},
  {Fiore}, {Franco}, {Fynbo}, {Galametz}, {Georgakakis}, {Giavalisco},
  {Grazian}, {Hathi}, {Jung}, {Kim}, {Koekemoer}, {Khusanova}, {Le F{\`e}vre},
  {Lotz}, {Mannucci}, {Maltby}, {Matsuoka}, {McLeod}, {Mendez-Hernandez},
  {Mendez-Abreu}, {Mignoli}, {Moresco}, {Mortlock}, {Nonino}, {Pannella},
  {Papovich}, {Popesso}, {Rosario}, {Salvato}, {Santini}, {Schaerer},
  {Schreiber}, {Stark}, {Tasca}, {Thomas}, {Treu}, {Vanzella}, {Wild},
  {Williams}, {Zamorani}, \& {Zucca}}]{vandels2018}
{Pentericci}, L., {McLure}, R.~J., {Garilli}, B., {et~al.} 2018,
  \href{http://dx.doi.org/10.1051/0004-6361/201833047}{\JournalTitle{\aap},
  616, A174}

\bibitem[{Perez \& Granger(2007)}]{ipython}
Perez, F., \& Granger, B.~E. 2007,
  \href{http://dx.doi.org/10.1109/MCSE.2007.53}{\JournalTitle{Computing in
  Science Engineering}, 9, 21}

\bibitem[{Petri {et~al.}(2017)Petri, Haiman, \& May}]{PhysRevD.95.123503}
Petri, A., Haiman, Z., \& May, M. 2017,
  \href{http://dx.doi.org/10.1103/PhysRevD.95.123503}{\JournalTitle{Phys. Rev.
  D}, 95, 123503}

\bibitem[{{Rahman} {et~al.}(2015){Rahman}, {M{\'e}nard}, {Scranton}, {Schmidt},
  \& {Morrison}}]{rahman2015}
{Rahman}, M., {M{\'e}nard}, B., {Scranton}, R., {Schmidt}, S.~J., \&
  {Morrison}, C.~B. 2015,
  \href{http://dx.doi.org/10.1093/mnras/stu2636}{\JournalTitle{\mnras}, 447,
  3500}

\bibitem[{{Rau} {et~al.}(2020){Rau}, {Wilson}, \&
  {Mandelbaum}}]{rau_et_al_2020}
{Rau}, M.~M., {Wilson}, S., \& {Mandelbaum}, R. 2020,
  \href{http://dx.doi.org/10.1093/mnras/stz3295}{\JournalTitle{\mnras}, 491,
  4768}

\bibitem[{{Reid} {et~al.}(2016){Reid}, {Ho}, {Padmanabhan}, {Percival},
  {Tinker}, {Tojeiro}, {White}, {Eisenstein}, {Maraston}, {Ross},
  {S{\'a}nchez}, {Schlegel}, {Sheldon}, {Strauss}, {Thomas}, {Wake}, {Beutler},
  {Bizyaev}, {Bolton}, {Brownstein}, {Chuang}, {Dawson}, {Harding}, {Kitaura},
  {Leauthaud}, {Masters}, {McBride}, {More}, {Olmstead}, {Oravetz}, {Nuza},
  {Pan}, {Parejko}, {Pforr}, {Prada}, {Rodr{\'\i}guez-Torres},
  {Salazar-Albornoz}, {Samushia}, {Schneider}, {Sc{\'o}ccola}, {Simmons}, \&
  {Vargas-Magana}}]{reid2016}
{Reid}, B., {Ho}, S., {Padmanabhan}, N., {et~al.} 2016,
  \href{http://dx.doi.org/10.1093/mnras/stv2382}{\JournalTitle{\mnras}, 455,
  1553}

\bibitem[{{Rozo} {et~al.}(2015){Rozo}, {Rykoff}, {Becker}, {Reddick}, \&
  {Wechsler}}]{Rozo15}
{Rozo}, E., {Rykoff}, E.~S., {Becker}, M., {Reddick}, R.~M., \& {Wechsler},
  R.~H. 2015,
  \href{http://dx.doi.org/10.1093/mnras/stv1560}{\JournalTitle{\mnras}, 453,
  38}

\bibitem[{{Rozo} {et~al.}(2016){Rozo}, {Rykoff}, {Abate}, {Bonnett}, {Crocce},
  {Davis}, {Hoyle}, {Leistedt}, {Peiris}, {Wechsler}, {Abbott}, {Abdalla},
  {Banerji}, {Bauer}, {Benoit-L{\'e}vy}, {Bernstein}, {Bertin}, {Brooks},
  {Buckley-Geer}, {Burke}, {Capozzi}, {Rosell}, {Carollo}, {Kind}, {Carretero},
  {Castander}, {Childress}, {Cunha}, {D'Andrea}, {Davis}, {DePoy}, {Desai},
  {Diehl}, {Dietrich}, {Doel}, {Eifler}, {Evrard}, {Neto}, {Flaugher},
  {Fosalba}, {Frieman}, {Gaztanaga}, {Gerdes}, {Glazebrook}, {Gruen},
  {Gruendl}, {Honscheid}, {James}, {Jarvis}, {Kim}, {Kuehn}, {Kuropatkin},
  {Lahav}, {Lidman}, {Lima}, {Maia}, {March}, {Martini}, {Melchior}, {Miller},
  {Miquel}, {Mohr}, {Nichol}, {Nord}, {O'Neill}, {Ogando}, {Plazas}, {Romer},
  {Roodman}, {Sako}, {Sanchez}, {Santiago}, {Schubnell}, {Sevilla-Noarbe},
  {Smith}, {Soares-Santos}, {Sobreira}, {Suchyta}, {Swanson}, {Thaler},
  {Thomas}, {Uddin}, {Vikram}, {Walker}, {Wester}, {Zhang}, \& {da
  Costa}}]{Rozo2016}
{Rozo}, E., {Rykoff}, E.~S., {Abate}, A., {et~al.} 2016,
  \href{http://dx.doi.org/10.1093/mnras/stw1281}{\JournalTitle{\mnras}, 461,
  1431}

\bibitem[{{Runco} {et~al.}(2021){Runco}, {Shapley}, {Sanders}, {Topping},
  {Kriek}, {Reddy}, {Coil}, {Mobasher}, {Siana}, {Freeman}, {Shivaei}, {Azadi},
  {Price}, {Leung}, {Fetherolf}, {de Groot}, {Zick}, {Fornasini}, \&
  {Barro}}]{Runco2021}
{Runco}, J.~N., {Shapley}, A.~E., {Sanders}, R.~L., {et~al.} 2021,
  \href{http://dx.doi.org/10.1093/mnras/stab119}{\JournalTitle{\mnras}, 502,
  2600}

\bibitem[{{Rykoff} {et~al.}(2012){Rykoff}, {Koester}, {Rozo}, {Annis},
  {Evrard}, {Hansen}, {Hao}, {Johnston}, {McKay}, \& {Wechsler}}]{rykoff2012}
{Rykoff}, E.~S., {Koester}, B.~P., {Rozo}, E., {et~al.} 2012,
  \href{http://dx.doi.org/10.1088/0004-637X/746/2/178}{\JournalTitle{\apj},
  746, 178}

\bibitem[{{Rykoff} {et~al.}(2014){Rykoff}, {Rozo}, {Busha}, {Cunha},
  {Finoguenov}, {Evrard}, {Hao}, {Koester}, {Leauthaud}, {Nord}, {Pierre},
  {Reddick}, {Sadibekova}, {Sheldon}, \& {Wechsler}}]{rykoff2014}
{Rykoff}, E.~S., {Rozo}, E., {Busha}, M.~T., {et~al.} 2014,
  \href{http://dx.doi.org/10.1088/0004-637X/785/2/104}{\JournalTitle{\apj},
  785, 104}

\bibitem[{{Rykoff} {et~al.}(2016){Rykoff}, {Rozo}, {Hollowood}, {Bermeo-Hernand
  ez}, {Jeltema}, {Mayers}, {Romer}, {Rooney}, {Saro}, {Vergara Cervantes},
  {Wechsler}, {Wilcox}, {Abbott}, {Abdalla}, {Allam}, {Annis},
  {Benoit-L{\'e}vy}, {Bernstein}, {Bertin}, {Brooks}, {Burke}, {Capozzi},
  {Carnero Rosell}, {Carrasco Kind}, {Castander}, {Childress}, {Collins},
  {Cunha}, {D'Andrea}, {da Costa}, {Davis}, {Desai}, {Diehl}, {Dietrich},
  {Doel}, {Evrard}, {Finley}, {Flaugher}, {Fosalba}, {Frieman}, {Glazebrook},
  {Goldstein}, {Gruen}, {Gruendl}, {Gutierrez}, {Hilton}, {Honscheid}, {Hoyle},
  {James}, {Kay}, {Kuehn}, {Kuropatkin}, {Lahav}, {Lewis}, {Lidman}, {Lima},
  {Maia}, {Mann}, {Marshall}, {Martini}, {Melchior}, {Miller}, {Miquel},
  {Mohr}, {Nichol}, {Nord}, {Ogando}, {Plazas}, {Reil}, {Sahl{\'e}n},
  {Sanchez}, {Santiago}, {Scarpine}, {Schubnell}, {Sevilla-Noarbe}, {Smith},
  {Soares-Santos}, {Sobreira}, {Stott}, {Suchyta}, {Swanson}, {Tarle},
  {Thomas}, {Tucker}, {Uddin}, {Viana}, {Vikram}, {Walker}, {Zhang}, \& {DES
  Collaboration}}]{rykoff2016}
{Rykoff}, E.~S., {Rozo}, E., {Hollowood}, D., {et~al.} 2016,
  \href{http://dx.doi.org/10.3847/0067-0049/224/1/1}{\JournalTitle{\apjs}, 224,
  1}

\bibitem[{Sadeh {et~al.}(2016)Sadeh, Abdalla, \& Lahav}]{Sadeh:16}
Sadeh, I., Abdalla, F.~B., \& Lahav, O. 2016,
  \href{http://dx.doi.org/10.1088/1538-3873/128/968/104502}{\JournalTitle{PASP},
  128, 104502}

\bibitem[{{Safonova} {et~al.}(2020){Safonova}, {Norberg}, \&
  {Cole}}]{safonova2020}
{Safonova}, S., {Norberg}, P., \& {Cole}, S. 2020, \JournalTitle{arXiv
  e-prints}, arXiv:2009.00005

\bibitem[{{Salvato} {et~al.}(2019){Salvato}, {Ilbert}, \&
  {Hoyle}}]{Salvato2019}
{Salvato}, M., {Ilbert}, O., \& {Hoyle}, B. 2019,
  \href{http://dx.doi.org/10.1038/s41550-018-0478-0}{\JournalTitle{Nature
  Astronomy}, 3, 212}

\bibitem[{{S{\'a}nchez} \& {Bernstein}(2019)}]{Sanchez:19}
{S{\'a}nchez}, C., \& {Bernstein}, G.~M. 2019,
  \href{http://dx.doi.org/10.1093/mnras/sty3222}{\JournalTitle{\mnras}, 483,
  2801}

\bibitem[{Sanchez {et~al.}(2021)Sanchez, Mendoza, Kirkby, \&
  Burchat}]{Sanchez2021}
Sanchez, J., Mendoza, I., Kirkby, D.~P., \& Burchat, P.~R. 2021, Effects of
  overlapping sources on cosmic shear estimation: Statistical sensitivity and
  pixel-noise bias, \href{http://arxiv.org/abs/2103.02078}{{\sffamily
  arXiv:2103.02078 [astro-ph.CO]}}

\bibitem[{{Sheldon} {et~al.}(2020){Sheldon}, {Becker}, {MacCrann}, \&
  {Jarvis}}]{sheldon2020}
{Sheldon}, E.~S., {Becker}, M.~R., {MacCrann}, N., \& {Jarvis}, M. 2020,
  \href{http://dx.doi.org/10.3847/1538-4357/abb595}{\JournalTitle{\apj}, 902,
  138}

\bibitem[{{Sheldon} {et~al.}(2004){Sheldon}, {Johnston}, {Frieman}, {Scranton},
  {McKay}, {Connolly}, {Budav{\'a}ri}, {Zehavi}, {Bahcall}, {Brinkmann}, \&
  {Fukugita}}]{sheldon2004}
{Sheldon}, E.~S., {Johnston}, D.~E., {Frieman}, J.~A., {et~al.} 2004,
  \href{http://dx.doi.org/10.1086/383293}{\JournalTitle{\aj}, 127, 2544}

\bibitem[{{Shirasaki} {et~al.}(2019){Shirasaki}, {Hamana}, {Takada},
  {Takahashi}, \& {Miyatake}}]{shirasaki2019}
{Shirasaki}, M., {Hamana}, T., {Takada}, M., {Takahashi}, R., \& {Miyatake}, H.
  2019, \href{http://dx.doi.org/10.1093/mnras/stz791}{\JournalTitle{\mnras},
  486, 52}

\bibitem[{{Shirasaki} \& {Takada}(2018)}]{2018MNRAS.478.4277S}
{Shirasaki}, M., \& {Takada}, M. 2018,
  \href{http://dx.doi.org/10.1093/mnras/sty1327}{\JournalTitle{\mnras}, 478,
  4277}

\bibitem[{{Singh} \& {Mandelbaum}(2016)}]{Singh2016}
{Singh}, S., \& {Mandelbaum}, R. 2016,
  \href{http://dx.doi.org/10.1093/mnras/stw144}{\JournalTitle{\mnras}, 457,
  2301}

\bibitem[{{Singh} {et~al.}(2015){Singh}, {Mandelbaum}, \& {More}}]{Singh2015}
{Singh}, S., {Mandelbaum}, R., \& {More}, S. 2015,
  \href{http://dx.doi.org/10.1093/mnras/stv778}{\JournalTitle{\mnras}, 450,
  2195}

\bibitem[{{Somerville} {et~al.}(2018){Somerville}, {Behroozi}, {Pandya},
  {Dekel}, {Faber}, {Fontana}, {Koekemoer}, {Koo}, {P{\'e}rez-Gonz{\'a}lez},
  {Primack}, {Santini}, {Taylor}, \& {van der Wel}}]{somerville2018}
{Somerville}, R.~S., {Behroozi}, P., {Pandya}, V., {et~al.} 2018,
  \href{http://dx.doi.org/10.1093/mnras/stx2040}{\JournalTitle{\mnras}, 473,
  2714}

\bibitem[{{Stoughton} {et~al.}(2002){Stoughton}, {Lupton}, {Bernardi},
  {Blanton}, {Burles}, {Castander}, {Connolly}, {Eisenstein}, {Frieman},
  {Hennessy}, {Hindsley}, {Ivezi{\'c}}, {Kent}, {Kunszt}, {Lee}, {Meiksin},
  {Munn}, {Newberg}, {Nichol}, {Nicinski}, {Pier}, {Richards}, {Richmond},
  {Schlegel}, {Smith}, {Strauss}, {SubbaRao}, {Szalay}, {Thakar}, {Tucker},
  {Vanden Berk}, {Yanny}, {Adelman}, {Anderson}, {Anderson}, {Annis},
  {Bahcall}, {Bakken}, {Bartelmann}, {Bastian}, {Bauer}, {Berman},
  {B{\"o}hringer}, {Boroski}, {Bracker}, {Briegel}, {Briggs}, {Brinkmann},
  {Brunner}, {Carey}, {Carr}, {Chen}, {Christian}, {Colestock}, {Crocker},
  {Csabai}, {Czarapata}, {Dalcanton}, {Davidsen}, {Davis}, {Dehnen},
  {Dodelson}, {Doi}, {Dombeck}, {Donahue}, {Ellman}, {Elms}, {Evans}, {Eyer},
  {Fan}, {Federwitz}, {Friedman}, {Fukugita}, {Gal}, {Gillespie}, {Glazebrook},
  {Gray}, {Grebel}, {Greenawalt}, {Greene}, {Gunn}, {de Haas}, {Haiman},
  {Haldeman}, {Hall}, {Hamabe}, {Hansen}, {Harris}, {Harris}, {Harvanek},
  {Hawley}, {Hayes}, {Heckman}, {Helmi}, {Henden}, {Hogan}, {Hogg}, {Holmgren},
  {Holtzman}, {Huang}, {Hull}, {Ichikawa}, {Ichikawa}, {Johnston}, {Kauffmann},
  {Kim}, {Kimball}, {Kinney}, {Klaene}, {Kleinman}, {Klypin}, {Knapp},
  {Korienek}, {Krolik}, {Kron}, {Krzesi{\'n}ski}, {Lamb}, {Leger},
  {Limmongkol}, {Lindenmeyer}, {Long}, {Loomis}, {Loveday}, {MacKinnon},
  {Mannery}, {Mantsch}, {Margon}, {McGehee}, {McKay}, {McLean}, {Menou},
  {Merelli}, {Mo}, {Monet}, {Nakamura}, {Narayanan}, {Nash}, {Neilsen},
  {Newman}, {Nitta}, {Odenkirchen}, {Okada}, {Okamura}, {Ostriker}, {Owen},
  {Pauls}, {Peoples}, {Peterson}, {Petravick}, {Pope}, {Pordes}, {Postman},
  {Prosapio}, {Quinn}, {Rechenmacher}, {Rivetta}, {Rix}, {Rockosi}, {Rosner},
  {Ruthmansdorfer}, {Sandford}, {Schneider}, {Scranton}, {Sekiguchi}, {Sergey},
  {Sheth}, {Shimasaku}, {Smee}, {Snedden}, {Stebbins}, {Stubbs}, {Szapudi},
  {Szkody}, {Szokoly}, {Tabachnik}, {Tsvetanov}, {Uomoto}, {Vogeley}, {Voges},
  {Waddell}, {Walterbos}, {Wang}, {Watanabe}, {Weinberg}, {White}, {White},
  {Wilhite}, {Wolfe}, {Yasuda}, {York}, {Zehavi}, \& {Zheng}}]{Stoughton2002}
{Stoughton}, C., {Lupton}, R.~H., {Bernardi}, M., {et~al.} 2002,
  \href{http://dx.doi.org/10.1086/324741}{\JournalTitle{\aj}, 123, 485}

\bibitem[{{Sullivan} {et~al.}(2010){Sullivan}, {Conley}, {Howell}, {Neill},
  {Astier}, {Balland}, {Basa}, {Carlberg}, {Fouchez}, {Guy}, {Hardin}, {Hook},
  {Pain}, {Palanque-Delabrouille}, {Perrett}, {Pritchet}, {Regnault}, {Rich},
  {Ruhlmann-Kleider}, {Baumont}, {Hsiao}, {Kronborg}, {Lidman}, {Perlmutter},
  \& {Walker}}]{sullivan2010}
{Sullivan}, M., {Conley}, A., {Howell}, D.~A., {et~al.} 2010,
  \href{http://dx.doi.org/10.1111/j.1365-2966.2010.16731.x}{\JournalTitle{\mnras},
  406, 782}

\bibitem[{{The LSST Dark Energy Science
  Collaboration}(2018{\natexlab{a}})}]{lsst-srd}
{The LSST Dark Energy Science Collaboration}. 2018{\natexlab{a}},
  \JournalTitle{arXiv e-prints}, arXiv:1809.01669

\bibitem[{{The LSST Dark Energy Science
  Collaboration}(2018{\natexlab{b}})}]{SRD}
---. 2018{\natexlab{b}}, \JournalTitle{arXiv e-prints}, arXiv:1809.01669

\bibitem[{{To} {et~al.}(2021){To}, {Krause}, {Rozo}, {Wu}, {Gruen}, {Wechsler},
  {Eifler}, {Rykoff}, {Costanzi}, {Becker}, {Bernstein}, {Blazek}, {Bocquet},
  {Bridle}, {Cawthon}, {Choi}, {Crocce}, {Davis}, {DeRose}, {Drlica-Wagner},
  {Elvin-Poole}, {Fang}, {Farahi}, {Friedrich}, {Gatti}, {Gaztanaga},
  {Giannantonio}, {Hartley}, {Hoyle}, {Jarvis}, {MacCrann}, {McClintock},
  {Miranda}, {Pereira}, {Park}, {Porredon}, {Prat}, {Rau}, {Ross}, {Samuroff},
  {S{\'a}nchez}, {Sevilla-Noarbe}, {Sheldon}, {Troxel}, {Varga}, {Vielzeuf},
  {Zhang}, {Zuntz}, {Abbott}, {Aguena}, {Amon}, {Annis}, {Avila}, {Bertin},
  {Bhargava}, {Brooks}, {Burke}, {Carnero Rosell}, {Carrasco Kind},
  {Carretero}, {Chang}, {Conselice}, {da Costa}, {Davis}, {Desai}, {Diehl},
  {Dietrich}, {Everett}, {Evrard}, {Ferrero}, {Flaugher}, {Fosalba}, {Frieman},
  {Garc{\'\i}a-Bellido}, {Gruendl}, {Gutierrez}, {Hinton}, {Hollowood},
  {Honscheid}, {Huterer}, {James}, {Jeltema}, {Kron}, {Kuehn}, {Kuropatkin},
  {Lima}, {Maia}, {Marshall}, {Menanteau}, {Miquel}, {Morgan}, {Muir}, {Myles},
  {Palmese}, {Paz-Chinch{\'o}n}, {Plazas}, {Romer}, {Roodman}, {Sanchez},
  {Santiago}, {Scarpine}, {Serrano}, {Smith}, {Suchyta}, {Swanson}, {Tarle},
  {Thomas}, {Tucker}, {Weller}, {Wester}, {Wilkinson}, \& {DES
  Collaboration}}]{DES_Y1_WL}
{To}, C., {Krause}, E., {Rozo}, E., {et~al.} 2021,
  \href{http://dx.doi.org/10.1103/PhysRevLett.126.141301}{\JournalTitle{\prl},
  126, 141301}

\bibitem[{{To} {et~al.}(2020){To}, {Reddick}, {Rozo}, {Rykoff}, \&
  {Wechsler}}]{to2020}
{To}, C.-H., {Reddick}, R.~M., {Rozo}, E., {Rykoff}, E., \& {Wechsler}, R.~H.
  2020, \href{http://dx.doi.org/10.3847/1538-4357/ab9636}{\JournalTitle{\apj},
  897, 15}

\bibitem[{{van Daalen} {et~al.}(2016){van Daalen}, {Henriques}, {Angulo}, \&
  {White}}]{vanDaalen2016}
{van Daalen}, M.~P., {Henriques}, B. M.~B., {Angulo}, R.~E., \& {White}, S.
  D.~M. 2016,
  \href{http://dx.doi.org/10.1093/mnras/stw405}{\JournalTitle{\mnras}, 458,
  934}

\bibitem[{van~der Walt {et~al.}(2011)van~der Walt, Colbert, \&
  Varoquaux}]{numpy}
van~der Walt, S., Colbert, S.~C., \& Varoquaux, G. 2011,
  \href{http://dx.doi.org/10.1109/MCSE.2011.37}{\JournalTitle{Computing in
  Science Engineering}, 13, 22}

\bibitem[{{van der Wel} {et~al.}(2014){van der Wel}, {Franx}, {van Dokkum},
  {Skelton}, {Momcheva}, {Whitaker}, {Brammer}, {Bell}, {Rix}, \&
  {Wuyts}}]{2014ApJ...788...28V}
{van der Wel}, A., {Franx}, M., {van Dokkum}, P.~G., {et~al.} 2014,
  \href{http://dx.doi.org/10.1088/0004-637X/788/1/28}{\JournalTitle{\apj}, 788,
  28}

\bibitem[{{Varga} {et~al.}(2019){Varga}, {DeRose}, {Gruen}, {McClintock},
  {Seitz}, {Rozo}, {Costanzi}, {Hoyle}, {MacCrann}, {Plazas}, {Rykoff},
  {Simet}, {von der Linden}, {Wechsler}, {Annis}, {Avila}, {Bertin}, {Brooks},
  {Buckley-Geer}, {Burke}, {Carnero Rosell}, {Carrasco Kind}, {Carretero},
  {Cunha}, {D'Andrea}, {da Costa}, {De Vicente}, {Desai}, {Diehl}, {Dietrich},
  {Doel}, {Evrard}, {Flaugher}, {Fosalba}, {Frieman}, {Garc{\'\i}a-Bellido},
  {Gaztanaga}, {Gerdes}, {Gruendl}, {Gschwend}, {Gutierrez}, {Hartley},
  {Hollowood}, {Honscheid}, {James}, {Jeltema}, {Kuehn}, {Kuropatkin}, {Lima},
  {Maia}, {March}, {Marshall}, {Melchior}, {Menanteau}, {Miller}, {Miquel},
  {Ogando}, {Romer}, {Sanchez}, {Scarpine}, {Schubnell}, {Serrano},
  {Sevilla-Noarbe}, {Smith}, {Sobreira}, {Suchyta}, {Swanson}, {Tarle},
  {Thomas}, {Tucker}, {Zhang}, \& {DES Collaboration}}]{varga19}
{Varga}, T.~N., {DeRose}, J., {Gruen}, D., {et~al.} 2019,
  \href{http://dx.doi.org/10.1093/mnras/stz2185}{\JournalTitle{\mnras}, 489,
  2511}

\bibitem[{{Velander} {et~al.}(2014){Velander}, {van Uitert}, {Hoekstra},
  {Coupon}, {Erben}, {Heymans}, {Hildebrandt}, {Kitching}, {Mellier}, {Miller},
  {Van Waerbeke}, {Bonnett}, {Fu}, {Giodini}, {Hudson}, {Kuijken}, {Rowe},
  {Schrabback}, \& {Semboloni}}]{Velander2014}
{Velander}, M., {van Uitert}, E., {Hoekstra}, H., {et~al.} 2014,
  \href{http://dx.doi.org/10.1093/mnras/stt2013}{\JournalTitle{\mnras}, 437,
  2111}

\bibitem[{{von der Linden} {et~al.}(2014){von der Linden}, {Mantz}, {Allen},
  {Applegate}, {Kelly}, {Morris}, {Wright}, {Allen}, {Burchat}, {Burke},
  {Donovan}, \& {Ebeling}}]{vonderlinden2014}
{von der Linden}, A., {Mantz}, A., {Allen}, S.~W., {et~al.} 2014,
  \href{http://dx.doi.org/10.1093/mnras/stu1423}{\JournalTitle{\mnras}, 443,
  1973}

\bibitem[{{Wang} {et~al.}(2013){Wang}, {Brunner}, \& {Dolence}}]{tpcf_wang}
{Wang}, Y., {Brunner}, R.~J., \& {Dolence}, J.~C. 2013,
  \href{http://dx.doi.org/10.1093/mnras/stt450}{\JournalTitle{\mnras}, 432,
  1961}

\bibitem[{{Weaver} {et~al.}(2021){Weaver}, {Kauffmann}, {Ilbert}, {McCracken},
  {Moneti}, {Toft}, {Brammer}, {Shuntov}, {Davidzon}, {Hsieh}, {Laigle},
  {Anastasiou}, {Jespersen}, {Vinther}, {Capak}, {Casey}, {McPartland},
  {Milvang-Jensen}, {Mobasher}, {Sanders}, {Zalesky}, {Arnouts}, {Aussel},
  {Dunlop}, {Faisst}, {Franx}, {Furtak}, {Fynbo}, {Gould}, {Greve}, {Gwyn},
  {Kartaltepe}, {Kashino}, {Koekemoer}, {Kokorev}, {Le Fevre}, {Lilly},
  {Masters}, {Magdis}, {Mehta}, {Peng}, {Riechers}, {Salvato}, {Sawicki},
  {Scarlata}, {Scoville}, {Shirley}, {Sneppen}, {Smolcic}, {Steinhardt},
  {Stern}, {Tanaka}, {Taniguchi}, {Teplitz}, {Vaccari}, {Wang}, \&
  {Zamorani}}]{cosmos2021}
{Weaver}, J.~R., {Kauffmann}, O.~B., {Ilbert}, O., {et~al.} 2021,
  \JournalTitle{arXiv e-prints}, arXiv:2110.13923

\bibitem[{{Wechsler} {et~al.}(2021){Wechsler}, {DeRose}, {Busha}, {Becker},
  {Rykoff}, \& {Evrard}}]{wechsler2021}
{Wechsler}, R.~H., {DeRose}, J., {Busha}, M.~T., {et~al.} 2021,
  \JournalTitle{arXiv e-prints}, arXiv:2105.12105

\bibitem[{{Weinberg} {et~al.}(2013){Weinberg}, {Mortonson}, {Eisenstein},
  {Hirata}, {Riess}, \& {Rozo}}]{weinberg2013}
{Weinberg}, D.~H., {Mortonson}, M.~J., {Eisenstein}, D.~J., {et~al.} 2013,
  \href{http://dx.doi.org/10.1016/j.physrep.2013.05.001}{\JournalTitle{\physrep},
  530, 87}

\bibitem[{{White}(2001)}]{white2001}
{White}, M. 2001,
  \href{http://dx.doi.org/10.1051/0004-6361:20000357}{\JournalTitle{\aap}, 367,
  27}

\bibitem[{{Willmer}(2018)}]{willmer2018}
{Willmer}, C. N.~A. 2018,
  \href{http://dx.doi.org/10.3847/1538-4365/aabfdf}{\JournalTitle{\apjs}, 236,
  47}

\bibitem[{{Zehavi} {et~al.}(2011){Zehavi}, {Zheng}, {Weinberg}, {Blanton},
  {Bahcall}, {Berlind}, {Brinkmann}, {Frieman}, {Gunn}, {Lupton}, {Nichol},
  {Percival}, {Schneider}, {Skibba}, {Strauss}, {Tegmark}, \&
  {York}}]{zehavi2011}
{Zehavi}, I., {Zheng}, Z., {Weinberg}, D.~H., {et~al.} 2011,
  \href{http://dx.doi.org/10.1088/0004-637X/736/1/59}{\JournalTitle{\apj}, 736,
  59}

\bibitem[{{Zhang} \& {Yang}(2019)}]{zhang2019}
{Zhang}, Y.-C., \& {Yang}, X.-H. 2019,
  \href{http://dx.doi.org/10.1088/1674-4527/19/1/6}{\JournalTitle{Research in
  Astronomy and Astrophysics}, 19, 006}

\bibitem[{Zhou {et~al.}(2019)Zhou, Cooper, Newman, Ashby, Aird, Conselice,
  Davis, Dutton, Faber, Fang, Fazio, Guhathakurta, Kocevski, Koo, Nandra,
  Phillips, Rosario, Schlafly, Trump, Weiner, Willmer, \& Yan}]{Zhou2019}
Zhou, R., Cooper, M.~C., Newman, J.~A., {et~al.} 2019,
  \href{http://dx.doi.org/10.1093/mnras/stz1866}{\JournalTitle{\mnras}, 488,
  4565}

\bibitem[{{Zuntz} {et~al.}(2018){Zuntz}, {Sheldon}, {Samuroff}, {Troxel},
  {Jarvis}, {MacCrann}, {Gruen}, {Prat}, {S{\'a}nchez}, {Choi}, {Bridle},
  {Bernstein}, {Dodelson}, {Drlica-Wagner}, {Fang}, {Gruendl}, {Hoyle}, {Huff},
  {Jain}, {Kirk}, {Kacprzak}, {Krawiec}, {Plazas}, {Rollins}, {Rykoff},
  {Sevilla-Noarbe}, {Soergel}, {Varga}, {Abbott}, {Abdalla}, {Allam}, {Annis},
  {Bechtol}, {Benoit-L{\'e}vy}, {Bertin}, {Buckley-Geer}, {Burke}, {Carnero
  Rosell}, {Carrasco Kind}, {Carretero}, {Castander}, {Crocce}, {Cunha},
  {D'Andrea}, {da Costa}, {Davis}, {Desai}, {Diehl}, {Dietrich}, {Doel},
  {Eifler}, {Estrada}, {Evrard}, {Fausti Neto}, {Fernandez}, {Flaugher},
  {Fosalba}, {Frieman}, {Garc{\'\i}a-Bellido}, {Gaztanaga}, {Gerdes},
  {Giannantonio}, {Gschwend}, {Gutierrez}, {Hartley}, {Honscheid}, {James},
  {Jeltema}, {Johnson}, {Johnson}, {Kuehn}, {Kuhlmann}, {Kuropatkin}, {Lahav},
  {Li}, {Lima}, {Maia}, {March}, {Martini}, {Melchior}, {Menanteau}, {Miller},
  {Miquel}, {Mohr}, {Neilsen}, {Nichol}, {Ogando}, {Roe}, {Romer}, {Roodman},
  {Sanchez}, {Scarpine}, {Schindler}, {Schubnell}, {Smith}, {Smith},
  {Soares-Santos}, {Sobreira}, {Suchyta}, {Swanson}, {Tarle}, {Thomas},
  {Tucker}, {Vikram}, {Walker}, {Wechsler}, {Zhang}, \& {DES
  Collaboration}}]{Zuntz2018}
{Zuntz}, J., {Sheldon}, E., {Samuroff}, S., {et~al.} 2018,
  \href{http://dx.doi.org/10.1093/mnras/sty2219}{\JournalTitle{\mnras}, 481,
  1149}

\end{thebibliography}

\appendix

\section{Snapshots of Readiness Tests}
\label{app:ready}

We present here selected snapshots of the web page that show the output of the readiness tests performed on the synthetic galaxy catalog, cosmoDC2. As discussed in \autoref{sec:test:readiness}, the test outputs include both distributions and summary statistics for the catalog data. We show here some examples of each.

\begin{figure}
  \centering \includegraphics[width=8.5cm]{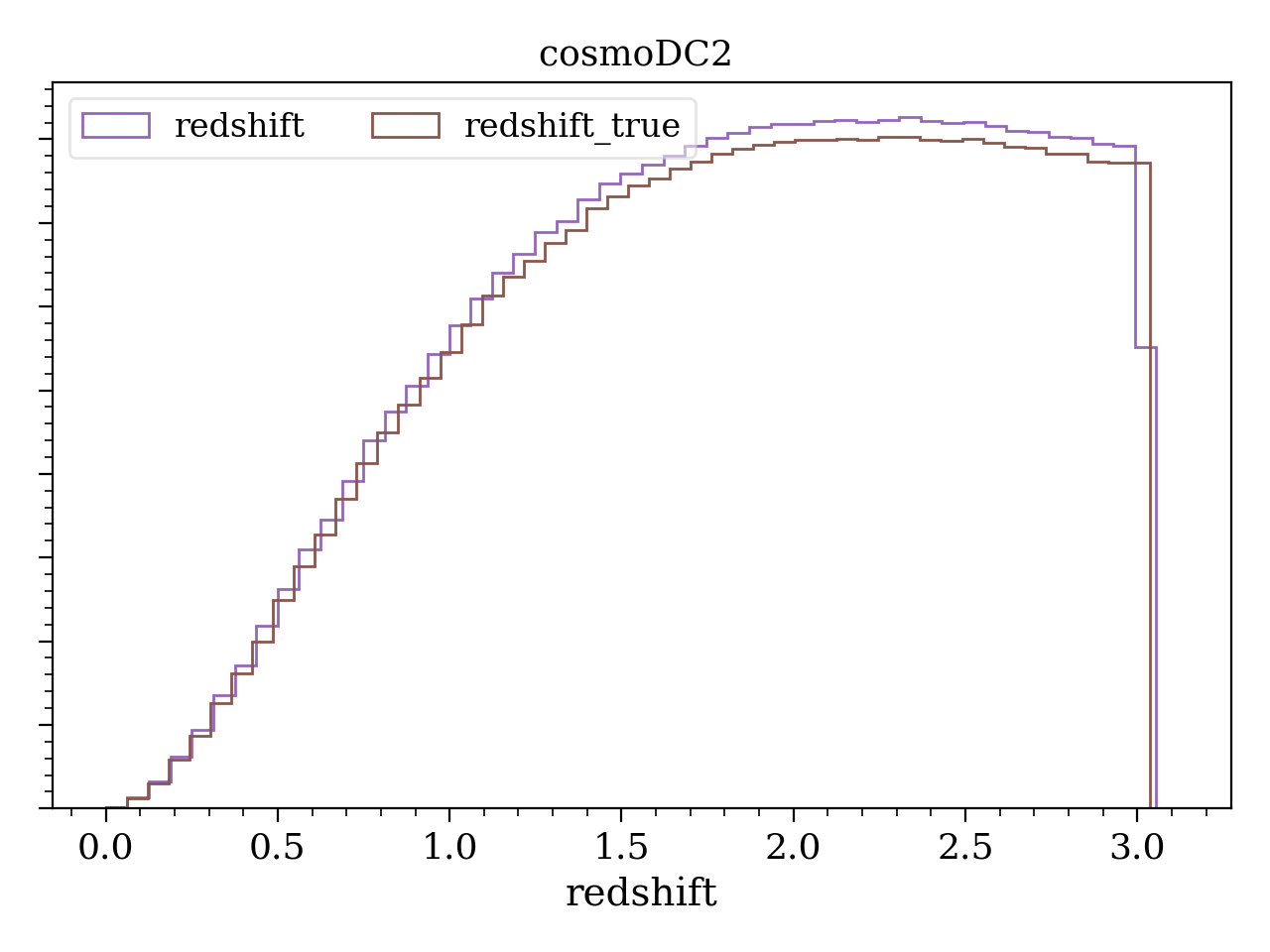}
  \centering \includegraphics[width=8.5cm]{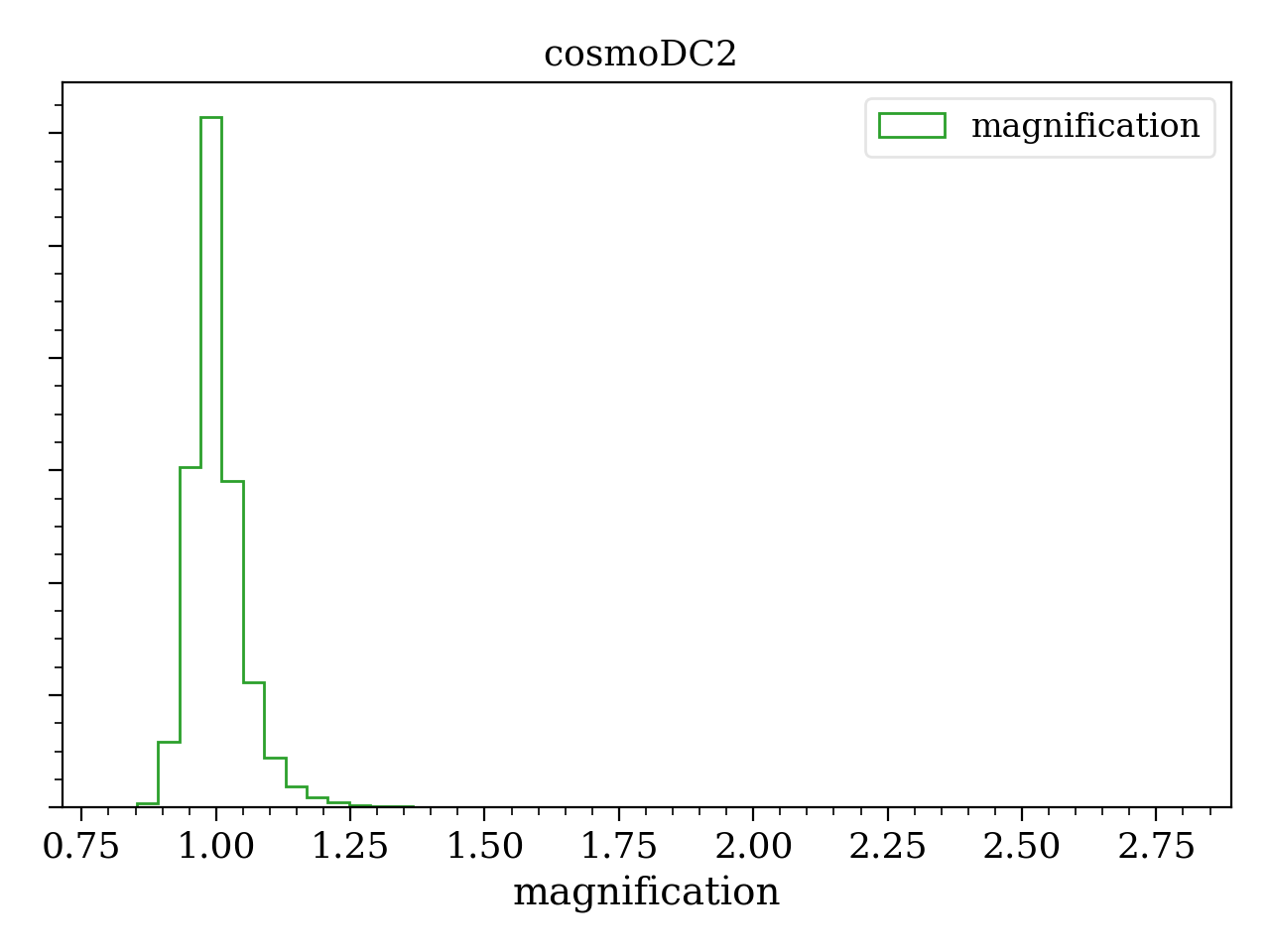}
  \caption{Histograms of the redshift (top) and  magnification (bottom) for cosmoDC2 galaxies. These distributions are easily examined for problems with the data. In this case of the magnification distribution for example, we do not expect to find any values $\gtrsim 2$. Similar plots are produced for all of the other user-selected quantities examined in the tests. The plots are shown here exactly as they appear on the web interface. Limits for the horizontal axes can be set in the configuration files for running the test. The vertical scale can be either linear or logarithmic.
  }
  \label{fig:readiness_dist}
\end{figure}

In \autoref{fig:readiness_dist} we show two examples of the distributions examined in the tests. These histograms provide a quick visual aid to check the distributions of catalog quantities for readily identifiable problems such as missing values or other issues. The plots here are shown exactly as they appear on the web interface. Limits for the horizontal axes can be set in the configuration files for running the tests. The vertical scale can be either linear or logarithmic.

\begin{figure*}
  \centering \includegraphics[width=7.0in]{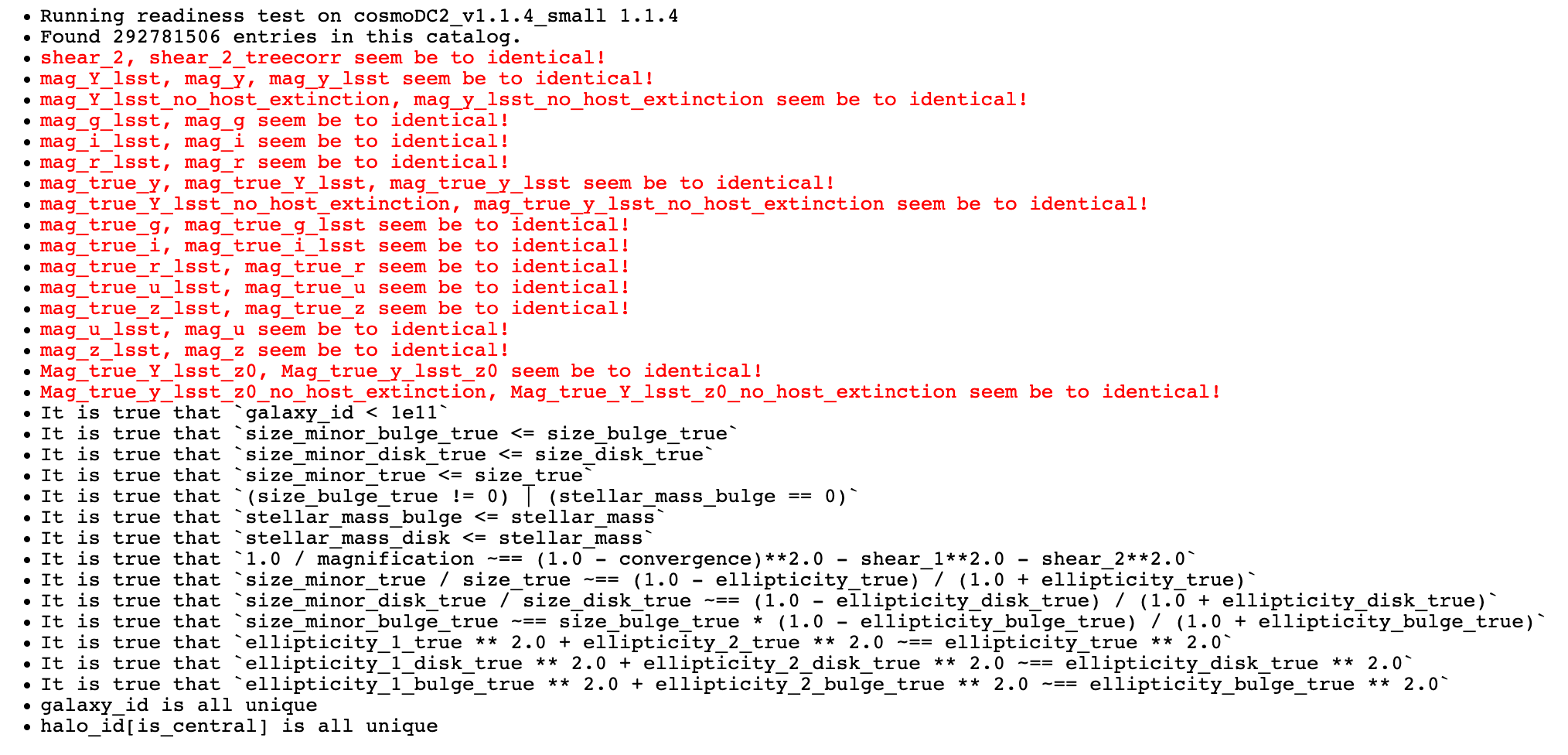}
  \caption{An excerpt of the readiness-test summary web page that shows the quantity properties and relationships checked by the tests. Red text flags catalog data columns that are identical. In this case, all of the identical columns are expected. Unexpected results, such as non-unique halo\_ids, are also flagged in red, if they occur.
  }
  \label{fig:readiness_check}
\end{figure*}

\begin{figure*}
  \centering \includegraphics[width=7.0in]{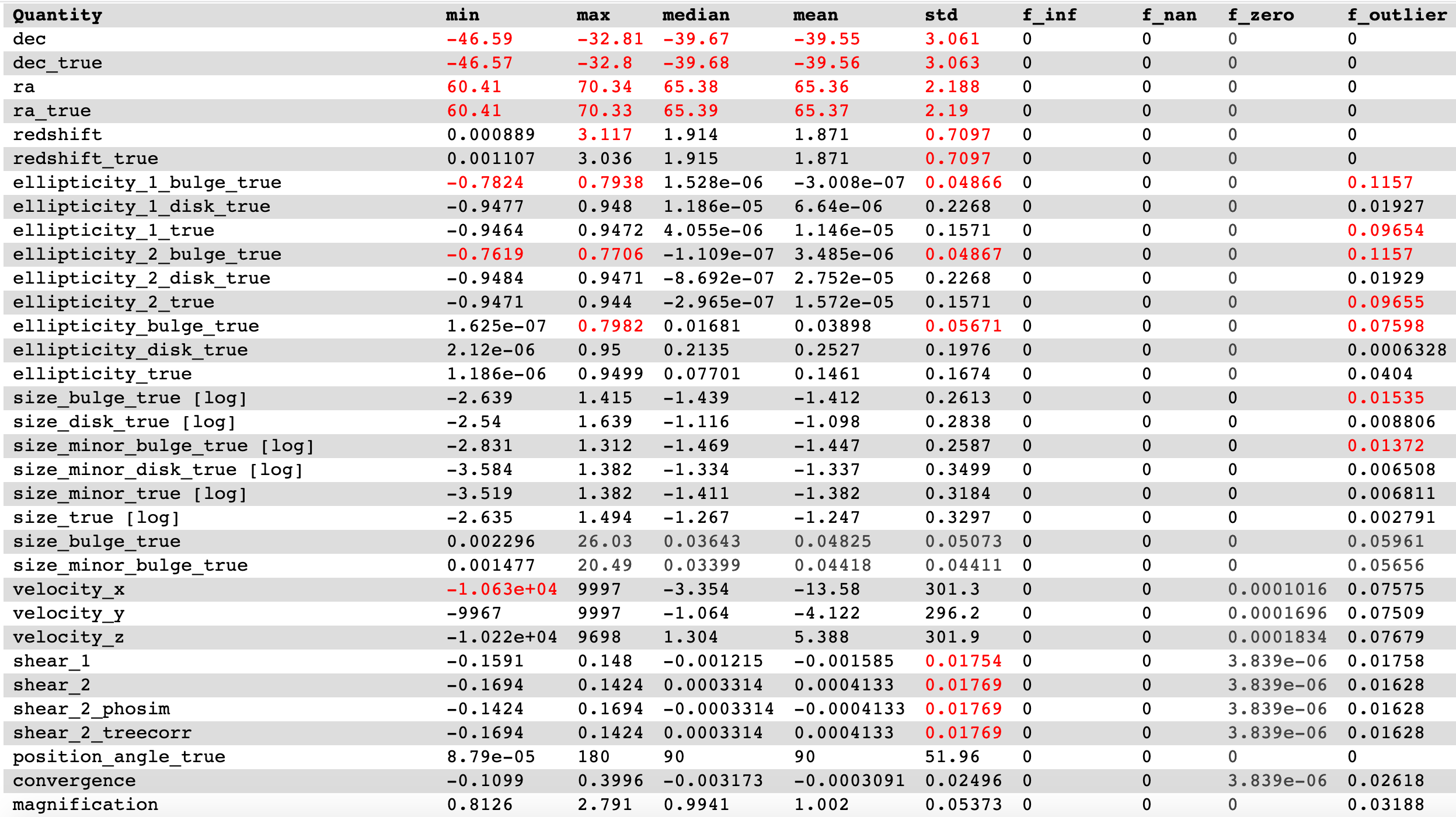}
  \caption{An excerpt of the summary web page that shows the summary statistics for the distributions of catalog quantities that are examined by the tests. The user supplies the expected range for each statistic. Values that lie outside these ranges are flagged in red. In this example, the test is run on a sub-volume of the catalog so that the statistics for the RA and Dec distributions do not match the expected values for the entire catalog. The remaining values flagged in red are only marginally outside the expected range and do not signify an underlying problem with the catalog data. 
  }
  \label{fig:readiness_summary}
\end{figure*}

In \autoref{fig:readiness_check} and \autoref{fig:readiness_summary}, we show excerpts of the summary web page produced by the readiness tests. \autoref{fig:readiness_check} shows the quantity relationships and properties that are checked and \autoref{fig:readiness_summary} shows an excerpt of the statistics evaluated for the distributions that are examined by the test. The quantities and relationships that are checked and the distributions that are examined are selected by the user by adjusting the test configuration file. The user also supplies the expected ranges for the statistics of the distributions so that values lying outside these ranges can be flagged. Items requiring attention are flagged in red. The summary page thus  provides an at-a-glance verification test of the catalog data that visually identifies any potential problems requiring further inspection.

\section{Mass Conversions}
\label{app:mass_conversion}

In this appendix, we present relationships between the various mass definitions mentioned in the tests in \autoref{sec:test:cl_density_profile}, \autoref{sec:test:mass_richness} and \autoref{sec:test:cl_shear}. The mass definition that is used in the cosmoDC2 catalog is based on the  mass of a halo identified by an FoF halo finder with a linking length, $b=0.168$. Previous studies~\citep{white2001, lukic2009} have examined the relationship between the SO mass $M_{\rm 200c}$ (defined with respect to the critical density $\rho_c$) and the FoF mass defined with a linking length $b=0.2$.  \citet{lukic2009} find the ratio of $M_{\rm FoF}/M_{\rm 200c}$ varies with $c$, the halo concentration,~\citep[and references therein]{child2018} and more weakly with the number of particles in the halo. For halos containing $>1000$ particles (which corresponds to a halo mass $\gtrsim 10^{12} M_\odot/h$ for the cosmoDC2 catalog), the ratio of these masses varies from 1.5 to 1.1 for halos with $3 < c <20$. See Figure~3 in \citet{lukic2009}. These concentration values span the range of concentrations found for most halos in $N$-body simulations. The authors find the tightest relationship following the mean ratios for so-called ``relaxed" halos whose potential centers and centers of mass lie within $0.4 h^{-1}{\rm Mpc}$. 

\begin{figure}
  \centering \includegraphics[width=8.5cm]{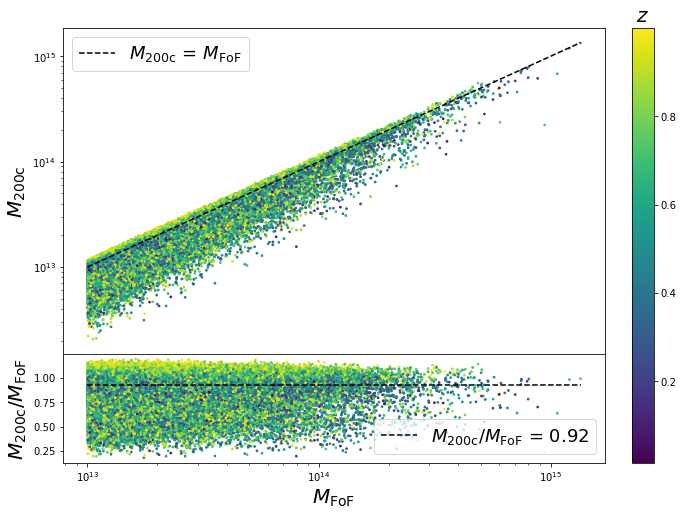}
  \caption{The scatter in the values (upper panel) and ratios (lower panel) of $M_{\rm 200c}$ versus $M_{\rm FoF}$ for halos from the Outer Rim simulation with FoF mass $>10^{13} h^{-1}{\rm M_\odot}$ and $z < 1$. The SO and FoF mass estimates are based on the identical halo centers, which are identified using the FoF halo finder using a linking length of $b=0.168$. The mean value of the ratio of masses $\langle M_{\rm 200c}/M_{\rm FoF}\rangle = 0.92$.}
  \label{fig:M200c_Mfof_OR}
\end{figure}

For cosmoDC2, the FoF linking length used to identify halos in the underlying Outer Rim simulation is smaller than that used in the~\citep{lukic2009} study. Hence we expect, on average, the values of $M_{\rm FoF}$ and $M_{\rm 200c}$ to be closer to one another.  At the time of production of the cosmoDC2 catalog, only FoF masses were available, so SO masses are not included in the cosmoDC2 catalog. Since then, further post-processing of the saved particle data from the Outer Rim simulation has produced SO masses for halos containing $>1000$ particles (or mass $\gtrsim 10^{12}~h^{-1}M_\odot$). These SO 
masses are measured by centering the SO halos on the FoF halo centers, so we can make a direct comparison of the halo masses in each case. We estimate the relationship between $M_{\rm 200c}$ and $M_{\rm FoF}$ for cluster-sized halos in the cosmoDC2 catalog by selecting halos from the Outer Rim simulation with $M_{\rm FoF} >10^{13} h^{-1}{\rm M_\odot}$ and $z < 1$ and compare their FoF and SO masses. In \autoref{fig:M200c_Mfof_OR} we show the scatter in the values (upper panel) and ratios (lower panel) of $M_{\rm 200c}$ and $M_{\rm FoF}$ for this halo sample.
The mean value of the ratio of masses $\langle M_{\rm 200c}/M_{\rm FoF}\rangle = 0.92$. We note that the scatter is quite large since we are including both relaxed and unrelaxed halos. 

\begin{figure}
  \centering \includegraphics[width=8.5cm]{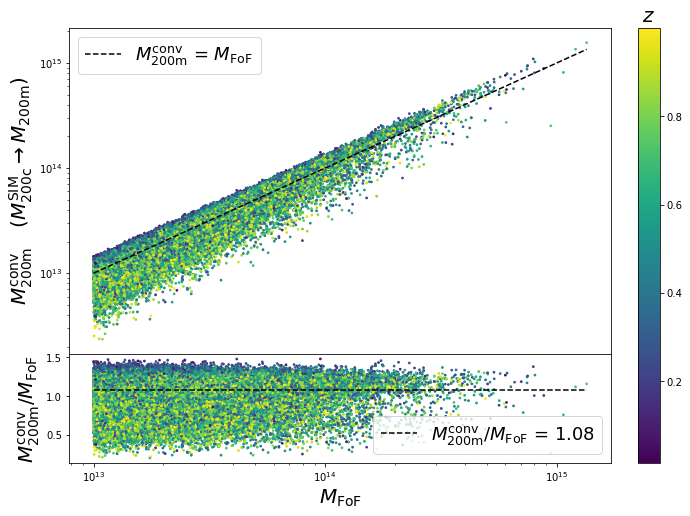}
  \caption{The scatter in the values (upper panel) and ratios (lower panel) of $M_{\rm 200m}$ versus $M_{\rm FoF}$ for Outer Rim halos with FoF mass $>10^{13} h^{-1}{\rm M_\odot}$ and $z < 1$. The values of $M_{\rm 200m}^{\rm conv}$ are obtained from the measured values of $M_{\rm 200c}$ by assuming an NFW halo profile, assigning a concentration based on the mean and scatter measured in \citet[see Equation 19]{child2018} and the \texttt{Colossus} code.~\citep{colossus_soft, colossus}.
  The mean value of the ratio of masses $\langle M_{\rm 200m}^{\rm conv}/M_{\rm FoF}\rangle = 1.08$}
  \label{fig:M200m_Mfof_OR}
\end{figure}

In addition to $M_{\rm 200c}$, the mass definition $M_{\rm 200m}$ is often used in the observational data. This mass has not been measured in the Outer Rim simulation, but it is possible to convert between these two mass definitions by assuming the halo profile is described by an NFW distribution and using either measured values of the halo concentration or a known concentration-mass ($c-M$) relation. We use the $c-M$ relation of \citet{child2018} (see Equation 19) to assign a concentration based on the measured $M_{\rm 200c}$ and add a scatter with variance $\bar{c}(M,z)/3$ around the mean. We then convert the measured values of $M_{\rm 200c}$ to $M_{\rm 200m}^{\rm conv}$ using the \texttt{Colossus} code~\citep{colossus_soft, colossus}. This conversion has a redshift dependence which is not apparent in the figure and which we discuss below.
In \autoref{fig:M200m_Mfof_OR}, we show the resulting values (upper panel) and ratios (lower panel) of the $M_{\rm 200c}^{\rm conv}$ and $M_{\rm FoF}$ for the same halo sample as \autoref{fig:M200c_Mfof_OR}. The mean value of the ratio of masses $\langle M_{\rm 200m}^{\rm conv}/M_{\rm FoF}\rangle = 1.08$. 

\begin{figure}
  \centering \includegraphics[width=8.5cm]{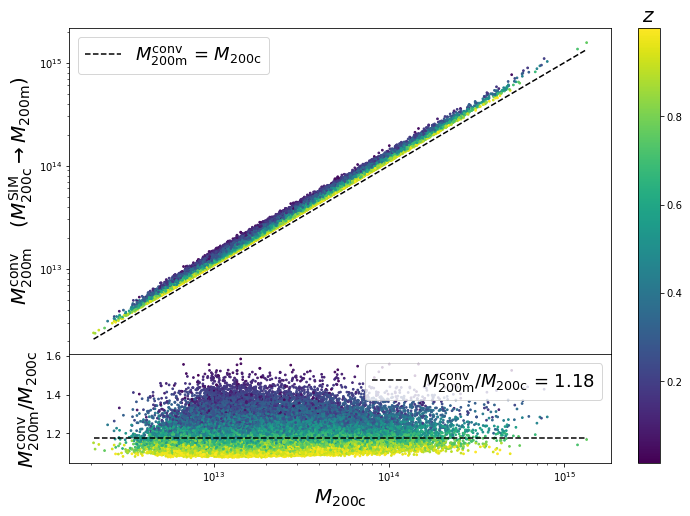}
  \caption{The relation between $M_{\rm 200m}^{\rm conv}$ and $M_{\rm 200c}$ obtained by assigning concentrations using the $c-M$ relationship from \citet{child2018}, assigning a scatter to the mean $c(M, z)$ using a variance $\sigma/\bar{c}(M, z) = 1/3$ and assuming an NFW halo profile. We use the \texttt{Colossus} code to convert the mass values.}
  \label{fig:m200m_from_m200c}
\end{figure}

In \autoref{fig:m200m_from_m200c}, we show the redshift dependence of the conversion between $M_{\rm 200m}^{\rm conv}$ and $M_{\rm 200c}$ obtained as described above. The mean value of the ratio of masses $\langle M_{\rm 200m}^{\rm conv}/M_{\rm 200c}\rangle = 1.18$, and varies with redshift from $\approx 1.1$ for halos with $z\approx 1$ to $\approx 1.3 - 1.4$ for halos with  $z\approx 0$.

\end{document}